\pdfoutput=1
\documentclass[cernpreprint,texlive=2011,txfonts,UKenglish]{latex/atlasdoc}
\usepackage{caption}

\usepackage[lineno=false,csquotes=false,subcaption=true,subfigure=false]{latex/atlaspackage}

\usepackage{latex/atlasphysics}
\usepackage{latex/atlascontribute}
\usepackage{latex/atlasbiblatex}
\usepackage{xspace}
\usepackage{definenumbers}
\usepackage{multirow}
\usepackage{accents}
\usepackage{booktabs}
\usepackage{rotating}

\newcommand*{\pbar}[1]{\accentset{(-)}{#1}}

\graphicspath{{logos/}{figures/}}

\def\WZtotXsecDDWjetTotUpee{3.1}
\def\WZtotXsecDDWjetTotUpmm{2.0}
\def\WZtotXsecDDWjetTotUpem{3.6}
\def\WZtotXsecDDWjetTotUp  {2.8}
\def\WZtotXsecBkgDDStatEEUpee{2.03}

\def\WZtotXsecBkgMCStatEEUpee{0.32}

\newcommand{\nWWSMCrossSectionNNLO}{63.2}

\def\WZtotXsecLumiErrUpee{1.5}

\def\WZtotXsecpileupUpee{2.00}

\def\WZtotXseceResoUpee{0.23}

\def\WZtotXseceScaleUpee{1.45}

\def\WZtotXseceIDSFUpee{2.19}

\def\WZtotXseceIsoUpee{0.47}

\def\WZtotXsecJVFUpee{0.26}

\def\WZtotXsecJERUpee{1.30}

\def\WZtotXsecMETResoSoftUpee{0.38}

\def\WZtotXsecMETScaleSoftUpee{2.07}

\def\WZtotXsecPtResoSoftUpee{0.19}

\def\WZtotXsecPtScaleSoftUpee{0.38}

\def\WZtotXsecDibosonCrossSectionUpee{1.01}

\def\WZtotXsecDDtopUpee{1.82}

\def\WZtotXsecDDZjetUpee{3.00}

\def\WZtotXsecMERGED_MCsigStatUpee{0.00}
\def\WZtotXsecMERGED_MCsigStatDwee{-0.00}
\def\WZtotXsecMERGED_MCbkgStatUpee{0.00}
\def\WZtotXsecMERGED_MCbkgStatDwee{-0.00}
\def\WZtotXsecMERGED_DataDrivenStatUpee{0.00}
\def\WZtotXsecMERGED_DataDrivenStatDwee{-0.00}
\def\WZtotXsecMERGED_BkgCrosthUpee{0.00}
\def\WZtotXsecMERGED_BkgCrosthDwee{-0.00}
\def\WZtotXsecMERGED_DataDrivenZjetUpee{0.00}
\def\WZtotXsecMERGED_DataDrivenZjetDwee{-0.00}

\def\WZtotXsecee{73.6}
\def\WZtotXsecStatErrUpee{4.2}
\def\WZtotXsecStatErrDwee{-4.1}

\def\WZtotXsecSysErrUpee{7.5}
\def\WZtotXsecSysErrDwee{-6.4}

\def\WZtotXsecmIsoUpmm{0.00}

\def\WZtotXsecmIDSFUpmm{0.00}

\def\WZtotXsecLumiErrUpmm{1.5}

\def\WZtotXsecpileupUpmm{2.03}

\def\WZtotXsecmuMSSmearUpmm{0.21}

\def\WZtotXsecmuIDSmearUpmm{1.67}

\def\WZtotXsecmuScaleUpmm{0.39}

\def\WZtotXsecmIDSFUpmm{0.82}

\def\WZtotXsecmIsoUpmm{1.20}

\def\WZtotXsecJVFUpmm{0.24}

\def\WZtotXsecJERUpmm{1.47}

\def\WZtotXsecMETResoSoftUpmm{0.53}

\def\WZtotXsecMETScaleSoftUpmm{1.85}

\def\WZtotXsecPtResoSoftUpmm{0.14}

\def\WZtotXsecPtScaleSoftUpmm{0.35}

\def\WZtotXsecDibosonCrossSectionUpmm{0.55}

\def\WZtotXsecDDtopUpmm{1.42}

\def\WZtotXsecDDZjetUpmm{2.26}

\def\WZtotXsecBkgDDStatMMUpmm{1.39}

\def\WZtotXsecBkgMCStatMMUpmm{0.18}

\def\WZtotXsecMERGED_MCsigStatUpmm{0.00}
\def\WZtotXsecMERGED_MCsigStatDwmm{-0.00}
\def\WZtotXsecMERGED_MCbkgStatUpmm{0.00}
\def\WZtotXsecMERGED_MCbkgStatDwmm{-0.00}
\def\WZtotXsecMERGED_DataDrivenStatUpmm{0.00}
\def\WZtotXsecMERGED_DataDrivenStatDwmm{-0.00}
\def\WZtotXsecMERGED_BkgCrosthUpmm{0.00}
\def\WZtotXsecMERGED_BkgCrosthDwmm{-0.00}
\def\WZtotXsecMERGED_DataDrivenZjetUpmm{0.00}
\def\WZtotXsecMERGED_DataDrivenZjetDwmm{-0.00}

\def\WZtotXsecmm{74.0}
\def\WZtotXsecStatErrUpmm{3.0}

\def\WZtotXsecSysErrUpmm{7.1}
\def\WZtotXsecSysErrDwmm{-5.9}

\def\WZtotXsecLumiErrUpem{1.4}

\def\WZtotXsecpileupUpem{1.35}

\def\WZtotXsecmuMSSmearUpem{0.09}

\def\WZtotXsecmuIDSmearUpem{0.56}

\def\WZtotXsecmuScaleUpem{0.10}

\def\WZtotXsecmIDSFUpem{0.41}

\def\WZtotXsecmIsoUpem{0.59}

\def\WZtotXseceResoUpem{0.04}

\def\WZtotXseceScaleUpem{0.42}

\def\WZtotXseceIDSFUpem{0.99}

\def\WZtotXseceIsoUpem{0.22}

\def\WZtotXsecJVFUpem{0.22}

\def\WZtotXsecJERUpem{1.35}

\def\WZtotXsecMETResoSoftUpem{0.31}

\def\WZtotXsecMETScaleSoftUpem{1.12}

\def\WZtotXsecPtResoSoftUpem{0.13}

\def\WZtotXsecPtScaleSoftUpem{0.23}

\def\WZtotXsecDibosonCrossSectionUpem{0.70}

\def\WZtotXsecDDtopUpem{1.35}

\def\WZtotXsecDDZjetUpem{0.46}

\def\WZtotXsecBkgDDStatEMUpem{0.61}

\def\WZtotXsecBkgMCStatEMUpem{0.10}

\def\WZtotXsecMERGED_MCsigStatUpem{0.00}
\def\WZtotXsecMERGED_MCsigStatDwem{-0.00}
\def\WZtotXsecMERGED_MCbkgStatUpem{0.00}
\def\WZtotXsecMERGED_MCbkgStatDwem{-0.00}
\def\WZtotXsecMERGED_DataDrivenStatUpem{0.00}
\def\WZtotXsecMERGED_DataDrivenStatDwem{-0.00}
\def\WZtotXsecMERGED_BkgCrosthUpem{0.00}
\def\WZtotXsecMERGED_BkgCrosthDwem{-0.00}
\def\WZtotXsecMERGED_DataDrivenZjetUpem{0.00}
\def\WZtotXsecMERGED_DataDrivenZjetDwem{-0.00}

\def\WZtotXsecem{70.6}
\def\WZtotXsecStatErrUpem{1.3}

\def\WZtotXsecSysErrUpem{5.8}
\def\WZtotXsecSysErrDwem{-5.1}

\def\WZtotXsecTriggerUpem{0.43}

\def\WZfidXsecLumiErrUpee{1.5}

\def\WZfidXsecMERGED_MCsigStatUpee{0.00}
\def\WZfidXsecMERGED_MCsigStatDwee{-0.00}
\def\WZfidXsecMERGED_MCbkgStatUpee{0.00}
\def\WZfidXsecMERGED_MCbkgStatDwee{-0.00}
\def\WZfidXsecMERGED_DataDrivenStatUpee{0.00}
\def\WZfidXsecMERGED_DataDrivenStatDwee{-0.00}
\def\WZfidXsecMERGED_BkgCrosthUpee{0.00}
\def\WZfidXsecMERGED_BkgCrosthDwee{-0.00}
\def\WZfidXsecMERGED_DataDrivenZjetUpee{0.00}
\def\WZfidXsecMERGED_DataDrivenZjetDwee{-0.00}

\def\WZfidXsecee{73.4}
\def\WZfidXsecStatErrUpee{4.2}
\def\WZfidXsecStatErrDwee{-4.1}

\def\WZfidXsecSysErrUpee{6.5}
\def\WZfidXsecSysErrDwee{-5.6}

\def\WZfidXsecLumiErrUpmm{1.6}

\def\WZfidXsecMERGED_MCsigStatUpmm{0.00}
\def\WZfidXsecMERGED_MCsigStatDwmm{-0.00}
\def\WZfidXsecMERGED_MCbkgStatUpmm{0.00}
\def\WZfidXsecMERGED_MCbkgStatDwmm{-0.00}
\def\WZfidXsecMERGED_DataDrivenStatUpmm{0.00}
\def\WZfidXsecMERGED_DataDrivenStatDwmm{-0.00}
\def\WZfidXsecMERGED_BkgCrosthUpmm{0.00}
\def\WZfidXsecMERGED_BkgCrosthDwmm{-0.00}
\def\WZfidXsecMERGED_DataDrivenZjetUpmm{0.00}
\def\WZfidXsecMERGED_DataDrivenZjetDwmm{-0.00}

\def\WZfidXsecmm{80.2}
\def\WZfidXsecStatErrUpmm{3.3}
\def\WZfidXsecStatErrDwmm{-3.2}

\def\WZfidXsecSysErrUpmm{6.4}
\def\WZfidXsecSysErrDwmm{-5.5}

\def\WZfidXsecLumiErrUpem{7.5}
\def\WZfidXsecLumiErrDwem{-7.3}

\def\WZfidXsecMERGED_MCsigStatUpem{0.00}
\def\WZfidXsecMERGED_MCsigStatDwem{-0.00}
\def\WZfidXsecMERGED_MCbkgStatUpem{0.00}
\def\WZfidXsecMERGED_MCbkgStatDwem{-0.00}
\def\WZfidXsecMERGED_DataDrivenStatUpem{0.00}
\def\WZfidXsecMERGED_DataDrivenStatDwem{-0.00}
\def\WZfidXsecMERGED_BkgCrosthUpem{0.00}
\def\WZfidXsecMERGED_BkgCrosthDwem{-0.00}
\def\WZfidXsecMERGED_DataDrivenZjetUpem{0.00}
\def\WZfidXsecMERGED_DataDrivenZjetDwem{-0.00}

\def\WZfidXsecem{374.4}
\def\WZfidXsecStatErrUpem{6.9}

\def\WZfidXsecSysErrUpem{25.0}
\def\WZfidXsecSysErrDwem{-22.5}

\def\WZtotXsecLumiErrUp{1.4}

\def\WZtotXsecpileupUp{1.48}

\def\WZtotXsecmuMSSmearUp{0.11}

\def\WZtotXsecmuIDSmearUp{0.67}

\def\WZtotXsecmuScaleUp{0.14}

\def\WZtotXsecmIDSFUp{0.43}

\def\WZtotXsecmIsoUp{0.62}

\def\WZtotXseceResoUp{0.05}

\def\WZtotXseceScaleUp{0.43}

\def\WZtotXseceIDSFUp{0.91}

\def\WZtotXseceIsoUp{0.21}

\def\WZtotXsecJVFUp{0.23}

\def\WZtotXsecJERUp{1.35}

\def\WZtotXsecMETResoSoftUp{0.35}

\def\WZtotXsecMETScaleSoftUp{1.28}

\def\WZtotXsecPtResoSoftUp{0.13}

\def\WZtotXsecPtScaleSoftUp{0.25}

\def\WZtotXsecDibosonCrossSectionUp{0.69}

\def\WZtotXsecDDtopUp{1.39}

\def\WZtotXsecDDZjetUp{0.86}

\def\WZtotXsecMERGED_MCsigStatUp{0.00}
\def\WZtotXsecMERGED_MCsigStatDw{-0.00}
\def\WZtotXsecMERGED_MCbkgStatUp{0.00}
\def\WZtotXsecMERGED_MCbkgStatDw{-0.00}
\def\WZtotXsecMERGED_DataDrivenStatUp{0.00}
\def\WZtotXsecMERGED_DataDrivenStatDw{-0.00}
\def\WZtotXsecMERGED_BkgCrosthUp{0.00}
\def\WZtotXsecMERGED_BkgCrosthDw{-0.00}
\def\WZtotXsecMERGED_DataDrivenZjetUp{0.00}
\def\WZtotXsecMERGED_DataDrivenZjetDw{-0.00}

\def\WZtotXsec{71.1}
\def\WZtotXsecStatErrUp{1.1}

\def\WZtotXsecSysErrUp{5.7}
\def\WZtotXsecSysErrDw{-5.0}

\def\WZtotXsecTriggerUp{0.75}

\def\WZtotXsecBkgDDStatAllUp{0.53}

\def\WZtotXsecBkgMCStatAllUp{0.09}

\hypersetup{pdftitle={ATLAS draft},pdfauthor={The ATLAS Collaboration}}
\AtlasTitle{Measurement of total and differential $W^+W^-$ production cross sections in proton-proton collisions at $\mathbf{\sqrt{s} = 8}$~\TeV\ with the ATLAS detector and limits on anomalous triple-gauge-boson couplings}

\author{The ATLAS Collaboration}

\PreprintIdNumber{CERN-PH-EP-2015-323}

\AtlasJournalRef{JHEP 09 (2016) 029}
\AtlasDOI{10.1007/JHEP09(2016)029}

\AtlasAbstract{
The production of $W$ boson pairs in proton-proton collisions at $\sqrt s = 8$~\TeV\ is studied using data corresponding to 20.3 \ifb\ of integrated luminosity collected by the ATLAS detector during 2012 at the CERN Large Hadron Collider. The $W$ bosons are reconstructed using their leptonic decays into electrons or muons and neutrinos. Events with reconstructed jets are not included in the candidate event sample. A total of 6636 $WW$ candidate events are observed. Measurements are performed in fiducial regions closely approximating the detector acceptance. The integrated measurement is corrected for all acceptance effects and for the $W$ branching fractions to leptons in order to obtain the total $WW$ production cross section, which is found to be \WZtotXsec $\pm\WZtotXsecStatErrUp$(stat) $^{+\WZtotXsecSysErrUp}_{\WZtotXsecSysErrDw}$(syst) $\pm\WZtotXsecLumiErrUp$(lumi) pb. This agrees
with the next-to-next-to-leading-order Standard Model prediction of 63.2$^{+1.6}_{-1.4}$(scale)$\pm1.2$(PDF) pb. Fiducial differential cross sections are measured as a function of each of six kinematic
variables. The distribution of the transverse momentum of the leading lepton is used to set limits on anomalous triple-gauge-boson couplings. 
}

\begin{document}
\nolinenumbers

\maketitle

\tableofcontents

\section{Introduction}
\label{sec:introduction}

The measurement of the production of pairs of electroweak gauge bosons plays a central role in 
tests of the Standard Model (SM) and in searches for new physics at the \TeV\ scale~\cite{Baak:2013fwa}. The $WW$ production cross section would grow arbitrarily large as a function of the centre-of-mass energy of the production process, $\sqrt{\hat{s}}$, were it not for the cancellations of $s$- and $t$-channel $W^+W^-$ (henceforth denoted $WW$) processes. New physics phenomena can occur as deviations from the gauge
structure of the Standard Model in the triple-gauge-boson couplings
$ZWW$ or $\gamma WW$~\cite{HISZ1993}, termed anomalous triple-gauge-boson couplings
(aTGCs). As the cross section for $WW$ production is one of the largest
among those involving a triple-gauge-boson vertex, 
it allows tests of the self-interaction of the gauge bosons to be made with high precision through
measurements of differential kinematic distributions. Studies of the $WW$ production process are particularly important as it constitutes a large irreducible background to searches for physics beyond the SM as well as to resonant $H\rightarrow W^+W^-$ production. 

A precise measurement of $WW$ production also tests the validity of the theoretical calculations. Perturbative quantum chromodynamics (pQCD) is the
essential ingredient in all these calculations and a recent
calculation of non-resonant $WW$ production has been performed up to
next-to-next-to-leading order (NNLO)~\cite{Gehrmann:2014fva}. However, fixed-order
calculations may fail to describe effects that arise from restrictions imposed on the phase space of the measurement. In this analysis, it is required that there be no jets above a certain transverse
momentum threshold, which introduces an additional momentum scale in
the theoretical calculation. Resumming the resulting large logarithms
can improve the accuracy of the prediction. Several calculations including resummation effects up to next-to-next-to-leading logarithms
(NNLL) have appeared recently in the literature~\cite{Grazzini:2015wpa,Meade:2014fca,Jaiswal:2014yba,Monni:2014zra}. Both the fixed-order and resummed predictions are compared to the measurements in this paper, except for Ref.~\cite{Grazzini:2015wpa} which coincides with the central prediction of the NNLO fixed-order prediction.

The existence of a non-zero self-coupling of the Standard Model gauge bosons has been proved by measurements of $WW$ production in electron--positron collisions at LEP~\cite{schael:2013ita}. The first measurement of the production of $W$ boson pairs at a hadron collider was conducted by the CDF experiment using Tevatron Run I data~\cite{Abe:1996dw}. Since then, more precise results have been published by the CDF~\cite{Aaltonen:2009aa} and D\O{} experiments~\cite{Abazov:2009ys}. The $WW$ production cross sections have already been measured at the LHC for a centre-of-mass energy of 
$\sqrt{s}=7$~\TeV\ by the ATLAS Collaboration~\cite{ww7tev} and for centre-of-mass energies of $\sqrt{s}=7$~\TeV\ and 8~\TeV\ by the CMS Collaboration~\cite{cms7tev,Khachatryan:2015sga}. Limits on anomalous couplings have been reported in these publications as well and, in several cases, are comparable to the most stringent aTGC limits set by the LEP\ experiments~\cite{schael:2013ita}. 

The present analysis uses a data sample with an integrated luminosity of 20.3 fb$^{-1}$ at a centre-of-mass energy of $\sqrt{s}=8$~\TeV. The total and fiducial $WW$ production cross sections are measured using $W\rightarrow e\nu$ and $W\rightarrow \mu\nu$ decays. Furthermore, measurements of differential cross sections are presented and limits on anomalous triple-gauge-boson couplings are reported.

\section{Analysis overview}
\label{sec:overview}

\begin{figure}[htbp]
\centering
\begin{subfigure}[t]{0.24\textwidth}
 \centering
 \includegraphics[width=\textwidth]{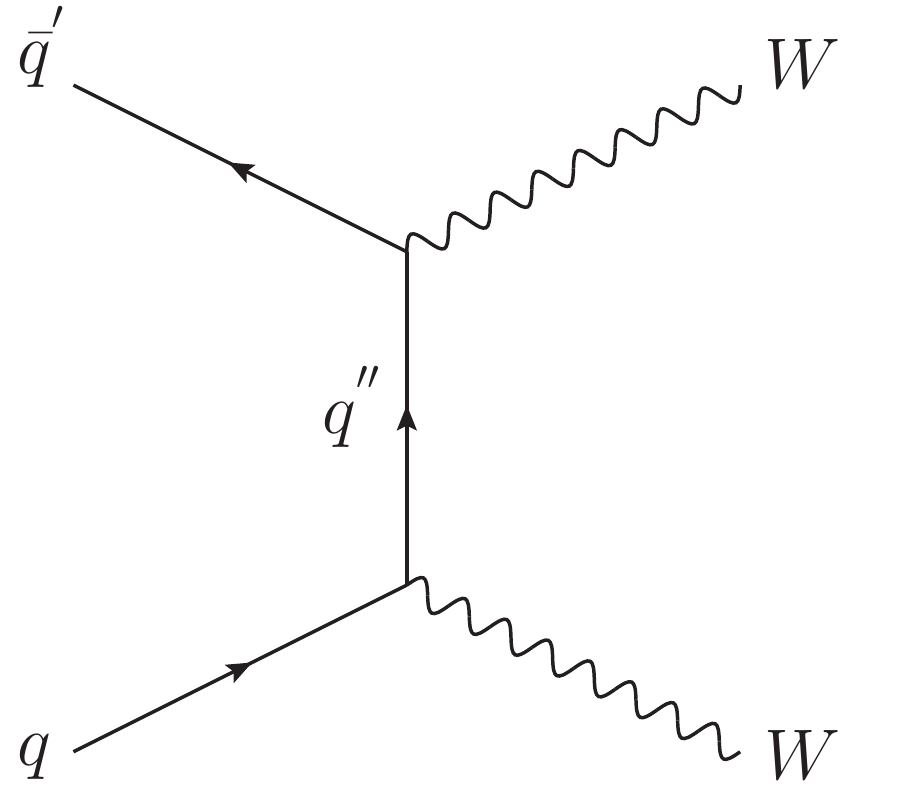}\caption{$t$- channel}
 \end{subfigure}
\begin{subfigure}[t]{0.24\textwidth}
 \centering
 \includegraphics[width=\textwidth]{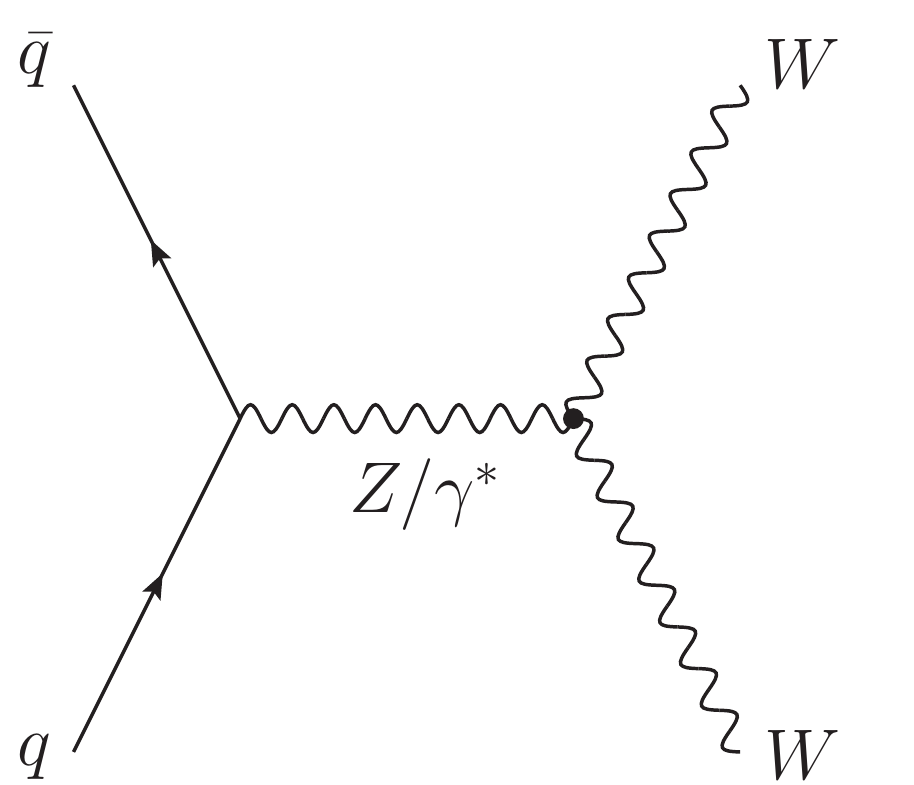}\caption{$s$- channel (TGC vertex)}
 \end{subfigure}
\begin{subfigure}[t]{0.24\textwidth}
 \centering
 \includegraphics[width=\textwidth]{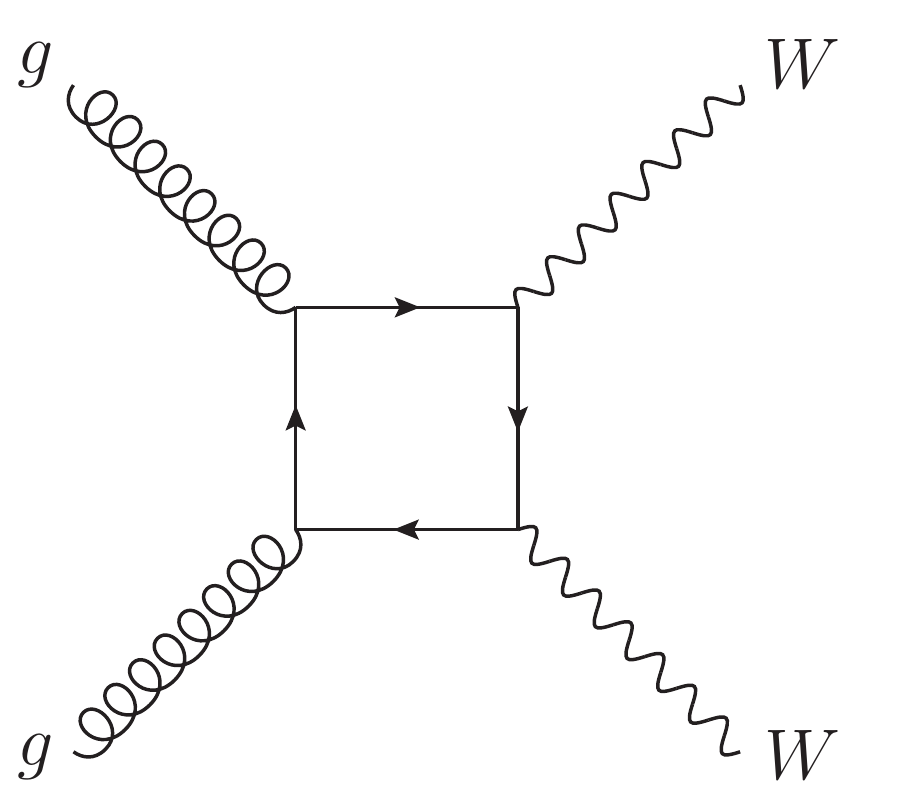}\caption{gluon fusion}
 \end{subfigure}
\begin{subfigure}[t]{0.24\textwidth}
 \centering
 \includegraphics[width=\textwidth]{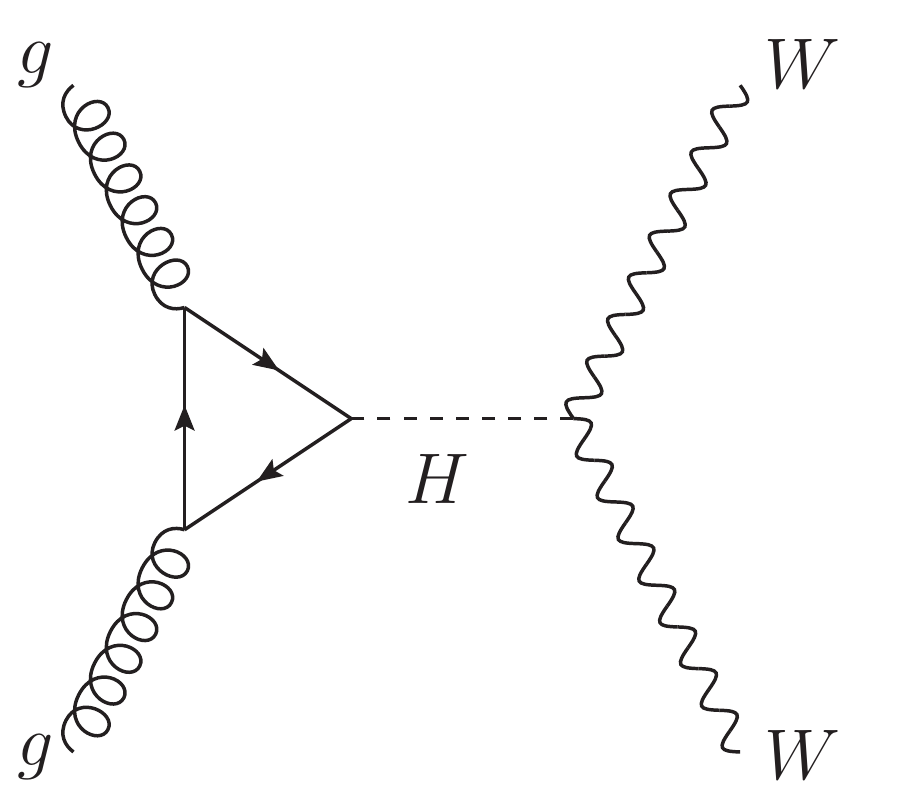}\caption{Higgs boson production}
 \end{subfigure}
\caption{(a) The SM tree-level Feynman diagram for \WW production through the $q\overline{q}$ initial state in the $t$-channel. (b)~The corresponding tree-level diagram in the $s$-channel, which contains the $WWZ$ and $WW\gamma$ TGC vertices. (c) The gluon fusion process, which is mediated by a quark loop. (d) The Higgs boson production process through gluon fusion and the subsequent decay of the Higgs boson to \WW.}
\label{fig:ww_prod}
\end{figure}

The production of $WW$ signal events takes place dominantly through quark--antiquark $t$-channel scattering and
$s$-channel annihilation, denoted by \qqWW,\footnote{In the following, \qqWW\ is taken to also include $qg$ initial states contributing to $t$-channel and
$s$-channel $WW$ production.} and are shown in Figures~\ref{fig:ww_prod}(a) and \ref{fig:ww_prod}(b), where the latter process involves a triple-gauge-boson vertex. In addition, $W$ boson pairs can be produced via gluon fusion through a quark loop;
these are the non-resonant $gg\rightarrow W^+W^-$ and the resonant Higgs boson  $gg\rightarrow H \rightarrow\ W^+W^-$ production processes in Figures ~\ref{fig:ww_prod}(c) and~\ref{fig:ww_prod}(d). All of these are considered as signal processes in this analysis. 

The $WW$ candidate events are selected in fully leptonic decay channels, resulting in final states of $e^\pm\pbar{\nu}_e \mu^\mp\pbar{\nu}_\mu$, $e^+\nu_e e^-\bar{\nu}_e$ and $\mu^+\nu_\mu \mu^-\bar{\nu}_\mu$. In the following, the different final states are referred to as $e\mu$, $ee$ and $\mu\mu$.

Backgrounds to these final states originate from a variety of processes.
Top-quark production ($t\bar t $ and the associated production of a single top quark and a $W$ boson) also results in events with $W$ pairs. In this case, the $W$ bosons are, however, accompanied by $b$-quarks that hadronise into jets. To enhance the purity of the signal candidates, events are rejected if any jets above a certain transverse momentum threshold are present in the final state. The Drell--Yan background is suppressed by requirements on missing transverse momentum, caused in $WW$ events by final-state neutrinos. For final states with same-flavour leptons, a veto on dilepton invariant masses close to the $Z$ pole mass is used. Other backgrounds stem from the $W$+jets or \mj\ production processes where one or more jets are misidentified as leptons. Diboson processes such as production of a heavy boson with an off- or on-shell photon or a Z boson, $WZ(\gamma^*)$, $W/Z+\gamma$ and $ZZ$ production, where one of the leptons falls outside the acceptance of the detector or a photon converts to an electron--positron pair, are additional sources of backgrounds. Backgrounds stemming from top-quark, Drell--Yan, $W$+jets and multijet production are evaluated using partially data-driven methods, where simulated event samples are only used to describe the shape of kinematic distributions or to
validate the methods. The background from diboson production
processes is modelled using Monte Carlo samples normalised to the
expected production cross section using theoretical calculations at
the highest available order. Other processes, such as double parton interactions, vector-boson fusion processes or associated $WH$ production, resulting in $e\mu$, $ee$ and $\mu\mu$ final states are not considered explicitly in the analysis as their contribution to the selected event sample is expected to be negligible ($<$0.6\%). 

The $e\mu$, $ee$ and $\mu\mu$ measurements of the total $WW$ production cross section are combined using a likelihood fit that includes the branching fractions into electrons or muons, whereas the fiducial cross sections are calculated per final state. Contributions from leptonic $\tau$-decays are not included in the definitions of the fiducial cross sections in order to allow comparisons with existing theoretical predictions. 
Because of its larger signal acceptance and smaller background, only the $e\mu$ final state is used to measure differential cross sections and to set limits on anomalous triple-gauge-boson-couplings.

The differential cross sections are reported as a function of the transverse momentum of the leading lepton, $\pt^{\rm lead}$, the transverse momentum of the dilepton system, $p_{\rm{T}}(\ell\ell)$, and the dilepton invariant mass, $m_{\ell\ell}$, all of which are correlated with the centre-of-mass energy of the interaction and thus sensitive to contributions from new physics processes at high values of $\sqrt{\hat{s}}$. Differential cross sections are also reported as a function of the azimuthal angle between the decay leptons, $\Delta \phi_{\ell\ell}$, which is correlated with the polarisation of the $W$ bosons and plays a special role in the extraction of the scalar Higgs boson signal. Additional measurements are presented as a function of the rapidity of the dilepton system, \absyll, and the observable \abscostheta, which is defined using the difference between the pseudorapidities of the leptons, $\Delta \eta_{\ell\ell}$, as follows:

\begin{equation}
\abscostheta = \left|\tanh\left(\frac{\Delta \eta_{\ell\ell}}{2} \right) \right|,
\label{eq:costheta}
\end{equation}
where the pseudorapidity is defined as $\eta = -\ln \tan(\theta/2)$ with $\theta$ being the polar angle.\footnote{The 
ATLAS experiment uses a right-handed coordinate system
with its origin at
the nominal $pp$ interaction point at the centre of the detector.
The positive $x$-axis is defined by the direction
from the interaction point to the centre of the LHC ring, with the
positive $y$-axis pointing upwards, while the
beam direction defines the $z$-axis. The azimuthal angle $\phi$ is measured
around the beam axis and the polar angle $\theta$ is the angle from
the $z$-axis. The distance in $\eta-\phi$ space between two objects is defined as
$\Delta R = \sqrt{({\Delta\eta})^2+({\Delta\phi})^2}$.
Transverse energy is computed as $\et = E \cdot \sin\theta$.} These variables are correlated with the rapidity and the boost of the $WW$ system along the $z$-axis. The $\abscostheta$ variable has been suggested for searches for new physics in $WW$ production in the low-$p_\mathrm{T}$ regime~\cite{Kim:2014eva}.

\section{The ATLAS detector}
\label{s:dectector}

The ATLAS detector is a general-purpose detector that is used to study collisions at the Large Hadron Collider (LHC). A detailed description can be found in Ref.~\cite{detectorpaper}. 

The inner detector (ID) is used to measure trajectories and momenta of charged particles within the central region of the ATLAS detector with pseudorapidities of $|\eta|<$ 2.5. The ID is located inside a solenoid that provides a 2 T axial magnetic field. The ID consists of three sub-detector systems: a three-layer silicon-pixel tracker, a four-layer silicon-strip detector built of modules with pairs of single-sided sensors glued back-to-back, and a transition radiation tracker consisting of straw tubes. In the central region these sub-detectors are constructed in the shape of cylinders, while in the forward and backward regions, they take the form of disks. The innermost pixel layer of the ID is located just outside the beam-pipe. 

Electromagnetic (EM) energy deposits are measured using a liquid-argon calorimeter with accordion-shaped electrodes and lead absorbers. The EM calorimeter is divided into a barrel part ($|\eta| <$ 1.475) and two end-cap components (1.375 $<|\eta|<$ 3.2). 
The transition region between the barrel and the end-caps of the calorimeter, 1.37 $< |\eta| <$ 1.52, has a large amount of material in front of the first active calorimeter layer; therefore electromagnetic objects measured in this region suffer from worse energy resolution and are not considered in this analysis.

For hadronic calorimetry, three different technologies are used. In the barrel region ($|\eta| <$ 1.7), scintillator tiles with steel absorbers are used. 
Liquid argon with copper absorber plates are used in the end-cap region ($1.5 < |\eta| < 3.2$). The forward calorimeter ($3.1 < |\eta| < 4.9$) consists of liquid argon with tungsten and copper absorbers and has separate electromagnetic and hadronic sections.

The muon spectrometer (MS) provides precise measurements of the momentum of muons within $|\eta| < 2.7$ using three layers of precision tracking stations, consisting of drift tubes and cathode strip chambers. Resistive plate and thin-gap chambers are used to trigger on muons in the region $|\eta| <$ 2.4. The magnetic fields for the MS are produced by one barrel and two end-cap air-core toroid magnets surrounding the calorimeter. Each magnet consists of eight superconducting coils arranged symmetrically in $\phi$. 

The ATLAS trigger system uses three consecutive stages to decide whether an event is selected to be read out for permanent storage. The first level of the trigger is implemented using custom-made electronics and operates at a design rate of at most 75 kHz. It is complemented by two software-based high-level triggers (HLT). The second level consists of fast online algorithms to inspect regions of interest flagged by the first trigger level. At the third level, the full event is reconstructed using algorithms similar to those in the offline event selection. 

\section{Data and Monte Carlo samples}
\label{sec:samples}
\label{sec:MCSamples}

The analysis is based on data collected by the ATLAS experiment during the 2012 data-taking period. Only runs with stable proton--proton ($pp$) beam collisions at $\sqrt{s}=8~\TeV$ in which all relevant detector components were operating normally are used. This data sample corresponds to an integrated luminosity of \(\IntL=20.3\,\ifb\), determined with an uncertainty of $\pm1.9\%$ and derived from beam-separation scans performed in November 2012~\cite{Aaboud:2016hhf}.

The kinematic distributions of both the signal and background processes are modelled using Monte Carlo (MC) samples. The additional $pp$ collisions accompanying the hard-scatter interactions (pile-up) are modelled by overlaying minimum-bias events generated using \Pythia~8~\cite{newpythia8}. To simulate the detector response, the MC events are passed through a detailed simulation of the ATLAS detector~\cite{Aad:2010ah} based on \textsc{GEANT4}~\cite{Agostinelli:2002hh}.

For the $WW$ signal events, three different MC samples are generated. The $q\bar{q}\rightarrow W^+W^-$ events are generated using the {\textsc{Powheg-Box 1.0}} generator (referred to as \Powheg~below)~\cite{Nason:2004rx, Frixione:2007vw, Alioli:2010xd, Alioli:2008gx}. It is interfaced to \Pythia~8.170 for the simulation of parton shower and hadronisation processes. The non-resonant $gg$-induced $WW$ signal events are generated using the \ggtwoww\ program (version 3.1.3)~\cite{gg2ww} interfaced to \Herwig~6.5 and \Jimmy 4.31~\cite{herwig,Butterworth:1996zw} for parton showering, hadronisation and underlying event simulation. The resonant $WW$ production via a Higgs boson with a mass of $m_H =$ 125 \GeV\ is modelled using \Powheg+\Pythia~8.170. For these three samples, the CT10 NLO~\cite{CTEQ10} parton distribution function (PDF) is employed in the event generation. \Photos~\cite{Golonka:2005pn} is used to model the radiation of photons, and AU2~\cite{ATLAS:2012uec} and AUET2~\cite{ATLAS:2010osr} are used as the parameter tunes for the underlying event in the \Powheg+\Pythia\ and the \Herwig+\Jimmy samples, respectively. To calculate acceptances (see Section~\ref{sec:theorymodelling}) or make differential predictions for the $WW$ signal process, these samples are combined according to their respective cross sections as listed in Table~\ref{tab:mc}. Next-to-leading-order electroweak (EW) contributions of $\mathcal{O}(\alpha_{\mathrm{EW}}^3)$ are described in Refs.~\cite{WWEWKCorr,WWrr,PhysRevD.88.113005,Bierweiler:2013dja} and the corrections derived in Ref.~\cite{Gieseke:2014gka} are applied as scale factors to \qqWW\ production in the acceptance calculation and in the setting of limits on anomalous triple-gauge-couplings (see Section~\ref{atgc}), but not for any other purpose or distribution shown in this paper. 

The $t\overline{t}$ background is modelled with \MCatNLO\ 4.03~\cite{MCatNLO} using the CT10 NLO PDF interfaced with \Herwig~6.5+\Jimmy~4.31 with the AUET2 tune. The same generators and settings are used to simulate $s$-channel single-top production  and the associated production of a top quark with a $W$ boson, while the \textsc{AcerMC} 3.7~\cite{Kersevan:2004yg} MC generator interfaced to \Pythia~6~\cite{pythia} with the AUET2B tune~\cite{ATL-PHYS-PUB-2011-009} and the CTEQ6L1 PDF~\cite{Pumplin:2002vw} is employed for the single-top-$t$-channel process. Alternative samples employing \Powheg+\Pythia~8\ and \Powheg+\Herwig~6.5+\Jimmy~4.31 are used to determine systematic uncertainties in the data-driven estimate. 

Drell--Yan and $W$+jets events are modelled using \Alpgen\ 2.14~\cite{Mangano:2002ea} which is either interfaced to \Pythia~6 ($W$+jets and $Z\rightarrow ee/\mu\mu$ samples with $m_{\ell\ell}>60$~\GeV) or to \Herwig~6.5~\cite{herwig} and \Jimmy 4.31 ($Z\rightarrow \tau\tau$ and remaining $Z\rightarrow ee/\mu\mu$ samples) for the simulation of parton showering, hadronisation and the underlying event modelling. The AUET2 tune is used for the underlying event. The CTEQ6L1 PDF is employed in the event generation and for the parton shower. 
The MLM~\cite{Mangano:2001xp} matching scheme is used to remove overlaps between events with the same parton multiplicity generated by the matrix element and the parton shower. 

The $ZZ$ and $WZ(\gamma^*)$ diboson background processes are generated with \Powheg\ interfaced to \Pythia~8\ using the AU2CT10 tune with the CT10 NLO PDF. The lower limit on the invariant mass of the decay leptons of the $Z(\gamma^*)$ in the $ZZ$ sample is set to $m_{\ell\ell}>4$ \GeV, while for the $WZ(\gamma^*)$ it is $m_{\ell\ell}>7$~\GeV. For $\gamma^*$ masses below 7~\GeV, dedicated $W\gamma^*$ samples are produced using the {\sc Sherpa} 1.4.2 generator~\cite{Sherpa} with its built-in parton shower and hadronisation using the CT10 NLO PDF. Events from $W\gamma$ production can also mimic the $WW$ signature when the photon is misidentified as an electron. These events are generated using \Alpgen\ interfaced to \Herwig+\Jimmy. The CTEQ6L1 PDF and the AUET2 tune are used for this sample.

The MC samples used in this analysis are summarised in Table~\ref{tab:mc}, where the total cross sections, $\sigma_{\mathrm{total}}$, times the branching fractions, $\mathcal{B}$, into leptons are reported. 
The total cross sections are taken from theoretical calculations and the perturbative order of each calculation is also given in the table.  The total cross sections are used to normalize the MC samples, which are essential for the modelling of kinematic distributions.

\begin{table}[ht]
\centering
\begin{tabular}{llllll r@{.}l}
\hline\hline
\multicolumn{3}{l}{\multirow{2}{*}{Process}}
& \multicolumn{2}{l}{{MC generator}}& Calculation
 & \multicolumn{2}{r}{$\sigma_{\mathrm{total}}{\cdot}\mathcal{B}$~~}
\\
& 
&
&\multicolumn{2}{l}{{+parton shower}}&
& \multicolumn{2}{r}{[pb]~~}
\\
& 
&
&\multicolumn{2}{l}{{+hadronisation}}&
& \multicolumn{2}{r}{  }
\\
\hline
\multicolumn{2}{l}{\textbf{$\WW$ Signal} }& &\multicolumn{2}{l}{} \\
\multicolumn{3}{l}{\quad $q\overline{q} \rightarrow W^+W^-$                           }& \POWHEG+\PYTHIA~8&   & NLO~\cite{Campbell:2011bn}   &5&58 \\
\multicolumn{3}{l}{\quad $gg \rightarrow W^+W^-$ (non-resonant)                             }& \GGTOVV+\HERWIG&      & LO\dag~\cite{gg2ww} &0&153 \\
\multicolumn{3}{l}{\quad $gg \rightarrow  H \rightarrow W^+W^-$                             }& \POWHEG+\PYTHIA~8&    & NNLO~\cite{HiggsXS}  &0&435 \\
\hline
\multicolumn{3}{l}{\textbf{Top quark} }& &\multicolumn{2}{l}{} \\
\multicolumn{3}{l}{\quad $\ttbar$                                                }& \MCatNLO+\HERWIG &  &NNLO+NNLL~\cite{Czakon:2011xx}  &26&6\\
\multicolumn{3}{l}{\quad $Wt$                                                    }& \MCatNLO+\HERWIG & &NNLO+NNLL~\cite{Kidonakis:2010ux}  &2&35 \\
\multicolumn{3}{l}{\quad Single top $t$-channel                                             }& \ACERMC+\PYTHIA~6&    &NNLO+NNLL~\cite{Kidonakis:2011wy}  &28&4\\
\multicolumn{3}{l}{\quad Single top $s$-channel                                              }& \MCatNLO+\HERWIG&  &NNLO+NNLL~\cite{Kidonakis:2010tc}  &1&82 \\
\hline
\multicolumn{4}{l}{\textbf{Drell--Yan }}& &\multicolumn{2}{l}{}  \\
\multicolumn{1}{l}{\quad $Z\rightarrow ee / \mu\mu$   } &\multicolumn{2}{l}{($\mll{~>~}60\GeV$)         }& \ALPGEN+\PYTHIA~6&  & \multirow{3}{*}{NNLO~\cite{Catani:2007vq}\hspace{4mm}$\left. \vphantom{\begin{tabular}{l}3\\3\\3\end{tabular}}\right\}$} \\

\multicolumn{1}{l}{\quad $Z\rightarrow \tau\tau$   } &\multicolumn{2}{l}{($\mll{~>~}60\GeV$)         }& \ALPGEN+\HERWIG&   &  & \multicolumn{1}{r@{\hphantom{.}}}{\hskip-6mm 16500} \\
  \multicolumn{1}{l}{\quad $Z\rightarrow \ell\ell$   } &\multicolumn{2}{l}{($10\GeV{~<~}\mll{~<~}60 \GeV$)         }& \ALPGEN+\HERWIG&   \\

\hline
\multicolumn{4}{l}{\textbf{Other dibosons ($VV$)}}& & \multicolumn{2}{l}{} \\
\multicolumn{1}{l}{\quad $\Wg$  } &\multicolumn{2}{l}{($\pT^{\gamma} > 8\GeV$) }& \ALPGEN+\HERWIG&    &NLO~\cite{Campbell:2011bn}   &369&0 \\
\multicolumn{1}{l}{\quad $\WZ(/\gamma^*)$  } &\multicolumn{2}{l}{($\mll > 7\GeV$)         }& \POWHEG+\PYTHIA~8&  &NLO~\cite{Campbell:2011bn}     &12&7\\
\multicolumn{1}{l}{\quad $\WZ(/\gamma^*)$ } &\multicolumn{2}{l}{($\mll < 7\GeV$)         }& \SHERPA      &            &NLO~\cite{Campbell:2011bn}  &12&9\\
\multicolumn{1}{l}{\quad $\ZZ \rightarrow 4\ell$  } &\multicolumn{2}{l}{($\mll > 4\GeV$)         }& \POWHEG+\PYTHIA~8&  &NLO~\cite{Campbell:2011bn}    &0&733 \\
\multicolumn{1}{l}{\quad $\ZZ \rightarrow \ell\ell\,\nu\nu$} &\multicolumn{2}{l}{($\mll > 4\GeV$)  }& \POWHEG+\PYTHIA~8&  &NLO~\cite{Campbell:2011bn}   &0&504 \\
    & \\
\hline\hline
\end{tabular}
\caption{
Monte Carlo samples used to model the signal and background processes. The total cross sections times branching fractions, $\sigma_{\mathrm{total}}{\cdot}\mathcal{B}$, are quoted at $\sqrt{s}=8~\TeV$ using higher-order calculations. The branching fractions $\mathcal{B}$ include the decays $t{\TO}Wq$, $W\rightarrow\ell\nu$, and $Z{\rightarrow}\ell\ell$, while the decay of one $Z$ boson to neutrinos is considered for the process $\ZZ{\TO}\ell\ell\,\nu\nu$. Here, $\ell$ refers to $e$, $\mu$, or $\tau$ for signal and background processes, and all three lepton flavors are considered in $\mathcal {B}$. The \qqWW\ process also includes $qg$ initial states contributing to $t$-channel and
$s$-channel $WW$ production. The Higgs mass is taken to be $m_H=125$~\GeV.  \newline \dag{\footnotesize The process itself is calculated at LO, however it contributes only at  NNLO to the total $WW$ cross section.}}
\label{tab:mc}
\end{table}

\section{Object reconstruction and event selection}
\label{sec:selection}

\subsection{Pre-selection of events}

Fast selection algorithms based on the detection of electrons or
muons are used to trigger the readout of the events~\cite{ATLAS-CONF-2012-048,Aad:2014sca}. The
trigger selection algorithms are based on the transverse momentum of the
leptons and use certain object quality criteria. These object
quality criteria vary for the different triggers and are generally
looser and more efficient for dilepton triggers as opposed to single-lepton triggers, which are designed to yield larger rate reductions. Another important
consideration is the coverage of the first-level muon trigger, which
is only about 80\% in the central region ($|\eta|<1.05$) of the detector~\cite{Aad:2014sca}. In the $ee$ and $\mu\mu$ final states, highly efficient dilepton
triggers are used, which impose loose identification criteria on both electrons for the dielectron trigger and for the dimuon trigger only a single muon in the first trigger level. In the
$e\mu$ final state the optimal signal yield is achieved by combining single-lepton triggers with the $e\mu$ dilepton trigger, as the latter is
affected by the limited coverage for muons at the first trigger level and, due also to the trigger requirements on the electrons, yields a low efficiency.

For the single-electron trigger, the HLT criterion for the transverse momentum is either $\pte>24$\,\GeV, accompanied by track-based isolation requirements, or $\pte>60$\,\GeV. The single-muon trigger has a transverse momentum threshold of $\ptmu=24$\,\GeV\ when a loose track-based isolation requirement is satisfied, or a transverse momentum threshold of $\ptmu=36$\,\GeV. The combined electron--muon trigger requires $\ptmu>8$\,\GeV\ for the muon and $\pte>12$\,\GeV~for the electron. The dielectron trigger requires two electrons with a transverse momentum of $\pte > 12$~\GeV\, while the dimuon trigger applies a transverse momentum requirement of $\pt^\mu > 18$\,\GeV\ for one and $\pt^\mu > 8$\,\GeV\ for the second muon. With the chosen trigger scheme, the trigger efficiency defined with respect to the offline selection criteria is 99-100\% for all three channels.

\subsection{Lepton selection}

Electrons are reconstructed from energy deposits in the EM calorimeter with an associated track.
Electrons must satisfy $|\eta^e|<2.47$, excluding the calorimeter transition region $1.37<|\eta^e|<1.52$. To  efficiently reject multijet background, they are required to pass a very tight likelihood-based identification criterion~\cite{ATLAS-CONF-2014-032} that uses discriminating variables based on calorimetric shower shapes and track parameters of the electron candidates. 
Electrons
are required to be unaffected by known instrumental problems such as coherent noise in the calorimeters. Stringent requirements are placed on track impact parameters and electron isolation to reject electrons from \mj\ background events. These isolation and tracking requirements are the same as those utilised in Ref.~\cite{ATLAS:2014aga}. To reject electrons reconstructed from a bremsstrahlung photon emitted by a muon traversing the calorimeter, any electron candidate reconstructed at a distance $\Delta R< 0.1$ from a selected muon is removed.

Muons are reconstructed by combining tracks reconstructed separately in the ID and the MS. Muons are required to be within the pseudorapidity region $|\eta^{\mu}| < 2.4$. To reject backgrounds, quality criteria are applied to the muon candidates as described in detail in Ref.~\cite{aad:2014rra}. As in the case of electrons, the track parameter and isolation selection criteria applied to muons follow that in Ref.~\cite{ATLAS:2014aga}. For the rejection of muons from heavy-flavour decays, muons are removed if they are found within a cone of $\Delta R=0.3$ to a selected jet. 

\subsection{Jet selection}
\label{sec:jet}

Jets are reconstructed using the anti-$k_t$ algorithm~\cite{bib:antikt2}
with radius parameter $R=0.4$
implemented in the FastJet package~\cite{bib:antikt3}.
The inputs to the jet-finding algorithm are calibrated topological clusters~\cite{bib:topo}. The calibration of topological clusters to the hadronic energy scale depends on their local energy density and total energy~\cite{Barillari:1112035}. A jet-area-dependent correction is applied to correct the jet energy for contributions from additional $pp$ collisions based on an estimate of the pile-up activity in a given event using the method proposed in Ref.~\cite{Cacciari:2007fd}. The reconstructed jets are further calibrated using
jet-energy-scale corrections from simulation. Their calibration is refined using data-driven corrections to account for residual differences between data and MC simulation~\cite{JES,ATLAS-CONF-2015-037}.

Jets are required to have a transverse momentum of $\pt^{\mathrm{jet}}>25$~\GeV\ and pseudorapidity $|\eta^{\mathrm{jet}}|<4.5$. Jets are removed if they are found within a cone of $\Delta R=0.3$ around a selected electron. 
For jets with $\pt^{\mathrm{jet}}<50$~\GeV\ and $|\eta^{\mathrm{jet}}|<2.4$, an additional requirement is applied to reject jets from pile-up interactions in the event. The scalar sum of the transverse momenta of the tracks associated with both the primary vertex and the jet must be larger than one-half of the scalar sum of the momenta of all the tracks associated with the jet; jets with no associated tracks are also removed~\cite{ATLAS-CONF-2013-083}. This selection criteria is henceforth denoted as requirement on the jet vertex fraction (JVF). 

Selected $b$-jets are used in the estimation of the top-quark background described in Section~\ref{sec:top}. Jets containing $b$-hadrons are identified within the central region of the detector, $|\eta^{\mathrm{jet}}|<2.5$, using a multivariate approach based on track impact-parameter significance, secondary vertex reconstruction and other tracking variables described in Refs.~\cite{ATLAS-CONF-2011-102,ATLAS-CONF-2014-046}. In this analysis the requirement on the multivariate discriminant is chosen to have a $b$-jet identification efficiency of 85\%, which has been verified using a $t\bar{t}$ data sample. This corresponds to a rejection factor of 10 for light-flavour jets~\cite{ATLAS-CONF-2014-046}. 

\subsection{Reconstruction of missing transverse momentum}

The reconstruction of missing transverse momentum is optimised to reject backgrounds without neutrinos in the final state.

Calorimeter-based missing transverse momentum, \met, is reconstructed as the magnitude of the negative vectorial sum of all measured and identified physics objects, denoted as \textbf{\met}, where the bold notation indicates a vector throughout this paper. Additionally, energy deposits in the calorimeter not associated with any high-\pt\ objects are also included as described in Ref.~\cite{ATLAS-CONF-2013-082}. 

The relative missing transverse momentum, \metrel, is defined as
\begin{equation} \label{eqn:MetRelDef}
	\metrel = \left\{
	\begin{array} {ll}
	\met \times \sin\left(\Delta\phi_{\ell}\right) & \mbox{if} \; \Delta\phi_{\ell} < \pi/2 \\
	\met                                             & \mbox{if} \; \Delta\phi_{\ell} \geq \pi/2,
	\end{array}
	\right.
\end{equation}
where $\Delta\phi_{\ell}$ is the difference in azimuthal angle $\phi$ between \textbf{\met}~and the nearest lepton. With this definition, \metrel\ is less affected by the mis-measurement of the energy of a lepton leading to spurious large calorimeter-based missing transverse momentum.

Additionally, track-based \ptmiss\ is used, which is the magnitude of the negative vectorial sum (\textbf{\ptmiss}) of all identified and calibrated leptons and all tracks not associated with any lepton in the event. These tracks are required to have $\pt> 0.5$~\GeV\ and be associated to the reconstructed primary vertex, which makes \ptmiss\ robust against additional pile-up interactions in the same bunch-crossing. A more detailed description of the \ptmiss\ reconstruction can be found in Ref.~\cite{ATLAS-CONF-2014-019}.

In events with genuine missing transverse momentum due to undetected neutrinos, \textbf{\met}\ and \textbf{\ptmiss}\ are complementary estimators of the total missing transverse momentum vector. A large difference between \textbf{\met}\ and \textbf{\ptmiss}\ indicates a mis-reconstruction of either of these two quantities in the context of this analysis.

\subsection{$WW$ selection}

The $WW$ candidate events are required to contain two
oppositely charged leptons fulfilling the identification criteria, isolation and
track impact-parameter requirements specified earlier. The leading and sub-leading leptons have to satisfy
transverse momentum requirements of $\pt^{\ell} > 25$~\GeV\ and $\pt^{\ell} > 20$~\GeV, respectively. To suppress other diboson backgrounds, events are rejected if additional leptons with $\pt^{\ell}>$ 7~\GeV\ fulfilling the above described selection criteria are present.

The event selection criteria are optimised to enhance the $WW$ signal purity. The invariant mass of the dilepton pair is required to be greater than 15~\GeV\ for $ee$/$\mu\mu$ final states to reject \jpsi, \Ups\ and other low-mass resonances, while $e\mu$ final states are required to have an invariant mass above 10~\GeV\ to remove \mj\ events. 
Figure~\ref{fig:mass-ll} shows the invariant mass distributions of these selected dilepton events 
for the same-flavour and $e\mu$ final states. The backgrounds shown here are based purely on MC predictions, which are normalised to \(\IntL=20.3\,\ifb\) using the cross section times branching fractions shown in Table~\ref{tab:mc}. In Figure~\ref{fig:mass-ll}, Drell--Yan production is the largest background for the $ee$ and $\mu\mu$ final states, and it is therefore further suppressed by rejecting events that are reconstructed with an invariant mass closer than 15~\GeV\ to the $Z$ boson mass $m_{Z}$~\cite{Beringer:1900zz}. 

 \begin{figure}[h!tbp]
 \begin{center}
 \includegraphics[width=0.49\textwidth]{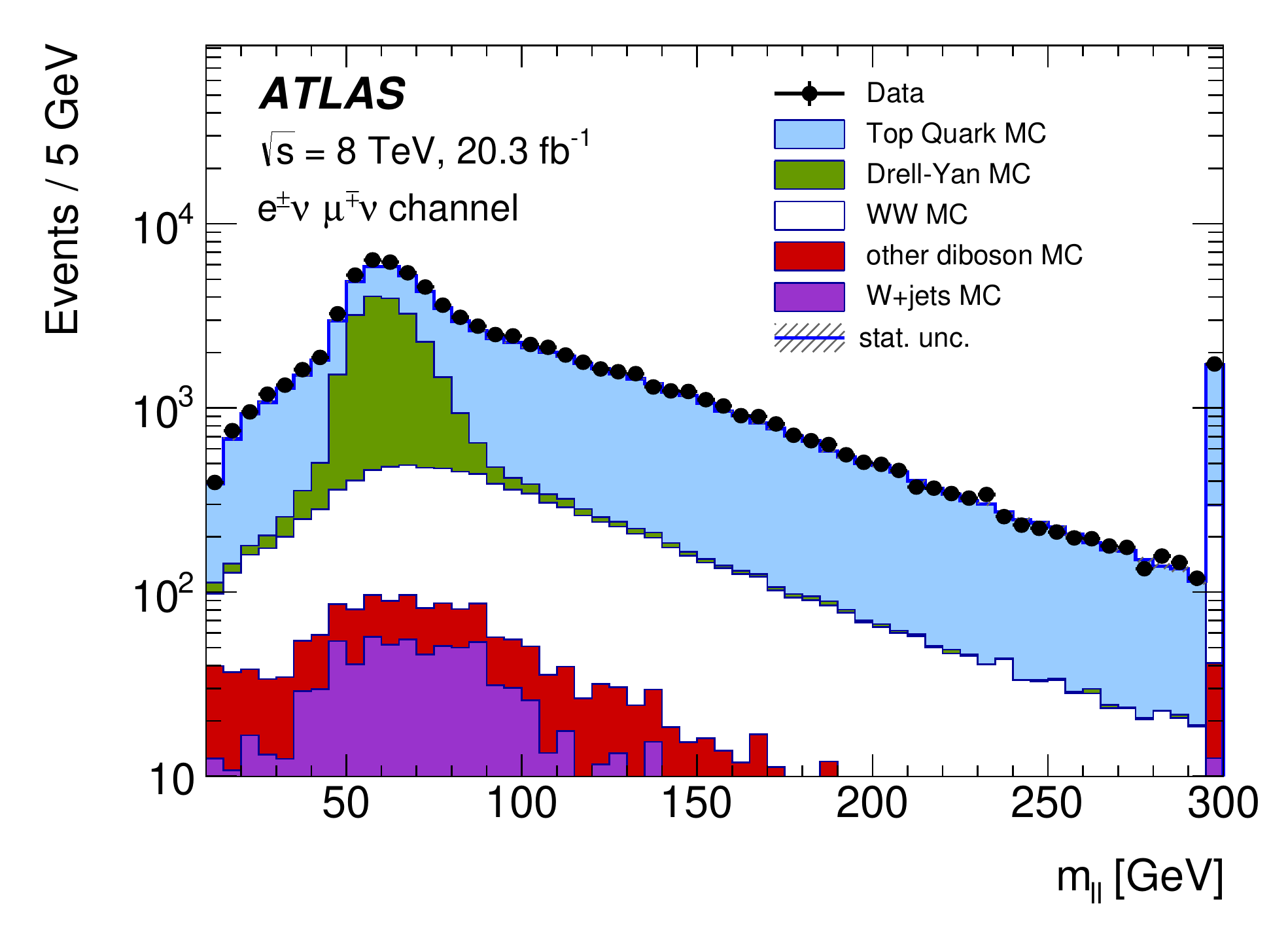}
 \includegraphics[width=0.49\textwidth]{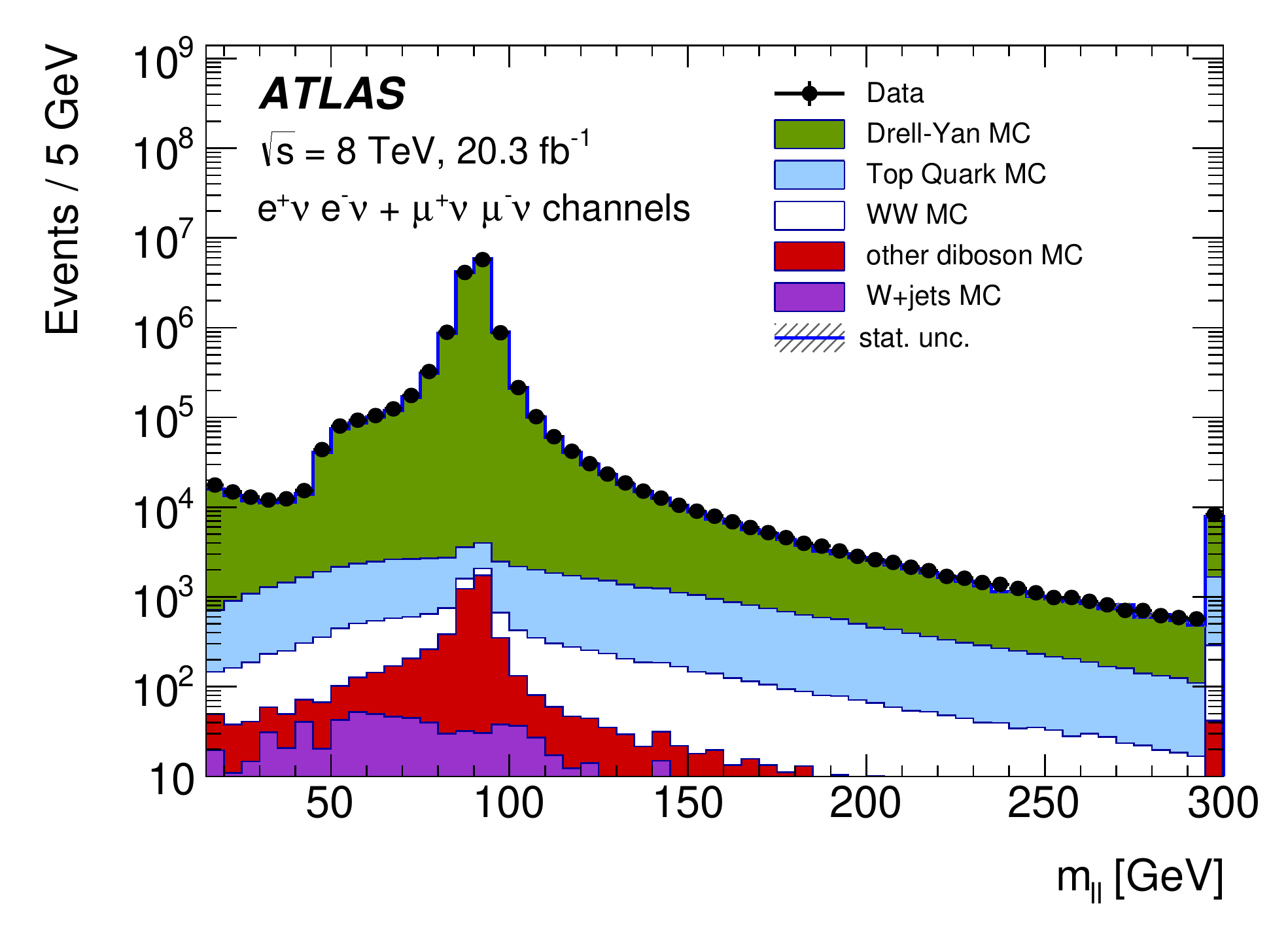}
 \caption{The invariant mass distributions are shown for dilepton pairs in selected events for $e\mu$ (left) and $ee+\mu\mu$ (right) final states after
    the dilepton selection and the $m_{\ell\ell}$ requirements described in the text.     The points represent data and the stacked histograms are
   the MC predictions, which are normalised to \(\IntL=20.3\,\ifb\) using the cross section times branching fractions shown in Table~\ref{tab:mc}. The last bin is an overflow bin.
   Only statistical uncertainties are shown. }
 \label{fig:mass-ll}
 \end{center}
 \end{figure}

The Drell--Yan background in the same-flavour channel is still
significant after this more restrictive invariant mass requirement,
so stringent conditions are imposed using selection criteria related to
missing transverse momentum. The requirements are less strict for $e\mu$ final states, where Drell--Yan production contributes only through $Z/\gamma^*\rightarrow \tau\tau$. The selection requirements are as follows. The relative missing transverse momentum, \metrel, is required to be larger than 15~\GeV\ for the \emu\ and larger than 45~\GeV\ for the $ee$ and $\mu\mu$ final states. Track-based missing transverse momentum, \ptmiss, is further required to be larger than 20~\GeV\ for the \emu\ and larger than 45~\GeV\ for the $ee$ and $\mu\mu$ final states. The azimuthal angle between \textbf{\met}\ and \textbf{\ptmiss}\ is calculated and the condition $\Delta\phi(\textbf{\met},\textbf{\ptmiss})< 0.6\,$ must be met in the $e\mu$ final state, while $\Delta\phi(\textbf{\met},\textbf{\ptmiss})< 0.3\,$ must be satisfied for the $ee$ and $\mu\mu$ final states. 
 
The jet multiplicity distributions for data, the signal MC simulation and the different background contributions
after applying these requirements are shown in Figure~\ref{fig:njet}. In order to suppress the dominant top-quark background, events are required to contain no selected jets. This requirement is referred to as the jet-veto requirement. 
The visible excess of events without selected jets at this stage is
still subject to changes from data-driven refinements in the
background estimate as discussed in Section~\ref{sec:backgrounds}. Furthermore, there is a significant uncertainty in $WW$ signal predictions as discussed in
Section~\ref{sec:results}.

A summary of all applied selection criteria is given in Table~\ref{recolevelcuts}. 

\begin{figure}[h!tbp]
 \begin{center}
\includegraphics[width=0.49\textwidth]{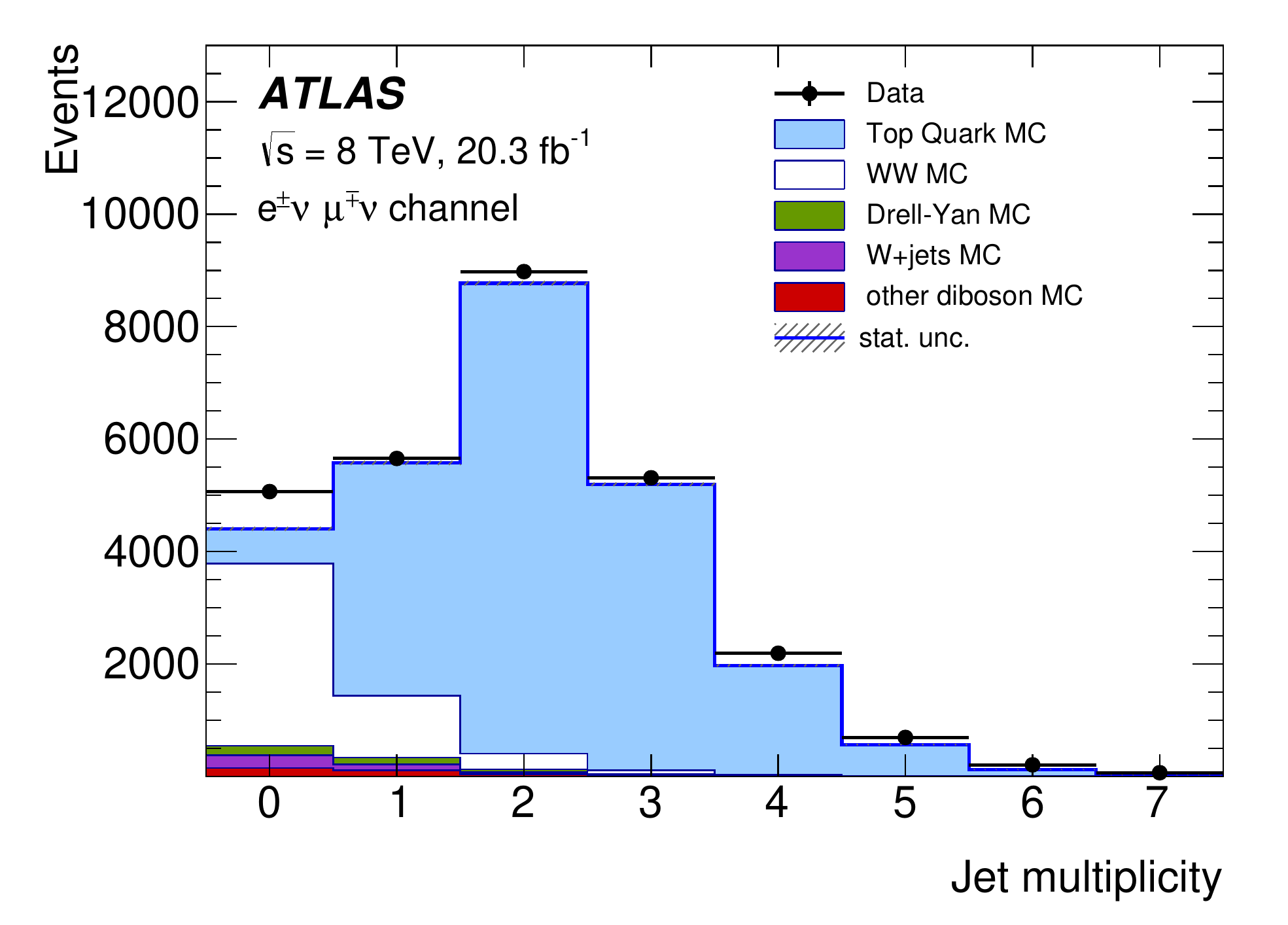}
\includegraphics[width=0.49\textwidth]{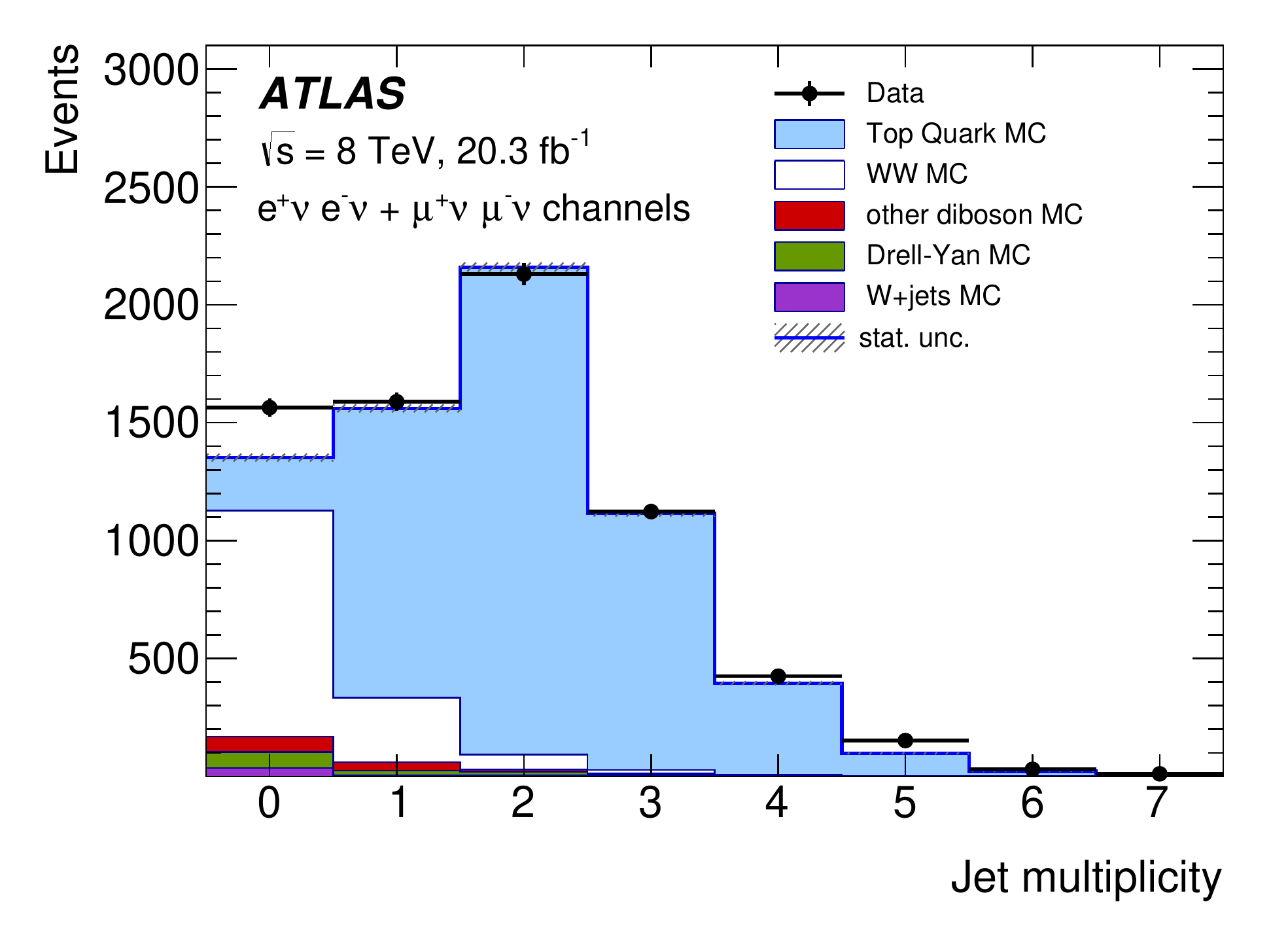}
 \caption{Jet multiplicity distributions for $e\mu$ (left) and $ee+\mu\mu$ (right) events
    before the jet-veto requirement is applied. The points represent data and the stacked histograms are
    the MC predictions, which are normalised to \(\IntL=20.3\,\ifb\) using the cross section times branching fractions shown in Table~\ref{tab:mc}. For the $t\bar{t}$ production process the NNLO+NNLL theoretical calculation from Ref.~\cite{Czakon:2011xx} is used. Only statistical uncertainties are shown. 
    }
 \label{fig:njet}
 \end{center}
 \end{figure}

\begin{table}[!ht]\renewcommand{\arraystretch}{1.2}
  \begin{center}
   \begin{tabular}{l|c|c}
\hline\hline
  											& $e\mu$ 							&    $ee/\mu\mu$       				\\\hline  
$ p_{\rm T}^\ell$ (leading/sub-leading)    						&  \multicolumn{2}{c}{$>25\,/\,20$ \GeV} 									\\\hline 
$ |\eta^{\ell}|$     									&   \multicolumn{2}{c}{ $|\eta^{\mu}|<2.4$ and $|\eta^{e}|<2.47$,}   		 	\\
      											&    \multicolumn{2}{c}{ excluding $1.37<|\eta^{e}|<1.52$}					\\\hline
Number of additional leptons with									& 								&  						\\       
$p_{\rm T}>$ 7 \GeV 							& 0 								& 0 						\\\hline

$ m_{\ell\ell}$     									&    $>10$ \GeV 							& $> 15$ \GeV 					\\\hline  
$|m_{Z} - m_{\ell\ell}|$     								&    --- 							& $> 15$ \GeV 					\\\hline  
$\metrel$    										&  $>15$ \GeV 							& $>45$ \GeV 					\\\hline  
$\ptmiss$    										&  $>20$ \GeV 							& $>45$ \GeV 					\\\hline  
$\Delta \phi (\textbf{\met},\textbf{\ptmiss})$ 						&  $<0.6$ 							& $<0.3$ 					\\\hline  
Number of jets with 									& 								&  						\\       
$p_{\rm T}>$ 25 \GeV, $|\eta| < 4.5$ 							& 0 								& 0 						\\\hline  

\hline\hline
  \end{tabular}
  \end{center}
  \caption{Criteria used to select $WW$ candidate events in data.}
  \label{recolevelcuts}
\end{table}

\section{Determination of backgrounds}
\label{sec:backgrounds}

After applying all selection requirements, the resulting \WW\ candidate sample has significant background contributions from top-quark ($t \bar{t}$ and single top) production, which is the dominant background. In the $e\mu$ final state, $W+$jets production and Drell--Yan production of $\tau$-leptons have similar contributions. Drell--Yan production is much larger than $W+$jets for the same-flavour final states. Diboson ($WZ(\gamma^*)$, $ZZ$, $W\gamma$) production constitutes a smaller background contribution for all final states. 

\subsection{Background from top-quark production}\label{sec:top}

The dominant background contribution to the selected $WW$ candidate events originates from top-quark ($t\bar t $ and single top) production. Top quarks decay into a real $W$ boson and a $b$ quark, such that top-quark events contain a pair of $W$-bosons accompanied by typically two jets. Even after rejecting events with reconstructed jets with \pt$>$ 25~\GeV, a small fraction of top-quark events remains if the jets fall outside the acceptance. This small fraction however still constitutes the largest background to the selected $WW$ candidate events. Background from top-quark production is estimated using a data-driven method first suggested in Ref.~\cite{TopJVSP}, in which the top-quark contribution is extrapolated from a control region ($\mathrm{CR}$) to the signal region ($\mathrm{SR}$). The method does not rely on the possibly imperfect theoretical modelling of the low-\pt\ spectrum of jets in top-quark production, reducing significantly the uncertainty in the top-quark background estimate compared to MC-driven estimates.

The $\mathrm{CR}$ is selected by applying the $WW$ signal selection with the sole exception of the jet-veto requirement, hence the $\mathrm{SR}$ is a subsample of the $\mathrm{CR}$. The majority of events in the $\mathrm{CR}$ stem from top-quark production, while the dominant non-top-quark contribution originates from the $WW$ signal process. In order to reduce the signal contamination and to reduce the overlap between the $\mathrm{SR}$ and $\mathrm{CR}$, an additional control region, $\mathrm{CR}+H_{\mathrm{T}}$ is selected by requiring the scalar sum of the transverse momenta of leptons and jets, $H_\mathrm{T}>130$~\GeV. In the resulting sample, the signal contamination is only about 9\%, while top-quark events contribute about 90\%. The number of top-quark events in the $\mathrm{CR}$, $N^{\mathrm{top}}_{\mathrm{CR}}$, is calculated as the number of data events in the $\mathrm{CR}+H_{\mathrm{T}}$ region from which the non-top-quark contribution, estimated using MC, is subtracted, $N^{\mathrm{data}}_{\mathrm{CR}+H_{\mathrm{T}}}-N^{\mathrm{non-top}}_{\mathrm{CR}+H_\mathrm{T}}$. Then ($N^{\mathrm{data}}_{\mathrm{CR}+H_{\mathrm{T}}}-N^{\mathrm{non-top}}_{\mathrm{CR}+H_\mathrm{T}}$) is corrected for the $H_{\mathrm{T}}$ cut efficiency $\epsilon_{H_\mathrm{T}}$ estimated from top-quark MC samples. With the efficiency $\epsilon_{\text{jet-veto}}$ for top-quark events to pass the jet-veto requirement, the top-quark background contribution in the $\mathrm{SR}$ can be calculated as:

\begin{linenomath}
\begin{equation}
\label{eq:topII}
N^{\mathrm{top}}_{\mathrm{SR}} = \frac{\left(N^{\mathrm{data}}_{\mathrm{CR}+H_{\mathrm{T}}}-N^{\mathrm{non-top}}_{\mathrm{CR}+H_{\mathrm{T}}}\right)}{ \epsilon_{H_{\mathrm{T}}}} \times \epsilon_{\text{jet-veto}}.
\end{equation}
\end{linenomath}

The jet-veto efficiency $\epsilon_{\text{jet-veto}}$ is calculated as the MC efficiency $\epsilon^{\mathrm{MC}}_{\text{jet-veto}}$ multiplied by a correction factor defined in Eq.~(\ref{eq:top1a}) and obtained from events with two leptons, the same requirements on missing transverse momentum, \metrel and \ptmiss, as for the signal selection, and at least one $b$-tagged jet in the central region of the detector, $|\eta^{\mathrm{jet}}|<2.5$. This $b$-tagged sample has a high purity of top-quark events and the small contribution from non-top-quark processes is subtracted. The probability $p$ that a jet in a top-quark event fails the jet-selection requirements can be evaluated as the fraction of top-quark events that contain no jets other than the $b$-tagged jet. The correction factor takes into account the difference between $p^{\mathrm{data}}$ and $p^{\mathrm{MC}}$, and the square of the ratio of data to MC probabilities accounts for the presence of on average two $b$-jets within the acceptance for the selected top-quark events. The jet-veto efficiency can be calculated as

\begin{linenomath}
\begin{equation}
\label{eq:top1a}
\epsilon_{\text{jet-veto}} = \epsilon^{\mathrm{MC}}_{\text{jet-veto}} \times \left(\frac{p^{\mathrm{data}}}{p^{\mathrm{MC}}}\right)^2.
\end{equation}
\end{linenomath}

The systematic uncertainty in $N^{\mathrm{top}}_{\mathrm{SR}}$ in Eq. \ref{eq:topII} is studied using MC simulation. The largest contribution to the total uncertainty in the top-quark background estimate arises from the MC ratio ${\epsilon^{\mathrm{MC}}_{\text{jet-veto}}}/{(p^{\mathrm{MC}})^2}$. The uncertainty from the reconstruction of objects and events for the MC ratio is about $\pm5\%$, dominated by the systematic uncertainties in the determination of jet energy scale, jet energy resolution and $b$-tagging efficiency. The modelling uncertainty for the MC ratio is around $\pm7\%$ and the dominant contribution comes from comparing the estimates from different parton shower and hadronisation models and different generators, while the PDF uncertainty and QCD scale variations have smaller effects. Further effects have been studied, but were found to be negligible. Among these effects is the uncertainty on the fraction due to single-top production which accounts for almost 40\% of the total top-quark background contribution after the jet-veto requirement. To account for potential differences between the single-top and $t\bar{t}$ processes, additional uncertainties are assigned by scaling the single-top cross section by a conservative 30\% (measurements of this cross-section at the LHC have uncertainties just below 20\%~\cite{Aad:2015eto}). However, the resulting effect on the MC ratio and consequently on the top-quark background estimate are very small. Interference effects between $Wt$ and $t\bar{t}$ have also been considered and similarly to the variations of the single-top cross section, the impact is found to be minor. An additional cross-check is performed by changing the exponent in the correction factor $\left({p^{\mathrm{data}}}/{p^{\mathrm{MC}}}\right)^2$ to be 1.5 or 2.5, which reflects the average jet multiplicity in top-quark background events (see Figure~\ref{fig:njet}). The resulting change in the estimated yield of top-quark background is found to be less than 1\%. This indicates that the result does not strongly depend on how one assumes the correction factor should account for the two jets in the final state. The value of $\left({p^{\mathrm{data}}}/{p^{\mathrm{MC}}}\right)$ ranges between 0.982 and 1.009 with an uncertainty of $1.5$--$5\%$ for the different final states, thus indicating good modelling of top-quark events in MC simulation. The uncertainty on this ratio is propagated to the total uncertainty on the top-quark background estimate. 
Apart from the MC ratio, further terms play a role in Eq. \ref{eq:topII} and need to be studied: The $H_{\mathrm{T}}$ cut efficiency $\epsilon_{H_{\mathrm{T}}}$ is 95\% with about $\pm1\%$ uncertainty taken as the difference between the efficiencies determined in data and MC simulation. Uncertainties that range from $\pm15\%$ (diboson production) to $\pm50\%$ ($Z/W$+jets) are assigned to the subtracted non-top-quark contributions in the $\mathrm{CR}$, $N^{\mathrm{non-top}}_{\mathrm{CR}+H_\mathrm{T}}$. The systematic effect on $N^{\mathrm{top}}_{\mathrm{SR}}$ resulting from $N^{\mathrm{non-top}}_{\mathrm{CR}+H_\mathrm{T}}$ and $\epsilon_{H_{\mathrm{T}}}$ is found to be about 2\%, and the statistical uncertainties of $N^{\mathrm{data}}_{\mathrm{CR}+H_{\mathrm{T}}}$ and $p^{\mathrm{data}}$ are negligible.

While the normalisation of the top-quark background is determined from data, the shape information used in the differential measurements relies on MC modelling. The bin-by-bin uncertainties in the differential distributions are evaluated by propagating (1) the uncertainties of the jet energy scale and resolution, (2) the uncertainties determining by taking the difference in the differential distributions found with different MC generators and parton shower models and (3) the uncertainties due to the QCD scale and parton distribution functions. All these uncertainties are added in quadrature and are treated as uncorrelated with the uncertainties for the top-quark background normalisation.

\subsection{Background from $W$+jets production}
\label{sec:wj_bkg}

In this paper, the $W$+jets background contribution also includes backgrounds from \mj\ production since they are determined together as explained below. 
The determination of background from $W$+jets production relies on comparing in data the number of events with leptons satisfying either of two alternative sets of selection requirements, namely the so-called \textit{loose} (L) and \textit{tight} (T) selection criteria, where the tight sample is a subset of the loose sample. The tight selection criteria are the same as those used for the signal selection. Loose electrons are selected by relaxing some of the particle identification criteria placed on tracking variables and calorimetric shower shapes and also by removing the requirements made on the electron isolation and impact parameters in the tight selection. For loose muons, the requirements on isolation and impact parameters are removed. Leptons satisfying the tight selection criteria can originate from \textit{real }prompt leptons or \textit{fake} background leptons, which are either due to non-prompt leptons from semileptonic decays of heavy-flavour hadrons, hadrons misidentified as electrons, or photon conversions producing electrons. The same applies to leptons satisfying the loose selection criteria. The total number of events with two leptons satisfying different combinations of loose and
tight criteria is the sum of four terms:

\begin{eqnarray}\label{eq:wjets}\nonumber
N^{\mathrm{LL}} &=& N^{\mathrm{LL}}_{\mathrm{fake,fake}} + N^{\mathrm{LL}}_{\mathrm{real,fake}} + N^{\mathrm{LL}}_{\mathrm{fake,real}}  + N^{\mathrm{LL}}_{\mathrm{real,real}} \\\nonumber
N^{\mathrm{LT}} &=& \epsilon_{\mathrm{fake}} N^{\mathrm{LL}}_{\mathrm{fake,fake}} + \epsilon_{\mathrm{fake}} N^{\mathrm{LL}}_{\mathrm{real,fake}} + \epsilon_{\mathrm{real}} N^{\mathrm{LL}}_{\mathrm{fake,real}} + \epsilon_{\mathrm{real}} N^{\mathrm{LL}}_{\mathrm{real,real}} \\\nonumber
N^{\mathrm{TL}} &=& \epsilon_{\mathrm{fake}} N^{\mathrm{LL}}_{\mathrm{fake,fake}} + \epsilon_{\mathrm{real}} N^{\mathrm{LL}}_{\mathrm{real,fake}} + \epsilon_{\mathrm{fake}} N^{\mathrm{LL}}_{\mathrm{fake,real}} + \epsilon_{\mathrm{real}} N^{\mathrm{LL}}_{\mathrm{real,real}} \\
N^{\mathrm{TT}} &=& \epsilon_{\mathrm{fake}}^2 N^{\mathrm{LL}}_{\mathrm{fake,fake}} + \epsilon_{\mathrm{real}}\epsilon_{\mathrm{fake}} N^{\mathrm{LL}}_{\mathrm{real,fake}} + \epsilon_{\mathrm{fake}}\epsilon_{\mathrm{real}} N^{\mathrm{LL}}_{\mathrm{fake,real}} + \epsilon_{\mathrm{real}}^2 N^{\mathrm{LL}}_{\mathrm{real,real}}\label{eq:last}.
\end{eqnarray}

Here, the number of events that have exactly one loose lepton and one tight lepton ($N^{\mathrm{LT}}$ and $N^{\mathrm{TL}}$), two loose leptons ($N^{\mathrm{LL}}$), or two tight leptons ($N^{\mathrm{TT}}$) are used. The first and second indices correspond to the qualities of the highest-\pt\ lepton and the lepton sub-leading in \pt\ respectively; $\epsilon_{\mathrm{real}}$ and $\epsilon_{\mathrm{fake}}$ in the above formulae are the probabilities for prompt and fake background leptons selected with the loose criteria to satisfy the tight selection criteria. The sample with two tight leptons, described by Eq.~(\ref{eq:last}), consists of contributions from \mj\ events with two fake leptons, $W$+jets events with one fake and one prompt lepton and finally events with two prompt leptons including the $WW$ signal events. If the numbers of events with loose and tight leptons as well as the efficiencies $\epsilon_{\mathrm{real}}$ and $\epsilon_{\mathrm{fake}}$ are known, the numbers of events with one prompt and one fake lepton ($N^{\mathrm{LL}}_{\mathrm{real,fake}}$+$N^{\mathrm{LL}}_{\mathrm{fake,real}}$) and two fake leptons ($N^{\mathrm{LL}}_{\mathrm{fake,fake}}$) for the loose selection criteria can be obtained by solving the above system of equations. The numbers of $W$+jets and \mj\ events in the signal region, which are selected using the tight criteria, can then be extracted using the following relations:

\begin{eqnarray}\label{eq:wjetss}
  N_{W+\mathrm{jets}} & = & \epsilon_{\mathrm{real}} \epsilon_{\mathrm{fake}}   N^{\mathrm{LL}}_{\mathrm{real,fake}}   +   
  \epsilon_{\mathrm{fake}} \epsilon_{\mathrm{real}}  N^{\mathrm{LL}}_{\mathrm{fake,real}}\\
  N_{\mathrm{\mj}} & = & \epsilon_{\mathrm{fake}}^2  N^{\mathrm{LL}}_{\mathrm{fake,fake}}  
\end{eqnarray}

The efficiency for real prompt leptons, $\epsilon_{\mathrm{real}}$, is evaluated using MC simulation, where data-to-MC correction factors extracted from $Z\rightarrow\ell\ell$ events~\cite{ATLAS-CONF-2014-032,aad:2014rra} are applied. The efficiency for fake leptons, $\epsilon_{\mathrm{fake}}$, is measured using a data control region enriched with
fake leptons from \mj\ production. This control sample is selected using a lepton trigger which does not bias the loose selection. 
The sample must contain a jet that is opposite
in azimuthal angle ($\Delta\phi>$ 2.0) to a lepton satisfying the loose selection criteria to enhance the contribution of \mj\ events. The fraction of these selected loose leptons that satisfy
the tight selection criteria is $\epsilon_{\mathrm{fake}}$. Prompt leptons from $W$ and $Z$ decay contaminate this multijet sample. To remove these prompt leptons, which would bias the determination of $\epsilon_{\mathrm{fake}}$, it is required that the missing transverse momentum is small, \met$<$ 25~\GeV, and that the transverse mass of the lepton and \met\ is below 40 \GeV, $m_\mathrm{T}^W<$ 40~\GeV. Only one lepton is allowed in the event. Up to 35\% of the selected \mj\ control sample consists of prompt leptons from $W$+jets and Drell--Yan events, which are subtracted using MC simulation.

Both $\epsilon_{\mathrm{real}}$ and $\epsilon_{\mathrm{fake}}$ are determined separately for muons and electrons and also differentially as functions of \pt\ and $\eta$ of the lepton.
The main uncertainty in the fake-lepton efficiency comes from the fact that the composition of the various sources of fake leptons, e.g. heavy flavour decays, charged hadrons or conversions, might not be the same in the sample used to measure the fake-lepton efficiency as in the sample these fake-lepton efficiencies are applied to. 
The effect is estimated using a comparison between the fake-lepton efficiency predicted using the above described $W$+jets MC sample and a simulated \mj\ MC sample, generated and showered using {\sc Pythia~8}. The sample-dependence uncertainty is determined to be $\sim$30--$50\%$, depending on the lepton flavour and the event kinematics. Furthermore, systematic uncertainties from the prompt lepton subtraction and statistical uncertainties are propagated to the $W$+jets background estimate.
The total $W+$jets and \mj\ contribution to the final selected \WW candidate sample is summarised in Table~\ref{ta:selected_data_MC}.
A qualitative check of the estimated $W$+jets background and \mj\ yield is performed using events with two leptons of the same charge, as described in Section~\ref{sec:diboson} below.

The differential $W$+jets distributions needed for a differential cross-section measurement are also determined in a fully data-driven way,
by evaluating the system of linear equations Eqs.~(\ref{eq:wjets}) in each bin of the differential distributions.

\subsection{Other diboson processes and validation of diboson and $W$+jets backgrounds}\label{sec:diboson}

All backgrounds from diboson production are estimated using MC simulation. The main systematic uncertainties are due to the theoretical uncertainties of predicted cross sections used for normalisation and the description of the jet-veto requirement.

The predicted contributions for backgrounds from diboson production, $W+$jets and multijets are validated using a data control sample in which the two selected leptons are required to have the same electric charge (same-sign) and satisfy all the other selection requirements. The electron pseudorapidity is restricted to lie within $|\eta^e|<2.1$ to suppress contributions from $WW$ signal events where the electron is reconstructed with a wrong charge assignment, which become significant for the high-$|\eta^e|$ region due to the increase in material in the inner tracking detector. Since the rate of charge-misidentification is negligible for muons, they are accepted if $|\eta^\mu|<2.4$. This selection only yields a sufficient number of events for comparisons in the $e\mu$ channel. Figure~\ref{fig:SSplot} shows the \met~and $m_{\ell\ell}$ distributions for this same-sign control sample, which is dominated by $WZ(\gamma^*)$ production,  that is estimated using the simulated MC samples described above, and $W+$jets events that are estimated from data, as described in Section~\ref{sec:wj_bkg}. Both statistical and systematic uncertainties are shown for the $W$+jets estimate. For the diboson samples the theoretical uncertainty in the cross-section predictions are included but their experimental uncertainties have not been evaluated in this control region. The predictions and the data agree well. 

  \begin{figure}[!ht]
  \begin{center}
  \includegraphics[width=0.49\textwidth]{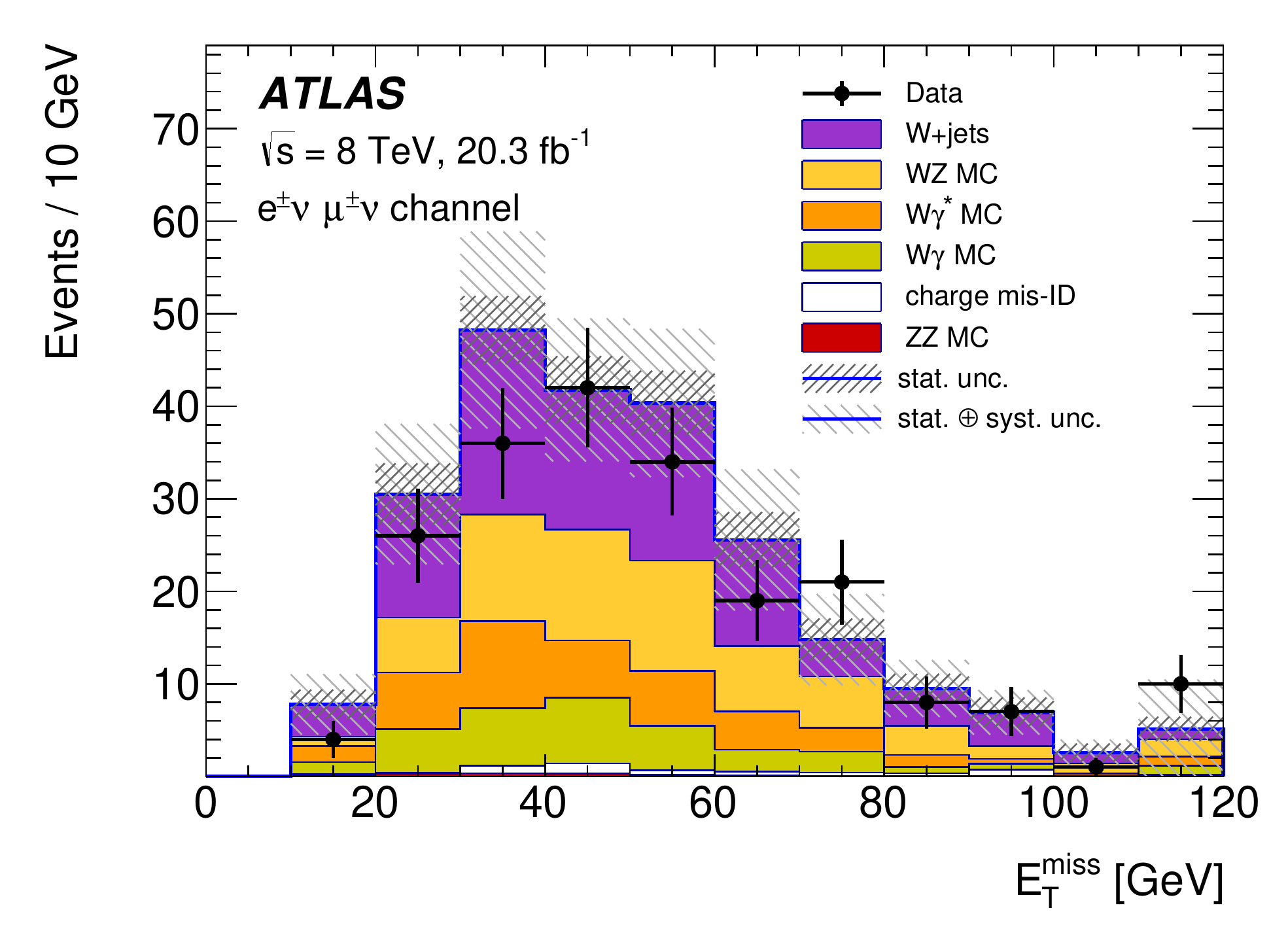}\hfill
  \includegraphics[width=0.49\textwidth]{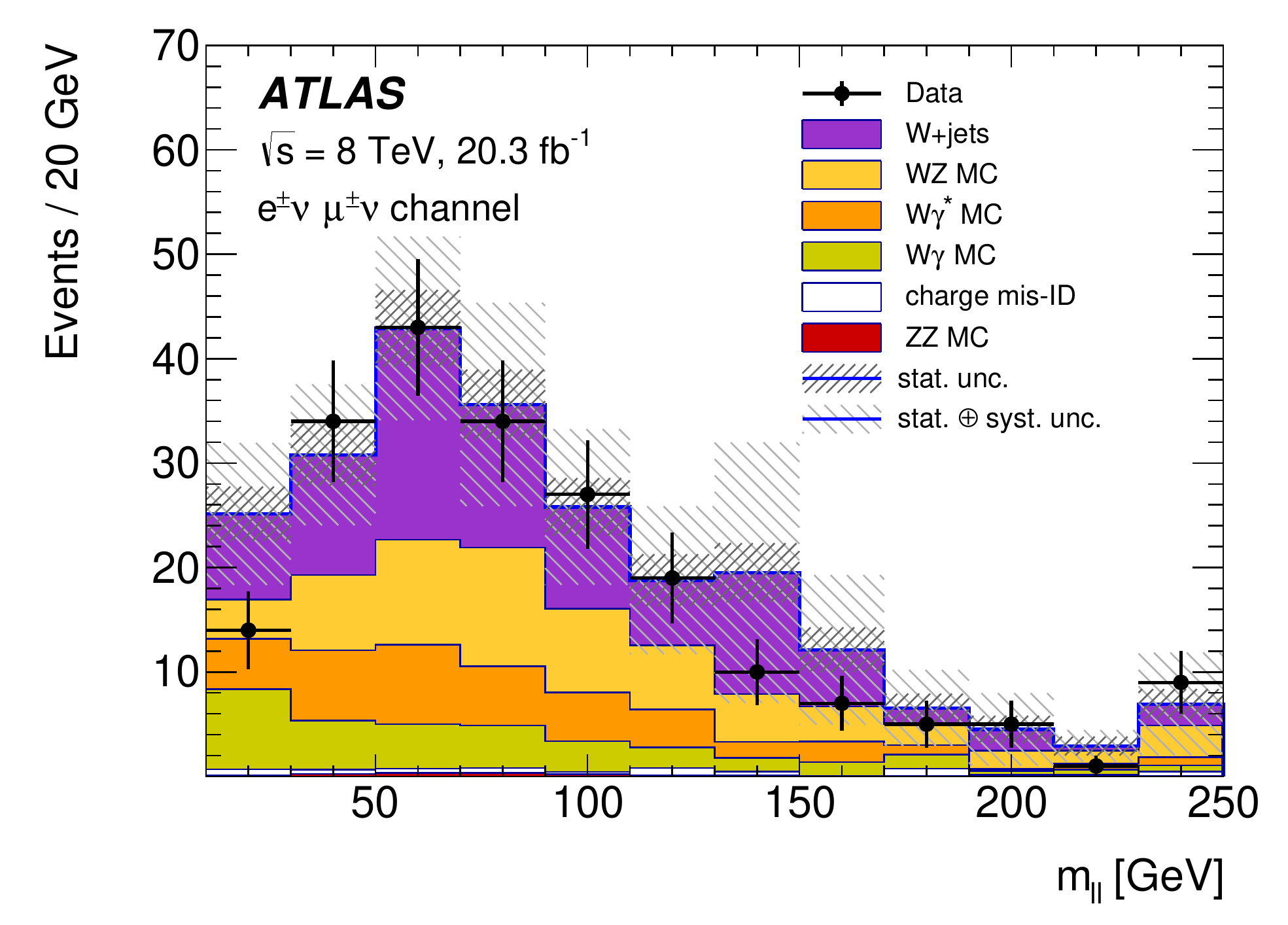}

  \caption{Distributions of \met and $m_{\ell\ell}$ for the same-sign control sample in the $e\mu$ channel. The last bin is an overflow bin. The selected leptons are required to have the same electric charge. 
  The uncertainties shown include statistical and systematic uncertainties in the $W$+jets estimate as well as statistical uncertainties in all MC predictions. For the diboson contributions, the theoretical uncertainties in the cross-section predictions are also included. The experimental systematic uncertainties for the diboson production processes are not included. Contributions from processes with two opposite-sign final-state leptons, where one of them is reconstructed with a wrong charge assignment, are denoted by ``charge mis-ID''. 
  }
  \label{fig:SSplot}
  \end{center}
  \end{figure}

\subsection{Background from Drell--Yan production}

The Drell--Yan background normalisation is constrained by an auxiliary fit. It is based on a profile likelihood approach
where the numbers of signal and background events in signal and control regions are described by a Poisson probability density function.
For the $W$+jets and multijet backgrounds, the normalisation and shape from the data-driven estimates described above are used.
Similarly, the top-quark and diboson contributions are obtained as described above.
The input template shapes for signal and Drell--Yan events are obtained from MC simulation.
The different sources of experimental and theoretical systematic uncertainties are included as nuisance parameters in the fit. Both the $WW$ signal and Drell--Yan normalisation factors are fitted, but only the Drell--Yan background normalisation is used further for the cross-section determination. The $WW$ cross section measured in Section~\ref{sec:results} is fully compatible with the $WW$ normalisation factor extracted here.

For the fit, a control region dominated by Drell--Yan events is defined
by inverting the \ptmiss requirement of $5< \ptmiss < 20 (45)$ \GeV\ for the $e\mu$ $(ee{+}\mu\mu)$ final states, where the minimum requirement of 5 \GeV\ ensures that there is a well-defined \ptmiss\ direction, and removing the $\Delta \phi (\textbf{\met},\textbf{\ptmiss})$ requirement.
The fit is performed on the $\Delta \phi (\textbf{\met},\textbf{\ptmiss})$ distribution in five bins of equal size for both the control region and the signal region simultaneously.
In addition, a validation region dominated by Drell--Yan events is defined by inverting both the calorimetric and the track-based missing
transverse momentum requirements but keeping the requirement on $\Delta \phi (\textbf{\met},\textbf{\ptmiss})$. The result of the fit is extrapolated to this validation region where good data--MC agreement is observed.

In addition to the experimental uncertainties,  theoretical uncertainties (QCD scale, PDF, parton-shower modelling in the simulation) are considered. For the uncertainties in the differential distributions of background events from Drell--Yan production, the constraints on the nuisance parameters from the likelihood fit are used. This information is propagated to the MC simulation, and predictions for Drell--Yan events are extracted for each bin with their uncertainties.

The largest uncertainties arise from the description of the jet and \met\ energy scale and resolution in the MC simulation and from the MC parton shower modelling. 
The latter is estimated by the difference between using the \herwig/\Jimmy and \pythia approaches in the MC simulation.

Figure~\ref{fig:Zjetsplot} shows the $\Delta \phi (\textbf{\met},\textbf{\ptmiss})$ distributions for $e\mu$ final states in the control and validation regions before and after the profile likelihood fit of the Drell--Yan background. 
Good agreement between the data and the post-fit prediction is seen.

  \begin{figure}[!ht]
  \begin{center}
  \includegraphics[width=0.49\textwidth]{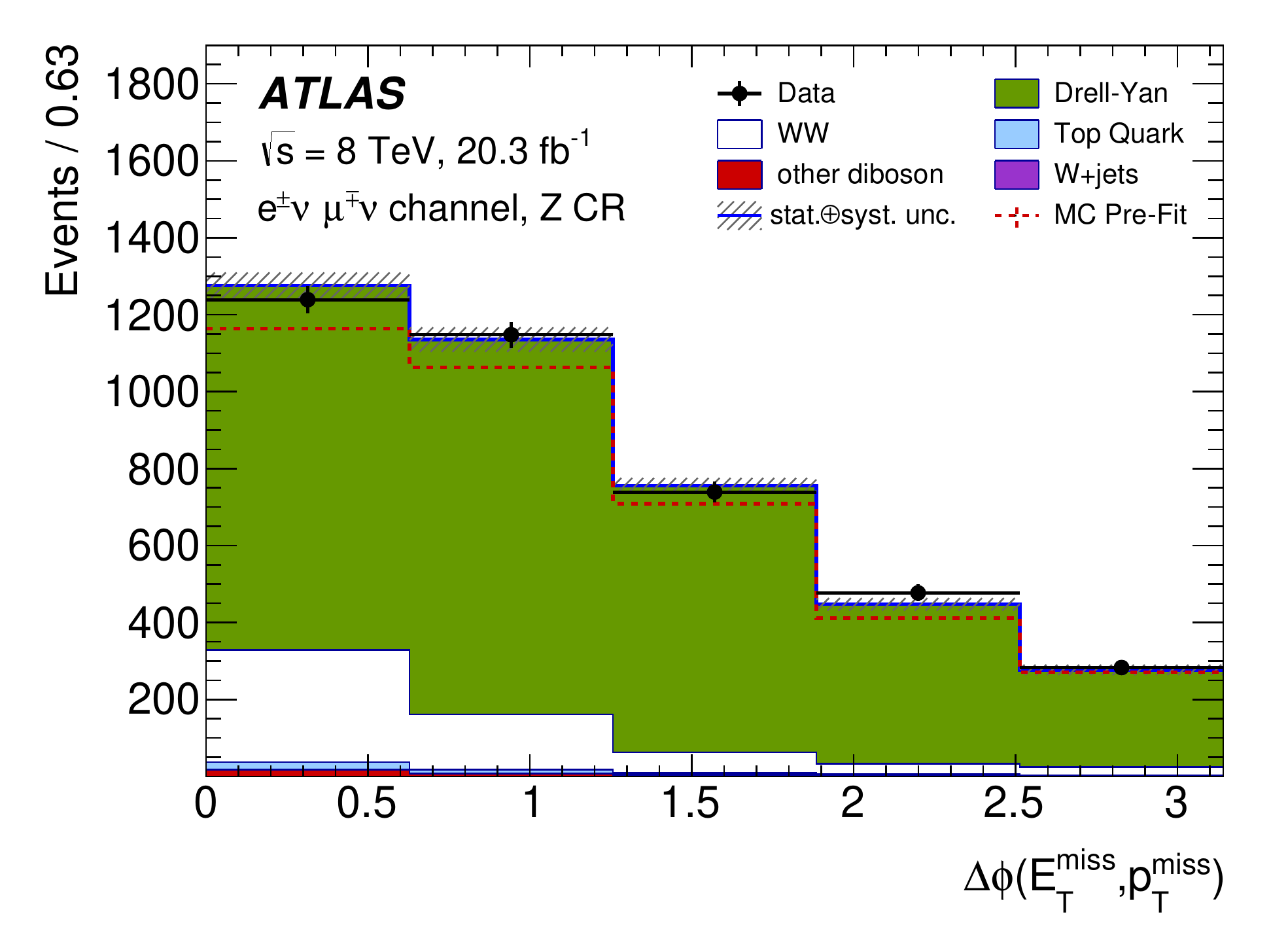}\hfill
  \includegraphics[width=0.49\textwidth]{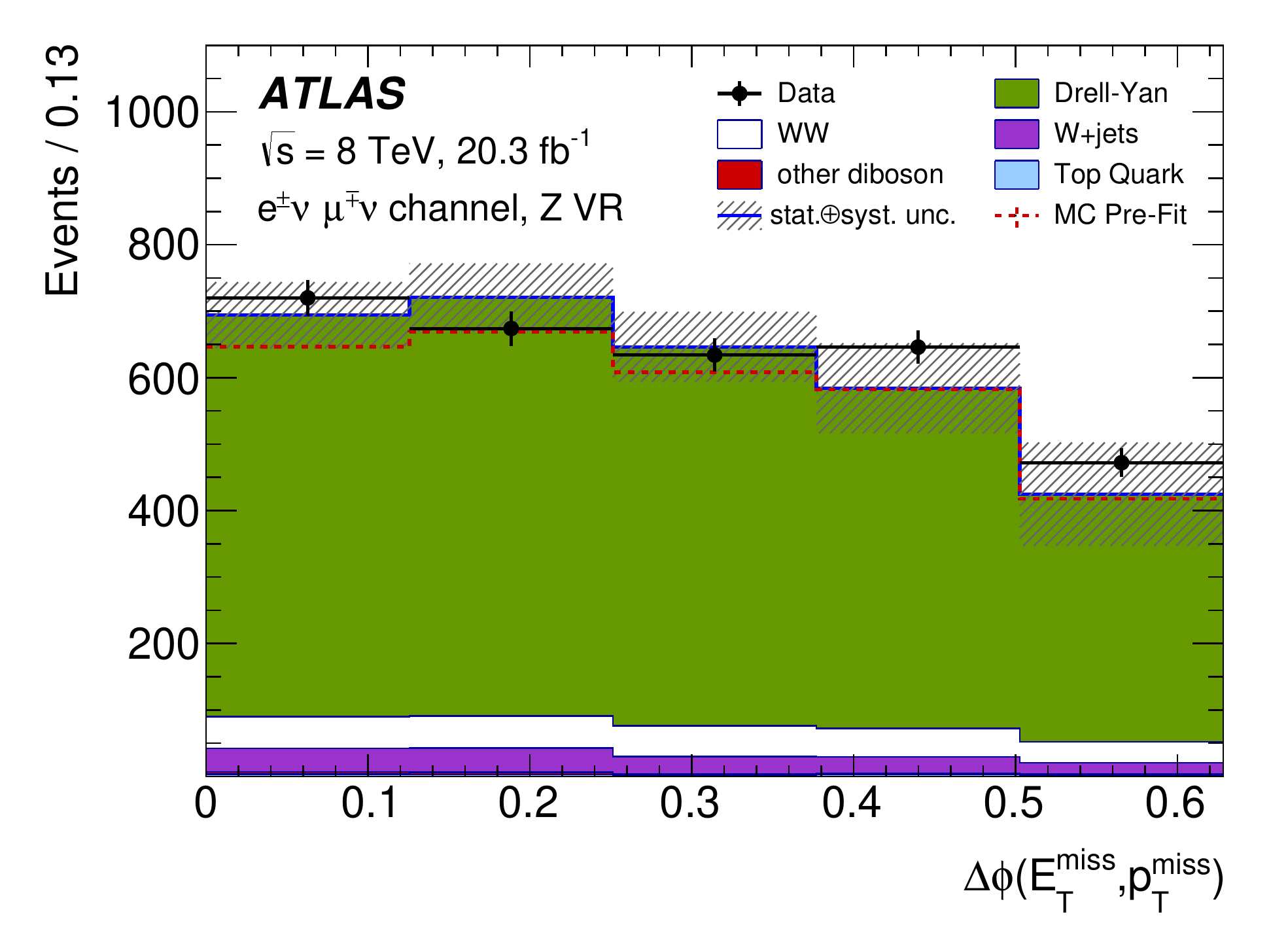}\hfill

  \caption{Distributions of $\Delta \phi (\textbf{\met},\textbf{\ptmiss})$ are shown for data and MC predictions for the Drell--Yan control region (left) and the validation region (right). The MC predictions for Drell--Yan and $WW$ signal have been scaled from the pre-fit predictions to reflect the results of the fit. The fit improves the description of the data by the simulated Drell--Yan events as compared to the MC pre-fit prediction.}
  \label{fig:Zjetsplot}
  \end{center}
  \end{figure}

\subsection{Other background contributions}
\label{ref:bkg_neglected}

The background contributed by $WW$ pairs from vector-boson scattering, Higgs boson production via vector-boson fusion, as well as $WH$ and $ZH$
associated production with $H\rightarrow WW$ is evaluated using MC
simulation. The sum of these processes contributes typically 
$0.3$--$0.6\%$ to each final state at detector level for the selection considered in this analysis and is therefore neglected. The contributions from these processes are neither subtracted as backgrounds nor included explicitly as signal in the calculation of the cross section.

The background contributed by $W$ pair production in double parton interactions is evaluated using a \Pythia~8 MC sample scaled to a theoretical cross section obtained by combining the NNLO prediction for single $W$ boson production and the measured effective-area parameter for double parton interactions~\cite{Aad:2013bjm}. The contribution in the signal region is found to be around 0.3\%. To increase the impact of double parton interactions on the dominant $e\mu$ channel beyond the percent level would require an increase of the effective cross section by more than ten times its uncertainty. This background contribution is neglected. 

\subsection{$WW$ candidate events and estimated background yields}
\label{ref:selected_data_MC}

The data event yields and the estimated background contributions are summarised in Table~\ref{ta:selected_data_MC}. The MC predicts that 93\% of all signal events selected in the sample are produced via the $q\bar{q}\rightarrow W^+W^-$ process, while 4\% stem from non-resonant $gg \rightarrow W^+W^-$ and 3\% from resonant $H\rightarrow WW$ production. Kinematic distributions comparing the selected data to the signal and backgrounds are shown in~Figures~\ref{fig:pre_3} and~\ref{fig:pre_1}. The $W$+jets and \mj\ backgrounds are determined using fully data-driven methods, while for top-quark and Drell--Yan production the normalisation is determined from data, but their differential shapes are taken from MC predictions. The diboson background and the \WW signal are taken from MC simulation. 

The signal contribution is normalised to the integrated luminosity using the nNLO cross-section prediction, which is defined in Section~\ref{sec:crosssections}.
The transverse momentum of the leading lepton, $\pt^{\mathrm{lead}}$,  invariant mass of the dilepton system, $m_{\ell\ell}$, and its transverse momentum, $\pt(\ell\ell)$, the difference in azimuthal angle between the decay leptons, $\Delta \phi_{\ell\ell}$, their combined rapidity, \absyll, as well as the observable \abscostheta, defined in Eq.~(\ref{eq:costheta}), are shown. For the same-flavour final states in Figure~\ref{fig:pre_1}, a discontinuity in the 
distribution of the invariant mass of the dilepton system, $m_{\ell\ell}$, is visible due to the rejection of events that are reconstructed with an invariant mass close to the $Z$ boson mass $m_{Z}$. For all distributions, an excess of the data over the signal and background is observed, and this is discussed in more detail in Section~\ref{sec:results}.

\begin{table}[!ht]
  \centering
  \sisetup{retain-explicit-plus,round-mode = places,group-separator = ., table-format = 9.9,
    table-number-alignment = right, round-mode=figures }
  {\small
    \begin{tabular}{llll}
      \hline \hline
      Final state     		& \multicolumn{1}{c}{\text{$e\mu$ }}           													& \multicolumn{1}{c}{\text{$ee$}}                															& \multicolumn{1}{c}{\text{$\mu\mu$ }} 																			\\
      \hline          		                                                                                                                                		                                                                                                                                                			                                                                                                                        								
      Observed events 		& \multicolumn{1}{l}{5067} 															& \multicolumn{1}{l}{594} 																		& \multicolumn{1}{l}{975} 																				\\
      \hline         
      \\
      Total expected events 	
      & \num[round-precision=3]{4419.7}\hphantom{0} 		$\pm$ 
      \hphantom{0}\num[round-precision=1]{25.9}\hphantom{0}	$\pm$ \hphantom{0}\num[round-precision=2]{319.2}\hphantom{.0}	
      
      & \num[round-precision=3]{507.3}\hphantom{0} 		$\pm$ 
      \hphantom{0}\num[round-precision=1]{9.4}\hphantom{0}	$\pm$ \hphantom{0}\num[round-precision=2]{39.2}\hphantom{.0} 
      
      & \num[round-precision=3]{817.2}\hphantom{0} 		$\pm$ 
      \hphantom{0}\num[round-precision=2]{11.6}\hphantom{0} 	$\pm$ \hphantom{0}\num[round-precision=2]{64.67}\hphantom{.0}					\\
      (Signal + background)   & 	& 	& \\                                                                                                                                                                                                                                                                                                       
      \hline                                                                                                                                                                                                                                                                                                                                                                                                                                                                                
	\\
      \WW\ signal (MC)	
      & \num[round-precision=3]{3238.1}\hphantom{0} 		$\pm$ 
      \hphantom{0}\num[round-precision=1]{10.2}\hphantom{0}	$\pm$ \hphantom{0}\num[round-precision=2]{284.0} \hphantom{.0}
      & \num[round-precision=3]{346.3}\hphantom{0}		$\pm$ 
      \hphantom{0}\num[round-precision=1]{3.3}\hphantom{0} 	$\pm$ \hphantom{0}\num[round-precision=2]{33.4}\hphantom{.0} 		
      & \num[round-precision=3]{612.5}\hphantom{0} 		$\pm$ 
      \hphantom{00}\num[round-precision=1]{4.5}\hphantom{0} 	$\pm$ \hphantom{0}\num[round-precision=2]{59.9}\hphantom{.0} 			\\
      \hline                                                                                                                                               	    \\                                                                                                                                                                                                                                                                                                                                           
      Top quark (data-driven) 	
      & \hphantom{0}\num[round-precision=3]{608.6}\hphantom{0} 	$\pm$ 
      \hphantom{0}\num[round-precision=2]{17.5}\hphantom{0}	$\pm$ \hphantom{00}\num[round-precision=2]{52.3} \hphantom{.0}
      
      & \hphantom{0}\num[round-precision=2]{91.8}\hphantom{0} 	$\pm$ 
      \hphantom{0}\num[round-precision=1]{7.3}\hphantom{0}	$\pm$ \hphantom{00}\num[round-precision=1]{7.9} 
      
      & \num[round-precision=3]{127.2}\hphantom{0} 		$\pm$ 
      \hphantom{00}\num[round-precision=1]{9.4}\hphantom{0} 	$\pm$ \hphantom{0}\num[round-precision=2]{10.9}\hphantom{.0} \\
      
      $W$+jets (data-driven) 	
      & \hphantom{0}\num[round-precision=2]{248.8}\hphantom{0}	$\pm$ 
      \hphantom{0}\num[round-precision=1]{15.3}\hphantom{0}	$\pm$ \hphantom{0}\num[round-precision=2]{138.7} \hphantom{.0}
      
      & \hphantom{0}\num[round-precision=2]{13.9}\hphantom{0} 	$\pm$ 
      \hphantom{0}\num[round-precision=1]{4.9}\hphantom{0}	$\pm$ \hphantom{0}\num[round-precision=2]{14.2}\hphantom{.0} 		
      & \hphantom{00}\num[round-precision=1]{6.1}\hphantom{0}   $\pm$ \hphantom{00}\num[round-precision=1]{5.0}\hphantom{0} $\pm$ \hphantom{0}\num[round-precision=2]{11.5}\hphantom{.0} \\
      
      Drell--Yan (data-driven)	
      & \hphantom{0}\num[round-precision=3]{174.5}\hphantom{0} 	$\pm$ 
      \hphantom{00}\num[round-precision=1]{3.4}\hphantom{0}	$\pm$ \hphantom{00}\num[round-precision=2]{17.7} \hphantom{.0}
      & \hphantom{0}\num[round-precision=2]{28.0}\hphantom{0} 	$\pm$ 
      \hphantom{0}\num[round-precision=0]{0.5}\hphantom{0}	$\pm$ \hphantom{0}\num[round-precision=2]{13.0}\hphantom{.0} 
      & \hphantom{0}\num[round-precision=2]{33.0}\hphantom{0} 	$\pm$ \hphantom{00}\num[round-precision=0]{0.5}\hphantom{0} $\pm$ \hphantom{0}\num[round-precision=2]{17.4}\hphantom{.0} 	\\	
      
      Other dibosons (MC) 	
      & \hphantom{0}\num[round-precision=3]{149.7}\hphantom{0} 	$\pm$ 
      \hphantom{00}\num[round-precision=1]{4.0}\hphantom{0}	$\pm$ \hphantom{00}\num[round-precision=2]{29.5} \hphantom{.0}
      & \hphantom{0}\num[round-precision=2]{27.3}\hphantom{0} 	$\pm$ 
      \hphantom{0}\num[round-precision=1]{1.4}\hphantom{0}	$\pm$ \hphantom{00}\num[round-precision=1]{4.9} 			
      & \hphantom{0}\num[round-precision=2]{38.4}\hphantom{0} 	$\pm$ 
      \hphantom{00}\num[round-precision=1]{1.3}\hphantom{0} 	$\pm$ \hphantom{00}\num[round-precision=1]{5.4} 		\\
      \hline             
      \\                                                                                                                                                       
      Total background 		
      & \num[round-precision=3]{1181.6}\hphantom{0} 		$\pm$ 
      \hphantom{0}\num[round-precision=1]{27.8}\hphantom{0}	$\pm$ \hphantom{0}\num[round-precision=2]{152.2} \hphantom{.0}
      & \num[round-precision=3]{161.0}\hphantom{0}  		$\pm$ 
      \hphantom{0}\num[round-precision=1]{8.8}\hphantom{0}	$\pm$ \hphantom{0}\num[round-precision=2]{21.0}\hphantom{.0} 
      & \num[round-precision=3]{204.7}\hphantom{0} 		$\pm$ 
      \hphantom{0}\num[round-precision=2]{10.7}\hphantom{0} 	$\pm$ \hphantom{0}\num[round-precision=2]{24.2}\hphantom{.0} 				\\
      \hline \hline
    \end{tabular}
  }
  \caption{\small Summary of observed events and expected signal and background contributions in three dilepton channels. The first uncertainty is statistical, the second one corresponds to the systematic uncertainty and includes the uncertainty due to the integrated luminosity (where used in the normalisation). The systematic uncertainties in the total background and total expectation are calculated as the sum in quadrature of the uncertainties of the individual components. The MC simulation of the $WW$ signal predicts that 93\% of the events are produced via the $q\bar{q}\rightarrow W^+W^-$ process, while 4\% stem from non-resonant $gg\rightarrow W^+W^-$ and 3\% from resonant $H\rightarrow WW$ production.
  }
  \label{ta:selected_data_MC}
\end{table}

\begin{figure}[p]
    \begin{center}
    \includegraphics[width=0.49\textwidth]{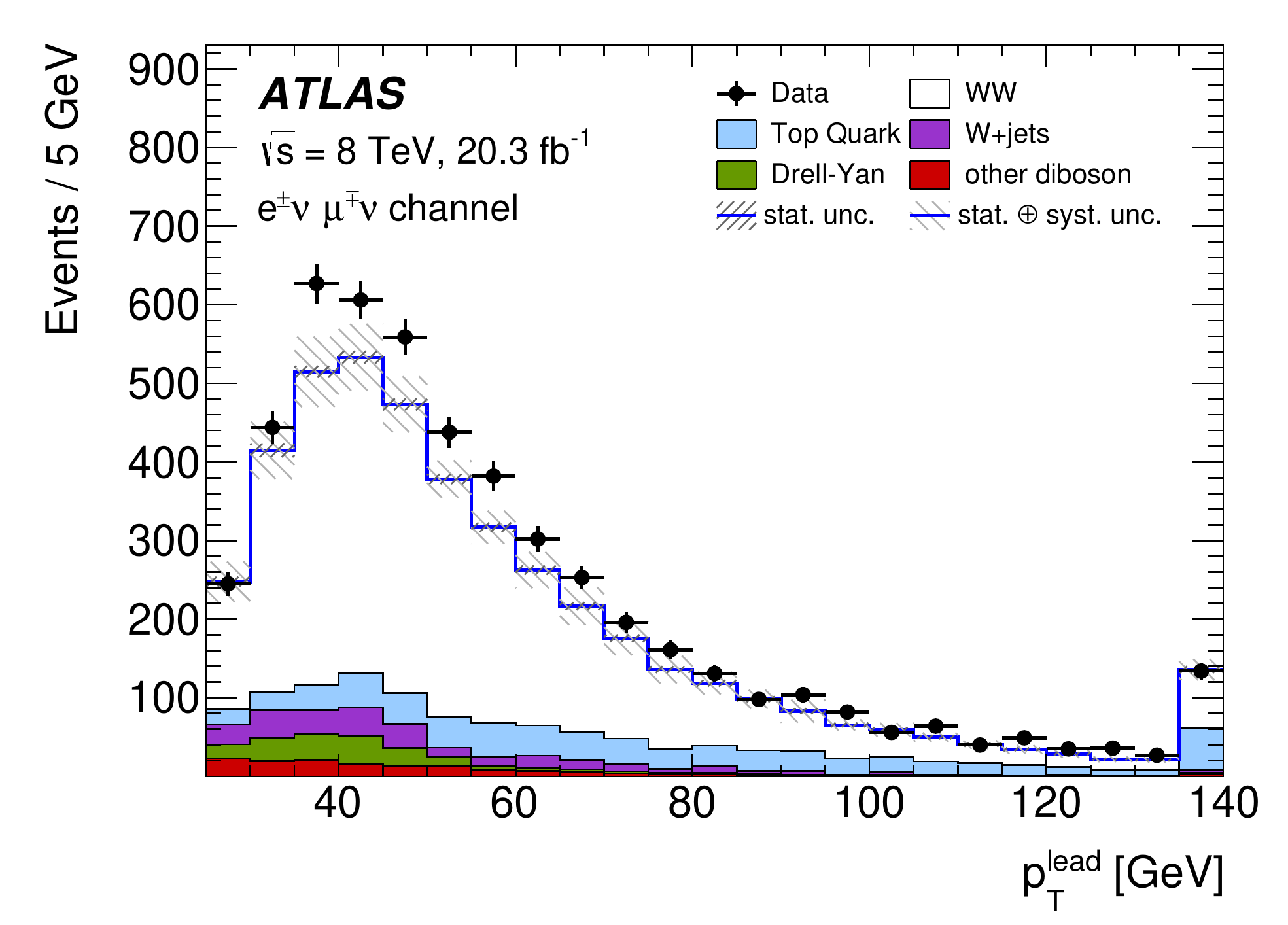}\hfill
    \includegraphics[width=0.49\textwidth]{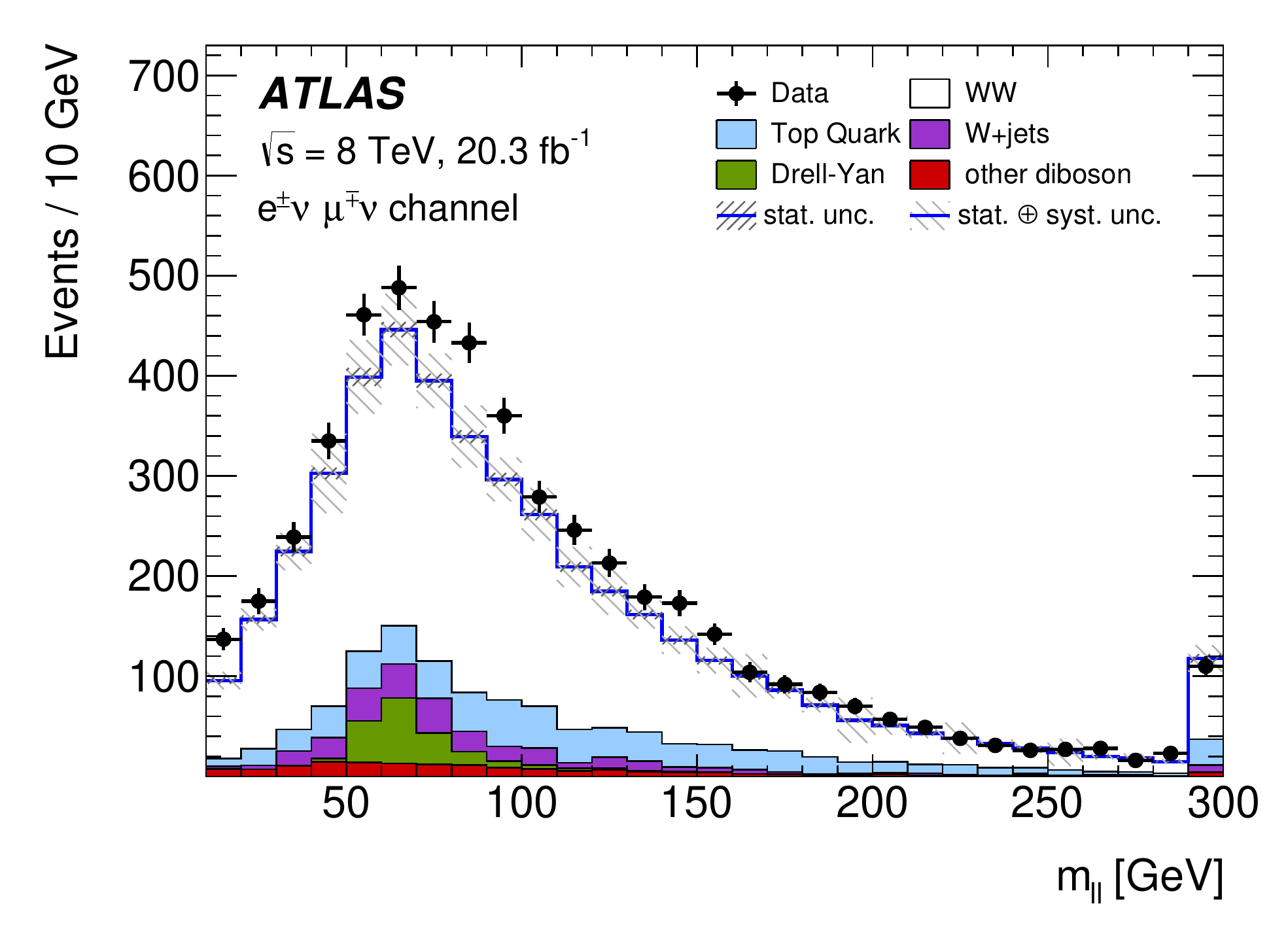}

    \includegraphics[width=0.49\textwidth]{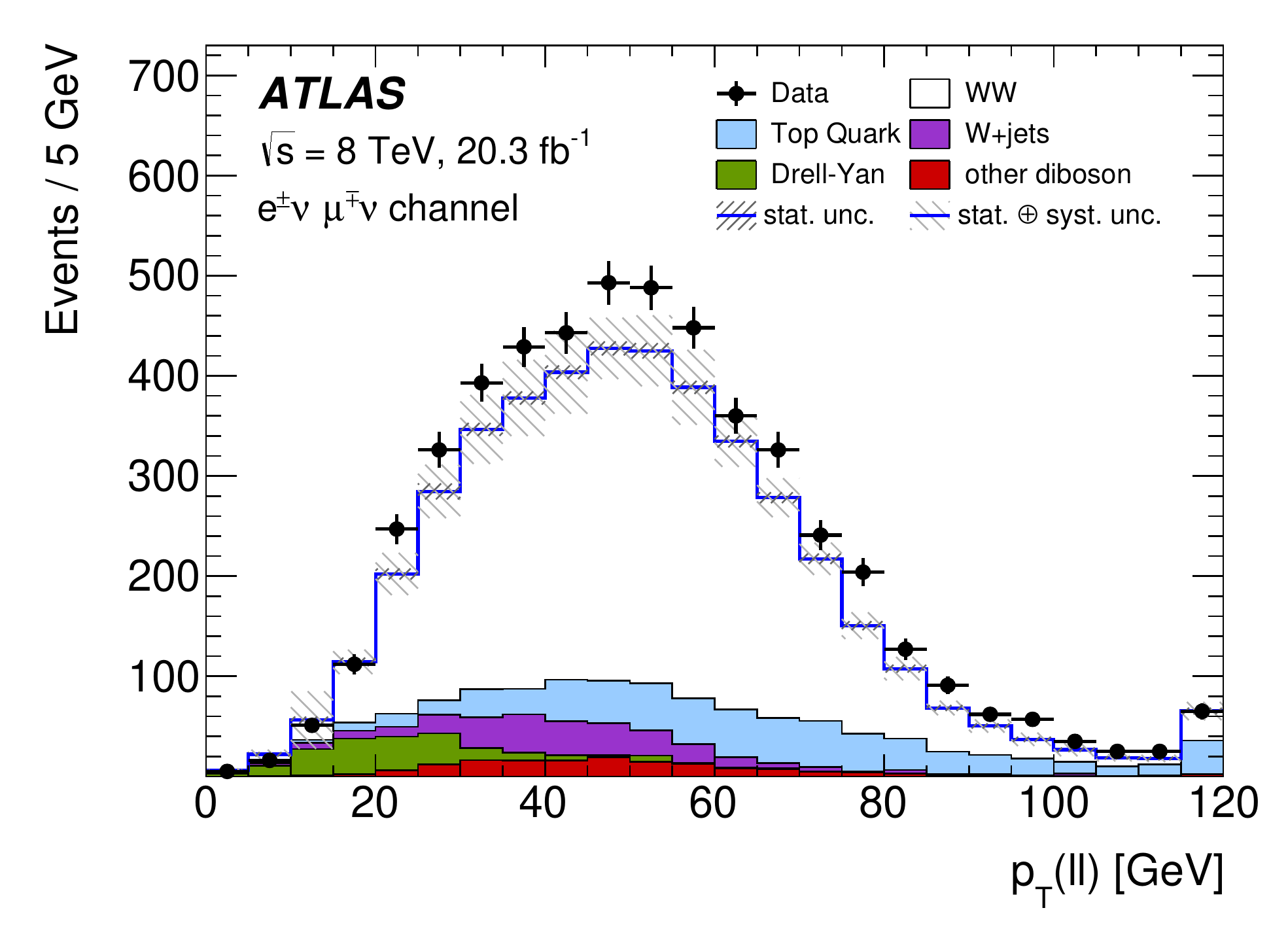}\hfill
    \includegraphics[width=0.49\textwidth]{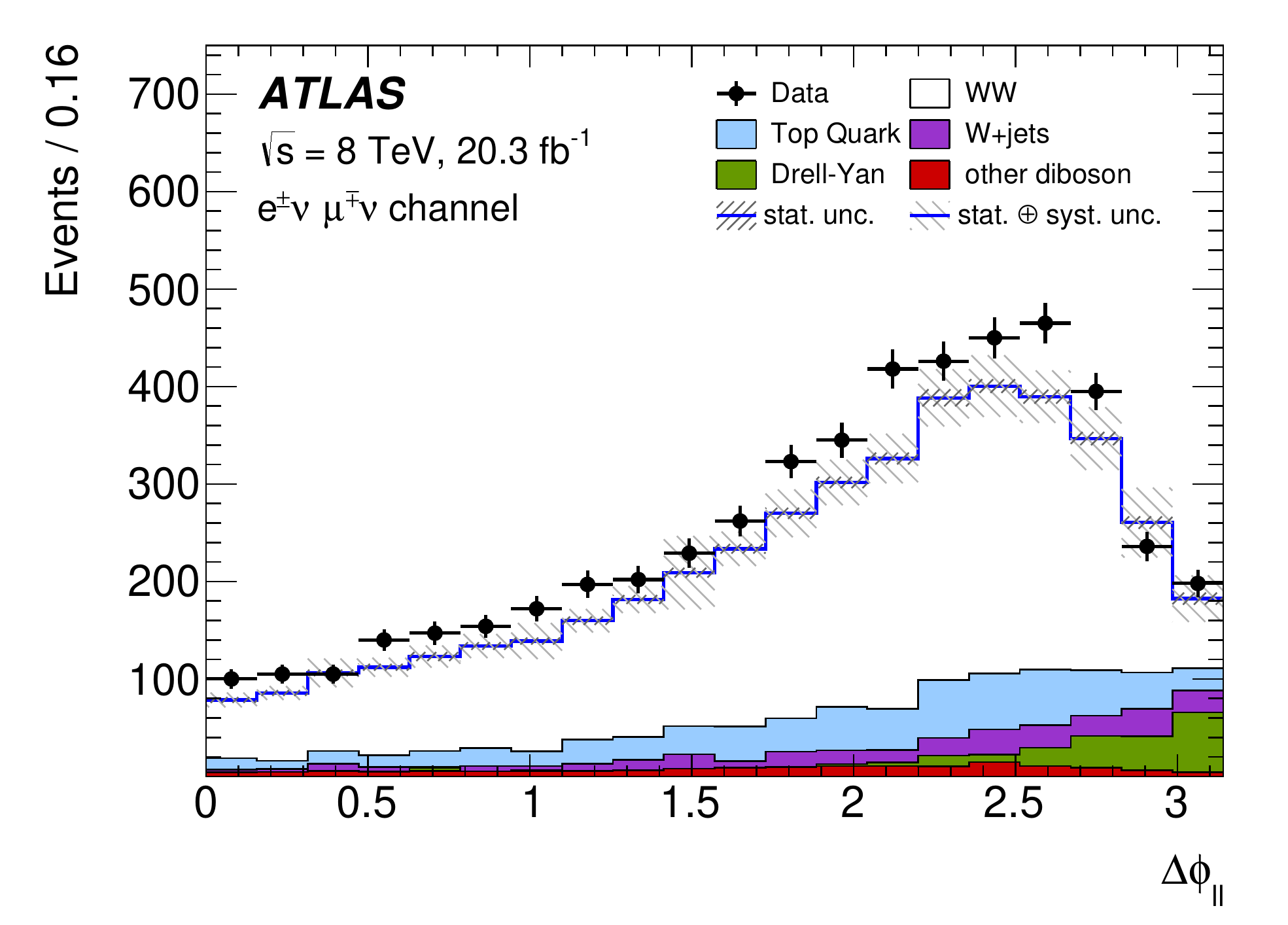}

   \includegraphics[width=0.49\textwidth]{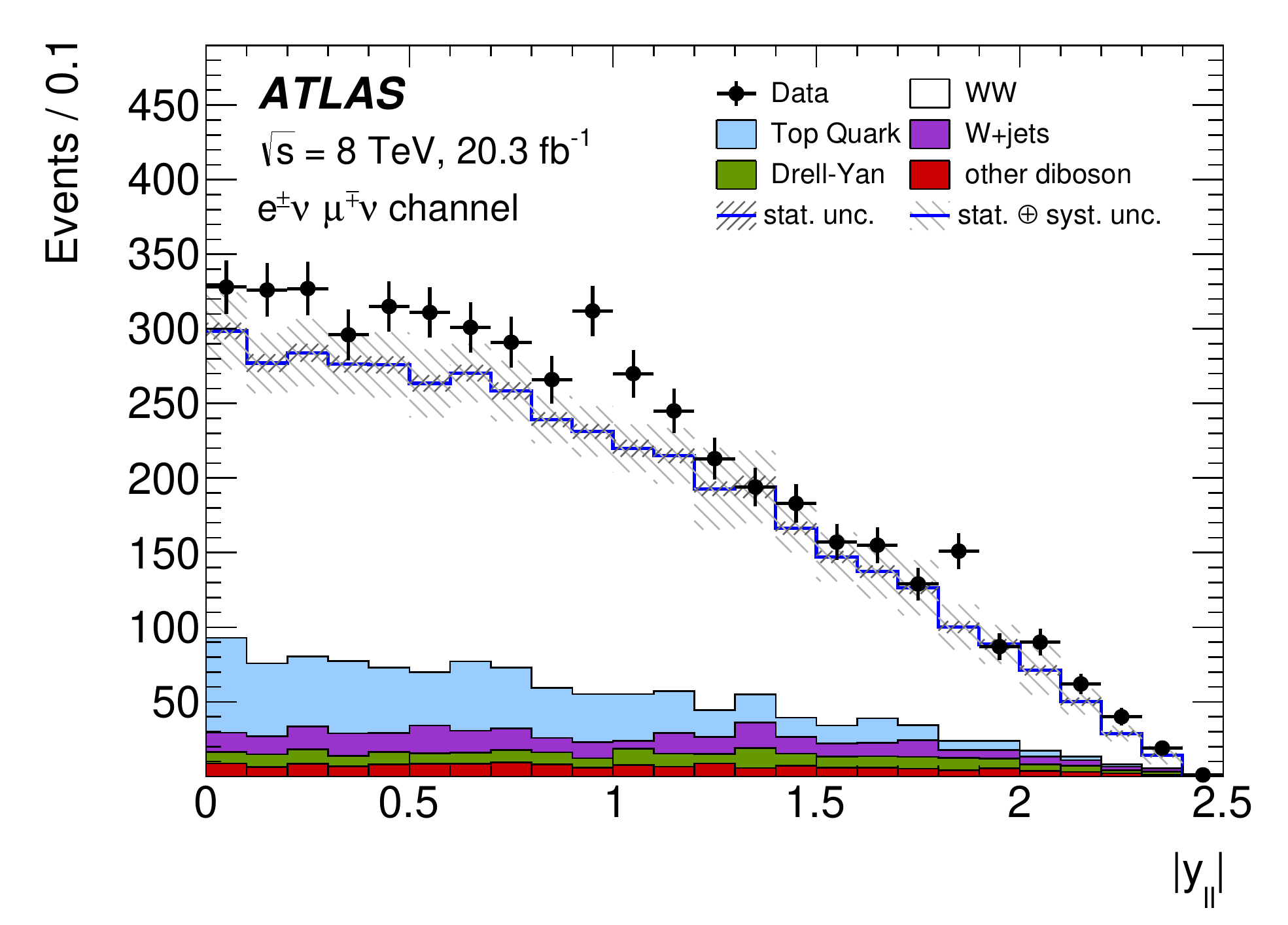}\hfill
    \includegraphics[width=0.49\textwidth]{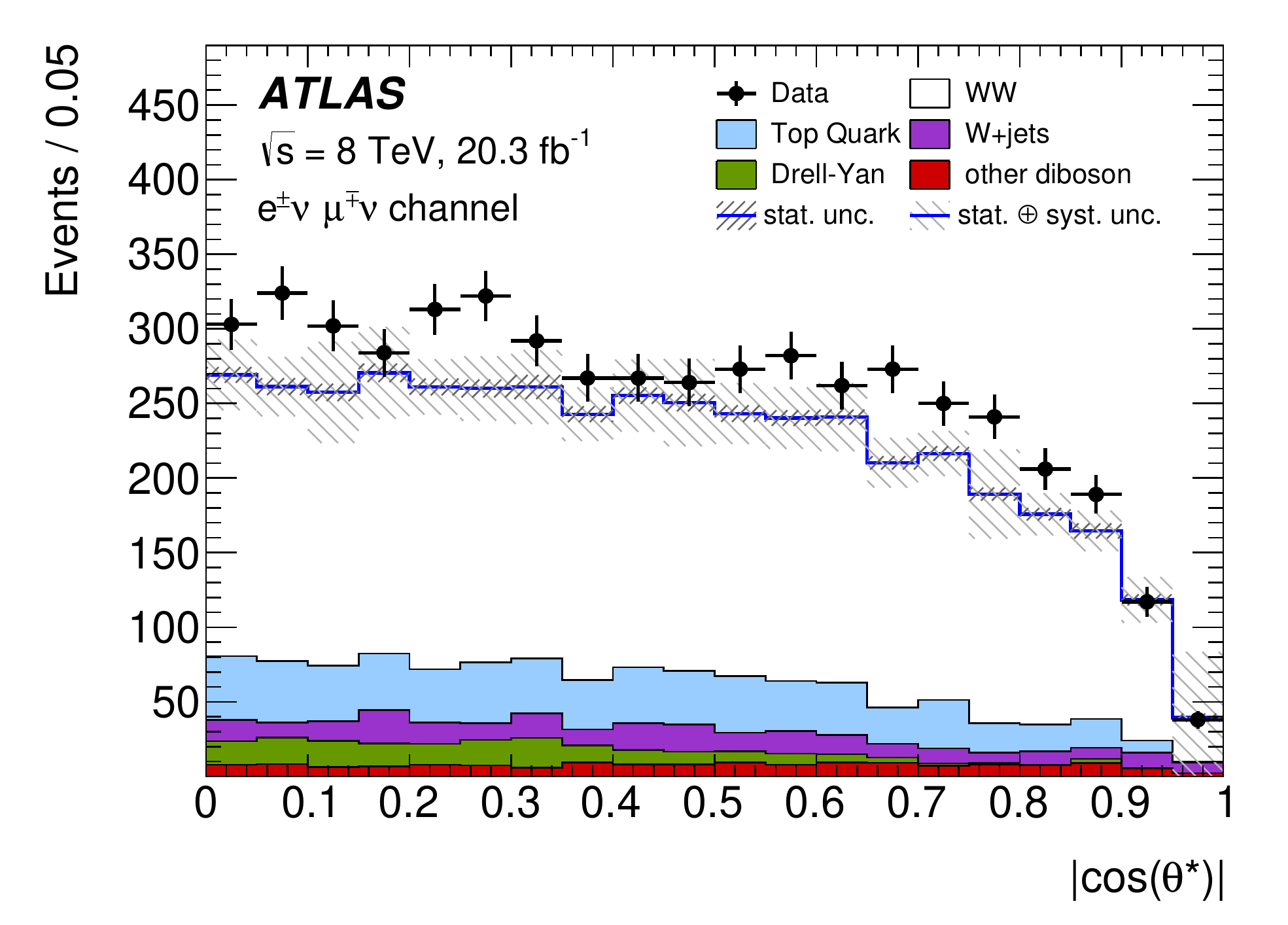}

\caption{Kinematic distributions of the selected data events after the full event selection for the $e\mu$ final state. Data are shown together with the predictions of the signal and background production processes. The transverse momentum  of the leading lepton, $\pt^{\mathrm{lead}}$, the invariant mass, $m_{\ell\ell}$, and the transverse momentum of the dilepton system, $\pt(\ell\ell)$, as well as the difference in azimuthal angle between the decay leptons, $\Delta \phi_{\ell\ell}$, the dilepton rapidity, \absyll, and the observable \abscostheta\ are shown (from left to right and top to bottom). The last bin of the $\pt^{\mathrm{lead}}$,  $m_{\ell\ell}$ and $\pt(\ell\ell)$ distributions is an overflow bin. Statistical and systematic uncertainties in the predictions are shown as bands in hatched style. }

    \label{fig:pre_3}
    \end{center}
\end{figure}
\begin{figure}[p]
    \begin{center}
    \includegraphics[width=0.49\textwidth]{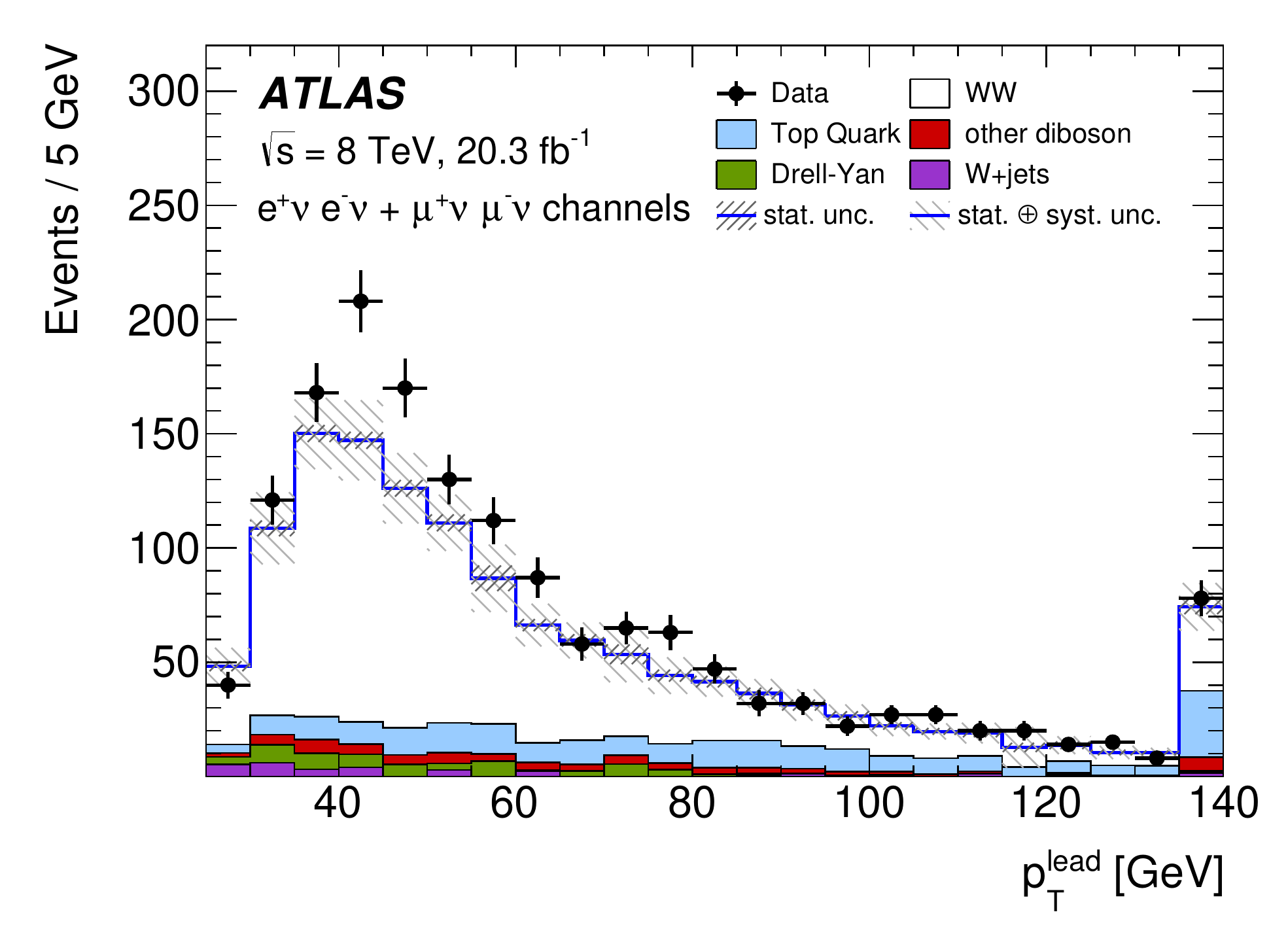}\hfill
    \includegraphics[width=0.49\textwidth]{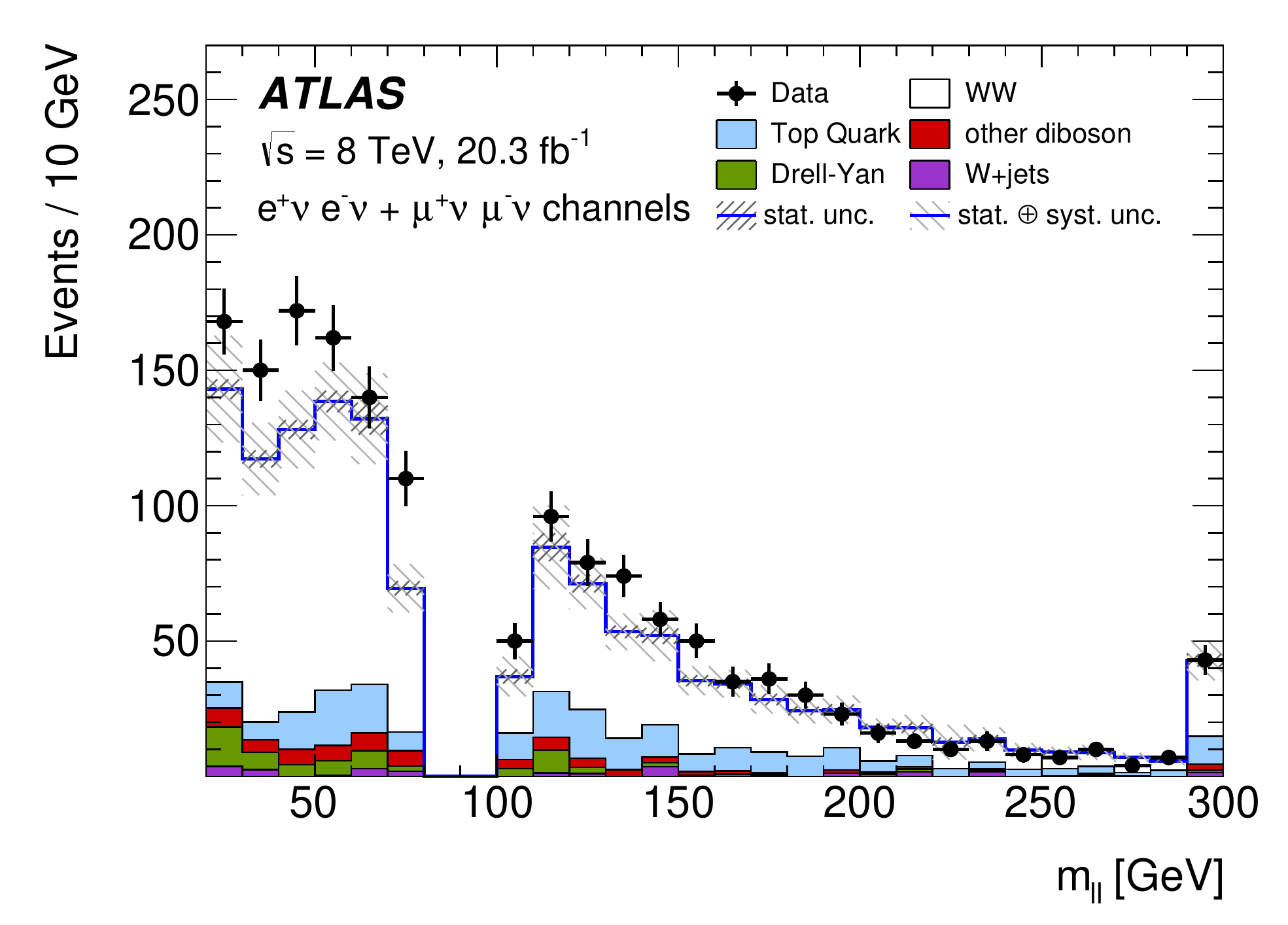}

    \includegraphics[width=0.49\textwidth]{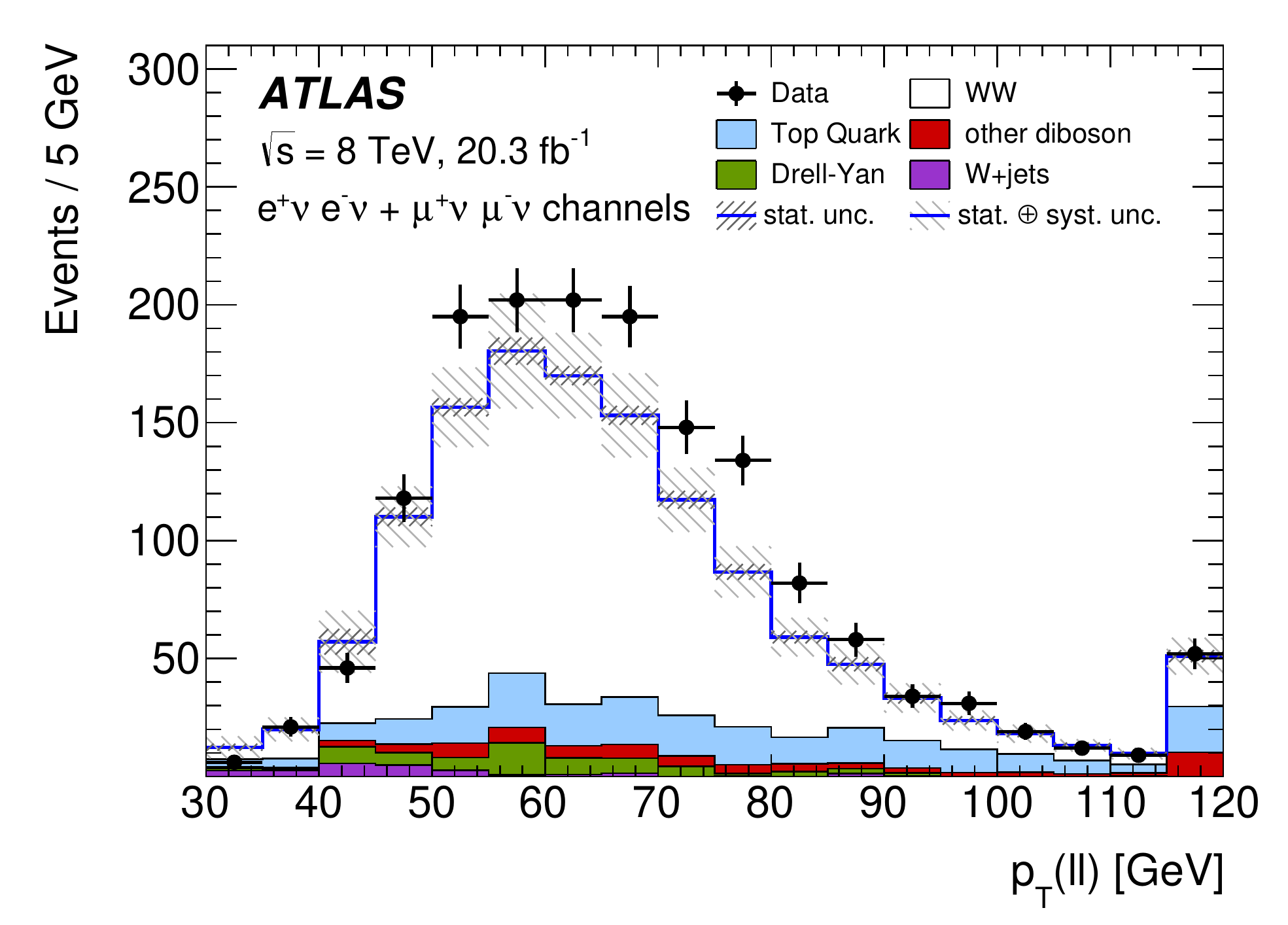}\hfill
    \includegraphics[width=0.49\textwidth]{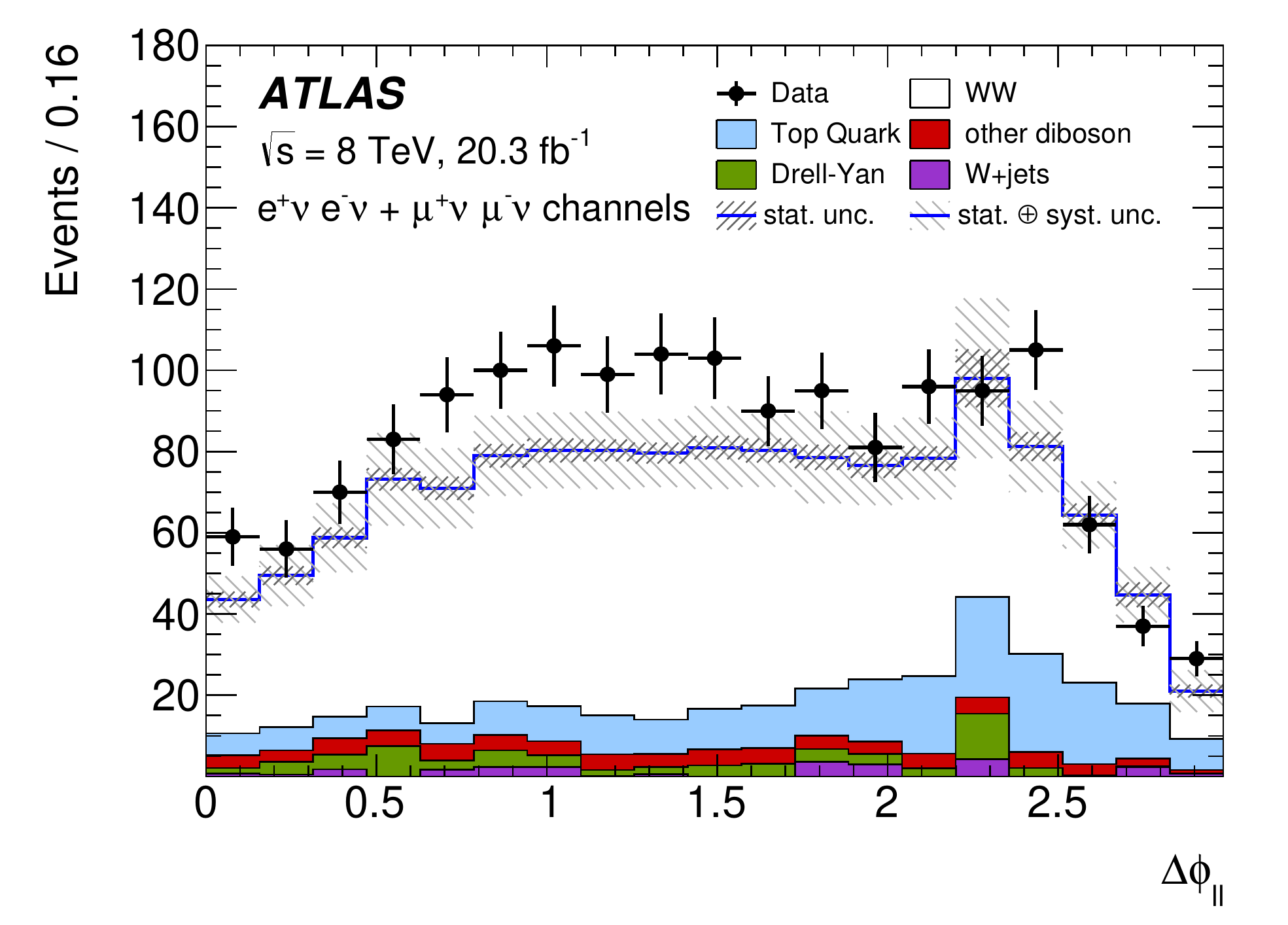}

   \includegraphics[width=0.49\textwidth]{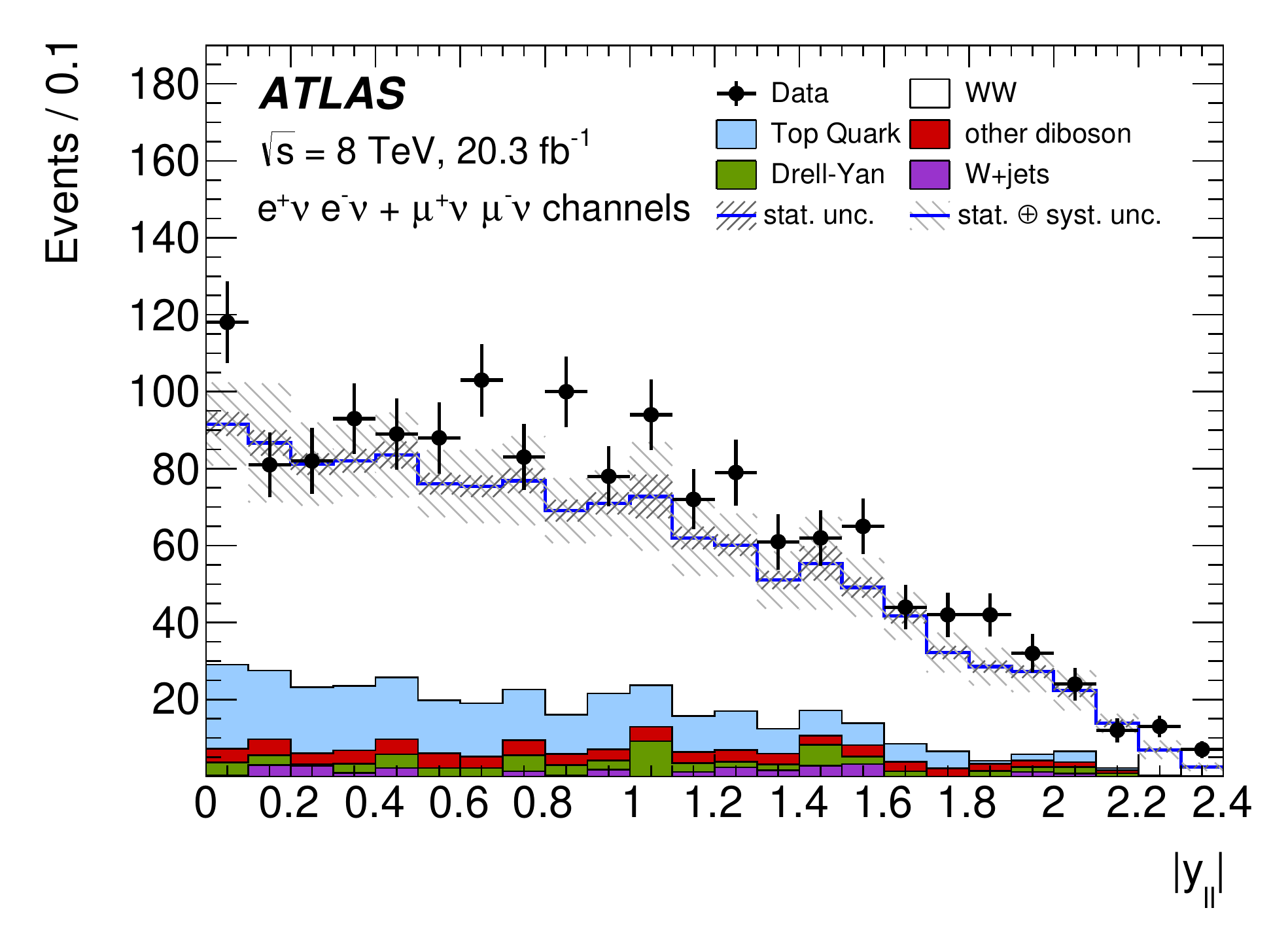}\hfill
    \includegraphics[width=0.49\textwidth]{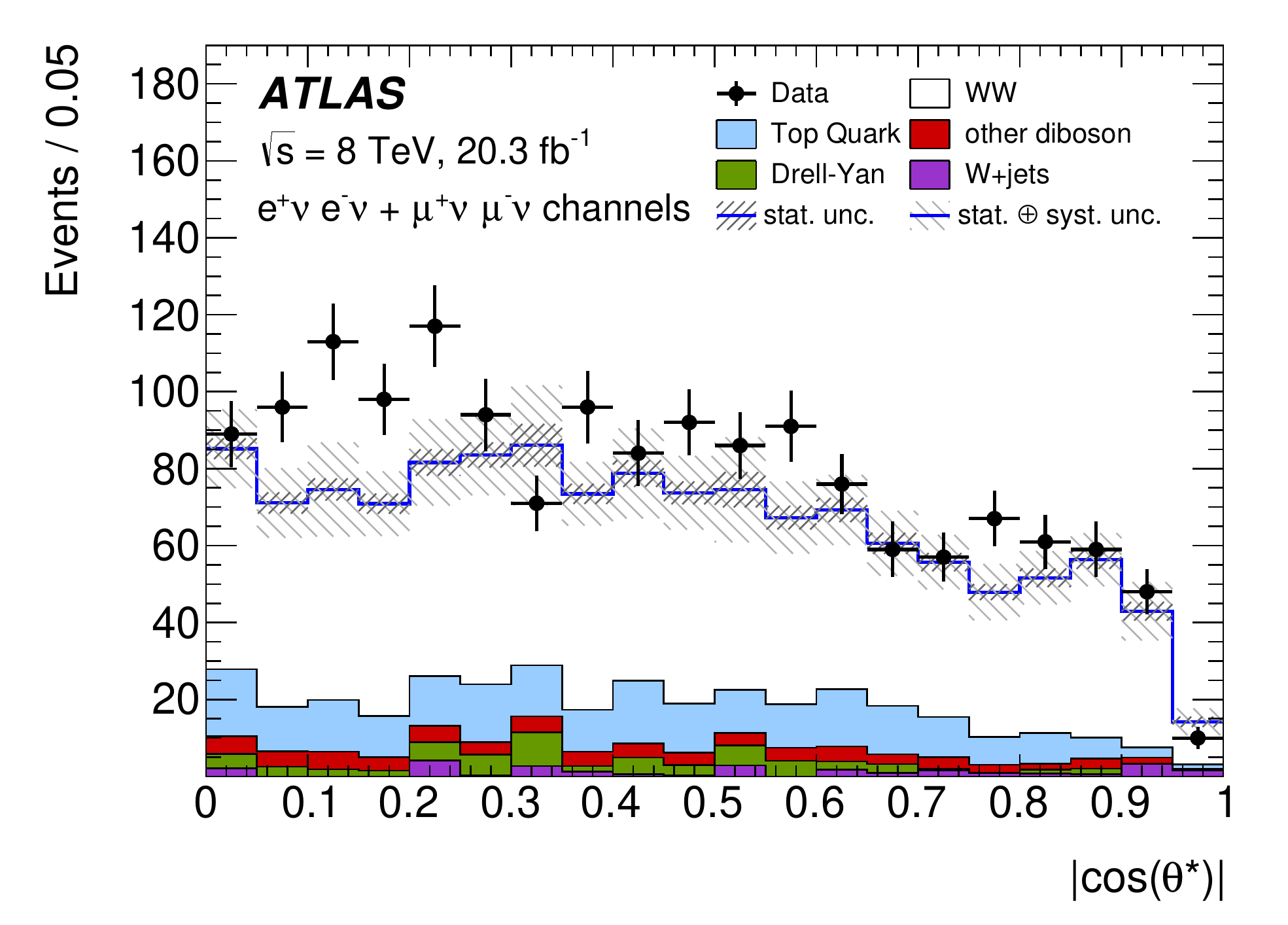}

\caption{Kinematic distributions of the selected data events after the full event selection for the combined $ee$ and $\mu\mu$ final states. Data are shown together with the predictions of the signal and background production processes. The transverse momentum of the leading lepton, $\pt^{\mathrm{lead}}$,  the invariant mass, $m_{\ell\ell}$, and the transverse momentum of the dilepton system, $\pt(\ell\ell)$,  as well as the difference in azimuthal angle between the decay leptons, $\Delta \phi_{\ell\ell}$, their combined rapidity, \absyll, and the observable \abscostheta\ are shown (from left to right and top to bottom). The last bin of the $\pt^{\mathrm{lead}}$,  $m_{\ell\ell}$ and $\pt(\ell\ell)$ distributions is an overflow bin. Statistical and systematic uncertainties in the predictions are shown as bands in hatched style. }

    \label{fig:pre_1}
    \end{center}
\end{figure}

\section{Cross-section determination}
\label{sec:xsec}

\subsection{Fiducial and total cross sections}
\label{s:phasespace}

After determining the background-subtracted number of signal candidate yields,  $N_{\mathrm{data}} - N_{\mathrm{bkg}}$, the fiducial \WW production cross section is extracted using a likelihood fit based on the following equation:

\begin{equation}
\label{eq:xsec}
\sigma_{\mathrm{fid}}^{\ell\ell^\prime}(\WW) = \frac{N_{\mathrm{data}} - N_{\mathrm{bkg}}} {C_{WW}  \times {\cal{L}} },
\end{equation}

\noindent
where 
${\cal{L}}$ is the integrated luminosity. The correction factor $C_{WW}$ is determined from MC simulation and accounts for detector efficiency, resolution effects and contributions from $\tau$-lepton decays. It is defined as the ratio of the number of reconstructed \WW events after the final selection with electrons or muons in the final state (including electrons or muons from $\tau$ decays) to the number of \WW events generated in the fiducial region where only direct decays of $W$ bosons to electrons and muons are allowed. The measured fiducial cross section $\sigma_{\mathrm{fid}}^{\ell\ell^\prime}$ thus describes $WW$ production with only prompt decays into $e\mu$, $ee$ and $\mu\mu$ final states. The correction for contributions with intermediate $W\rightarrow \tau\nu$ decays only relies on the correct relative acceptance and the well-known relative branching fractions~\cite{Beringer:1900zz}, not on the absolute normalization of the signal cross section.

\begin{table}[!t]\renewcommand{\arraystretch}{1.2}
  \begin{center}
  \begin{tabular}{l|c|c}
\hline\hline
     											& $e\mu$ 						&	$ee/\mu\mu$    				\\
\hline
$ p_{\rm T}^\ell$ (leading/sub-leading)							&	\multicolumn{2}{c}{ $>25\,/\,20$ \GeV} 								\\
\hline
    $ |\eta^{\ell}|$ 									&	\multicolumn{2}{c}{ $|\eta^{\mu}|<2.4$ and $|\eta^e|<2.47$,}\\
  											&	\multicolumn{2}{c}{ excluding $1.37<|\eta^e|<1.52$}	\\
\hline
$ m_{\ell\ell}$ 									&	$>10$ \GeV 					& $> 15$ \GeV 					\\
\hline
$|m_{Z} - m_{\ell\ell}|$ 								&	--- 						& $> 15$ \GeV 					\\
\hline
Number of jets with 									& 							&  						\\
$p_{\rm T}>$ 25 \GeV, $|\eta| < 4.5$ 							& 0 							& 0 						\\
\hline
$|\mathbf{\Sigma p^{\nu_i}_{\rm T}}|\;$ if $\Delta\phi_{\ell} > \pi/2 $&  $>15$ \GeV 						& $>45$ \GeV 					\\

$|\mathbf{\Sigma p^{\nu_i}_{\rm T}}| \times \sin\left(\Delta\phi_{\ell}\right)\;$ if $\Delta\phi_{\ell} < \pi/2 $	&   						& 					\\
(\metrel) &    							&  						\\ 
\hline
Transverse magnitude of the vectorial sum of all neutrinos, $|\mathbf{\Sigma p^{\nu_i}_{\rm T}}|$									&  $>20$ \GeV 						& $>45$ \GeV 					\\
(\ptmiss)		&   							&   						\\ 
\hline\hline
  \end{tabular}
  \end{center}
  \caption{
  Definitions of the respective fiducial regions used in the calculation of \xsecWWfidemu, \xsecWWfidee\ and \xsecWWfidmumu. In these definitions, $\ell$ is the charged lepton from the decays $W\rightarrow e\nu$ and $W\rightarrow \mu\nu$, and $\sin(\Delta\phi_{\ell})$ is the minimum difference in azimuthal angle between the vector sum of the momenta of the
neutrinos and any of the selected generator-level charged leptons.
  }
  \label{fiducialtruthlevelcuts}
\end{table}

The fiducial cross sections are measured separately in $e\mu$, $ee$ and $\mu\mu$ final states in regions closely approximating the experimental selection. The fiducial regions are summarised in Table~\ref{fiducialtruthlevelcuts}.

To define the fiducial region, the following selection is applied to events from the MC generator before passing them through the detector simulation. Leptons are required to originate directly from $W$ decays and be oppositely charged. They are recombined with any final-state photons from QED radiation that fall within $\Delta R = 0.1 $ of the respective
lepton to form so-called `dressed-leptons'. The lepton kinematic requirements are imposed on these dressed
leptons. Particle-level jets are constructed from stable particles with a lifetime of $\tau>30$ ps, excluding muons and neutrinos, using the anti-$k_t$ algorithm with a radius parameter of 0.4. To remove jets reconstructed from signal electrons, jets lying a distance $\Delta R < 0.3$ from any signal electrons are removed. The four-momentum sum of the neutrinos stemming from the $W$ boson decays is used for the calculation of both \ptmiss\ and \metrel.

The total cross section of $WW$ production is defined to include all
decay modes of the $W$ bosons and all jet multiplicities. It is obtained by
extrapolating the fiducial cross section for the effects of all
acceptance cuts listed in Table~\ref{fiducialtruthlevelcuts} with an additional acceptance
factor, $A_{WW}$, and correcting for the leptonic branching fraction of $W$
bosons $\mathcal{B}({W\rightarrow\ell\nu})=0.108$~\cite{Beringer:1900zz}:

\begin{equation}
\label{eq:totxsec}
\sigma_{\mathrm{tot}}(pp\rightarrow \WW) = \frac{\sigma_{\mathrm{fid}}^{\ell\ell^\prime}(pp\rightarrow \WW)}{A_{WW} \times \mathcal{B}^2({W\rightarrow\ell\nu})} = \frac{N_\mathrm{data}-N_\mathrm{bkg}}{C_{WW} \times A_{WW} \times \mathcal{B}^2({W\rightarrow\ell\nu})\times \mathcal{L}},
\end{equation}

where $A_{WW}$ is defined as the ratio of the MC signal event yield within the
fiducial region to the total number of generated signal MC events. The numerical values for the different final states are given in Table~\ref{ta:AWWCWW}. For the $e\mu$ final state, the right-hand side of Eq.~(\ref{eq:totxsec}) contains an
additional combinatorial factor of $1/2$.

The total cross sections for the individual final states, \emu, $ee$ and $\mu\mu$, are then combined. The combination procedure is based on a likelihood fit where the systematic uncertainties, including the uncertainties due to backgrounds, are included as nuisance parameters. The minimisation of the negative log-likelihood function and the error calculation are performed using the Minuit package~\cite{minuit}. Several independent sources of systematic uncertainty are treated as correlated among the different final states, while the
statistical uncertainties in the background estimates are treated as
uncorrelated.

The numerical values of the correction factors $C_{WW}$ and $A_{WW}$ are shown in Table~\ref{ta:AWWCWW}, while the uncertainties are listed in Table~\ref{ta:TheorySys}. These values are derived by adding the samples for all the $WW$ production processes according to their cross sections as detailed in Section~\ref{sec:samples}. The same holds for the determination of their uncertainties. Table~\ref{ta:AWWCWW} also gives the values of the correction factors for the different $WW$ production processes. The value for $C_{WW}$ is largest for the $e\mu$ final state because events with $W$ decays to $\tau$-leptons, which only contribute to the numerator, make up a larger fraction of events in the $e\mu$ channel. This is due to less stringent requirements on \met. The difference in the $C_{WW}$ values between $ee$ and $\mu\mu$ is due to the different lepton identification efficiencies.

\begin{table}[!ht]
\begin{center}
  \sisetup{group-separator = ., 
    table-number-alignment = left, round-mode=figures }
   \begin{tabular}{
    l|
    r@{.}l  
    r@{.}l  
    r@{.}l| 
    r@{.}l  
    r@{.}l  
    r@{.}l| 
    r@{.}l  
    r@{.}l  
    r@{.}l  
    }
    \hline \hline
    	&		 \multicolumn{6}{c|}{$C_{WW}$ [\%]} 	& \multicolumn{6}{c|}{$A_{WW}$ [\%]} 	& \multicolumn{6}{c}{$C_{WW} \times A_{WW}$ [\%]} \\
    \hline
      	&		 \multicolumn{2}{c}{\text{$e\mu$}} 	& \multicolumn{2}{c}{\text{$ee$}} 		& \multicolumn{2}{c|}{\text{$\mu\mu$}}		& \multicolumn{2}{c}{\text{$e\mu$}} 	& \multicolumn{2}{c}{\text{$ee$}} 	& \multicolumn{2}{c|}{\text{$\mu\mu$}} 	& \multicolumn{2}{c}{\text{$e\mu$}} 	& \multicolumn{2}{c}{\text{$ee$}} 	& \multicolumn{2}{c}{\text{$\mu\mu$}} 	\\
    \hline 

Total  &
   51&2&  29&1& 47&4 
   &   22&8 & 8&6 & 9&3&
  11&7 & 2&5 & 4&4\\ \hline

\qqWW\  &   
  51&4& 29&2 & 47&7 & 
    23&5 & 8&7 &9&5&
   12&1& 2&6 &4&5 \\ 
 
$gg{\rightarrow}W^+W^-$ (non-resonant) &  
   53&6& 33&4 & 48&2 & 
    30&6 & 14&7 &16&3 & 
   16&4& 4&9 & 7&8\\ 
  
$gg{\rightarrow} H \rightarrow W^+W^-$ &  
   43&5& 21&8 & 39&3 & 
  10&4 & 4&1 &4&6 &
    4&5& 0&9 &1&8 \\

    \hline \hline
  \end{tabular}
\end{center}
\caption{Central values of $C_{WW}$, $A_{WW}$ and $C_{WW}\times A_{WW}$ used in the calculation of the cross section. The numbers are derived using the weighted average of the numbers for the different $WW$ production processes which are weighted according to their cross sections as detailed in Section~\ref{sec:samples}. }
\label{ta:AWWCWW}
\end{table}

\subsection{Measurement of the differential cross sections}\label{sec:unfolding}

Differential cross sections are defined in the fiducial regions and are measured as a function of the kinematic variables described in Section~\ref{sec:overview}. The measurement is carried out in the $e\mu$ final state, which has a larger signal acceptance and lower relative background contamination compared to the same-flavour channels. The reconstructed spectra are
corrected for background contributions and then unfolded to the fiducial phase space
by correcting for detector resolution and reconstruction efficiencies. The iterative Bayesian approach~\cite{Adye:2011gm,DAgostini2010} with three iterations is employed in this analysis. The choice of three iterations is optimised to minimise the statistical uncertainties and the dependence on the prior Monte Carlo distribution in the unfolded spectra.

The measured differential cross sections are calculated from the unfolded signal spectra divided by the integrated luminosity and the corresponding bin widths. For the measurement, statistical uncertainties and the same systematic uncertainties as for the fiducial cross-section measurements are considered. In addition, the uncertainty due to the theoretical modelling is evaluated  
using a data-driven method that was introduced in Ref.~\cite{Malaescu:2011yg}, in which the kinematic distributions of the MC signal events are corrected to match those from data and the uncertainty is considered as the difference between the unfolded data spectra derived with the modified MC distributions and those using the original MC simulation. The modelling uncertainty is found to be small ($\pm$1\%) for most variables.
 
\section{Systematic uncertainties}
\label{sec:systematics}

Systematic uncertainties in the measured $WW$ cross section arise from the object reconstruction, the background determination, the procedures used to correct for detector effects, and the usage of theory predictions in correction and extrapolation procedures. 

\subsection{Experimental uncertainties}

The relative systematic uncertainties from the reconstruction of the events in the detector and the background subtraction are summarised in Table~\ref{tab:sys} for the fiducial and total cross-section measurements. The dominant systematic uncertainties in the combined measurement are the uncertainties due to the jet energy scale ($\sim$4\%), the $W$+jets background ($\sim$3\%) and the luminosity ($\sim$2\%).

The uncertainties due to pile-up are estimated by varying the reweighting procedure for the MC samples used to reproduce the distributions of the number of primary vertices in data. The uncertainties in the correction factors to match the simulated efficiencies to
the measured ones for the electron and muon trigger requirements~\cite{ATLAS-CONF-2012-048,Aad:2014sca} as well as for the  reconstruction,
identification and isolation requirements~\cite{ATLAS-CONF-2014-032,Aad:2014zya,aad:2014rra,Aad:2014fxa} are
propagated to the measurement. A similar procedure is used to assess the uncertainty due to the lepton momentum scale and resolution~\cite{aad:2014rra,Aad:2011mk}.

Uncertainties related to the selection and measurement of jets affect the measurement primarily via the definition of jets for the jet-veto requirement, but also via the \met\ reconstruction. The impact on the cross-section measurements is evaluated by varying each of these in the simulation within their respective uncertainties as determined from data~\cite{JES}. The main sources of uncertainty for jets are the jet energy scale (JES) and the jet energy resolution (JER). 
Uncertainties of the lepton and jet momentum scales and resolutions are propagated to the \met\ reconstruction. Additional uncertainties in the \met\ due to jets reconstructed with momenta below $\pt<20$\,\GeV\ and calorimeter cells not associated with any reconstructed objects are accounted for separately and denoted  ``soft terms'' in Table~\ref{tab:sys}. An uncertainty in the \ptmiss\ scale and resolution is estimated from a comparison between data and MC simulation in $Z$ boson events with muonic decays.

Backgrounds are determined as discussed in Section~\ref{sec:backgrounds}, and the uncertainties from the background subtraction are also given in Table~\ref{tab:sys}. For each of the top-quark and Drell--Yan background estimates, and the $W$+jets and the multijet background estimate, the total systematic uncertainty is given. The statistical uncertainties stem from the limited size of the MC samples used for the background estimates of the diboson production processes and from the limited size of the data samples used for data-driven estimations.  The uncertainty in the integrated luminosity is $\pm1.9$\% and affects the cross-section determination through Eqs.~(\ref{eq:xsec}) and~(\ref{eq:totxsec}) and the normalisation of background from other diboson processes.

\begin{table}[!ht]\renewcommand{\arraystretch}{1.2}
  \centering
  \sisetup{retain-explicit-plus}
  \sisetup{round-mode = places}
  \begin{tabular}{l
    S[round-mode = places, round-precision = 1,
    table-format = 1.1, table-number-alignment = left]
    @{\,}
    S[round-mode = places, round-precision = 1,
    table-format = 1.1, table-number-alignment = right]
    @{\,}
    S[round-mode = places, round-precision = 1,
    table-format = 1.1, table-number-alignment = left]
    @{\,}
    S[round-mode = places, round-precision = 1,
    table-format = 1.1, table-number-alignment = left]
    @{\,}
    }
    \hline\hline

    \multicolumn{1}{l}{Sources of uncertainty} 	& \multicolumn{1}{c}{~~~~~~~~$e\mu$~~~~~~~~} 	& \multicolumn{1}{c}{~~~~~~~~~$ee$~~~~~~~}	& \multicolumn{1}{c}{~~~~~~~~$\mu\mu$~~~~~~~~} 	& \multicolumn{1}{c}{Combined} \\ \hline

    \multicolumn{5}{l}{\textbf{Experimental uncertainties in fiducial and total cross sections [\%]}} \\
    \hline
	 Integrated luminosity                         &
	 ${{\pm 2.0}}$ &
	 ${{\pm 2.0}}$ &
	 ${{\pm 2.0}}$ &
	 ${{\pm 2.0}}$ \\
	 Pile-up                         &
	 ${{\pm\WZtotXsecpileupUpem}}$ &
	 ${\pm\WZtotXsecpileupUpee}$ &
	 ${{\pm\WZtotXsecpileupUpmm}}$ &
	 ${{\pm\WZtotXsecpileupUp}}$ \\
         Trigger                        &
         ${{\pm\WZtotXsecTriggerUpem}}$ &
         ${{\pm 2.8}}$ &
         ${{\pm 3.0}}$ &
         ${{\pm\WZtotXsecTriggerUp}}$ \\
    \hline
	  Electron energy scale            &
	  ${{\pm\WZtotXseceScaleUpem}}$ &
	  ${{\pm\WZtotXseceScaleUpee}}$ &
	  \multicolumn{1}{c}{---} &
	  ${{\pm\WZtotXseceScaleUp}}$ \\

	  Electron energy resolution       &
	  ${{\pm\WZtotXseceResoUpem}}$ &
	  ${{\pm\WZtotXseceResoUpee}}$ &
	  \multicolumn{1}{c}{---} &
	  ${{\pm\WZtotXseceResoUp}}$ \\
	  Electron identification and reconstruction &
	  ${{\pm\WZtotXseceIDSFUpem}}$ &
	  ${{\pm\WZtotXseceIDSFUpee}}$ &
	  \multicolumn{1}{c}{---} &
	  ${{\pm\WZtotXseceIDSFUp}}$ \\
	 Electron isolation  &
	 ${{\pm\WZtotXseceIsoUpem}}$ &
         ${{\pm\WZtotXseceIsoUpee}}$ &
	 \multicolumn{1}{c}{---} &
	 ${{\pm\WZtotXseceIsoUp}}$ \\

    \hline
   	  Muon momentum scale          &
	  ${{\pm\WZtotXsecmuScaleUpem}}$ &
 	  \multicolumn{1}{c}{---} &
	  ${{\pm\WZtotXsecmuScaleUpmm}}$ &
	  ${{\pm\WZtotXsecmuScaleUp}}$ \\

	  Muon momentum resolution (ID)        &
	  ${{\pm\WZtotXsecmuIDSmearUpem}}$ &
	  \multicolumn{1}{c}{---} &
	  ${{\pm\WZtotXsecmuIDSmearUpmm}}$ &
	  ${{\pm\WZtotXsecmuIDSmearUp}}$ \\

	  Muon momentum resolution (MS)         &
	  ${{\pm\WZtotXsecmuMSSmearUpem}}$ &
	  \multicolumn{1}{c}{---} &
	  ${{\pm\WZtotXsecmuMSSmearUpmm}}$ &
	  ${{\pm\WZtotXsecmuMSSmearUp}}$ \\

	 Muon identification and reconstruction&
	 ${{\pm\WZtotXsecmIDSFUpem}}$ &
	 \multicolumn{1}{c}{---} &
	 ${{\pm\WZtotXsecmIDSFUpmm}}$ &
	 ${{\pm\WZtotXsecmIDSFUp}}$ \\

	 Muon isolation  &
	 ${{\pm\WZtotXsecmIsoUpem}}$ &
	 \multicolumn{1}{c}{---} &
	 ${{\pm\WZtotXsecmIsoUpmm}}$ &
	 ${{\pm\WZtotXsecmIsoUp}}$ \\

    \hline
	  Jet vertex fraction (JVF)        &
	  ${{\pm\WZtotXsecJVFUpem}}$ &
	  ${{\pm\WZtotXsecJVFUpee}}$ &
	  ${{\pm\WZtotXsecJVFUpmm}}$ &
	  ${{\pm\WZtotXsecJVFUp}}$ \\

	  Jet energy scale        &
	  ${{\pm 4.1}}$ &
	  ${{\pm 3.9}}$ &
	  ${{\pm 4.4}}$ &
	  ${{\pm 4.1}}$ \\

    Jet energy resolution    &
${{\pm\WZtotXsecJERUpem}}$ &
${{\pm\WZtotXsecJERUpee}}$ &
${{\pm\WZtotXsecJERUpmm}}$ &
${{\pm\WZtotXsecJERUp}}$ \\

\met\ scale soft terms   &
${\pm\WZtotXsecMETScaleSoftUpem}$ &
${{\pm\WZtotXsecMETScaleSoftUpee}}$&
${{\pm\WZtotXsecMETScaleSoftUpmm}}$ &
${{\pm\WZtotXsecMETScaleSoftUp}}$ \\

\met\ resolution soft terms    &
 ${{\pm\WZtotXsecMETResoSoftUpem}}$ &
 ${{\pm\WZtotXsecMETResoSoftUpee}}$ &
 ${{\pm\WZtotXsecMETResoSoftUpmm}}$ &
 ${{\pm\WZtotXsecMETResoSoftUp}}$ \\

 \ptmiss\ scale soft terms            &
${{\pm\WZtotXsecPtScaleSoftUpem}}$ &
${{\pm\WZtotXsecPtScaleSoftUpee}}$ &
${{\pm\WZtotXsecPtScaleSoftUpmm}}$ &
${{\pm\WZtotXsecPtScaleSoftUp}}$ \\

\ptmiss\ resolution soft terms &
${{\pm\WZtotXsecPtResoSoftUpem}}$ &
${{\pm\WZtotXsecPtResoSoftUpee}}$ &
${{\pm\WZtotXsecPtResoSoftUpmm}}$ &
${{\pm\WZtotXsecPtResoSoftUp}}$ \\

\hline
\multicolumn{5}{l}{\textbf{Background uncertainties in fiducial and total cross sections [\%]}} \\
\hline

Top-quark background 	   &
${{\pm\WZtotXsecDDtopUpem}}$ &
${{\pm\WZtotXsecDDtopUpee}}$ &
${{\pm\WZtotXsecDDtopUpmm}}$ &
${{\pm\WZtotXsecDDtopUp}}$ \\

$W+$jets \& \mj\ background &
${{\pm\WZtotXsecDDWjetTotUpem}}$ &
${{\pm\WZtotXsecDDWjetTotUpee}}$ &
${{\pm\WZtotXsecDDWjetTotUpmm}}$ &
${{\pm\WZtotXsecDDWjetTotUp}}$  \\

Drell--Yan background &
${{\pm\WZtotXsecDDZjetUpem}}$ &
${{\pm\WZtotXsecDDZjetUpee}}$ &
${{\pm\WZtotXsecDDZjetUpmm}}$ &
${{\pm\WZtotXsecDDZjetUp}}$ \\

MC statistics (top-quark, $W$+jets, Drell--Yan) &
${{\pm\WZtotXsecBkgDDStatEMUpem}}$ &
${{\pm\WZtotXsecBkgDDStatEEUpee}}$ &
${{\pm\WZtotXsecBkgDDStatMMUpmm}}$ &
${{\pm\WZtotXsecBkgDDStatAllUp}}$ \\

Other diboson cross sections  &
${{\pm\WZtotXsecDibosonCrossSectionUpem}}$ &
${{\pm\WZtotXsecDibosonCrossSectionUpee}}$ &
${{\pm\WZtotXsecDibosonCrossSectionUpmm}}$ &
${{\pm\WZtotXsecDibosonCrossSectionUp}}$ \\

MC statistics (other diboson) &
${{\pm\WZtotXsecBkgMCStatEMUpem}}$ &
${{\pm\WZtotXsecBkgMCStatEEUpee}}$ &
${{\pm\WZtotXsecBkgMCStatMMUpmm}}$ &
${{\pm\WZtotXsecBkgMCStatAllUp}}$ \\
\hline\hline
   \end{tabular}
    \caption{Uncertainty sources and associated relative systematic uncertainties for the reconstruction and background subtraction for the \WW\ cross sections measured in the $e\mu$, $ee$ and $\mu\mu$ final states as well as for the combined cross section. The uncertainties apply to both the fiducial and total cross sections. 
    In cases where no uncertainties are quoted they do not affect the specific final state (e.g. electron energy scale uncertainties for muon final states).
        }
   \label{tab:sys}
\end{table}

\subsection{Modelling uncertainties}\label{sec:theorymodelling}

The modelling uncertainty in the $WW$ signal enters the cross-section determination through the calculation of the correction factor $C_{WW}$ and the acceptance correction factor $A_{WW}$. The dominant uncertainties in $A_{WW}$ stem from the jet-veto requirement (3.4\%), the parton shower model and the choice of generators ($\sim$2.5\%), while the uncertainties due to other sources (PDF choice, NLO electroweak correction, residual QCD scale dependence) are found to be small ($<$1\%). In contrast to $A_{WW}$, which has a sizeable theory dependence, $C_{WW}$ is mainly affected by the detector resolution and only has a small theoretical uncertainty ($\sim$1\%). The product of $C_{WW} \times A_{WW}$ and its corresponding uncertainties are evaluated taking into account correlations between the uncertainties in $C_{WW}$ and $A_{WW}$. Table~\ref{ta:TheorySys} gives an overview of the theoretical uncertainties used in the cross-section calculation. They are determined independently for the different $WW$ signal processes as detailed below and added according to the respective contribution of \qqWW\ and resonant and non-resonant $gg$-induced $W^+W^-$ production. 

The PDF uncertainty is estimated as the envelope of the CT10~\cite{CTEQ10} prediction and the predictions obtained using MSTW2008~\cite{MSTW} and NNPDF2.3~\cite{Ball:2012cx} and their respective PDF uncertainties at 68\% confidence level following Ref.~\cite{Alekhin:2011sk}. The PDF uncertainties for $q\bar{q}$ and $gg$-induced $WW$ production are combined assuming full correlation.

The effect of the next-to-leading-order EWK contributions of $\mathcal{O}(\alpha_{EW}^3)$ is considered for the \qqWW\ process only, but the resulting uncertainty in $C_{WW}$ and $A_{WW}$ is minor.

The uncertainties of the perturbative calculations can be estimated by varying the choice of renormalisation and factorisation scales and quoting the maximum difference between results from the central value and the alternative scale choices. However, it is suggested in Ref.~\cite{Stewart:2011cf} that this approach may underestimate the uncertainty in the case a jet-veto requirement is applied and therefore a more conservative approach is proposed. In this approach, the scale variation
uncertainties are treated as uncorrelated between cross sections for
different inclusive jet multiplicities and predictions for exclusive
jet multiplicities are obtained from the difference between the inclusive
predictions. The uncertainty in the difference is then used to estimate the perturbative uncertainty due to the jet-veto requirement. This uncertainty is found to be about $\pm2.9\%$ for the \qqWW\ process when using the NNLO predictions~\cite{Gehrmann:2014fva}. The uncertainty due to the jet-veto requirement for the $H\rightarrow W^+W^-$ production process has been determined to be $\pm11\%$~\cite{ATLAS:2014aga} and the same uncertainty is assumed for the non-resonant $gg\rightarrow W^+W^-$ process. The combined uncertainty due to the jet-veto requirement for the $WW$ signal is about $\pm3.4\%$, assuming full correlation of the $q\bar{q}$- and $gg$-induced processes. Without the jet-veto requirement, the residual perturbative uncertainty in $A_{WW}$ due to the scale choice for the perturbative calculations is estimated using an approach in which the renormalisation ($\mu_{\rm R}$) and factorisation ($\mu_{\rm F}$) scales are varied independently by a factor of two or one-half. The uncertainty in $A_{WW}$ due to the scale choice is calculated when not applying the jet-veto requirement to avoid double counting with the above described perturbative uncertainty due to the jet-veto requirement. It is found to be $\pm0.2\%$. The uncertainty in $C_{WW}$ due to the perturbative calculations, including the jet-veto requirement, is 
found to be about $\pm0.6\%$.

Parton shower, hadronisation and underlying-event uncertainties (collectively called ``soft QCD'' in the following), and generator uncertainties, have effects that are estimated by comparing various MC signal samples. For $q\overline{q}\rightarrow W^+W^-$ production, the generator uncertainty is evaluated by comparing samples generated using \Powheg\ and \MCatNLO\, both interfaced with \herwig+\Jimmy, whilst the soft QCD uncertainties are calculated using two samples generated with \Powheg\ and interfaced with \herwig+\Jimmy and \Pythia~8, respectively. This yields an uncertainty of 1.3\% dominated by the uncertainty in $A_{WW}$ due to the choice of generator. The $C_{WW}$ uncertainties are estimated to be $0.4$--$0.9\%$ for the \qqWW\ process and are applied to all production channels. The uncertainties in $A_{WW}$ for non-resonant $gg \rightarrow W^+W^-$ production due to soft QCD effects and the choice of the generator are evaluated by comparing the results obtained from samples generated with \ggtwoww and \MCFM\, both interfaced to \herwig+\Jimmy, and two samples generated with \MCFM\, interfaced either to \herwig+\Jimmy\ or \Pythia~8. This gives an uncertainty of about 30\% in $A_{WW}$, which is dominated by a $28$--$29\%$ uncertainty coming from the parton shower. Uncertainties in the modelling of resonant $H \rightarrow W^+W^-$ production are taken from Ref.~\cite{ATLAS:2014aga} and amount to 6.9\% for $A_{WW}$. 

\begin{table}[!ht]
\begin{center}
  \sisetup{group-separator = ., 
    table-number-alignment = left, round-mode=figures }
   \begin{tabular}{l|
    l  
    l  
    l| 
    l  
    l  
    l| 
    l  
    l  
    l  
    }
    \hline \hline
    			& \multicolumn{3}{c|}{$\sigma (C_{WW})$ [\%]} 	& \multicolumn{3}{c|}{$\sigma (A_{WW})$ [\%]} 	& \multicolumn{3}{c}{$\sigma (C_{WW} \times A_{WW})$ [\%]} \\
    \hline
    			& \multicolumn{1}{c}{\text{$e\mu$}} 	& \multicolumn{1}{c}{\text{$ee$}} 		& \multicolumn{1}{c|}{\text{$\mu\mu$}}		& \multicolumn{1}{c}{\text{$e\mu$}} 	& \multicolumn{1}{c}{\text{$ee$}} 	& \multicolumn{1}{c|}{\text{$\mu\mu$}} 	& \multicolumn{1}{c}{\text{$e\mu$}} 	& \multicolumn{1}{c}{\text{$ee$}} 	& \multicolumn{1}{c}{\text{$\mu\mu$}} 	\\
    \hline                                                                                                                                                                                                                                                                                                                      
    PDF     			& \num[round-precision=2]{0.10} 	& \num[round-precision=2]{0.34} 	& \num[round-precision=2]{0.13} 	 
    & \num[round-precision=2]{0.81} 	& \num[round-precision=2]{0.94} 		& \num[round-precision=2]{0.93} 	    &
    \num[round-precision=2]{0.85}  	& \num[round-precision=2]{1.3 } 	& \num[round-precision=2]{0.98} 	\\
    EWK corrections ($\mathrm{SF}_{\rm EW}$) 		& \num[round-precision=1]{0.01} 	& \num[round-precision=1]{0.06} 	& \num[round-precision=1]{0.04} 	
    & \num[round-precision=2]{0.46} 	& \num[round-precision=2]{0.41} 		& \num[round-precision=2]{0.43} 	
    & \num[round-precision=2]{0.47}  	& \num[round-precision=2]{0.34} 	& \num[round-precision=2]{0.40} 	\\
    Jet veto 		& \multicolumn{1}{c}{ --- }    		& \multicolumn{1}{c}{ ---   }   	& \multicolumn{1}{c|}{  ---   }  	
    & \num[round-precision=2]{3.40} 	& \num[round-precision=2]{3.40} 		& \num[round-precision=2]{3.40} 	
    & \num[round-precision=2]{3.40}  	& \num[round-precision=2]{3.40} 	& \num[round-precision=2]{3.40} 	\\
    Scale   			& \num[round-precision=2]{0.62}  	& \num[round-precision=2]{0.62}  	& \num[round-precision=2]{0.62}  	
    & \num[round-precision=2]{0.22}  	& \num[round-precision=2]{0.22}  		& \num[round-precision=2]{0.22}  	
    & \num[round-precision=2]{0.66}   	& \num[round-precision=2]{0.66}  	& \num[round-precision=2]{0.66}  	\\
    Soft QCD  			& \num[round-precision=2]{0.35} 	& \num[round-precision=2]{0.92} 	& \num[round-precision=2]{0.80} 
    & \num[round-precision=2]{2.50} 	& \num[round-precision=2]{2.60} 		& \num[round-precision=2]{2.70} 	
    & \num[round-precision=2]{2.50}  	& \num[round-precision=2]{3.00} 	& \num[round-precision=2]{2.90} 	\\
    \hline                                              	                                		                                		                                	                                	                                	                                 	                                	                                	
    Total    				& \num[round-precision=2]{0.70} 	& \num[round-precision=2]{1.2 }		& \num[round-precision=2]{1.0 } 
    & \num[round-precision=2]{4.30} 	& \num[round-precision=2]{4.40} 		& \num[round-precision=2]{4.50} 
    & \num[round-precision=2]{4.4 }  	& \num[round-precision=2]{4.8 } 	& \num[round-precision=2]{4.6 } 	\\
    \hline \hline
  \end{tabular}
\end{center}
\caption{Relative uncertainties of $C_{WW}$, $A_{WW}$ and $C_{WW}\times A_{WW}$, due to the theoretical modelling of the $WW$ signal processes, which include $q\overline{q}\rightarrow W^+W^-$ and non-resonant and resonant $gg\rightarrow W^+W^-$ production. The total uncertainties are calculated as the quadratic sum of the uncertainties from each source. The combined $C_{WW}\times A_{WW}$ uncertainties take into account the correlation between $C_{WW}$ and $A_{WW}$.}
\label{ta:TheorySys}
\end{table}

\clearpage
\section{Cross-section results}
\label{sec:results}

\subsection{Theoretical predictions}
\label{sec:crosssections}

Various calculations for $WW$ diboson production, involving higher loop QCD corrections, electroweak corrections, or involving resummation of soft gluon terms are available, with different approaches taken to approximate these effects~\cite{Gehrmann:2014fva,Meade:2014fca,Monni:2014zra}. Hence, when comparing experimental results with theoretical predictions, these corrections play a relevant role.

Figure~\ref{fig:ww_prod} illustrates the diagrams of the leading processes for $WW$ production. The cross section of the \qqWW\ process is known to $\mathcal{O}(\alpha_{\rm s}^2)$~\cite{Gehrmann:2014fva}. The loop-induced $gg$ processes that include the non-resonant $gg$ and resonant Higgs boson production processes start contributing at $\mathcal{O}(\alpha_{\rm s}^2)$. The non-resonant $gg$ process is only predicted at lowest order, $\mathcal{O}(\alpha_{\rm s}^2)$, while the Higgs boson production cross section has been calculated to $\mathcal{O}(\alpha_{\rm s}^4)$~\cite{HiggsXS}.\footnote{A more recent 3-loop calculation of the theoretical prediction for the resonant $gg$ process to $\mathcal{O}(\alpha_{\rm s}^5)$~\cite{Anastasiou:2015ema} yields a 0.15\% increase of the total $WW$ production cross section with respect to the calculation to $\mathcal{O}(\alpha_{\rm s}^4)$ but it is not considered here.} Neither of the loop processes interferes at $\mathcal{O}(\alpha_{\rm s}^2)$ with the \qqWW\ process and the interference between the $gg$-induced processes is small and can be neglected. The resonant Higgs boson production process makes up 6.6\% of the total cross section, while non-resonant $gg$ production contributes 2.2\%. 

The combination of the above processes calculated to $\mathcal{O}(\alpha_{\rm s}^2)$ for the $q\bar{q}$ and non-resonant $gg$ processes and to $\mathcal{O}(\alpha_{\rm s}^4)$ for the resonant $gg$ contribution yields a prediction for the total $WW$ production that is valid to NNLO accuracy in perturbative QCD.

These theoretical calculations are available only for the total production cross section; the kinematic distributions for the \qqWW\ process are predicted only at order $\mathcal{O}(\alpha_{\rm s})$~\cite{Campbell:2011bn}. Therefore, fiducial cross sections are estimated by applying the acceptance correction $A_{WW}$ as calculated from the MC samples described in Section~\ref{sec:samples} to the total cross sections. The first prediction for the fiducial cross section is that for $WW$ production predicted partially at NNLO in $\alpha_{\rm s}$ (nNLO) neglecting $\mathcal{O}(\alpha_{\rm s}^2)$ contributions to the $q\bar{q}$ process. The nNLO prediction for the total cross section is about 8\% lower than the NNLO prediction, and as a consequence, the nNLO fiducial cross section, $\sigma$(nNLO$_{\mathrm{fid}}$), is also lowered by the same amount.

An approximate NNLO fiducial cross section, $\sigma$(approx. NNLO$_{\mathrm{fid}}$), can be derived by using the total NNLO prediction~\cite{Gehrmann:2014fva,HiggsXS} and multiplying it by the fiducial acceptance and the branching fractions, $\sigma^{\mathrm{NNLO}}_{\mathrm{tot}} \times A_{WW}\times\mathcal{B}^2$. 

These calculations can be refined by considering further effects. Logarithmic contributions from soft gluon emission from the initial state can be resummed at next-to-next-to-leading logarithmic (NNLL) accuracy, yielding a theoretical prediction with reduced scale uncertainties and a more accurate description of the transverse momentum of the $WW$ system, $p_{\mathrm T}^{WW}$, and the jet multiplicity.

An approximate NNLO+NNLL prediction of fiducial cross sections, $\sigma$($\mathrm{approx.~NNLO+NNLL}_{\mathrm{fid}}$), is provided in Ref.~\cite{Monni:2014zra}, and is about $15$--$18\%$ higher than the nNLO prediction and around $7$--$10\%$ higher than the approximate NNLO prediction.

The use of resummation calculations also affects the calculation of the acceptance $A_{WW}$ and this effect was investigated in Ref.~\cite{Meade:2014fca}. Based on Ref.~\cite{Meade:2014fca} an alternative approximation for the fiducial cross section including NNLL effects, $\sigma$(NNLO $p_{\mathrm T}$-Resum$_\mathrm{fid}$), is calculated as the product of $\sigma($NNLO$_\mathrm{tot})$, the branching fraction, and a corrected fiducial acceptance factor $A_{\mathrm{p_{\mathrm T}-Resum}_{WW}}$. The corrected $A_{\mathrm{p_{\mathrm T}-Resum}_{WW}}$ is derived by reweighting the shape of the $p_{\mathrm T}^{WW}$ distribution in the MC samples described in Section~\ref{sec:samples} to the predicted NLO+NNLL shape given in Ref.~\cite{Meade:2014fca}. 

The uncertainty in the NNLO calculation of the total cross section accounts for the effects of QCD scale and PDF uncertainties, which are added linearly, while the theoretical uncertainties of the fiducial cross sections include the parton shower uncertainty in addition to the QCD scale and PDF uncertainties, the effects of which are again added linearly.

\subsection{Cross-section measurements and comparisons with theoretical predictions}
\label{sec:distributions}

The measured fiducial and total cross sections are compared to theoretical calculations evaluated at different orders in $\alpha_{\rm s}$, as summarised in  Tables~\ref{table:xsecMeasFid2} and~\ref{table:xsecMeasTot}, respectively.

The cross-section measurements in their fiducial phase spaces, defined in Table~\ref{fiducialtruthlevelcuts}, are minimally dependent on theory corrections, since the fiducial volumes correspond closely to the detector-level selection. The measured fiducial cross sections are summarised in Table~\ref{table:xsecMeasFid2}, including statistical and systematic uncertainties. They are about two standard deviations higher than the nNLO prediction, and about 1.4 standard deviations larger than  $\sigma$(approx. NNLO$_{\mathrm{fid}}$), while they are found to be compatible with the predictions that consider resummation corrections. A graphical comparison between the fiducial cross-section measurements and various theoretical predictions is shown in Figure~\ref{fig:fidplots}.

The total cross sections are measured separately in the three different final states and then combined. The results for the individual measurements and the combined cross section are summarised in Table~\ref{table:xsecMeasTot}. The combined cross section is found to be compatible with the NNLO prediction within about 1.4 standard deviations. A graphical comparison between the individual measurements and their combination is shown in Figure~\ref{fig:NNLOverviewPlot}. The result is fully compatible with the recently published measurement by the CMS Collaboration~\cite{Khachatryan:2015sga}.

\begin{table}[!t]
  \centering
  \sisetup{retain-explicit-plus,round-mode = places,group-separator = ., table-format = 9.9,
    table-number-alignment = right, round-mode=figures }
  \begin{tabular}{llc}
    \hline
    \hline
    Prediction &   &  Fiducial cross section  \\ 
    &   &  $pp \rightarrow WW \rightarrow \ell\ell\nu\nu$ [fb]  \\ \hline
    \vspace{-0.9em}  &  \\ [2pt]
    Measured \xsecWWfidemu& & \num[round-precision=3]{\WZfidXsecem} $\pm \num[round-precision=1]{\WZfidXsecStatErrUpem}$(stat) $^{+\num[round-precision=2]{\WZfidXsecSysErrUpem}}_{\num[round-precision=2]{\WZfidXsecSysErrDwem}}$(syst) $^{+\num[round-precision=1]{\WZfidXsecLumiErrUpem}}_{\num[round-precision=1]{\WZfidXsecLumiErrDwem}}$(lumi)  \\[2pt]
    
    $\sigma$(nNLO$_{\mathrm{fid,}e\mu}$)& = ($\sigma^{\mathrm{nNLO}}_{\mathrm{tot}} \times A_{WW}\times\mathcal{B}^2$) ~\cite{Campbell:2011bn}+\cite{HiggsXS} &  \num[round-precision=3]{311.4}  $\pm$ \num[round-precision=2]{15.1} \\[2pt]
    $\sigma$(approx. NNLO$_{\mathrm{fid,}e\mu}$)&= ($\sigma^{\mathrm{NNLO}}_{\mathrm{tot}} \times A_{WW}\times\mathcal{B}^2$) ~\cite{Gehrmann:2014fva}+\cite{HiggsXS}&   \num[round-precision=3]{335.3} $\pm$ \num[round-precision=2]{17.7} \\[2pt]
    $\sigma$($\mathrm{approx.~(NNLO+NNLL)}_{\mathrm{fid,}e\mu}$) &~\cite{Monni:2014zra} & \num[round-precision=3]{357.9} $\pm$ \num[round-precision=2]{14.4} \\[2pt]
    $\sigma$(NNLO $p_{\mathrm T}$-Resum$_{\mathrm{fid,}e\mu}$)&= ($\sigma^{\mathrm{NNLO}}_{\mathrm{tot}} \times A_{\mathrm{p_{\mathrm T}-Resum}_{WW}} \times \mathcal{B}^2$)~\cite{Meade:2014fca}&  \num[round-precision=3]{348.7} $\pm$ \num[round-precision=2]{19.1} \\[2pt]
    \hline
    \vspace{-0.9em}  &  \\[2pt]
    
    Measured \xsecWWfidee &	 & \WZfidXsecee $^{+\WZfidXsecStatErrUpee}_{\WZfidXsecStatErrDwee}$(stat) $^{+\WZfidXsecSysErrUpee}_{\WZfidXsecSysErrDwee}$(syst) $\pm\WZfidXsecLumiErrUpee$(lumi)   \\[2pt]
    
    $\sigma$(nNLO$_{\mathrm{fid,}ee}$)& ~\cite{Campbell:2011bn}+\cite{HiggsXS} & 58.5  $\pm$ 2.8 \\[2pt]
    $\sigma$(approx. NNLO$_{\mathrm{fid,}ee}$)& ~\cite{Gehrmann:2014fva}+\cite{HiggsXS}& 63.0  $\pm$ 3.4 \\[2pt]
    $\sigma$($\mathrm{approx.~(NNLO+NNLL)}_{\mathrm{fid,}ee}$) &~\cite{Monni:2014zra} & 69.0 $\pm$ 2.7 \\[2pt]
    $\sigma$(NNLO $p_{\mathrm T}$-Resum$_{\mathrm{fid,}ee}$) &~\cite{Meade:2014fca} &                                                                                                                                                                                             65.5 $\pm$ 3.6 \\[2pt]
    \hline
    \vspace{-0.9em}  &  \\[2pt]
    
    Measured \xsecWWfidmumu&	& \WZfidXsecmm $^{+\WZfidXsecStatErrUpmm}_{\WZfidXsecStatErrDwmm}$(stat) $^{+\WZfidXsecSysErrUpmm}_{\WZfidXsecSysErrDwmm}$(syst) $\pm\WZfidXsecLumiErrUpmm$(lumi)  \\[2pt]
    
    $\sigma$(nNLO$_{\mathrm{fid,\mu\mu}}$)& ~\cite{Campbell:2011bn}+\cite{HiggsXS}&  63.7  $\pm$ 3.1 \\[2pt]
    $\sigma$(approx. NNLO$_{\mathrm{fid,\mu\mu}}$)&~\cite{Gehrmann:2014fva}+\cite{HiggsXS}&   68.6 $\pm$ 3.7 \\[2pt]
    $\sigma$($\mathrm{approx.~(NNLO+NNLL)}_{\mathrm{fid,\mu\mu}}$)& ~\cite{Monni:2014zra}&  75.1 $\pm$ 3.0 \\[2pt]
    $\sigma$(NNLO $p_{\mathrm T}$-Resum$_\mathrm{fid,\mu\mu}$)& ~\cite{Meade:2014fca} &  71.2 $\pm$ 4.0 \\[2pt]

    \hline
    \hline
  \end{tabular}
  \caption{Measured cross sections in the fiducial region for each channel as defined in Table~\ref{fiducialtruthlevelcuts}, 
  compared with various theoretical predictions described in the text of Section~\ref{sec:crosssections}. }
  \label{table:xsecMeasFid2}
\end{table}

\begin{table}[t]
  \centering
  \begin{tabular}{ll}
    \hline
    \hline
    Final state & Total cross section $pp \rightarrow WW$ [pb]
     \\ 
     \vspace{-0.9em}  &  \\ [2pt]
      \hline 
     \vspace{-0.9em}  &  \\[2pt]
     $e\mu$ 	& \WZtotXsecem $\pm \WZtotXsecStatErrUpem$(stat) $^{+\WZtotXsecSysErrUpem}_{\WZtotXsecSysErrDwem}$(syst) $\pm\WZtotXsecLumiErrUpem$(lumi)  \\  [2pt]
    \vspace{-0.9em}  &  \\[2pt]
    $ee$ 	& \WZtotXsecee $^{+\WZtotXsecStatErrUpee}_{\WZtotXsecStatErrDwee}$(stat) $^{+\WZtotXsecSysErrUpee}_{\WZtotXsecSysErrDwee}$(syst) $\pm\WZtotXsecLumiErrUpee$(lumi)  \\  [2pt]
        \vspace{-0.9em}  &  \\[2pt]
    \vspace{2 mm}	  $\mu\mu$ 	& \WZtotXsecmm $\pm \WZtotXsecStatErrUpmm$(stat) $^{+\WZtotXsecSysErrUpmm}_{\WZtotXsecSysErrDwmm}$(syst) $\pm\WZtotXsecLumiErrUpmm$(lumi)  \\ [2pt] 

    \vspace{2 mm}	  Combined 	& \WZtotXsec $\pm \WZtotXsecStatErrUp$(stat) $^{+\WZtotXsecSysErrUp}_{\WZtotXsecSysErrDw}$(syst) $\pm\WZtotXsecLumiErrUp$(lumi)  \\ [2pt] \hline 
\vspace{-0.9em}  &  \\[2pt]
    \centering  $\sigma($NNLO$_\mathrm{tot})$ theory prediction~\cite{Gehrmann:2014fva}+\cite{HiggsXS} 	&   $\nWWSMCrossSectionNNLO ^{+1.6}_{-1.4}$(scale)$\pm1.2$(PDF) \\[2pt]

    \hline
    \hline
  \end{tabular}
  \caption{Measured total $WW$ production cross sections in each final state together with the combined value, compared to the $\sigma($NNLO$_\mathrm{tot})$ theory prediction.}
  \label{table:xsecMeasTot}
\end{table}

\begin{figure}[p]
  \centering
  \includegraphics[width=.49\textwidth]{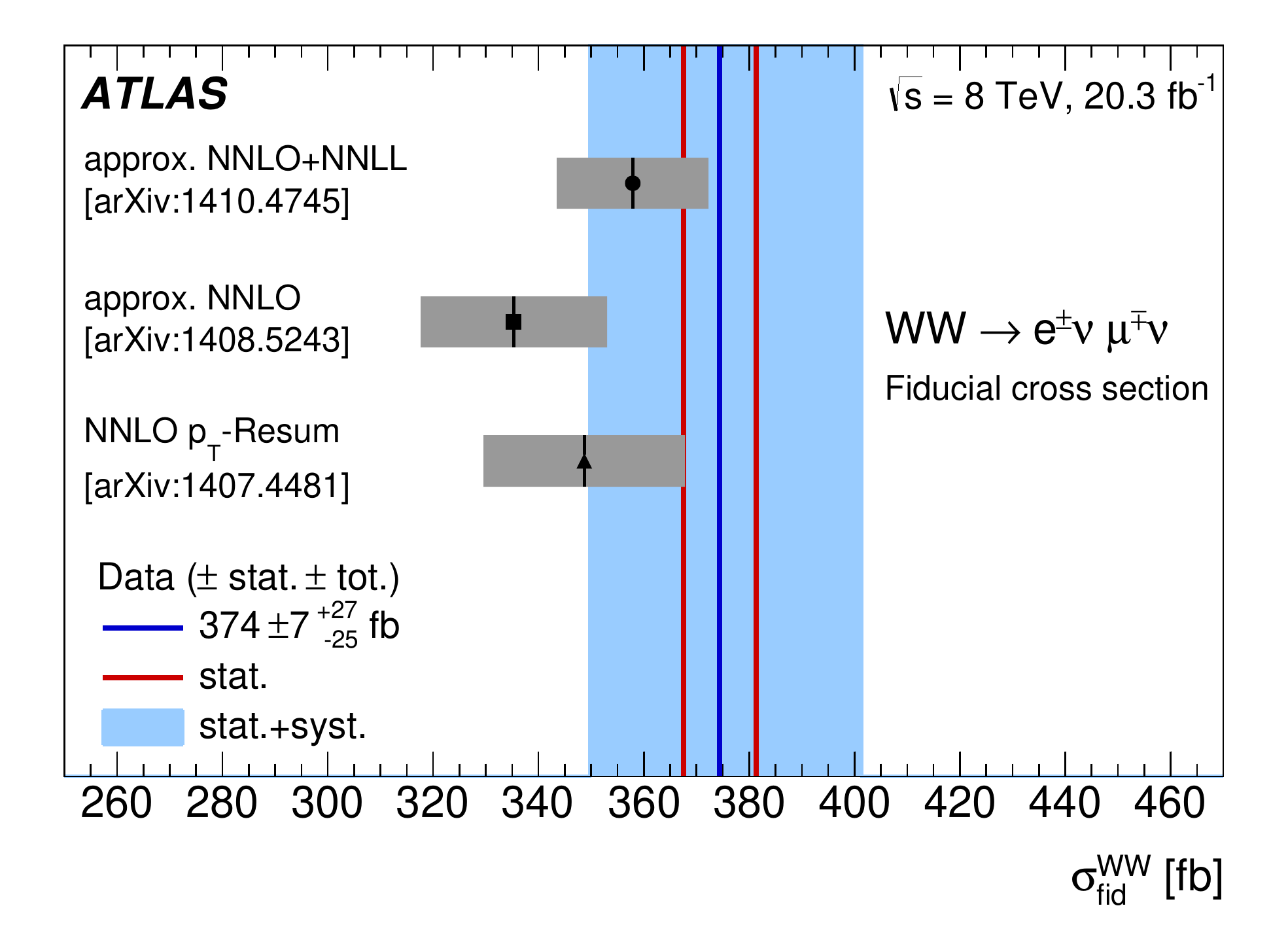}
  \includegraphics[width=.49\textwidth]{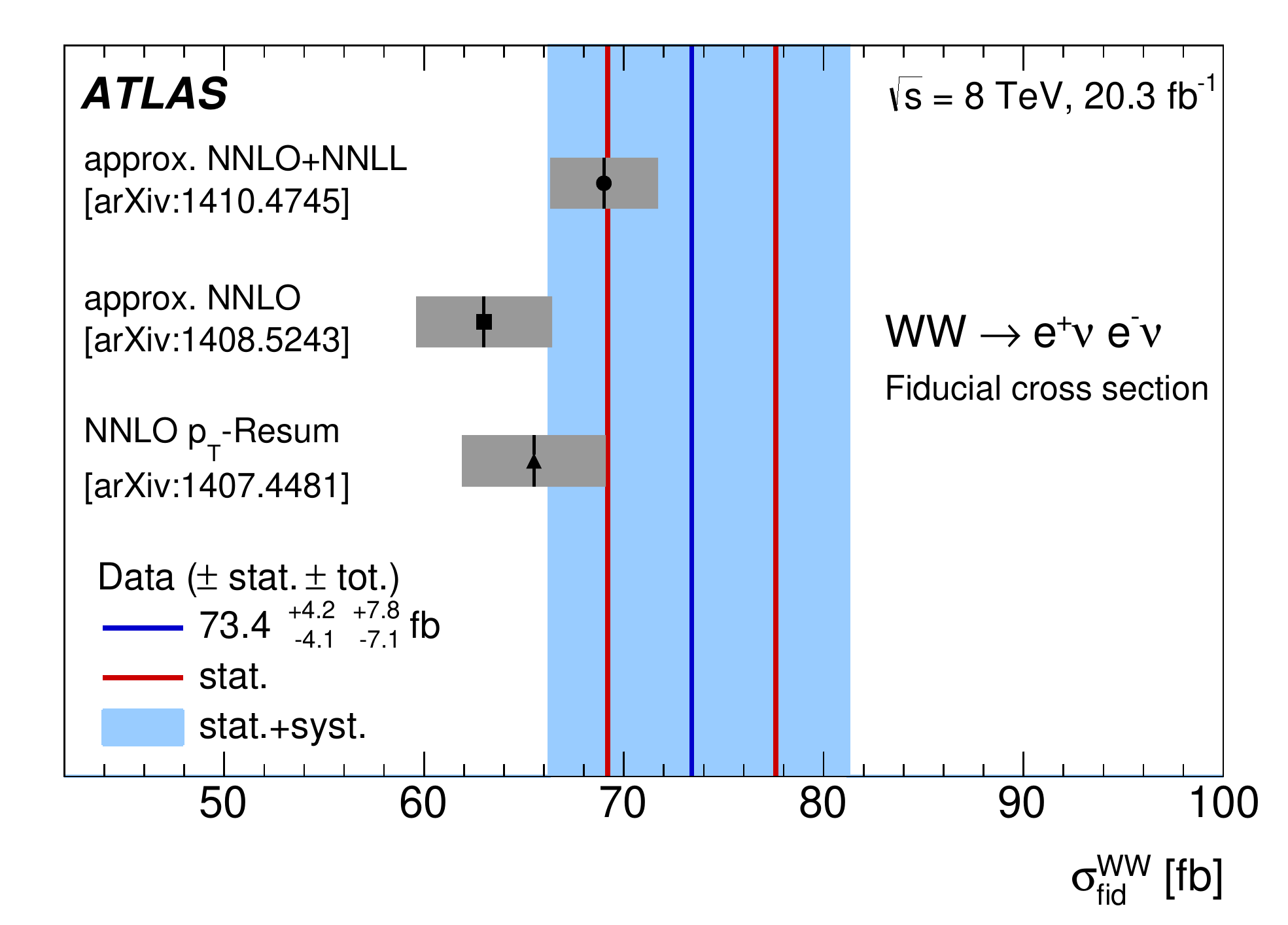}\\
  \includegraphics[width=.49\textwidth]{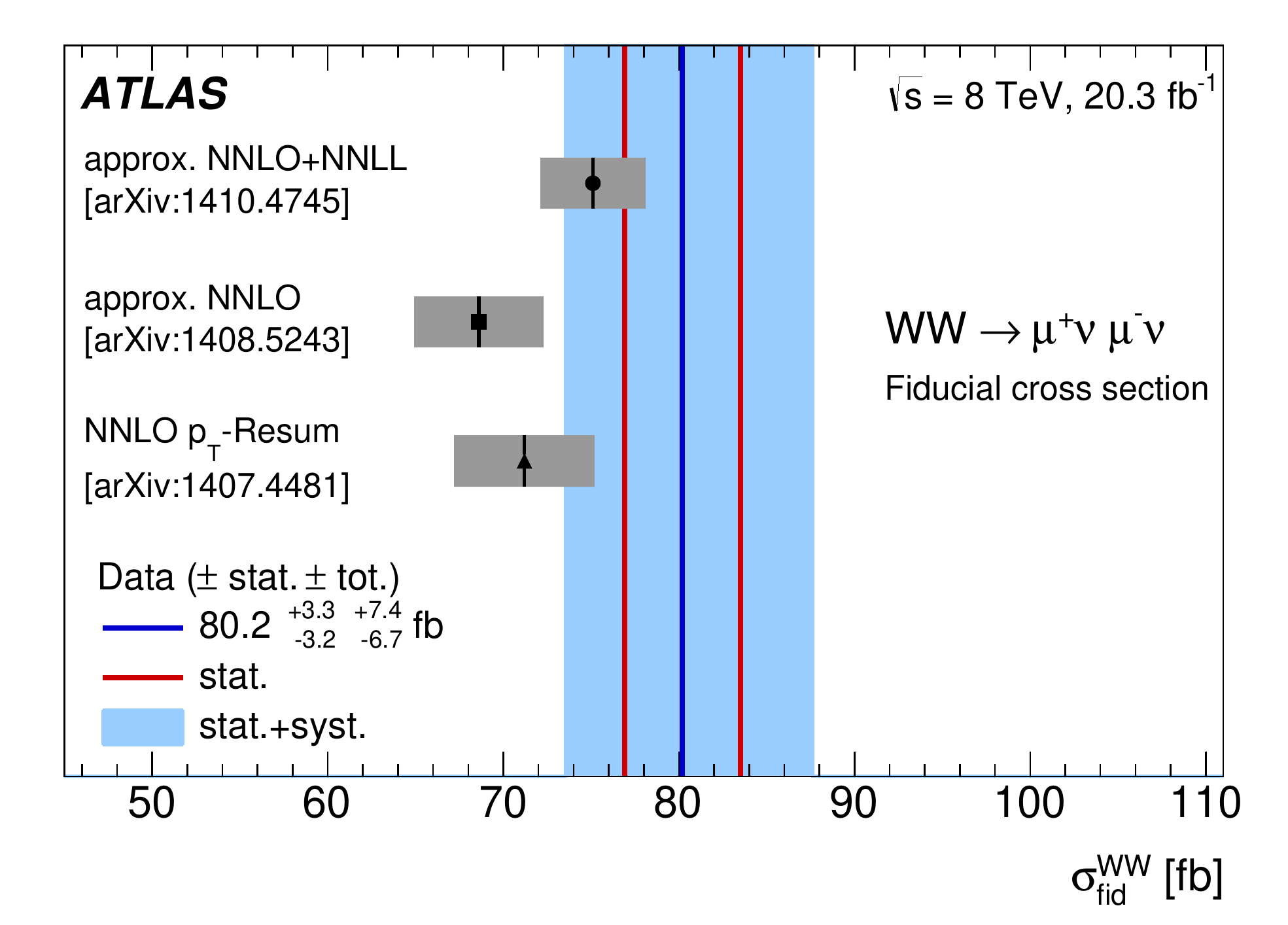}
  \includegraphics[width=.49\textwidth]{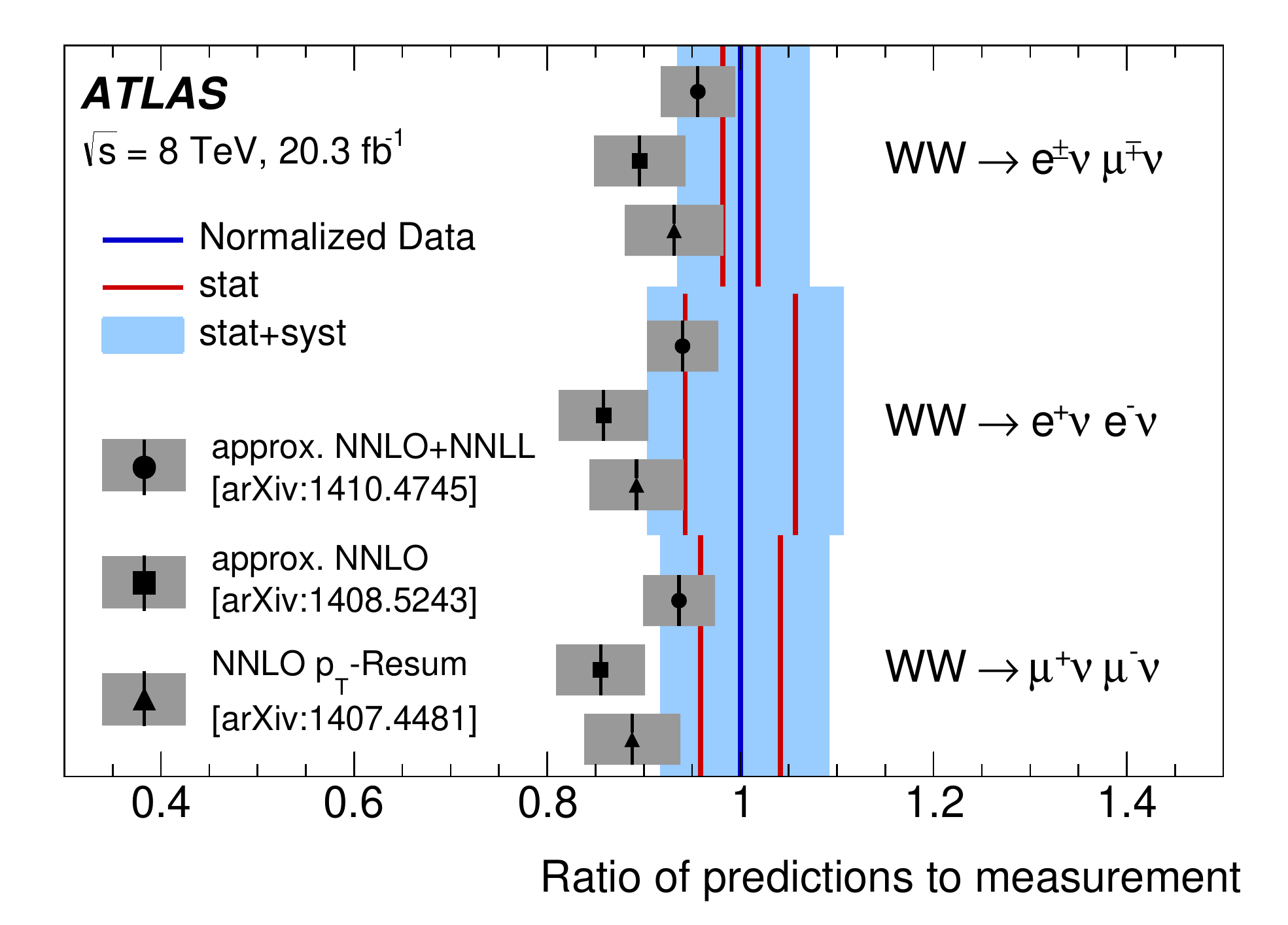}
  \caption{Comparison of the measured fiducial cross sections with various theoretical predictions. The comparison is made for all final states, $e\mu$ (top left), $ee$ (top right) and $\mu\mu$ (bottom left). The bottom right figure shows the measured and predicted fiducial cross sections normalised to the respective measured values for all final states. Theoretical predictions are indicated as black markers with grey error bands, while the central value of the measured cross sections is indicated by a blue line with red lines showing the statistical uncertainty and blue bands for the total uncertainty including statistical and systematic uncertainties.}
  \label{fig:fidplots}
\end{figure}

\begin{figure}[p]
  \centering
  \includegraphics[width=0.90\textwidth]{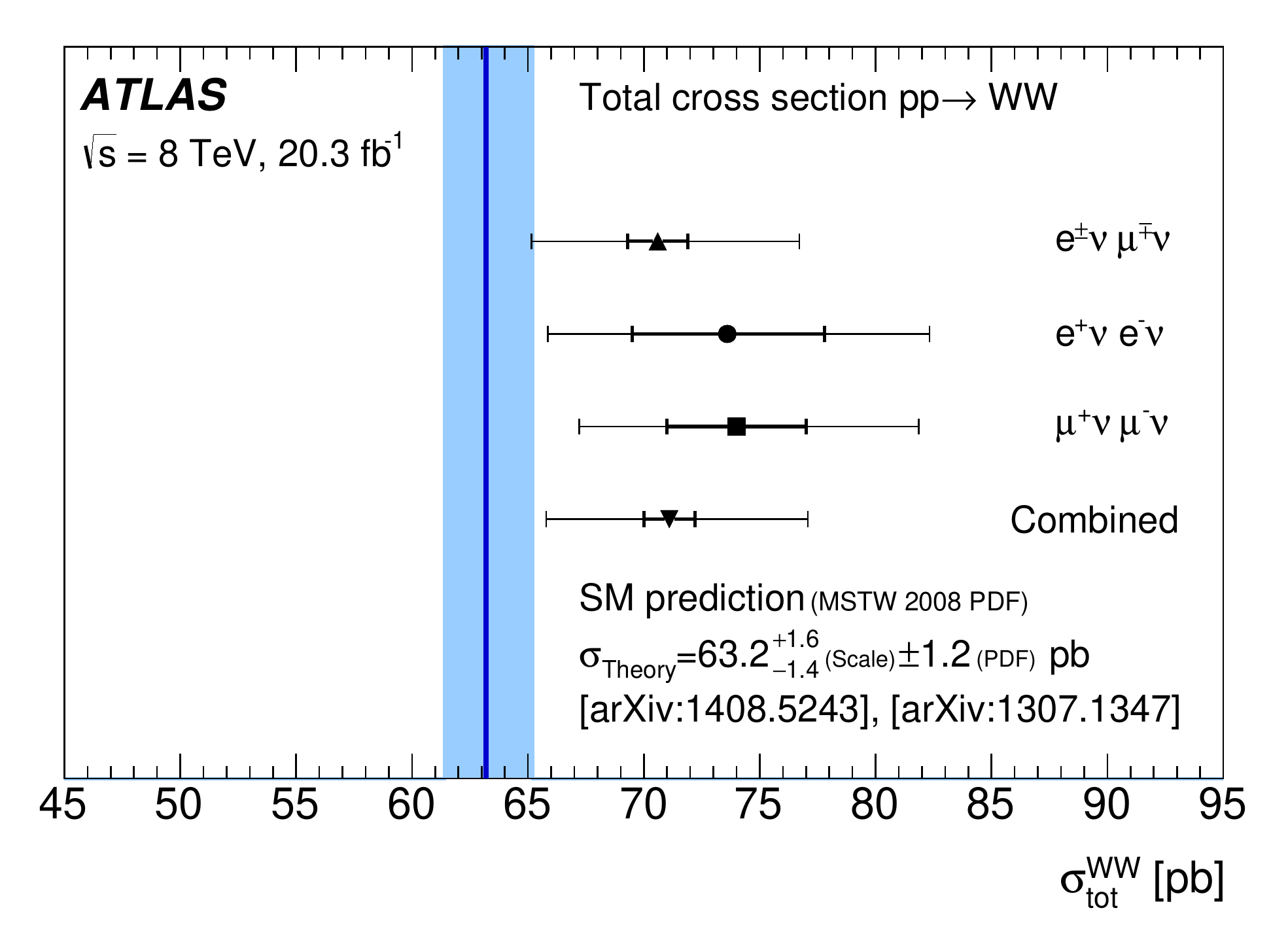}
  \caption
  {The $WW$ cross sections measured at $\sqrt{s}=8$~\TeV\ compared to the NNLO prediction. The uncertainties in the measurement show the statistical as well as the total uncertainty including the luminosity uncertainty. }
  \label{fig:NNLOverviewPlot}
\end{figure}

The measured unfolded differential cross sections are shown in Figure~\ref{fig::Analysis.Unfolding.Unnormalised.8TeV}. They are compared to predictions  
obtained using \Powheg+\Pythia\ for the \qqWW\ and $gg \rightarrow H \rightarrow W^+W^-$ production processes and \Ggtwoww+\Herwig+\Jimmy\ for non-resonant $gg$-induced $WW$ production. These predictions are added as described in Section~\ref{sec:MCSamples}. The data are also compared to an alternative prediction, where the \qqWW\ contribution is reweighted using the approximated resummed calculation from Ref.~\cite{Meade:2014fca} described above. For a third prediction, the \qqWW\ contribution is replaced using \MCatNLO+\herwig+\Jimmy. All three predictions are normalised to the NNLO theoretical prediction for the total cross section. 

The predictions generally undershoot the data, except for high \pt\ of the leading lepton, at high $m_{\ell\ell}$ and for high values of $\Delta\phi_{\ell\ell}$, where there is a small deficit in data compared to the expectation from either MC prediction. A small discrepancy between the MC predictions and the data is visible for the unfolded differential distribution of the transverse momentum of the leading lepton, $\pt^{\mathrm{lead}}$, which differs between \Powheg\ and \MCatNLO. The full difference between the two MC predictions is accounted for in Section~\ref{atgc}, when limits on anomalous couplings are extracted. The differences between data and MC simulation for the unfolded differential distributions in \absyll\ and \abscostheta\ are mostly constant as a function of these variables. 

In general, the shapes of the unfolded data distributions agree with either prediction at the level of $\pm$15\%. The $H\rightarrow  W^+W^-$ contribution to the
differential predictions is typically 2\% to at most 8.5\% per bin, and therefore has a small impact on the comparison.

\begin{figure}[p]
  \centering
  \includegraphics[width=0.4\textwidth]{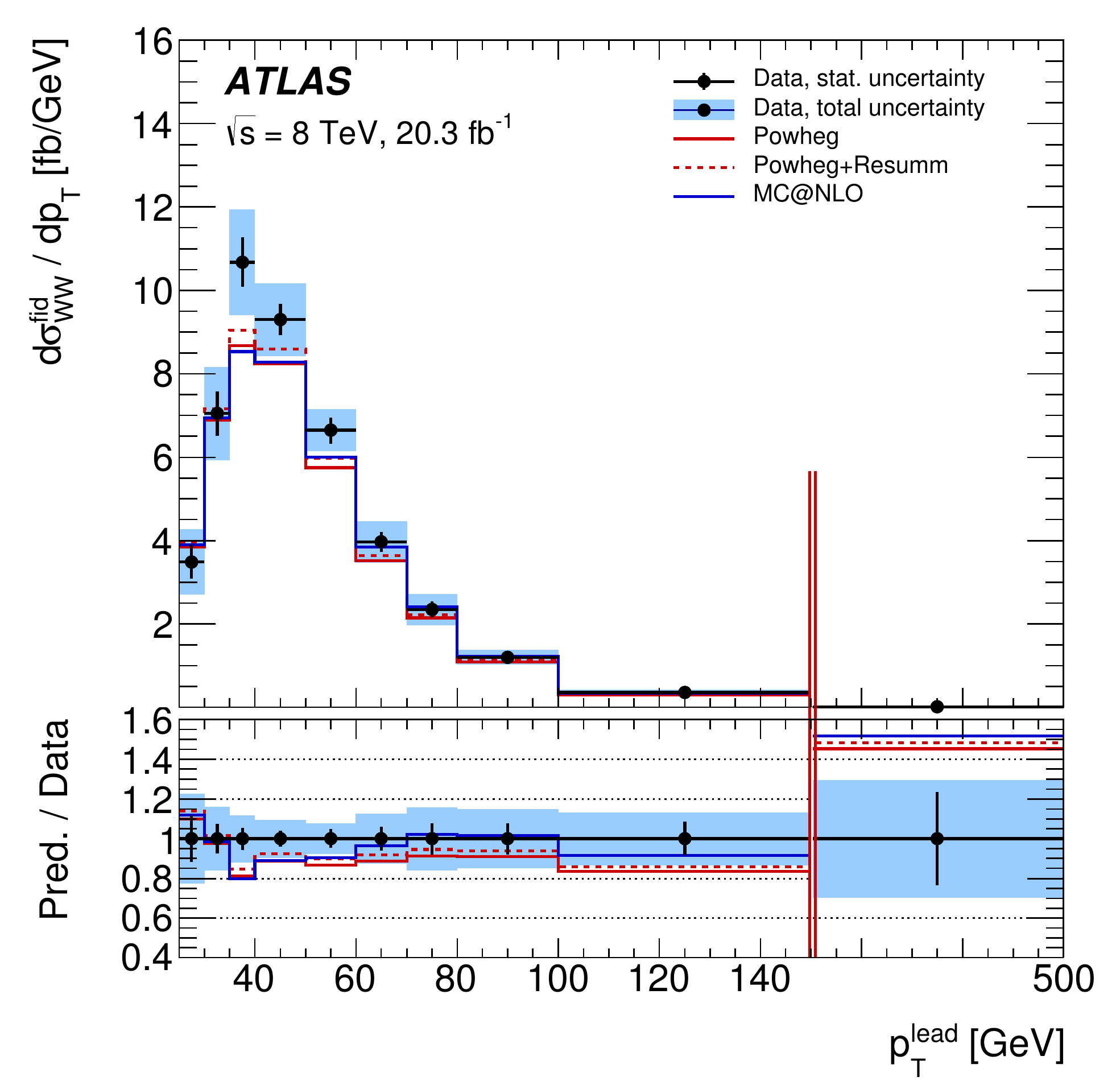}
  \includegraphics[width=0.4\textwidth]{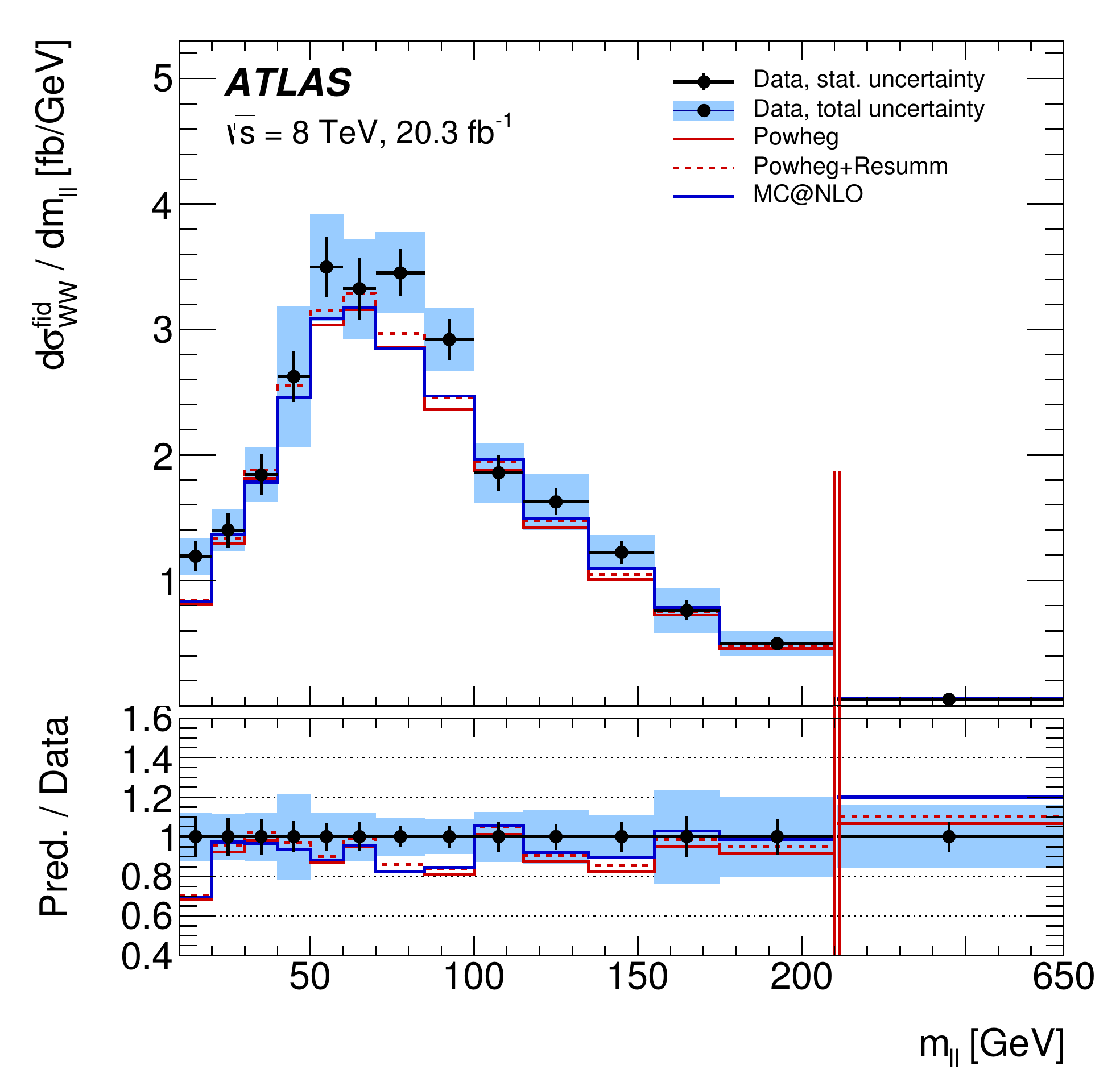}
  \includegraphics[width=0.4\textwidth]{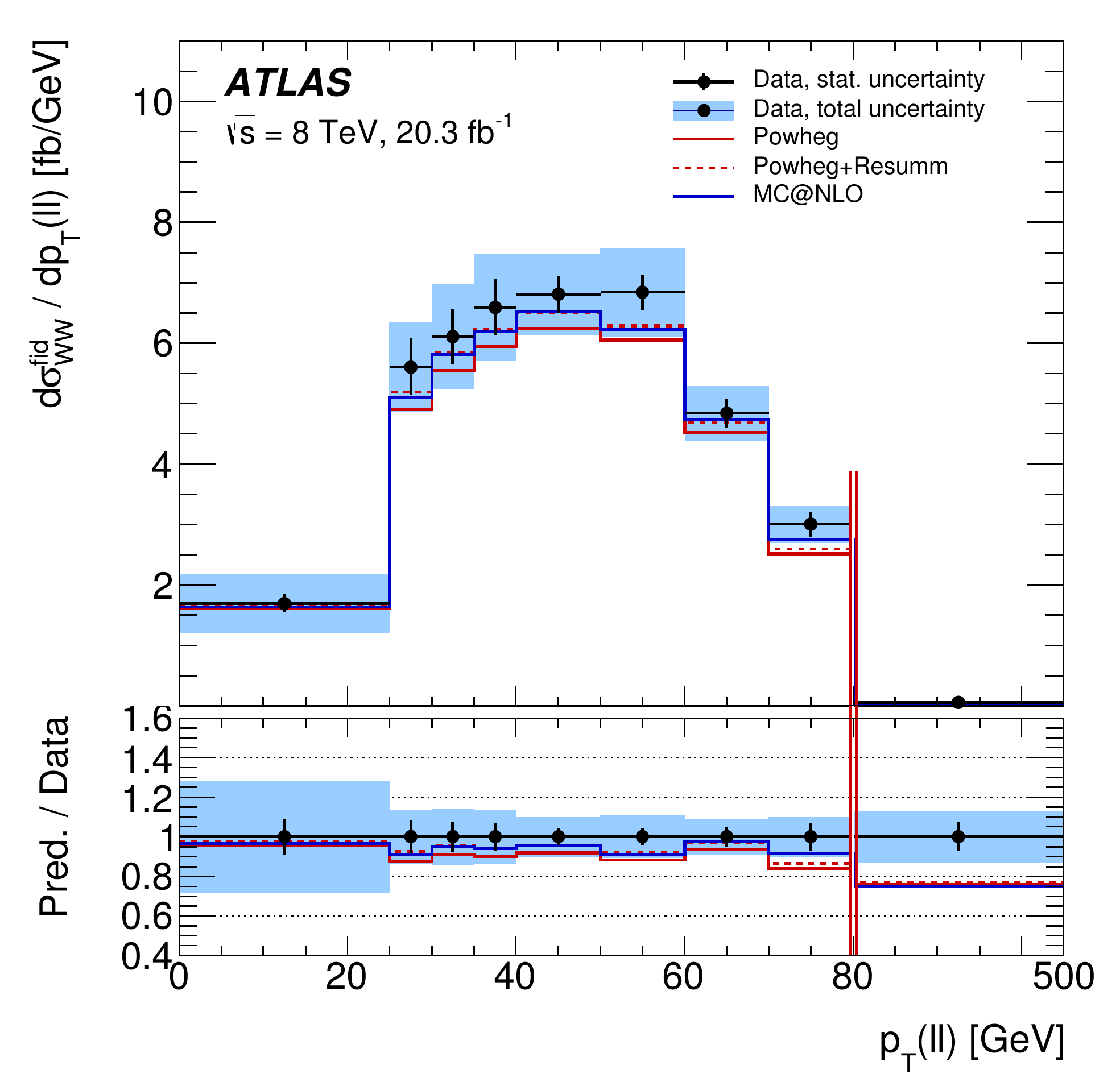}
  \includegraphics[width=0.4\textwidth]{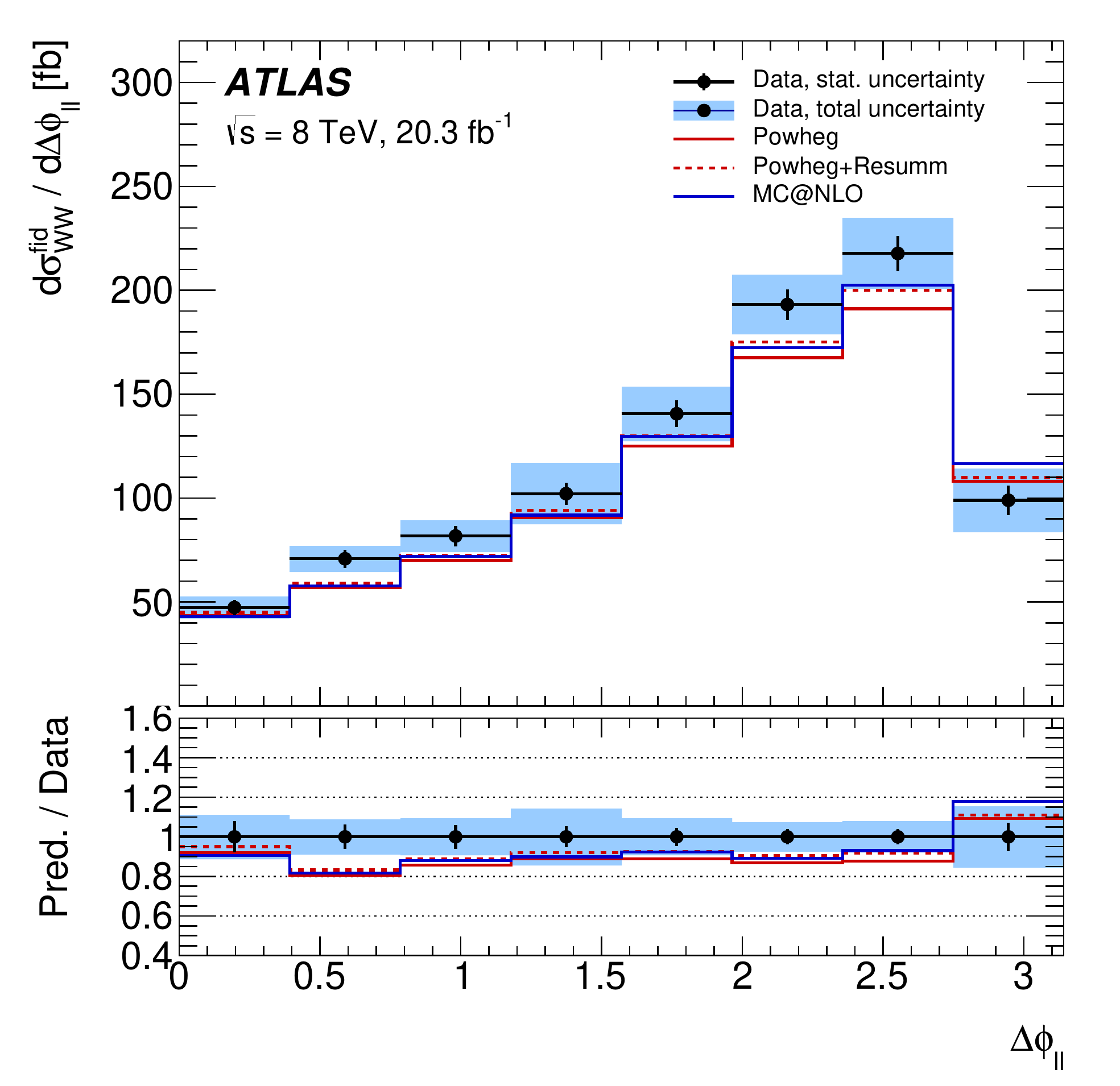}
  \includegraphics[width=0.4\textwidth]{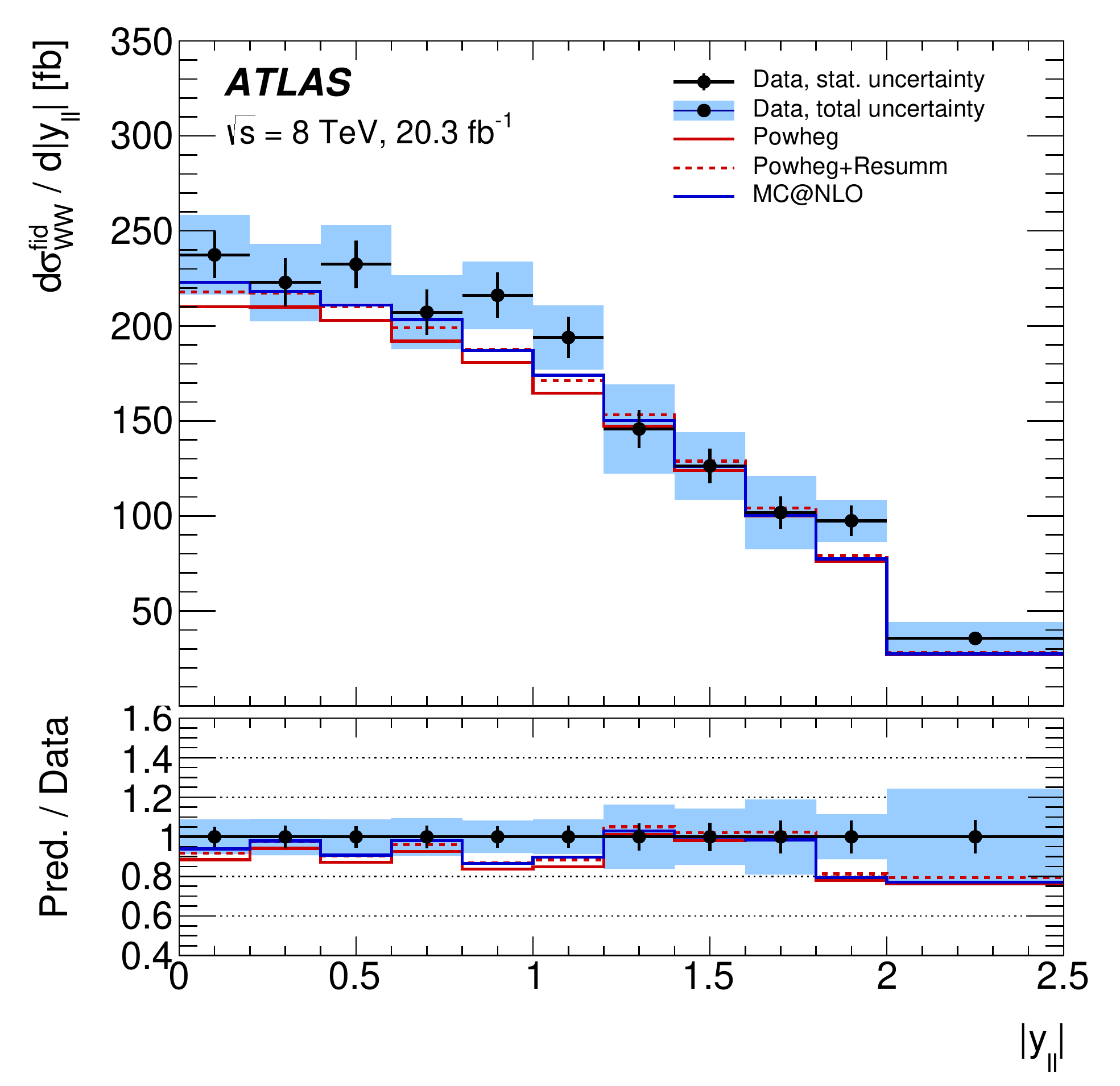}
  \includegraphics[width=0.4\textwidth]{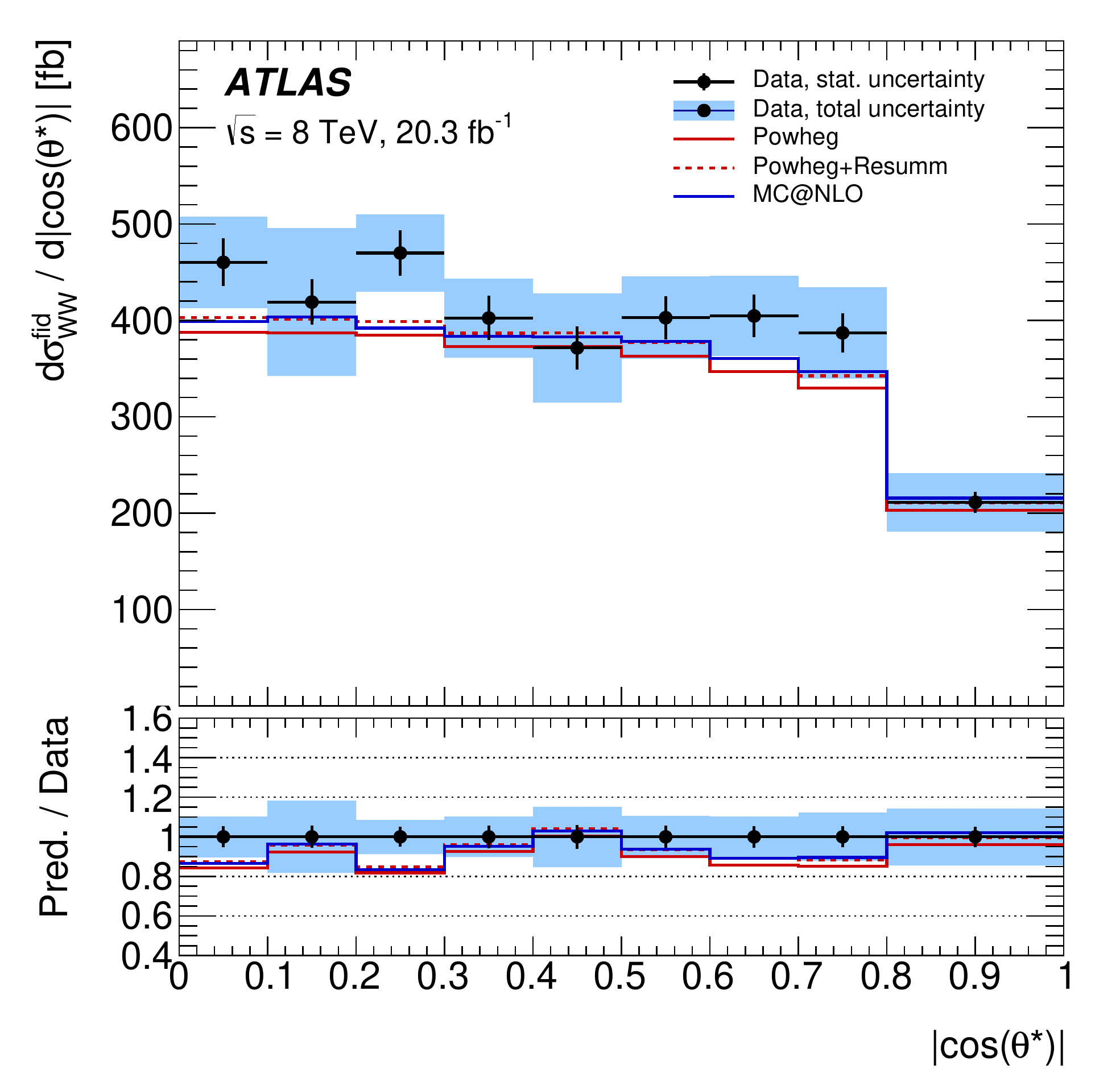}
  \caption 
  {Measured unfolded differential cross sections of $WW$ production in the $e\mu$ final state for 
   the transverse momentum of the leading lepton, $\pt^{\mathrm{lead}}$,  the invariant mass, $m_{\ell\ell}$, and the transverse momentum of the dilepton system, $\pt(\ell\ell)$,  as well as the difference in azimuthal angle between the decay leptons, $\Delta \phi_{\ell\ell}$, their combined rapidity, \absyll, and the observable \abscostheta. The measured cross-section values are shown as markers with error bars giving the statistical uncertainty and blue bands indicating the size of the total uncertainty. Three different MC predictions are compared to the measurement. The solid red line shows the nominal prediction, whilst the dashed red line shows the prediction in case the \qqWW\ contribution is replaced by the \Powheg+\Pythia\ prediction reweighted to the resummed calculation of Ref.~\cite{Meade:2014fca}. The blue line depicts a prediction obtained using \MCatNLO+\herwig+\Jimmy for the \qqWW\ contribution. All three predictions are normalised to the NNLO theoretical prediction for the total cross section. For the top three histograms, double red lines indicate changes in the $x$-axis scale.
  }
  \label{fig::Analysis.Unfolding.Unnormalised.8TeV}
\end{figure}

Tables~\ref{tab::Analysis.Results.Unfolding.8TeV.pt1} to \ref{tab::Analysis.Results.Unfolding.8TeV.AbsCosTheta} in the appendix give an overview of the measured unfolded differential cross sections and the statistical, experimental and background uncertainties in the measurement. The bin-to-bin correlations are preserved for each source of systematic uncertainty and the correlation matrices are made available in the appendix. The systematic uncertainties are treated as fully correlated. This includes the background uncertainties, except the uncertainties due to the limited statistics of the MC simulation and the uncertainties related to the $W$+jets estimate, specifically the uncertainties on the measured fake lepton efficiencies and the sample dependence, since both these uncertainties have a large statistical component. The background uncertainties are added in quadrature to the statistical and experimental uncertainties to obtain the total uncertainty in each bin. The total uncertainties range from 10\% to 30\%. Normalised unfolded differential cross-section distributions are also measured, as these are more suited to analysis of the shapes of differential distributions. Results and tables are made available in the appendix.

\clearpage
 
\section{Limits on anomalous triple-gauge-boson couplings}
\label{atgc}

\subsection{Theoretical parameterisation}
\label{atgc_par}

The non-Abelian self-couplings of $W$ and $Z$ bosons and photons can be probed via the $WWV$ vertex, where $V$ = $Z$ or $\gamma$, present when the bosons are produced via the $s$-channel exchange of a $Z$ or $\gamma$ as shown in Figure~\ref{fig:ww_prod}(b). The SM, with its $\mathit{SU}(2)\times U(1)$ structure, makes definite predictions for these triple-gauge-boson couplings~\cite{Hagiwara:1986vm}. The SM Lagrangian can be extended with additional degrees of freedom that modify the couplings. Considering only terms that conserve charge conjugation ($C$) and parity ($P$) separately, the modified Lagrangian can be written as:

\begin{equation}
\mathcal{L} = i g_{WWV}\left[g_{1}^{V}(W_{\mu\nu}^{+} W^{-\mu} - W^{+\mu}W^{-}_{\mu\nu})V^{\nu}
	+ \kappa^{V} W^{+}_{\mu} W^{-}_{\nu} V^{\mu\nu}
	+ \frac{\lambda^{V}}{m_{W}^{2}} W^{+\nu}_{\mu} W^{-\rho}_{\nu} V^{\mu}_{\rho}\right],
\end{equation}
where $V$ = $Z$ or $\gamma$; $W^{\pm}_{\mu\nu} = \partial_{\mu}W^{\pm}_{\nu} - \partial_{\nu}W^{\pm}_{\mu}$;
$V_{\mu\nu} = \partial_{\mu}V_{\nu} - \partial_{\nu}V_{\mu}$. The overall coupling constants $g_{WWV}$ are given by $g_{WW\gamma} = -e$ and $g_{WWZ} = -e \cot\theta_{W}$, where $\theta_{W}$ is the weak mixing angle.

Electromagnetic gauge invariance requires that $g_{1}^{\gamma}=1$. The three other coupling parameters that are non-zero in the SM are $g_{1}^{Z}=1$, $\kappa^{Z}=1$, and $\kappa^{\gamma}=1$. Deviations from the SM are introduced as
\begin{equation}
\Delta g_{1}^{Z} = 1 - g_{1}^{Z};\qquad
\Delta \kappa^{Z} = 1 - \kappa^{Z};\qquad
\Delta \kappa^{\gamma} = 1 - \kappa^{\gamma}.
\end{equation}
The remaining couplings are zero in the SM, $\lambda^{\gamma}=\lambda^{Z}=0$. A significant non-zero value for any of the parameters $\Delta g_{1}^{Z}$, $\Delta \kappa^{Z}$, $\Delta \kappa^{\gamma}$, $\lambda^{\gamma}$ and $\lambda^{Z}$ would be evidence of new interactions not included in the SM.

If anomalous couplings occur, these extra terms in the Lagrangian would contribute and would induce a violation of unitarity at sufficiently high energies. Therefore, form factors are introduced to dampen the rise of the $WW$ production cross section so that it takes physical values even at the highest partonic centre-of-mass energies relevant for 8 \TeV\ $pp$ collisions:

\begin{equation}
\Delta g_{1}^{V} \rightarrow \cfrac{\Delta g_{1}^{V}}{\left(1+\cfrac{\hat{s}}{\Lambda^{2}}\right)^{2}} ~, \qquad
\Delta \kappa^{V} \rightarrow \cfrac{\Delta \kappa^{V}}{\left(1+\cfrac{\hat{s}}{\Lambda^{2}}\right)^{2}} ~, \qquad
\lambda^{V} \rightarrow \cfrac{\lambda^{V}}{\left(1+\cfrac{\hat{s}}{\Lambda^{2}}\right)^{2}} ~, \\
\end{equation}
where $\hat{s}$ is the square of the invariant mass of the vector boson pair. The form-factor scale, $\Lambda$, is 
typically taken to be in the \TeV\ range. Upper bounds on the size of the anomalous gauge boson couplings can be derived as a function of $\Lambda$ based on unitarity considerations~\cite{FERMILAB-Pub-95-031}.

Several restrictions can be put on the couplings and are explored in this paper in addition to the scenario where none of the couplings is restricted per se:  the \textit{Equal Couplings} constraint assumes the coupling parameters for the $WWZ$ and $WW\gamma$ vertices to be equal. Hence, $g_{1}^{Z} = g_{1}^{\gamma} = 1$, which leaves only two independent parameters: $\Delta \kappa^{\gamma} = \Delta\kappa^{Z}$ and $\lambda^{\gamma} = \lambda^{Z}$. Imposing $\mathit{SU}(2)\times U(1)$ symmetry for the effective field operators~\cite{Gounaris:1996rz} suggests the following constraint

\begin{align}
\label{eqn:LEP_constraint}
\Delta g_{1}^{Z} &= \Delta \kappa^{Z} + \tan^{2}\theta_{W} \Delta \kappa^{\gamma} \nonumber, \\
\lambda^{\gamma} &= \lambda^{Z},
\end{align}

where $\theta_{W}$ is the weak mixing angle. This constraint is called the \textit{LEP} constraint hereafter since it was introduced due to the limited statistics available at LEP for anomalous TGC studies.
Assuming the absence of cancellations between tree-level and one-loop contributions, leads to the \textit{Hagiwara--Ishihara--Szalapski--Zeppenfeld (HISZ) }constraint scenario~\cite{HISZ1993} with two free parameters where the following relations hold:

\begin{align}
\label{eqn:HISZ_constraint}
\Delta g_{1}^{Z} &= \frac{\Delta \kappa^{Z}}{\cos^{2}\theta_{W} - \sin^{2}\theta_{W}} \nonumber, \\
\Delta \kappa^{\gamma} &= 2\Delta\kappa^{Z}\frac{\cos^{2}\theta_{W}}{\cos^{2}\theta_{W} - \sin^{2}\theta_{W}} \nonumber, \\
\lambda^{\gamma} &= \lambda^{Z}.
\end{align}

An alternative way to parameterise new physics in diboson production processes is based on \textit{effective field theory (EFT)}~\cite{EFT}. It removes two complications of the generalised Lagrangian described above: it respects $\mathit{SU}(2)\times U(1)$ gauge invariance and does not introduce arbitrary form factors, though it assumes that higher-dimensional operators are suppressed by the mass scale of new physics. In the effective field theory approach, the effective Lagrangian is an expansion in operators that are $\mathit{SU}(2)\times U(1)$ gauge invariant and conserve charge conjugation and parity.
The dimensionless coefficients, $C_{i}$, parameterise the strength of the coupling between new physics and SM particles:
\begin{equation}
\mathcal{L} = \mathcal{L}_{\rm SM} + \sum_{i}{\frac{C_{i}}{\Lambda^{2}}\mathcal{O}_{i}}.
\end{equation}
There are five dimension-six operators, $\mathcal{O}_{i}$, but only three of those conserve $C$ and $P$ and are considered in the following:
\begin{align}
\mathcal{O}_{WWW} &=  \mathrm{Tr}[{W_{\mu\nu}W^{\nu\rho}W^{\mu}_{\rho}}] \nonumber, \\
\mathcal{O}_{W} &= \left(D_{\mu} \Phi \right)^{\dagger}W^{\mu\nu}\left(D_{\nu} \Phi \right) \nonumber, \\
\mathcal{O}_{B} &= \left(D_{\mu} \Phi \right)^{\dagger}B^{\mu\nu}\left(D_{\nu} \Phi \right),
\end{align}
where $\Phi$ is the Higgs boson doublet field and
\begin{align}
D_{\mu} &= \partial_{\mu} + \frac{i}{2}g\tau^{I}W_{\mu}^{I}
+\frac{i}{2}g'B_{\mu} \nonumber, \\
W_{\mu\nu} &= \frac{i}{2}g\tau^{I}\left( \partial_{\mu}W_{\nu}^{I}
-\partial_{\nu}W_{\mu}^{I} + g \epsilon_{IJK}W_{\mu}^{J}W_{\nu}^{K} \right)\nonumber, \\
B_{\mu\nu} &= \frac{i}{2}g' \left(\partial_{\mu}B_{\nu} -
\partial_{\nu}B_{\mu} \right),
\end{align}

with $I=1,2,3$ and similarly for $J$ and $K$. The free parameters of the effective field theory approach used here are $C_{WWW}/\Lambda^2 , C_{W}/\Lambda^2$ and $C_{B}/\Lambda^2$.
The parameter $C_{W}/\Lambda^2$ also affects the Higgs production processes predicted to contribute at the 3\% level in the SM. Possible enhancements of the Higgs production process are neglected in the subtraction of the resonant $gg$-induced $H\rightarrow WW$ production process as background and are fully attributed to the \qqWW\ process in the limit setting. 

\subsection{Confidence intervals for the aTGC parameters}
\label{sec:atgc_results}

Anomalous triple-gauge-boson couplings (aTGCs) can modify the $WW$ production cross section at large $\hat{s}$. A number of variables were investigated, and the transverse momentum of the leading lepton, $p_{\rm T}^{\mathrm{lead}}$, was found to be particularly sensitive to aTGCs and is therefore used to extract limits on the aTGC parameters. The extraction is based on detector-level distributions. A profile-likelihood-ratio test statistic~\cite{frequentist} is used to check whether the data and predictions with aTGCs are compatible. Then a frequentist method~\cite{PhysRevD.57.3873} is used to determine the 95\% confidence interval for the aTGC parameters. The $e\mu$ final state constitutes a major fraction of the selected data sample and has a higher signal-to-background ratio than the $ee$ and $\mu\mu$ final states; therefore only the $e\mu$ events are used in the limit extraction.

The likelihood function used in the test statistic is the product of Poissonian probability density functions over the considered $p_{\rm T}^{\mathrm{lead}}$ bins (150--250 \GeV, 250--350 \GeV~and 350--1000 \GeV) where the binning extends to large $p_{\rm T}^{\mathrm{lead}}$ to maximise the sensitivity to anomalous couplings. However, the range shown in figure~\ref{fig::Analysis.Unfolding.Unnormalised.8TeV} is smaller as it has been optimised with respect to the uncertainties of the measured cross section based on the observed number of events. The binning for the aTGC analysis has been optimised using Asimov data~\cite{Cowan:2010js}. Events with $p_{\rm T}^{\mathrm{lead}}$ below 150 GeV are not considered because aTGC signals are not expected to contribute here significantly.

The number of observed data events and the prediction for the signal and background processes are used to construct the Poissonian probability density functions, in which systematic uncertainties are considered as nuisance parameters each constrained with a Gaussian distribution. A reweighting procedure implemented in \MCatNLO~\cite{MCatNLO} is used to obtain the signal predictions for arbitrary values of aTGCs; therefore in the study of aTGCs the SM \qqWW\ events are modelled using \MCatNLO~interfaced to \herwig+\Jimmy. Only the \qqWW\ process is considered as signal. 
The process $gg\rightarrow WW$, which includes resonant $H \rightarrow W^+W^-$ production, is considered as background, where the effects of possible anomalous couplings on the $H \rightarrow W^+W^-$ vertex are neglected. 

Next-to-leading-order electroweak corrections to SM $WW$ production are considered in the extraction of aTGC limits~\cite{WWEWKCorr,WWrr,PhysRevD.88.113005,Bierweiler:2013dja}. The correction is negative and becomes more significant in the high-$p_{\rm T}^{\mathrm{lead}}$ region.
Table~\ref{tab:ewcorrbinned} gives the relative size of the correction and its uncertainty. An additional systematic uncertainty in modelling the shape of the $p_{\rm T}^{\mathrm{lead}}$ distribution for the \qqWW\ process is estimated by comparing the predictions from \MCatNLO\ and \Powheg+\Pythia~8. The difference between these two MC predictions can be as large as 20\% in the high $p_{\rm T}^{\mathrm{lead}}$ bins. The shape of the $p_{\rm T}^{\mathrm{lead}}$ distribution is found to be less dependent on other theoretical modelling uncertainties that are described in Section~\ref{sec:systematics}. Experimental resolution and background uncertainties are fully accounted for. 

\begin{table}[!ht]
\centering
\begin{tabular}{lccccc}
\hline
 $p_{\rm T}^{\mathrm{lead}}$ [\GeV] &25--75 & 75--150 & 150--250 & 250--350 & 350--1000 \\
\hline
$\mathrm{SF}_{\rm EW}$ & $< 1\%$ & $-$4\% & $-$10\% & $-$16\% & $-$24\% \\
$\delta \mathrm{SF}_{\rm EW}$ & 0.1\% & $<$0.5\% & 2\% & 4\% & 7\% \\
\hline
\end{tabular}
\caption{Size of the next-to-leading-order EWK correction scale factor~\cite{Gieseke:2014gka}, $\mathrm{SF}_{\rm EW}$, and its systematic uncertainty ($\delta \mathrm{SF}_{\rm EW}$) in each bin of $p_{\rm T}^{\mathrm{lead}}$.
\label{tab:ewcorrbinned}
}
\end{table}

Figure~\ref{fig:atgc_reweighted_to_fitted} compares the detector-level $p_{\rm T}^{\mathrm{lead}}$ distribution with the SM prediction as well as the predictions for non-zero aTGC parameters, which are defined in the \textit{no constraints} scenario that assumes no correlation between the parameters. In the left plot the predictions with arbitrarily large aTGC parameters are shown to demonstrate the effect of anomalous-triple-gauge-boson couplings on the distribution. In comparison, the right plot shows the predicted shapes with the values of aTGC parameters corresponding to the upper bounds of the observed 95\% confidence interval.

\begin{figure}[!ht]
\centering 
	{\includegraphics[width=0.49\textwidth]{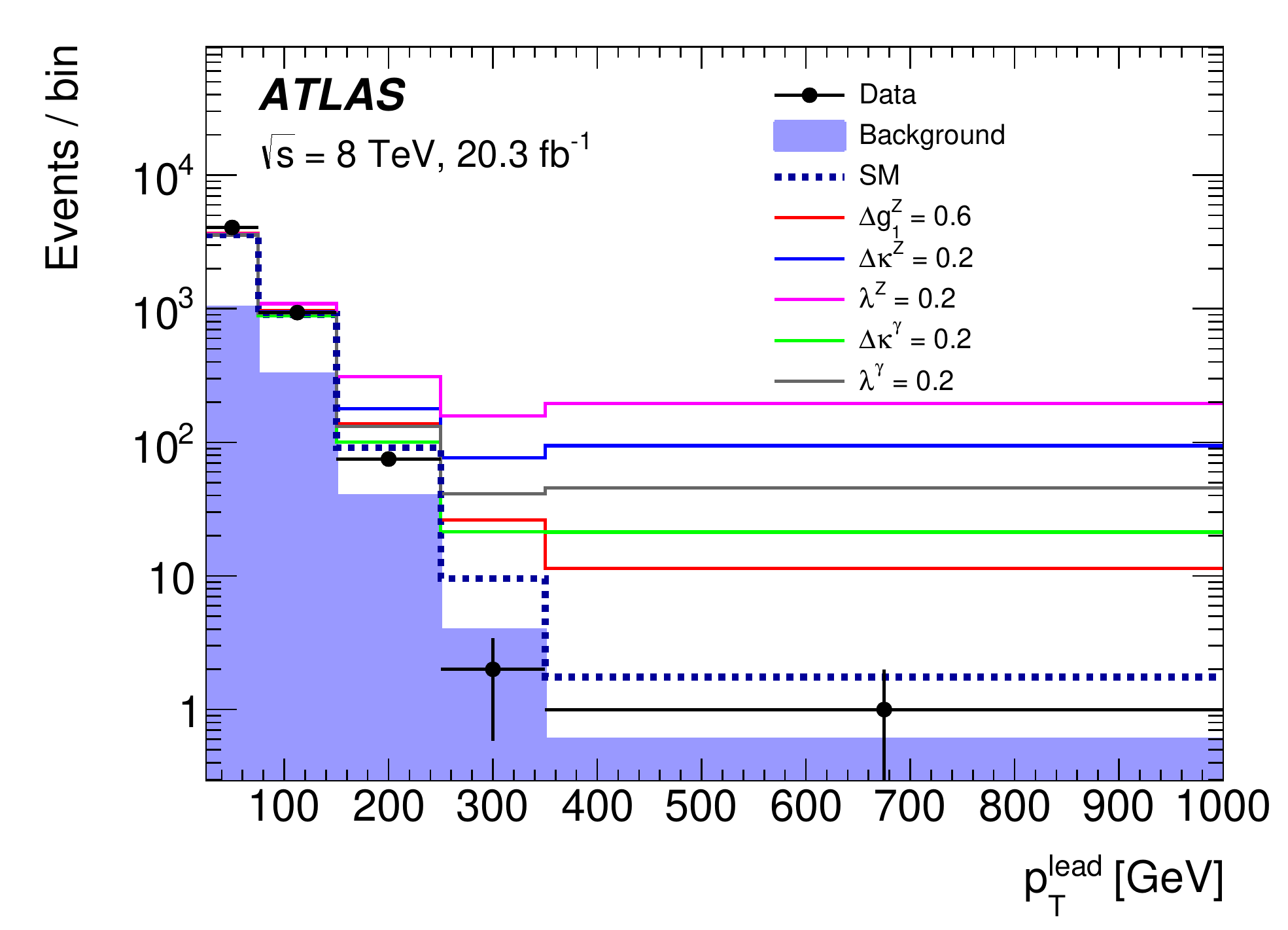}}
	{\includegraphics[width=0.49\textwidth]{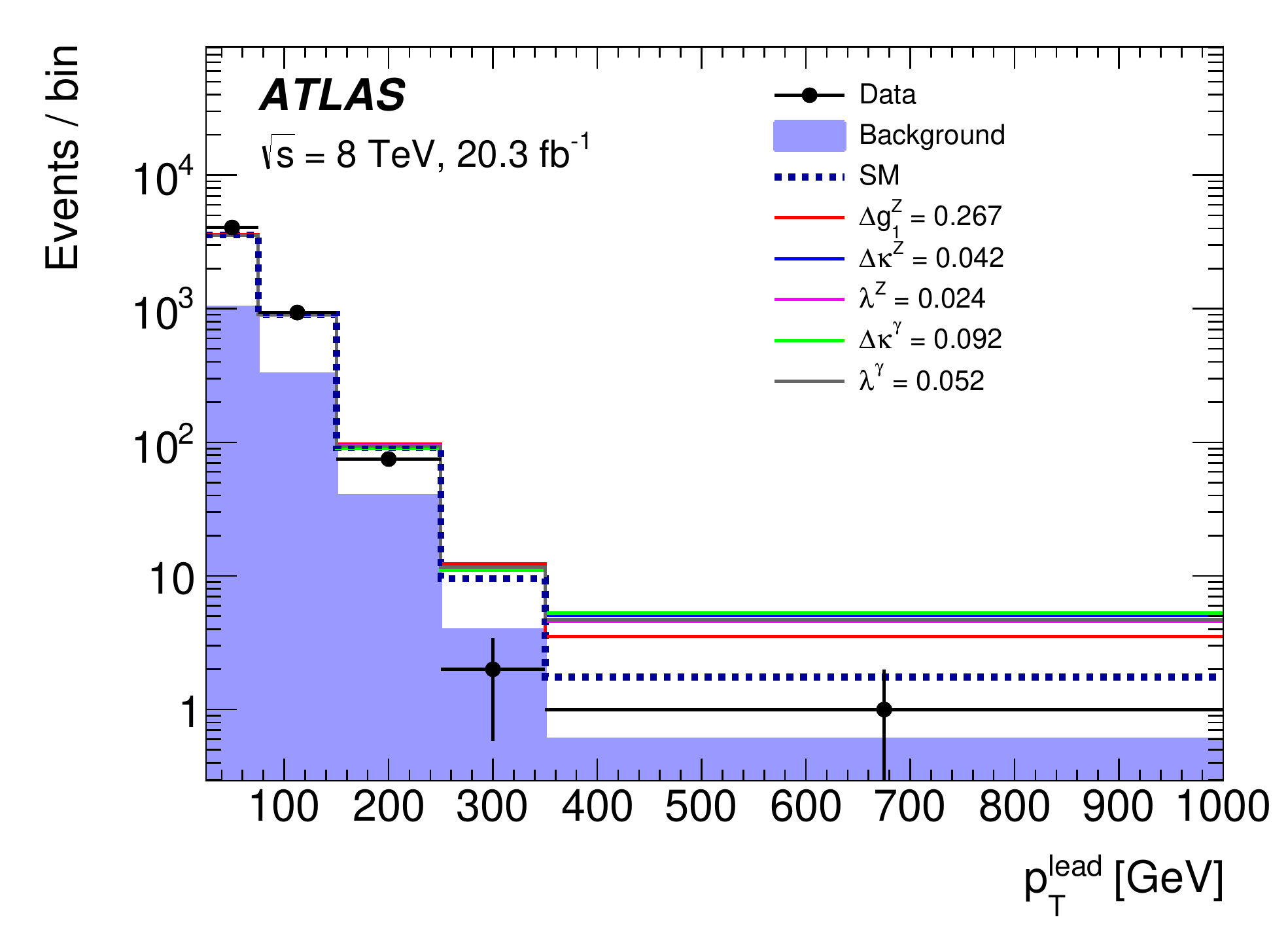}}
	\caption{The leading lepton transverse momentum, $p_{\rm T}^{\mathrm{lead}}$, for $e\mu$ final states is compared for data and MC-generated events using different arbitrary values for aTGC parameters (left). The detector-level distributions are shown using values of aTGC parameters corresponding to the 
	upper bounds of the observed 95\% confidence interval (right). 
    The aTGC parameters are defined in the \textit{no constraints} scenario, 
    and the form-factor scale is set to be infinity. The next-to-leading-order EWK correction scale factors from Table~\ref{tab:ewcorrbinned} have been applied here. Except for the anomalous coupling parameter under study, all others are set to zero.
    }
	\label{fig:atgc_reweighted_to_fitted}
\end{figure}

To derive the confidence interval for some specific anomalous coupling parameters in any of the described scenarios, the other parameters are set to their SM values. Table~\ref{tab:atgc_expected_observed_limits_skip2} gives the expected and observed 95\% confidence interval for each of the anomalous coupling parameters defined in the \textit{no constraints}, \textit{LEP}, \textit{HISZ} and \textit{Equal Couplings} scenarios. The limits are obtained with both $\Lambda=\infty$ and $\Lambda=7$~\TeV. A form-factor scale of 7~\TeV\ is chosen as the largest value allowed by the unitarity requirement~\cite{FERMILAB-Pub-95-031} for most aTGC parameters. The confidence intervals for the effective field theory approach are given in Table~\ref{tab:atgc_expected_observed_limits_skip2_EFT}.
Figure~\ref{fig:atgc_expected_limits_skip2_formfactors} shows the expected and observed limits at 95\% confidence level (C.L.), in red and black respectively, and the theoretical constraint due to the unitarity requirement (shown as blue dashed lines) as a function of form-factor scales from $\Lambda=2$~\TeV\ to $\Lambda=10$~\TeV.
The largest value of form-factor scales that can preserve unitarity is $\sim$7--$9$~\TeV\ for most aTGC parameters, while it is only about 3~\TeV\ for $\Delta g_{1}^{Z}$. All observed limits are more stringent than the expected limits because the data distribution falls more steeply than expected and a deficit of events is observed for the highest $p_\mathrm{T}^{\mathrm{lead}}$ bins.

\begin{table}[h!t]
\centering
\begin{tabular}{lccccc}
\hline
Scenario & Parameter            & Expected & Observed   & Expected & Observed \\
	 & 			& \multicolumn{2}{c}{$\Lambda=\infty$}  & \multicolumn{2}{c}{$\Lambda=7$~\TeV}  \\
\hline
\multirow{5}{*}{\parbox{2.31cm}{No constraints scenario}}
&$\Delta g_{1}^{Z}$       & [$-$0.498, 0.524]  & [$-0.215$, 0.267] 		           & [$-0.519$, 0.563]  & [$-0.226$, 0.279] \\
&$\Delta\kappa^{Z}$           & [$-$0.053, 0.059]  & [$-0.027$, 0.042] 	           & [$-0.057$, 0.064]  & [$-0.028$, 0.045] \\
&$\lambda^{Z}$            & [$-0.039$, 0.038]  & [$-0.024$, 0.024] 		           & [$-0.043$, 0.042]  & [$-0.026$, 0.025] \\
&$\Delta\kappa^{\gamma}$  & [$-0.109$, 0.124]  & [$-0.054$, 0.092]                       & [$-0.118$, 0.136]  & [$-0.057$, 0.099] \\
&$\lambda^{\gamma}$           & [$-$0.081, 0.082]  & [$-0.051$, 0.052]                   & [$-0.088$, 0.089]  & [$-0.055$, 0.055] \\
\hline
\multirow{3}{*}{LEP}
&$\Delta g_{1}^{Z}$           & [$-0.033$, 0.037]  & [$-0.016$, 0.027]                   & [$-0.035$, 0.041]  & [$-0.017$, 0.029] \\
&$\Delta\kappa^{Z}$           & [$-0.037$, 0.035]  & [$-0.025$, 0.020]                   & [$-0.041$, 0.038]  & [$-0.027$, 0.021] \\
&$\lambda^{Z}$            & [$-0.031$, 0.031]  & [$-0.019$, 0.019]                       & [$-0.033$, 0.033]  & [$-0.020$, 0.020] \\
\hline
\multirow{2}{*}{HISZ}
&$\Delta\kappa^{Z}$           & [$-0.026$, 0.030]  & [$-0.012$, 0.022]                   & [$-0.028$, 0.033]  & [$-0.013$, 0.024] \\
&$\lambda^{Z}$            & [$-0.031$, 0.031]  & [$-0.019$, 0.019]                       & [$-0.033$, 0.034]  & [$-0.020$, 0.020] \\
\hline
\multirow{2}{*}{Equal Couplings}
&$\Delta\kappa^{Z}$           & [$-0.041$, 0.048]  & [$-0.020$, 0.035]                   & [$-0.045$, 0.052]  & [$-0.021$, 0.037] \\
&$\lambda^{Z}$            & [$-0.030$, 0.030]  & [$-0.019$, 0.019]                       & [$-0.034$, 0.033]  & [$-0.020$, 0.020] \\
\hline
\end{tabular}
	\caption{ The expected and observed 95\% confidence intervals for the anomalous coupling parameters defined in the \textit{no constraints} scenario, \textit{LEP}, \textit{HISZ} and \textit{Equal Couplings} scenarios. The results are shown with $\Lambda=\infty$ and $\Lambda=7$~\TeV.}
	\label{tab:atgc_expected_observed_limits_skip2}
\end{table}

\begin{table}[h!t]
\centering
\begin{tabular}{lccc}
\hline
Scenario & Parameter            & Expected [TeV$^{-2}$] & Observed [TeV$^{-2}$] \\
\hline
\multirow{3}{*}{EFT}
&$C_{WWW}/\Lambda^{2}$& [$-7.62$, 7.38]  & [$-4.61$, 4.60]                               \\
&$C_{B}/\Lambda^{2} $& [$-35.8$, 38.4]  & [$-20.9$, 26.3]                                \\
&$C_{W}/\Lambda^{2}$& [$-12.58$, 14.32]  & [$-5.87$, 10.54]  					 \\
\hline
\end{tabular}
	\caption{ The expected and observed 95\% confidence intervals for the \textit{EFT} approach.}
	\label{tab:atgc_expected_observed_limits_skip2_EFT}
\end{table}

\begin{figure}[!ht]
\centering
	{\includegraphics[width=0.49\textwidth]{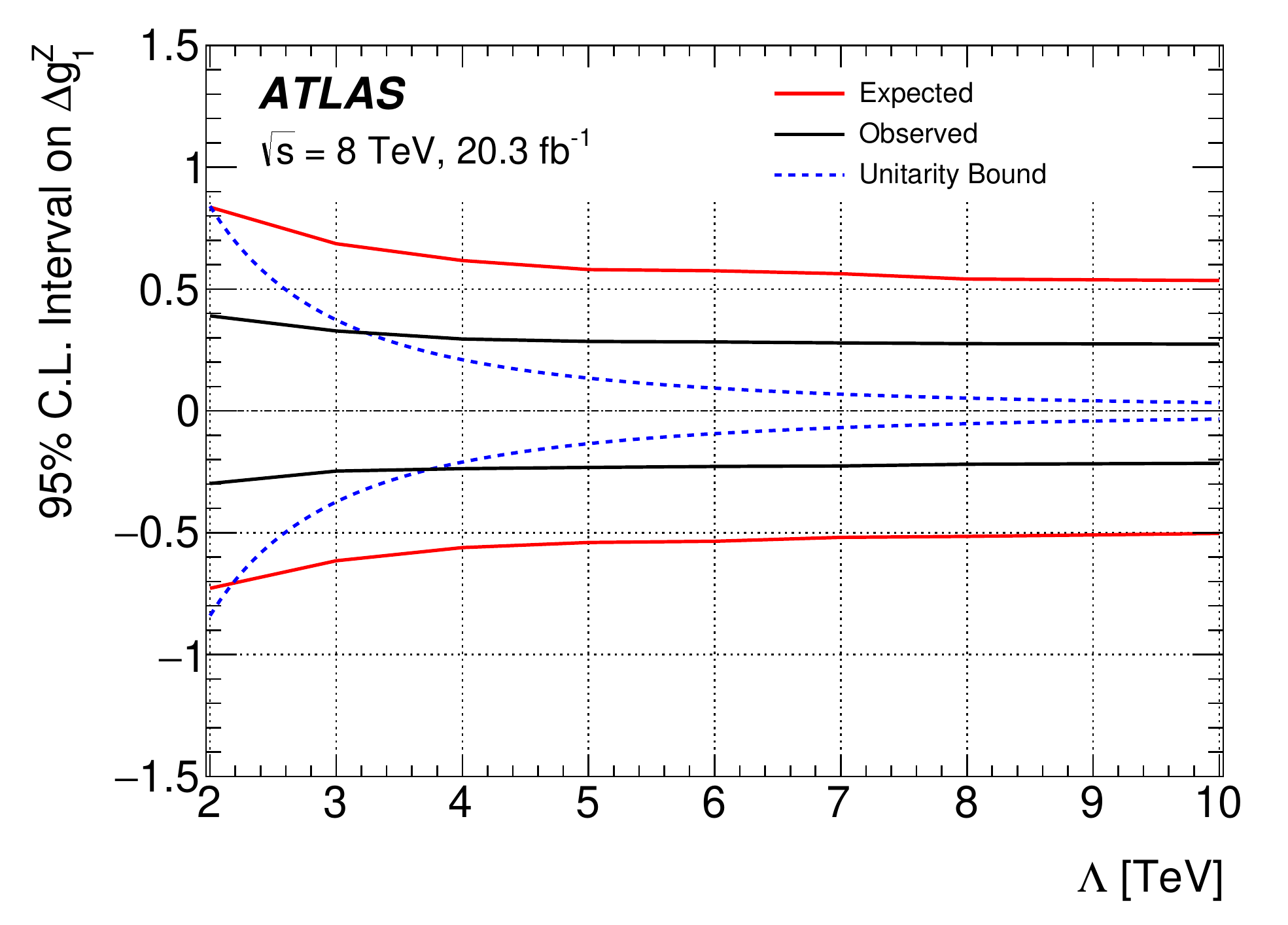}}
	{\includegraphics[width=0.49\textwidth]{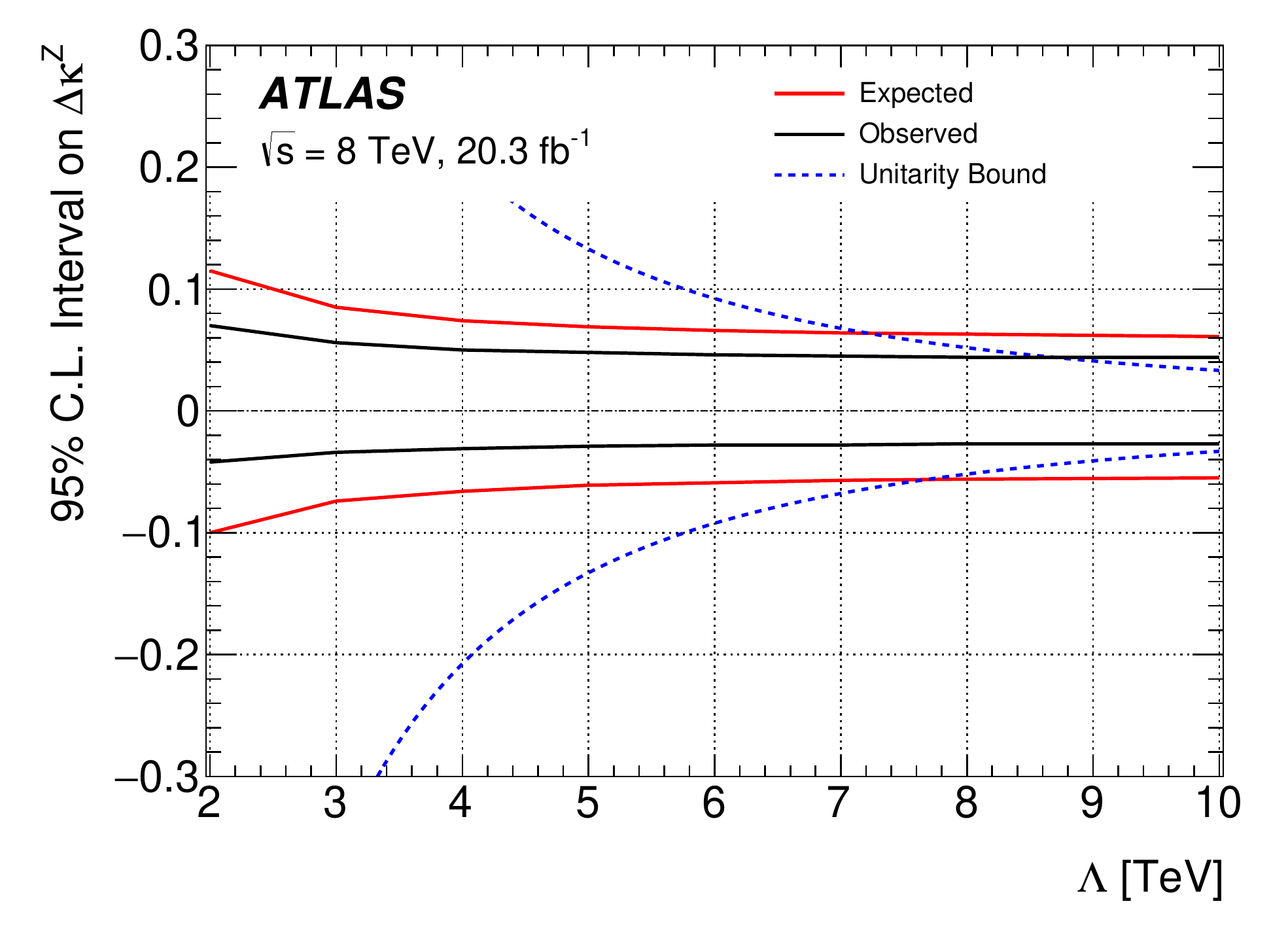}}
	{\includegraphics[width=0.49\textwidth]{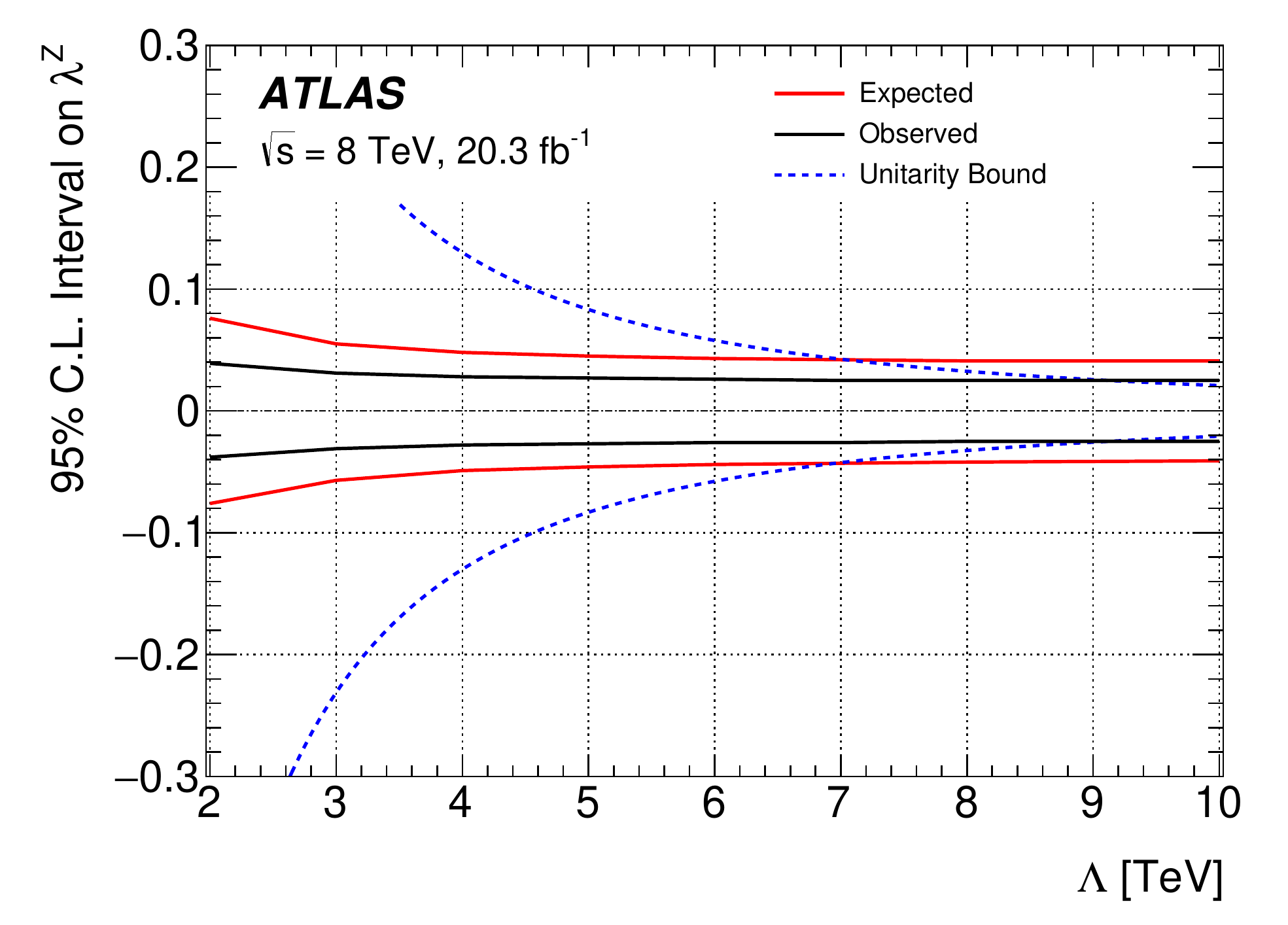}}
	{\includegraphics[width=0.49\textwidth]{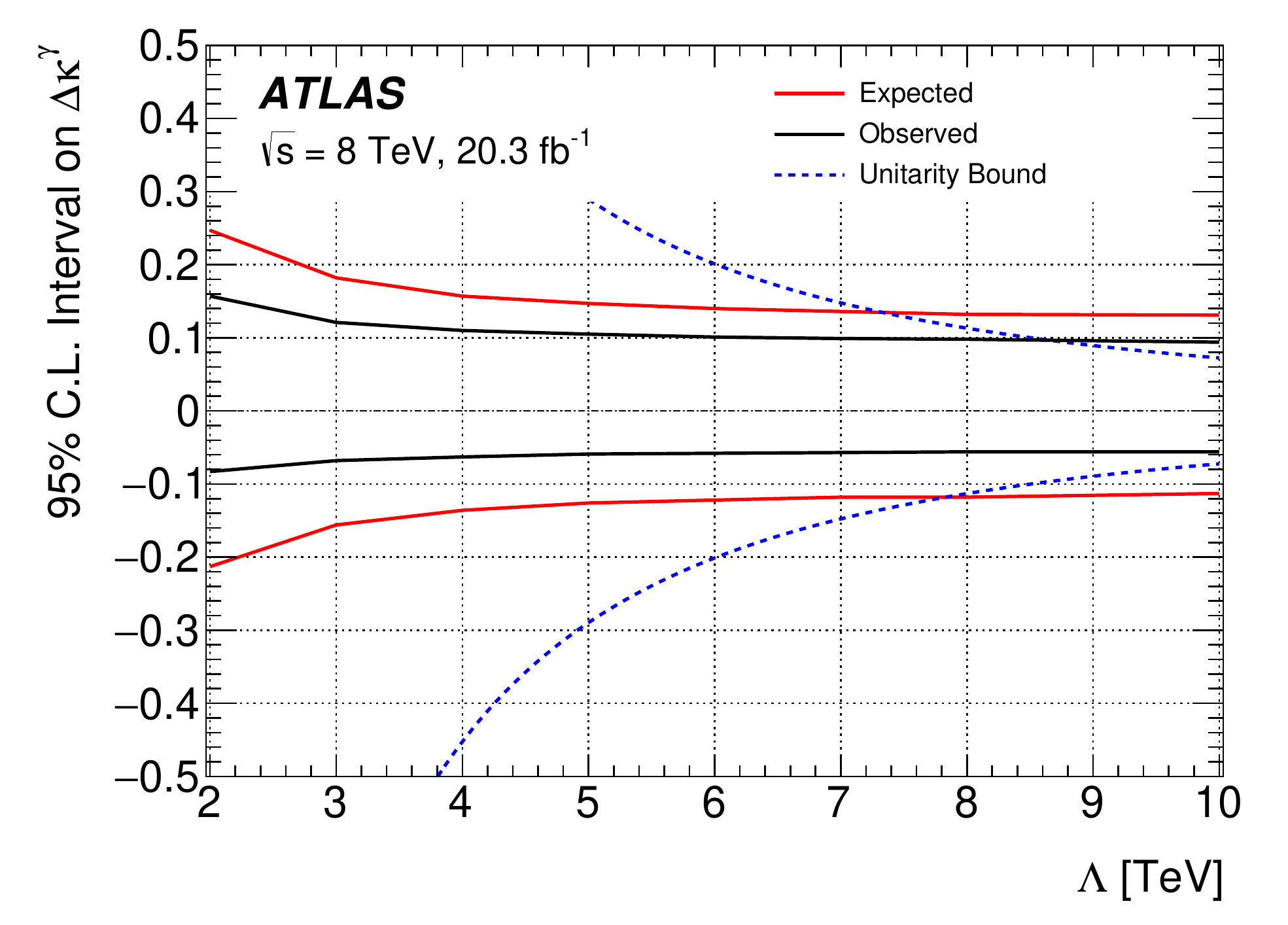}}
	{\includegraphics[width=0.49\textwidth]{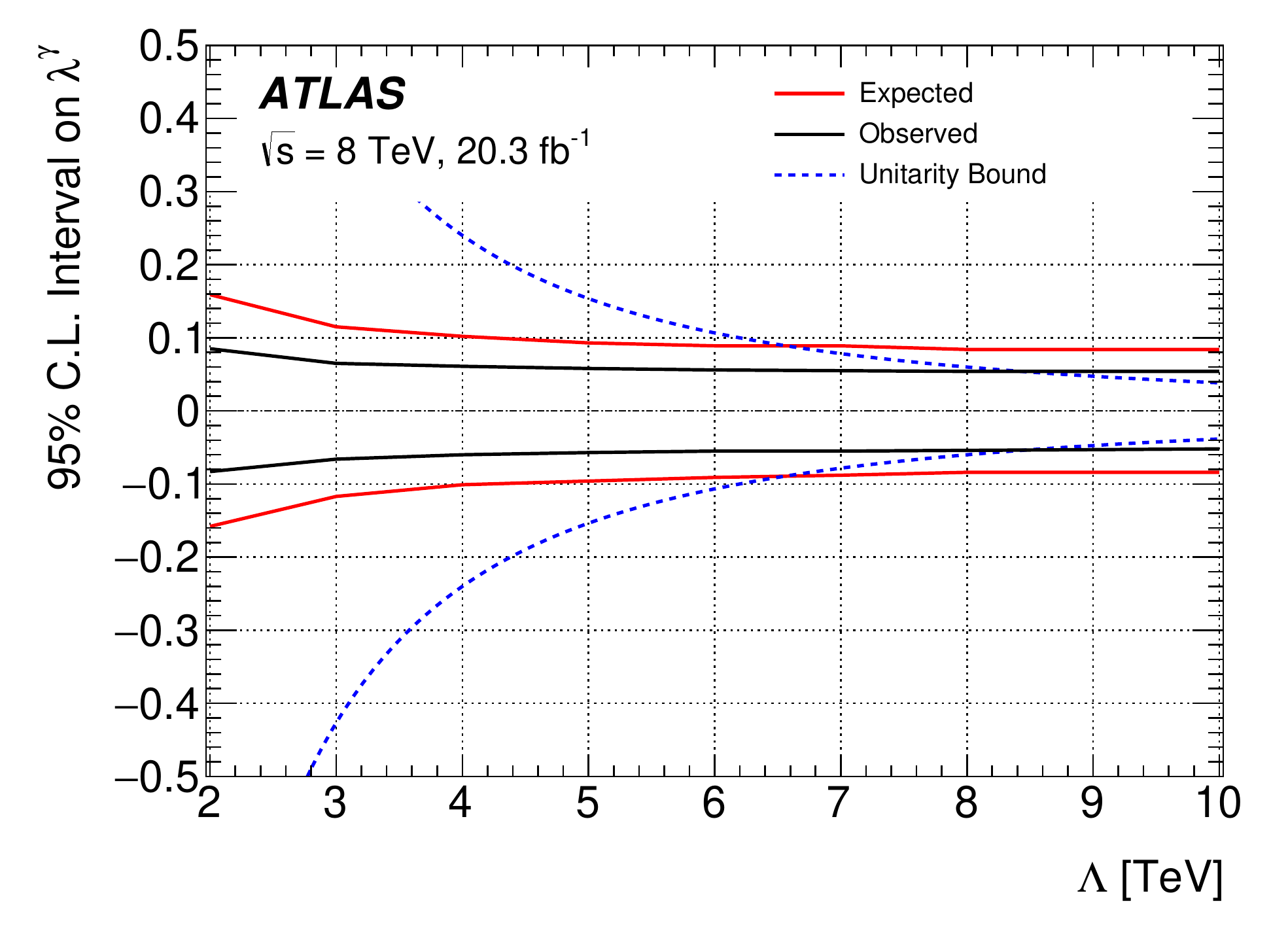}}
	\caption{
     The expected and observed intervals at 95\% confidence level (C.L.), in red and black respectively, and the theoretical constraint~\cite{FERMILAB-Pub-95-031} due to the unitarity requirement (shown as blue dashed lines) as a function of form-factor scales from $\Lambda=2$~\TeV\ to $\Lambda=10$~\TeV.  The plots are made for the aTGC parameters defined in the \textit{no constraints} scenario. Except for the anomalous coupling parameter shown, all others are set to zero.
    }\label{fig:atgc_expected_limits_skip2_formfactors}
\end{figure}

The limits in the plane of two coupling parameters are shown for the \textit{no constraints} and \textit{LEP} scenarios in Figure~\ref{fig:atgc_2D_limits_None} and Figure~\ref{fig:atgc_2D_limits_LEP}, respectively. Further limits obtained for the \textit{Equal Couplings} and \textit{HISZ} scenarios are shown in Figure~\ref{fig:atgc_2D_limits_HISZ_EqualCouplings}. Finally, the 95\% confidence-level contours for linear combinations of aTGC parameters defined in the effective field theory approach are shown in Figure~\ref{fig:atgc_2D_limits_EFT}.

Due to the increased integrated luminosity and the higher centre-of-mass energy, the new limits are more stringent by up to 50\% than those previously published by the ATLAS Collaboration using data taken at $\sqrt{s}=7$~\TeV~\cite{ww7tev}. The constraints derived in the LEP scenario are similar to the combined results of the LEP experiments and in a few cases the derived limits exceed the bounds placed by LEP. The 95\% confidence-level limits on $\Delta g_{1}^{Z}$ obtained in this analysis range from $-$0.016 to 0.027 whilst the limits from  LEP cover values from $-$0.021 to 0.054. The 95\% confidence intervals on $C_{WWW}/\Lambda^{2}$ and $C_{B}/\Lambda^{2}$ derived in this analysis are similar, or up to 20-30\% more restrictive than those obtained by the CMS Collaboration in Ref.~\cite{Khachatryan:2015sga}, which derives limits for the effective field theory approach only and uses the invariant dilepton mass distribution, $m_{\ell\ell}$. The limits derived on $C_{W}/\Lambda^{2}$ cover a complementary range around zero compared to the bounds from CMS, they have similar numerical values but opposite sign. Since the effects of EFT operators on distributions depend primarily on their absolute magnitude and not on their sign, these differences between the ATLAS and CMS constraints on $C_{W}/\Lambda^{2}$  can be considered insignificant. 

\begin{figure}[!ht]
	{\includegraphics[width=0.30\textwidth]{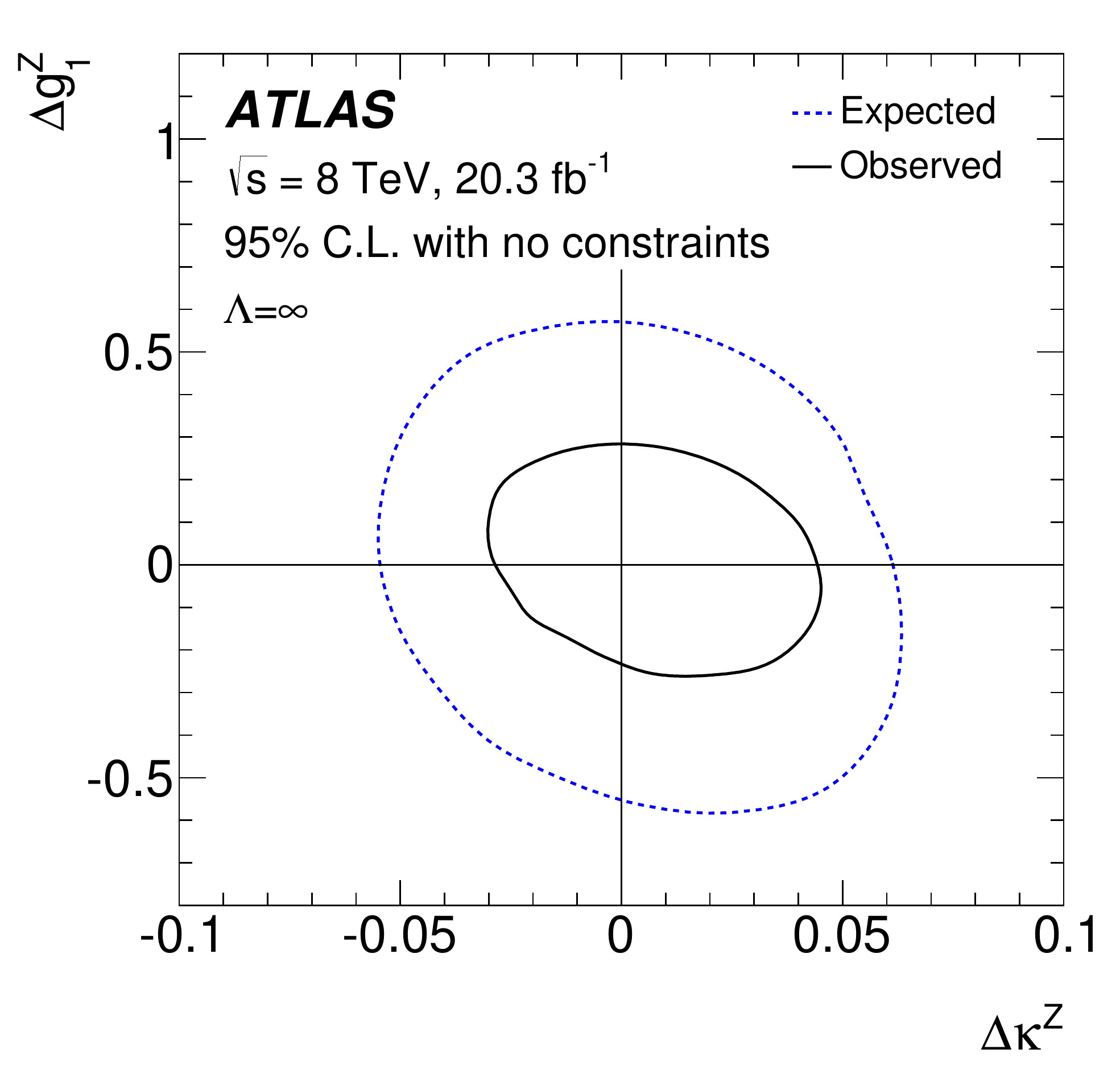}}
	{\includegraphics[width=0.30\textwidth]{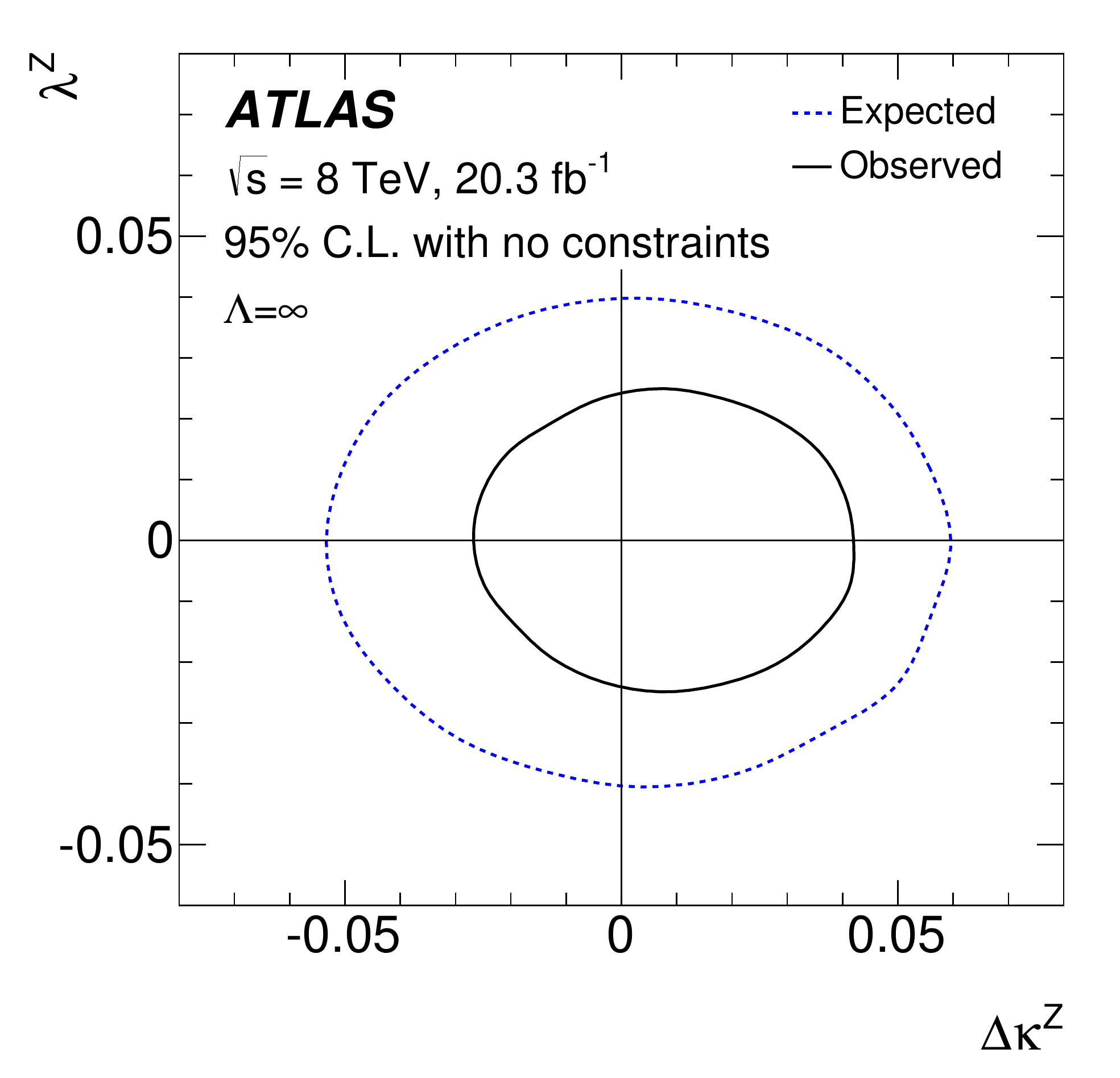}}
	{\includegraphics[width=0.30\textwidth]{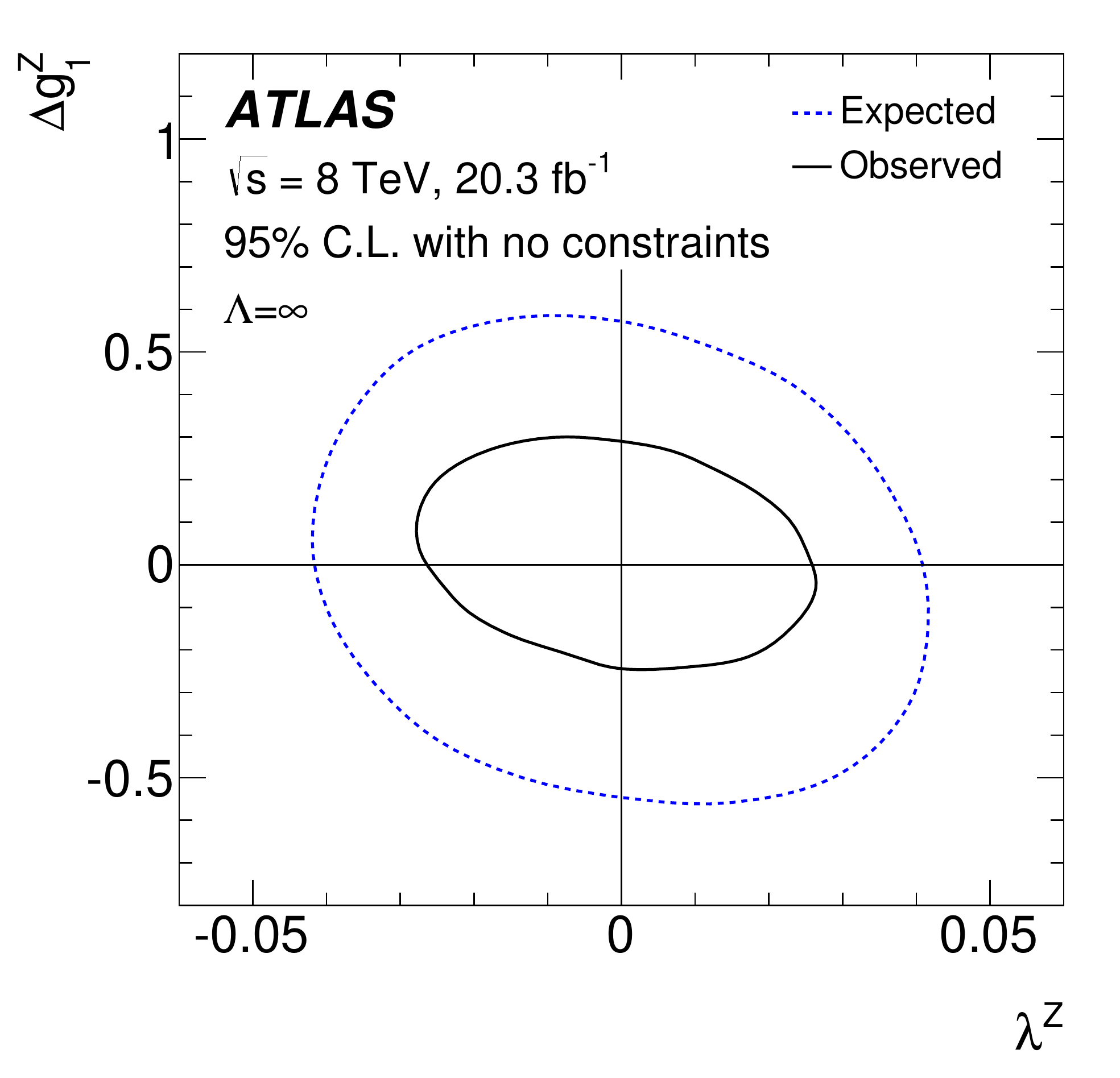}}
	{\includegraphics[width=0.30\textwidth]{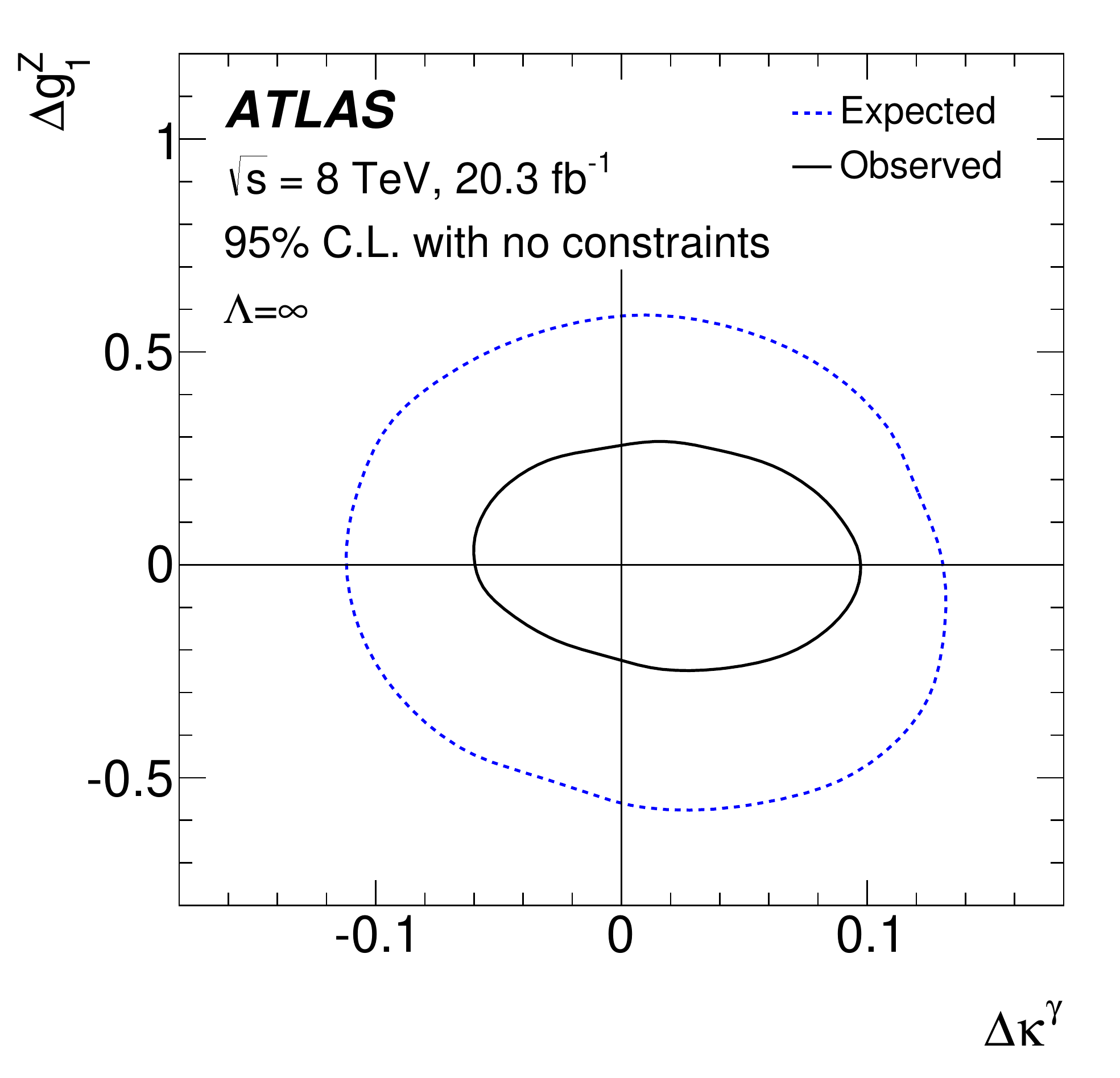}}
	{\includegraphics[width=0.30\textwidth]{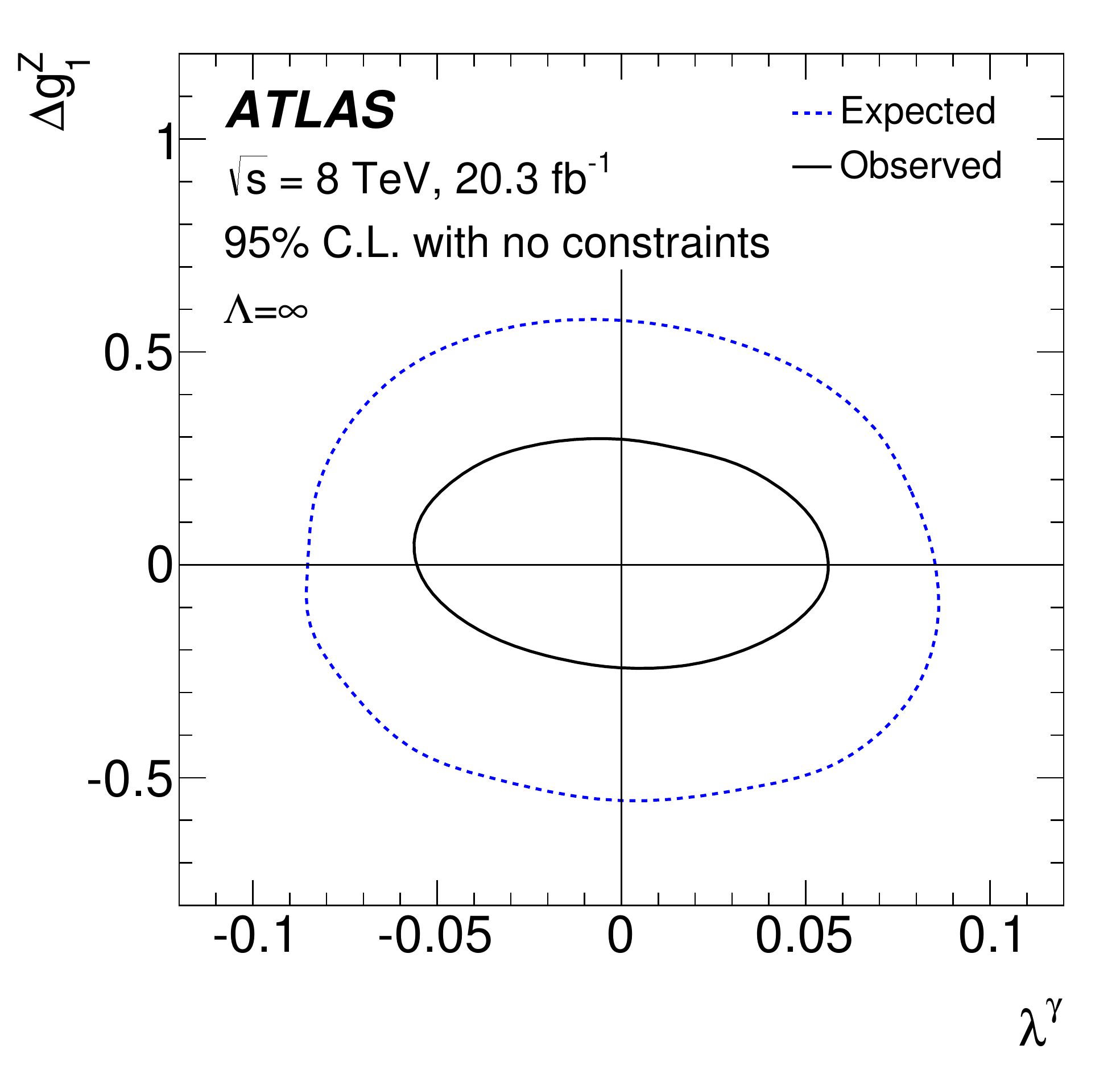}}
	{\includegraphics[width=0.30\textwidth]{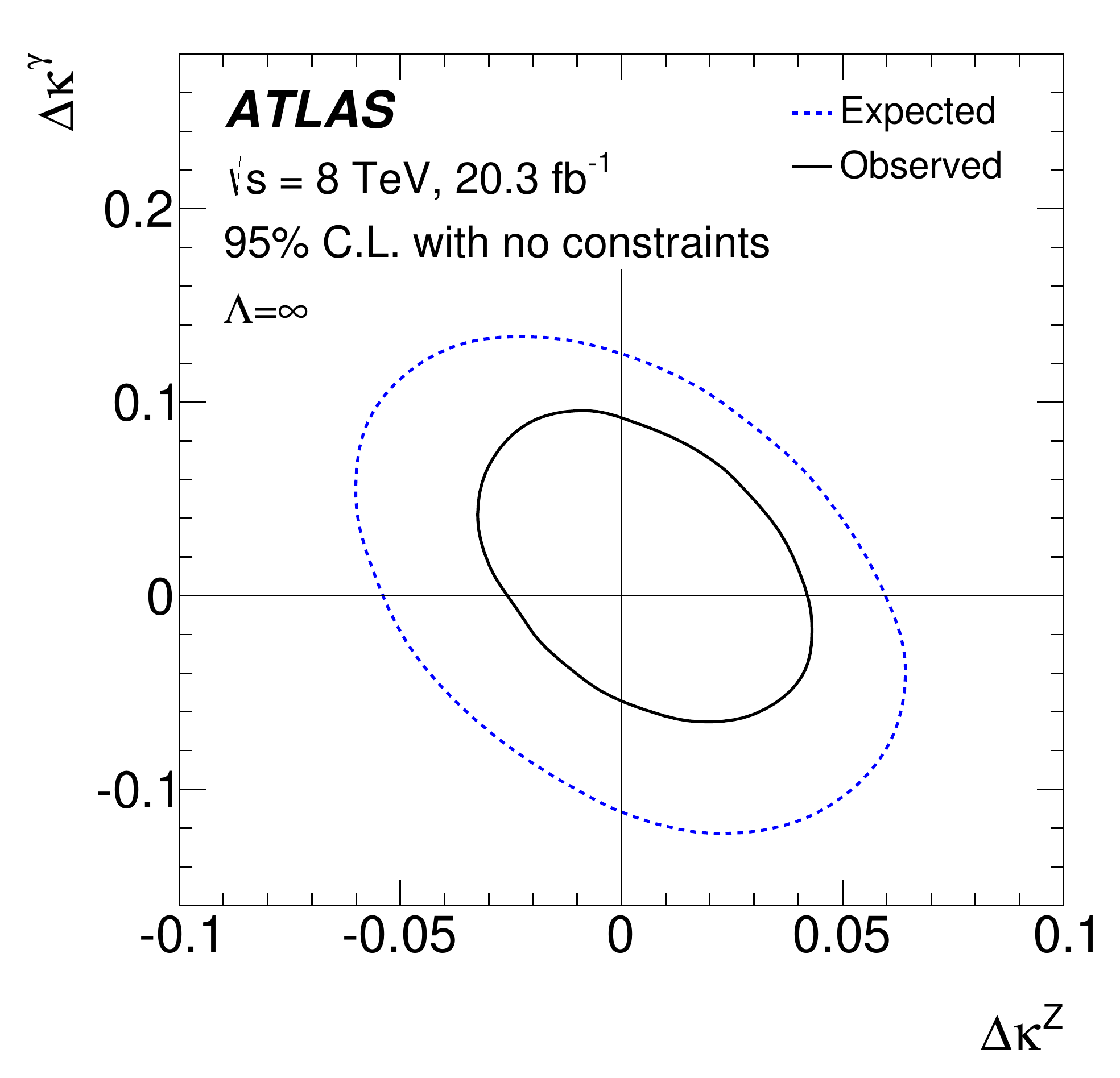}}
	{\includegraphics[width=0.30\textwidth]{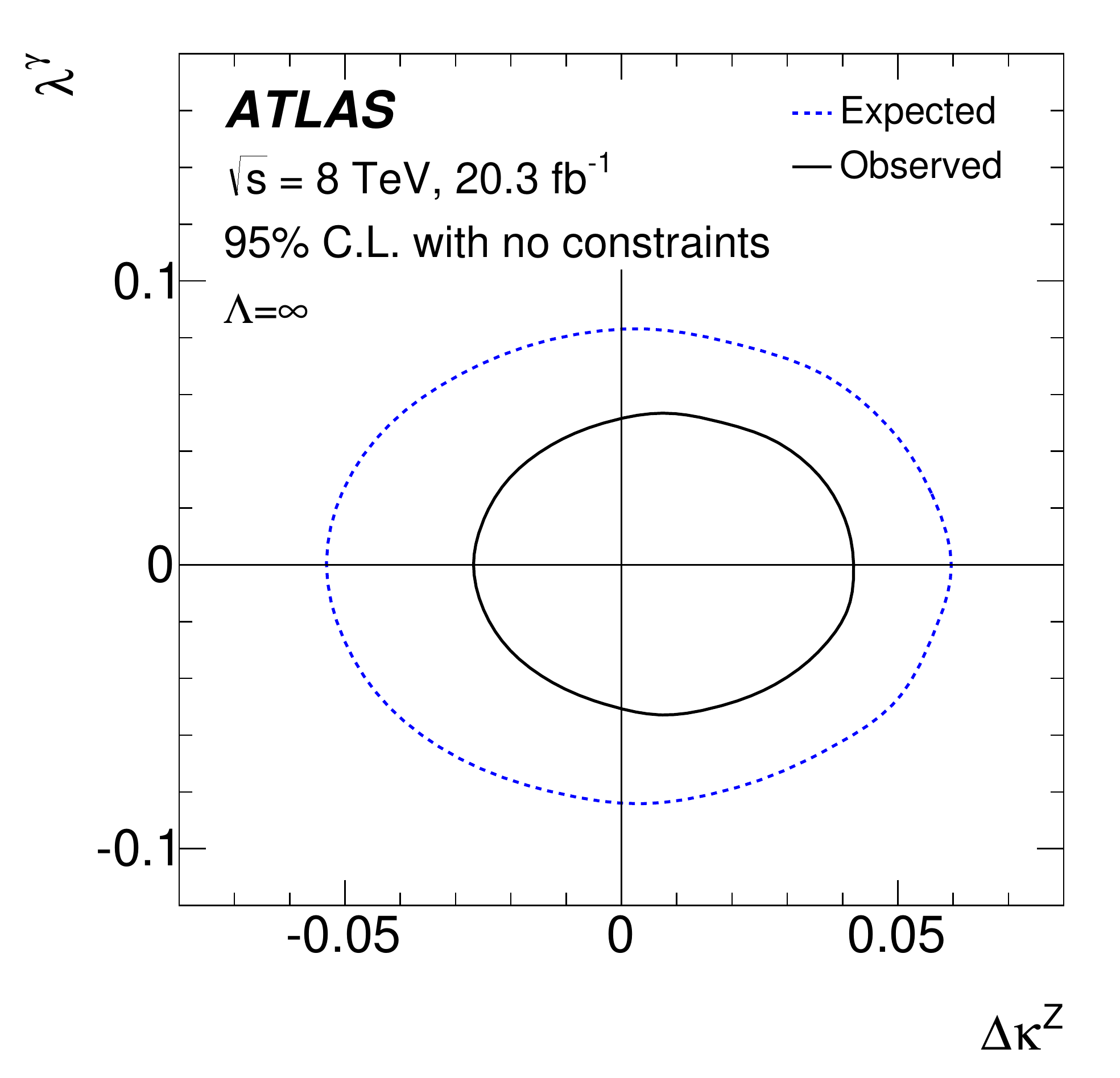}}
	{\includegraphics[width=0.30\textwidth]{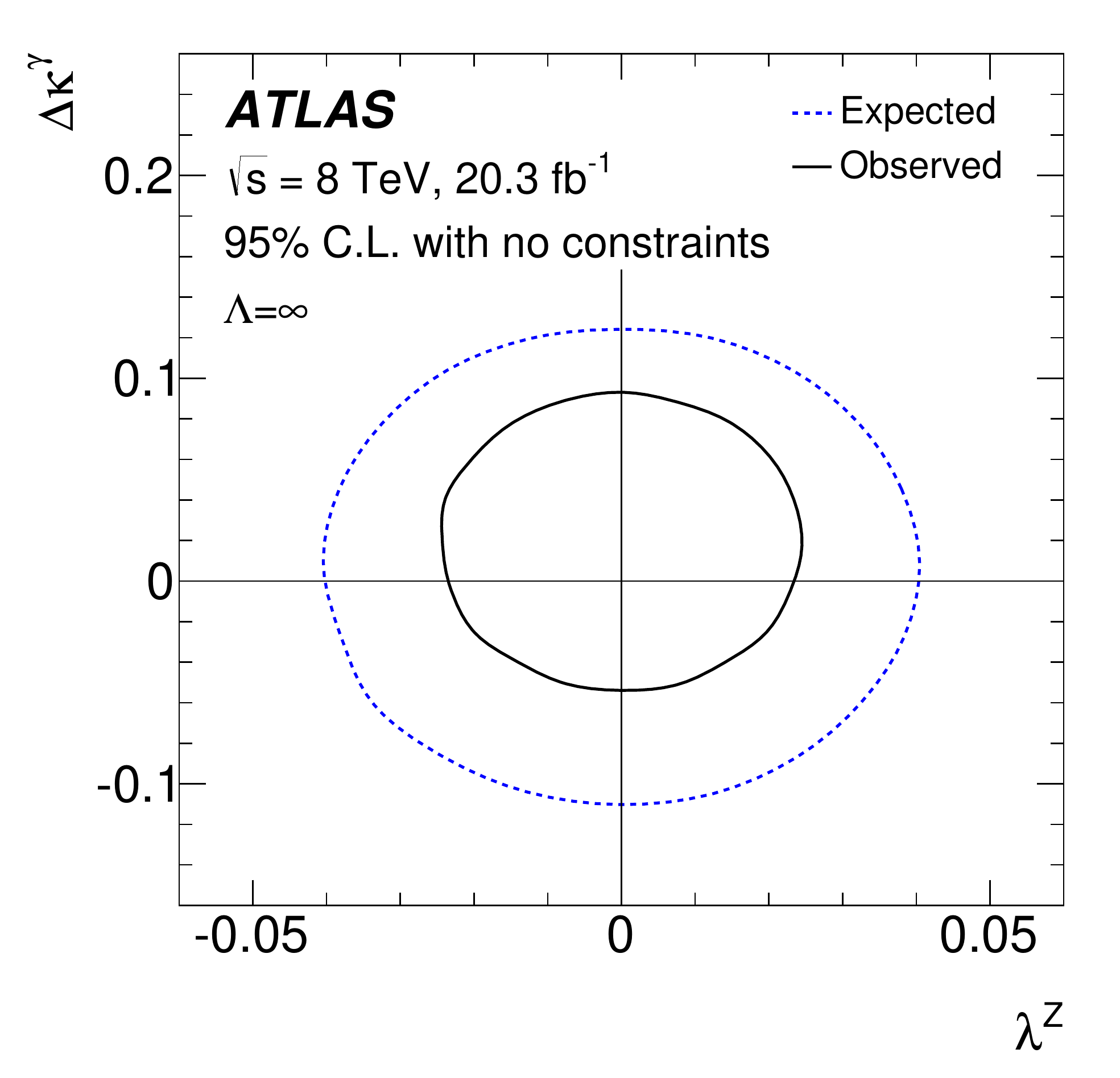}}
	{\includegraphics[width=0.30\textwidth]{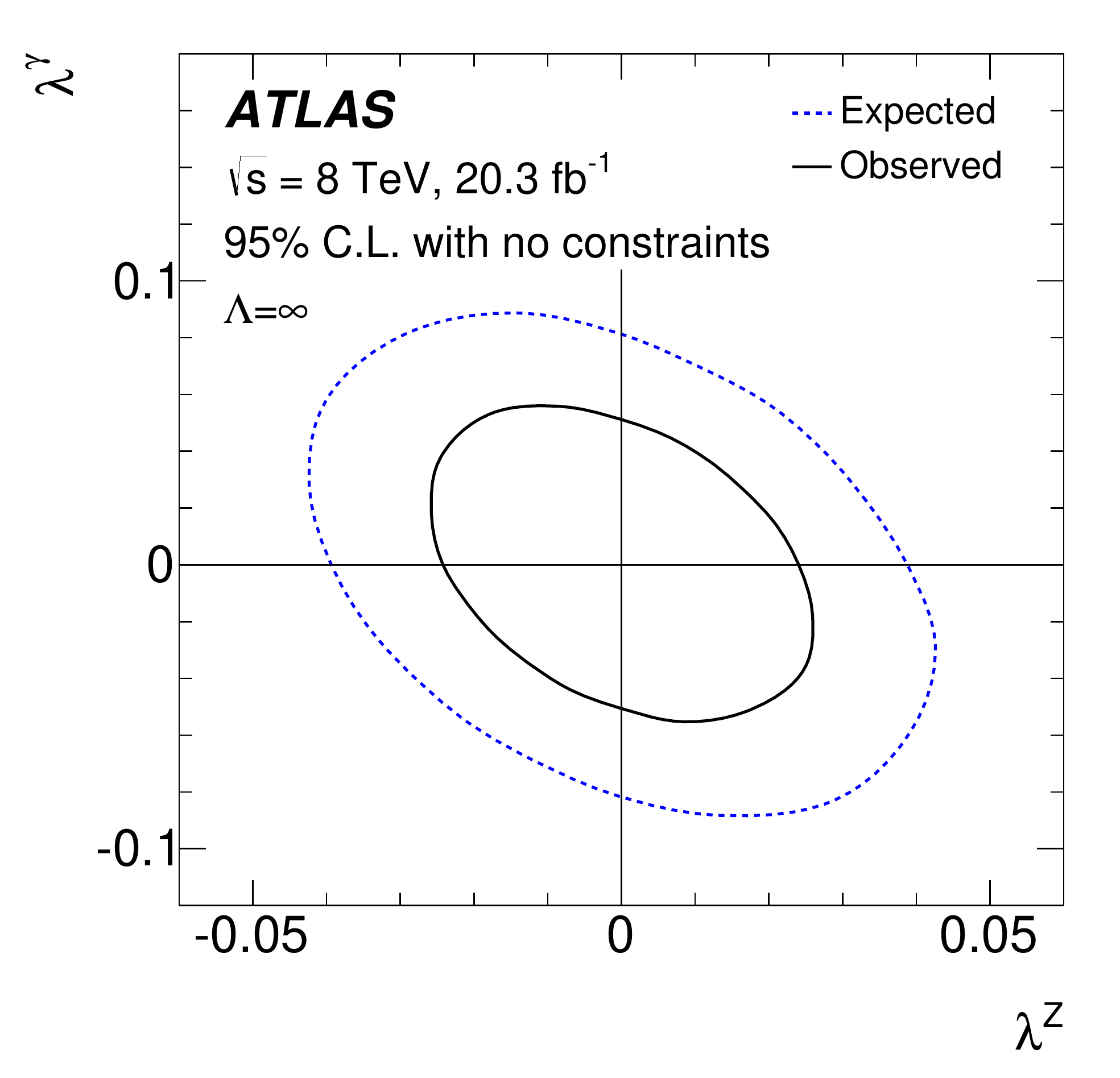}}
	{\includegraphics[width=0.30\textwidth]{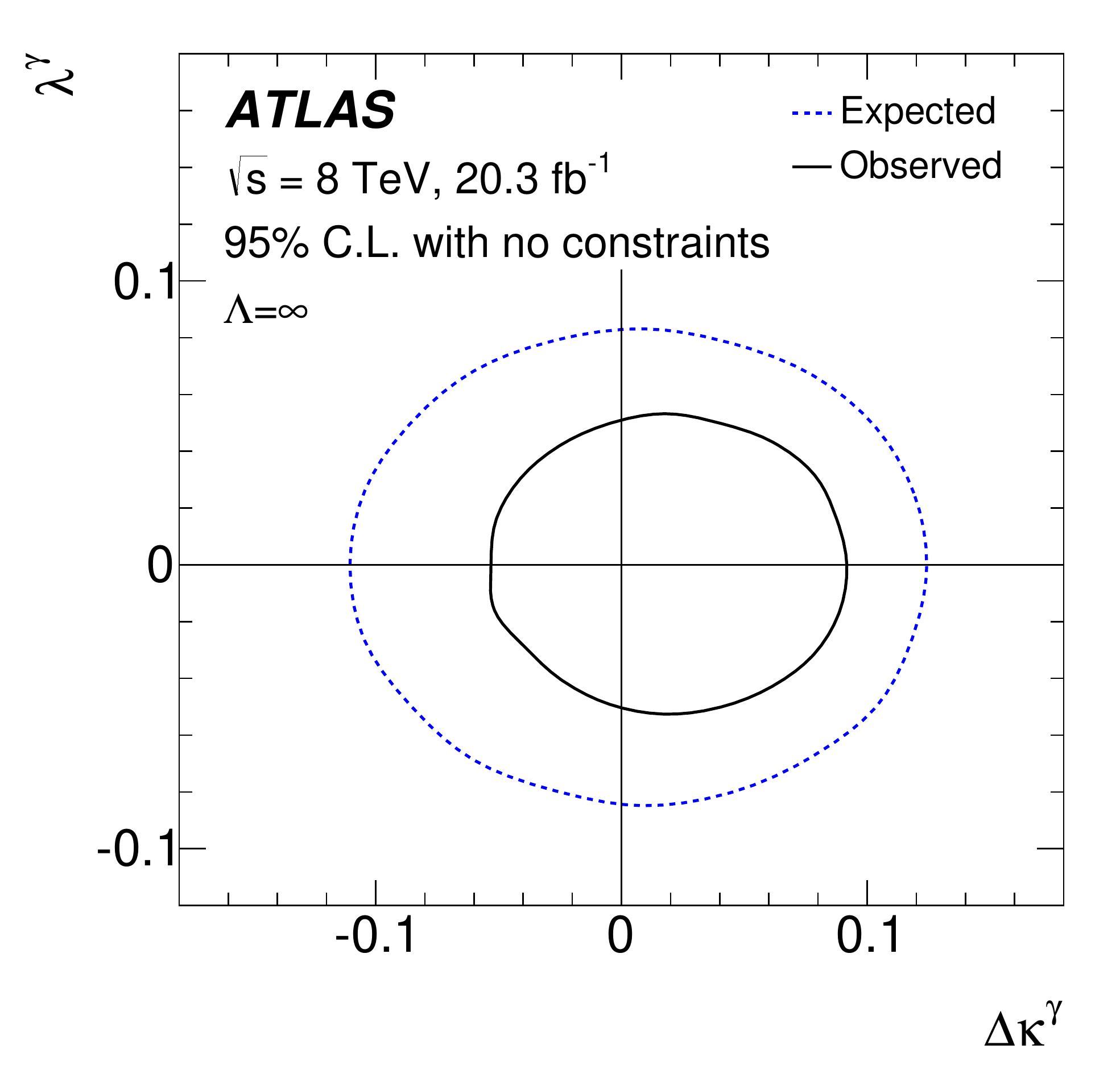}}
	\caption{The expected and observed 95\% confidence-level contours for limits in the plane of two simultaneously non-zero parameters in the \textit{no constraints} scenario. Except for the two anomalous coupling parameters under study, all others are set to zero.}
	\label{fig:atgc_2D_limits_None}
\end{figure}

\begin{figure}[!ht]
	{\includegraphics[width=0.49\textwidth]{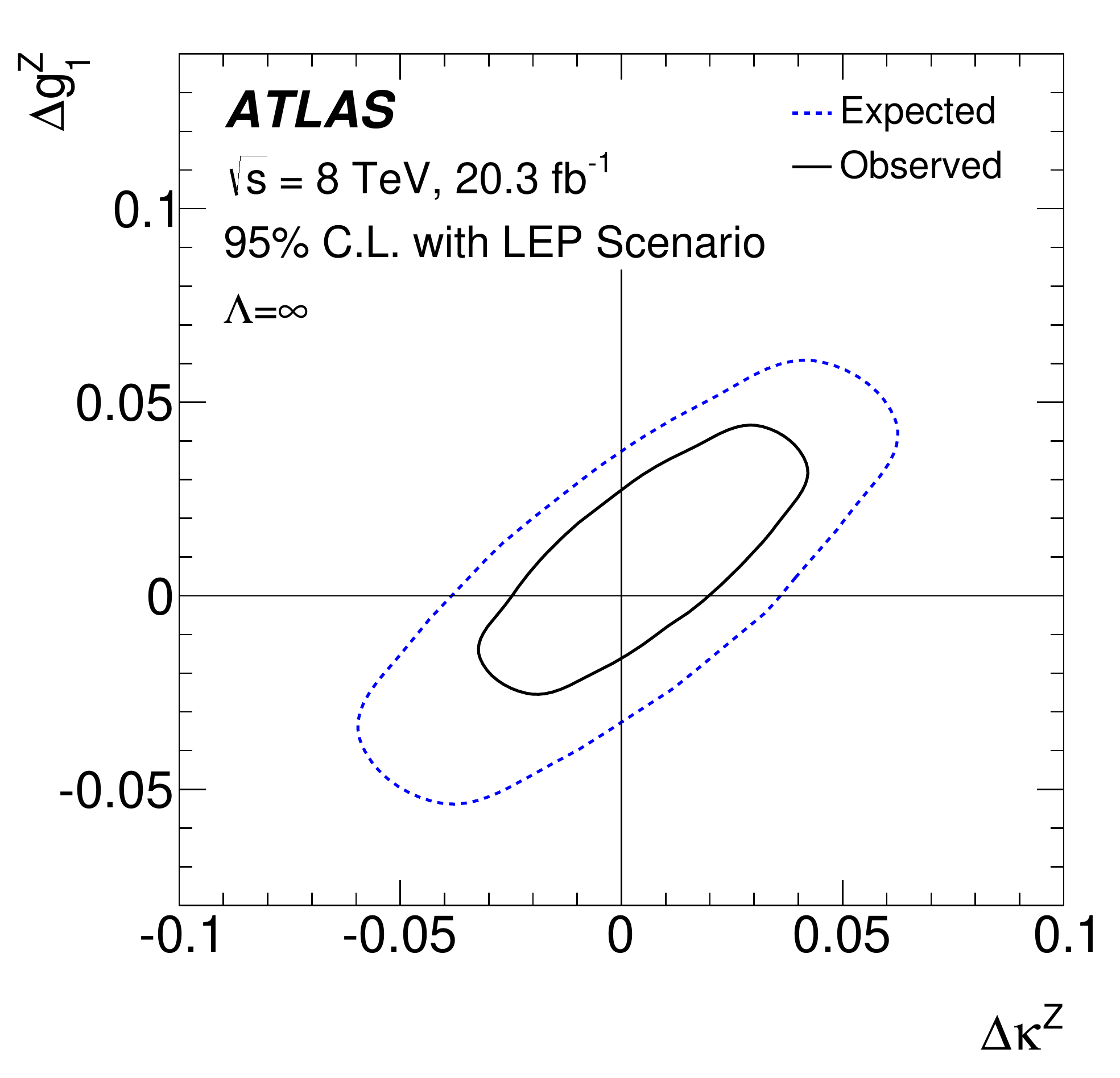}}
	{\includegraphics[width=0.49\textwidth]{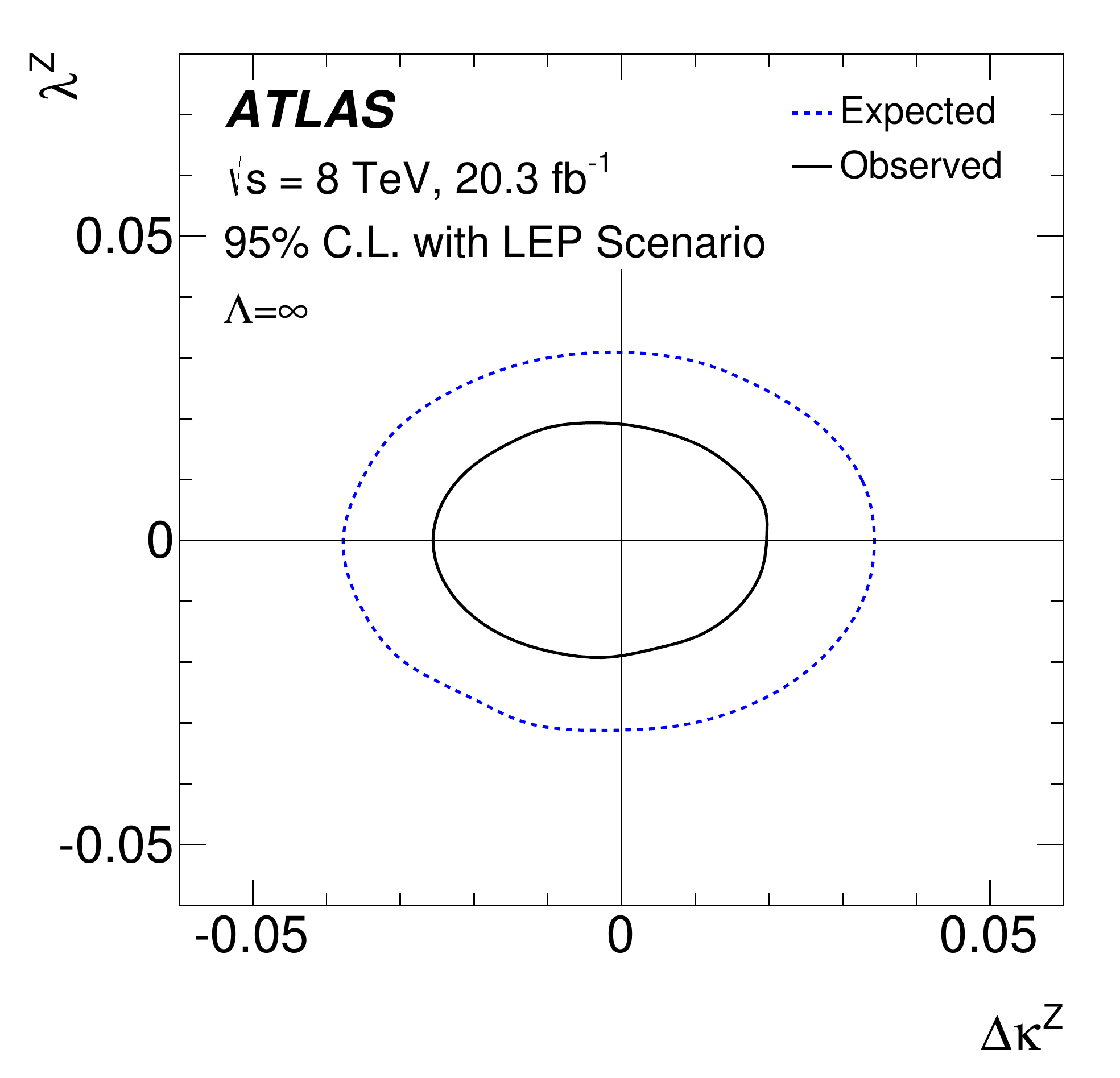}}
	{\includegraphics[width=0.49\textwidth]{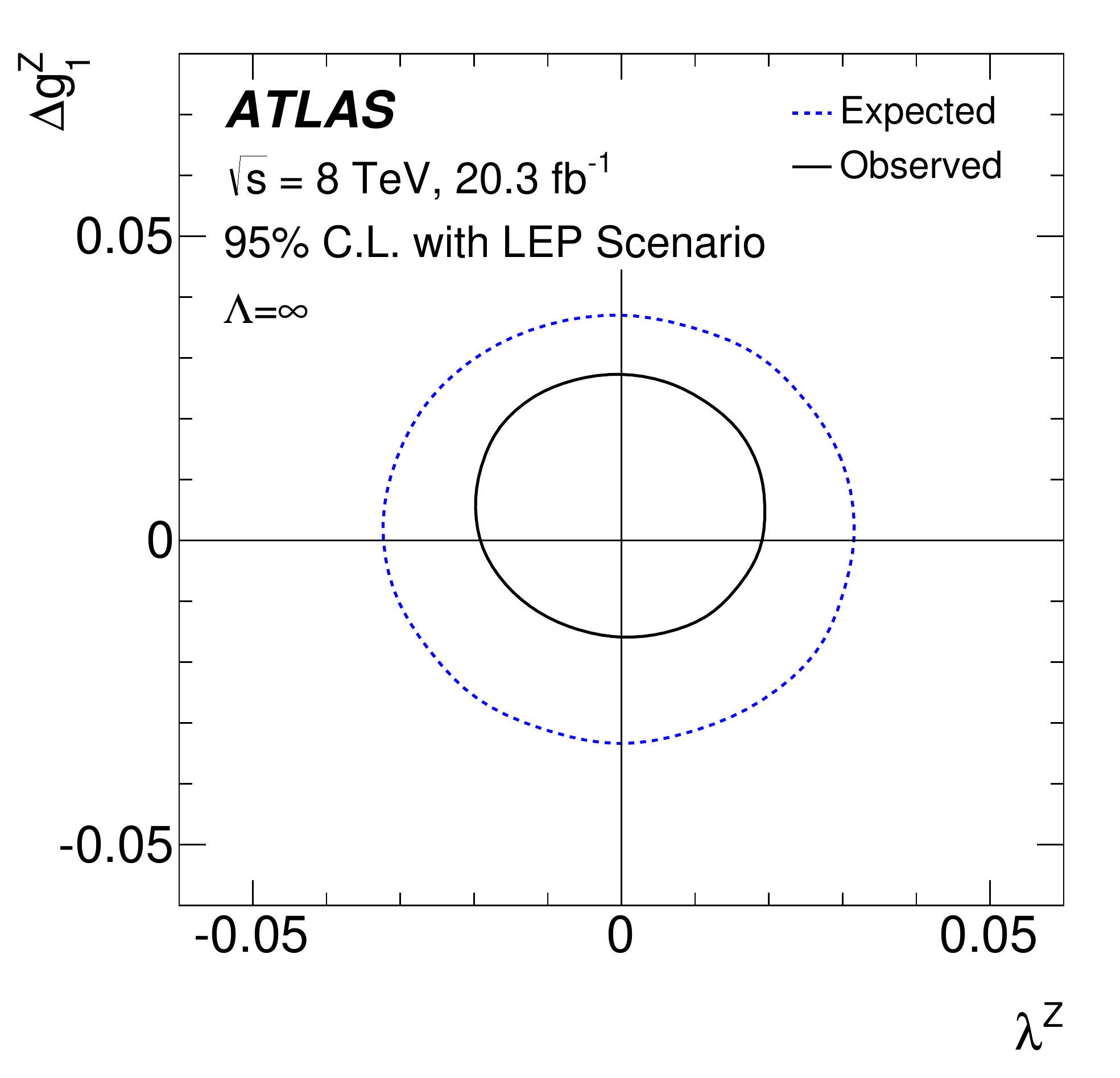}}
	\caption{The expected and observed 95\% confidence-level contours for limits in the plane of two simultaneously non-zero parameters in the \textit{LEP} scenario. Except for the two anomalous coupling parameters under study, all others are set to zero.}
	\label{fig:atgc_2D_limits_LEP}
\end{figure}

\begin{figure}[!ht]
	{\includegraphics[width=0.49\textwidth]{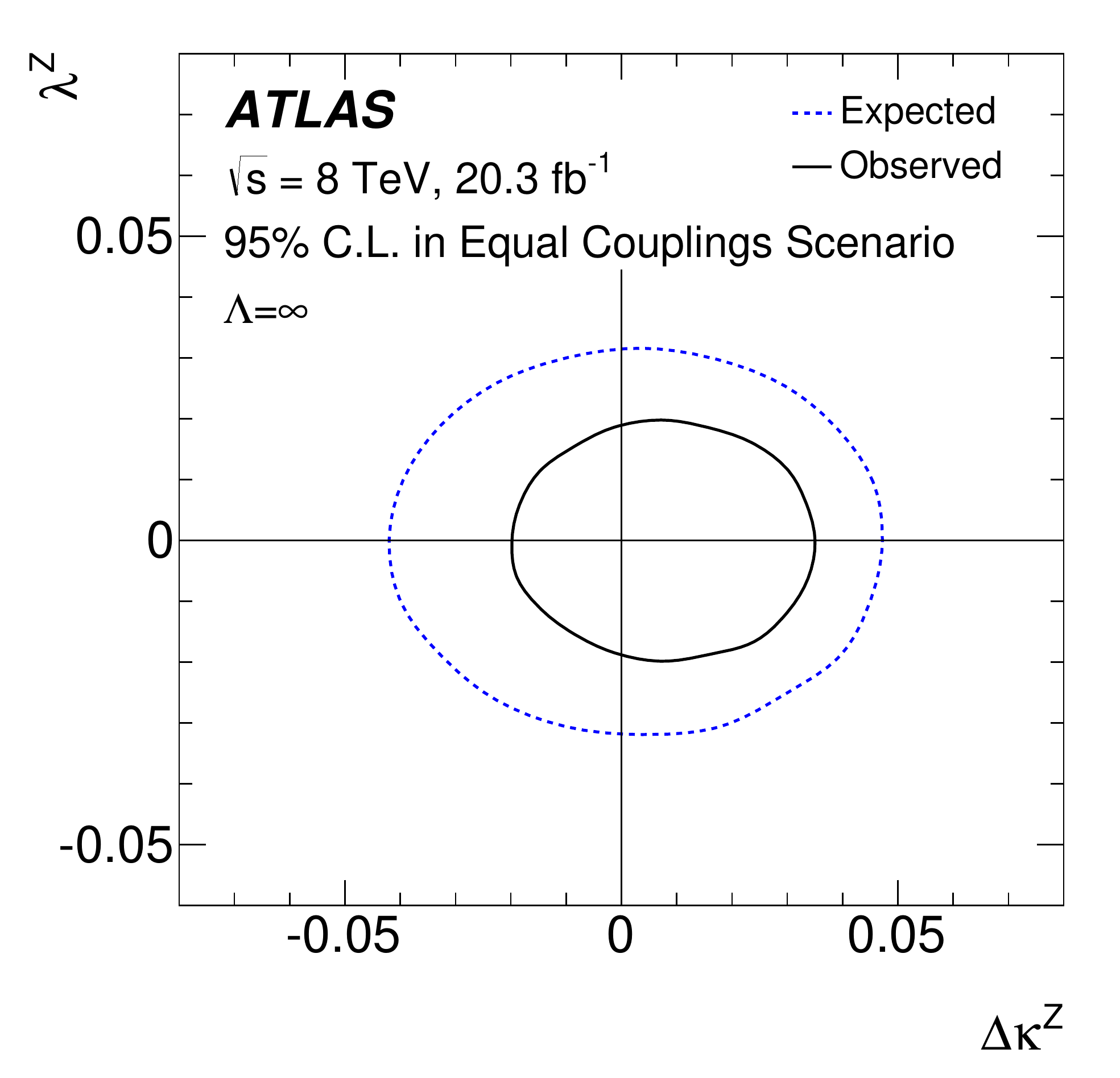}}
	{\includegraphics[width=0.49\textwidth]{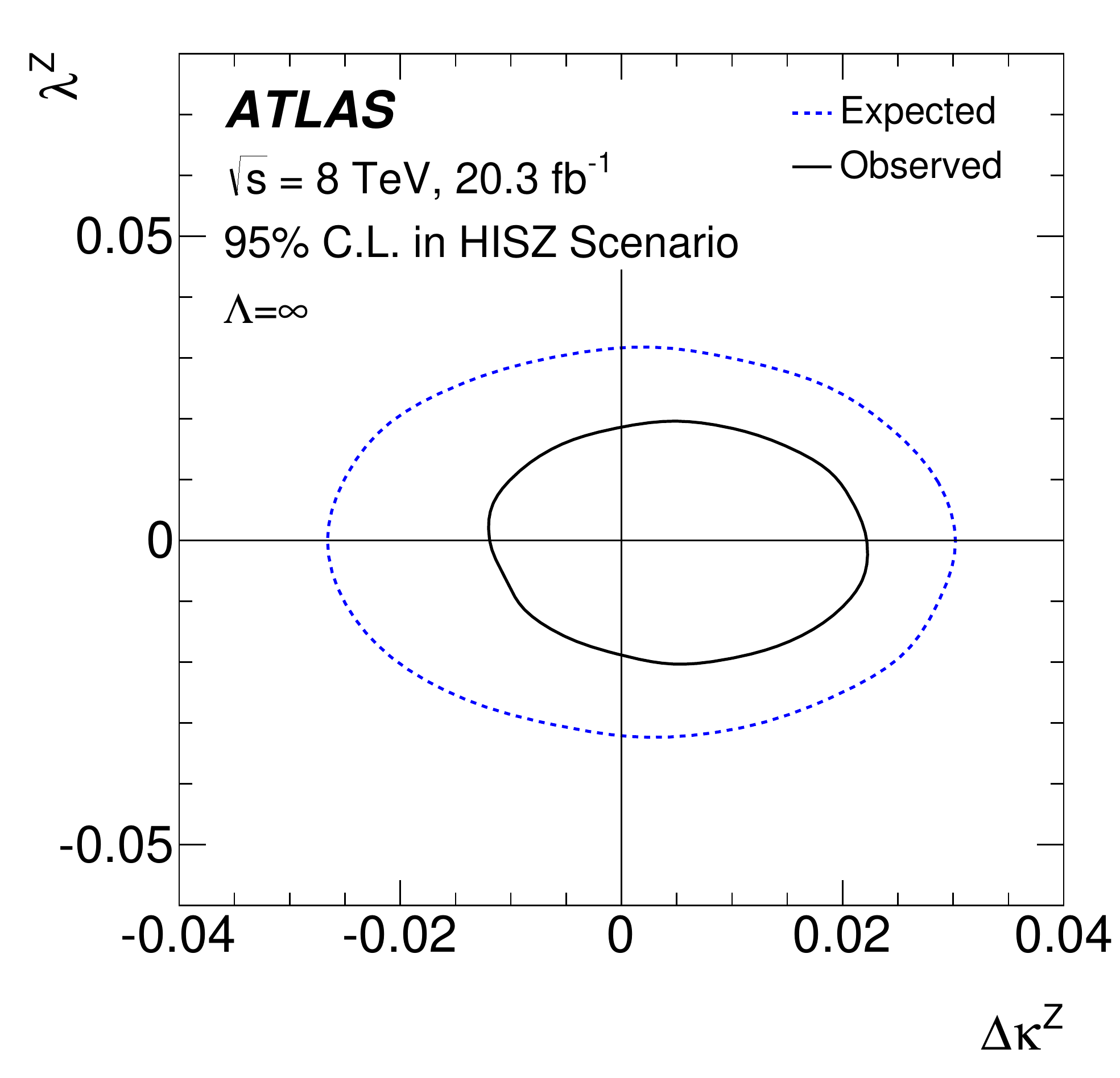}}
	\caption{The expected and observed 95\% confidence-level contours for limits in the plane of two simultaneously non-zero parameters in the \textit{Equal Couplings} (left) and the \textit{HISZ} scenario (right).}
	\label{fig:atgc_2D_limits_HISZ_EqualCouplings}
\end{figure}

\begin{figure}[!ht]
	{\includegraphics[width=0.49\textwidth]{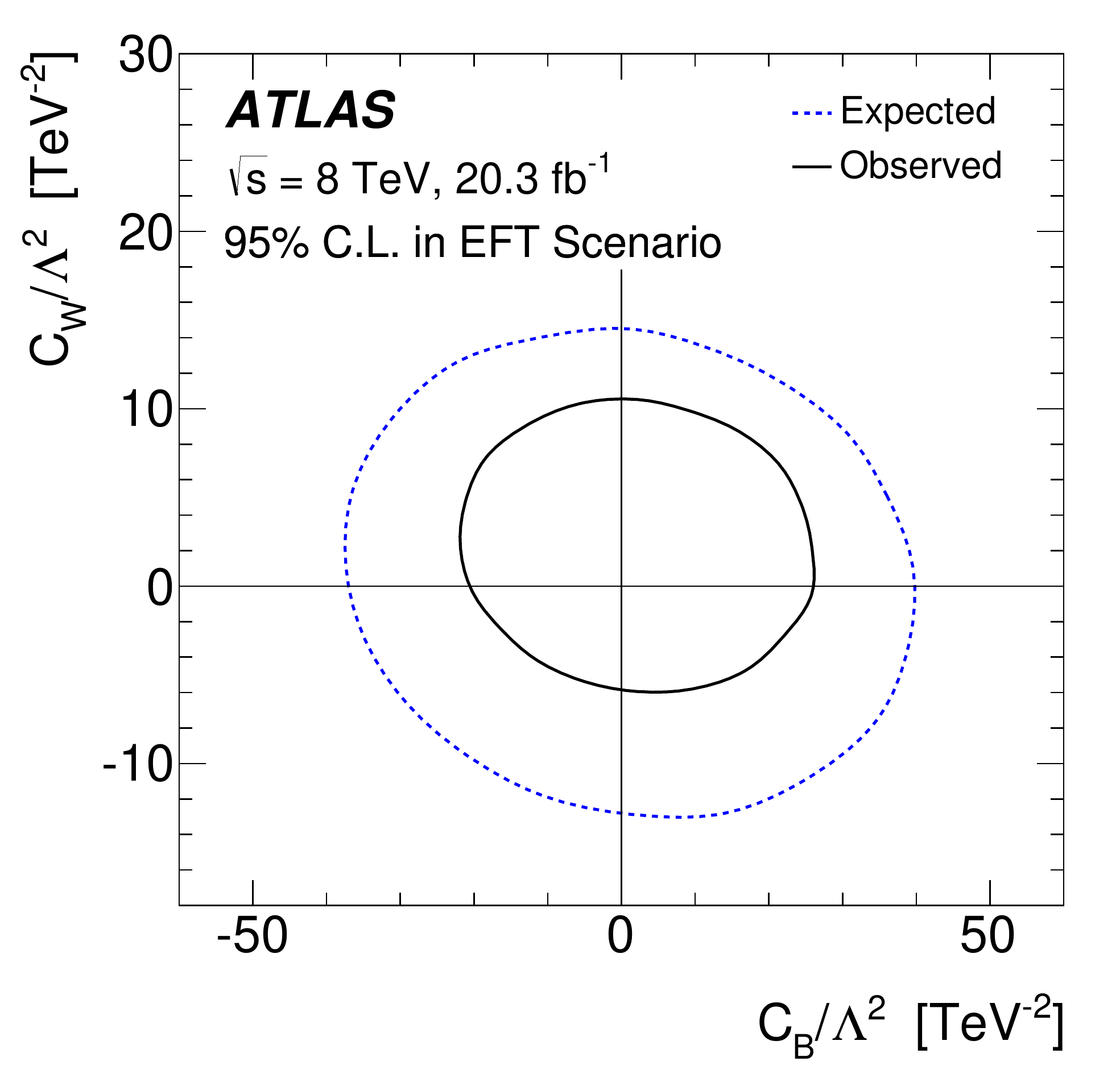}}

	{\includegraphics[width=0.49\textwidth]{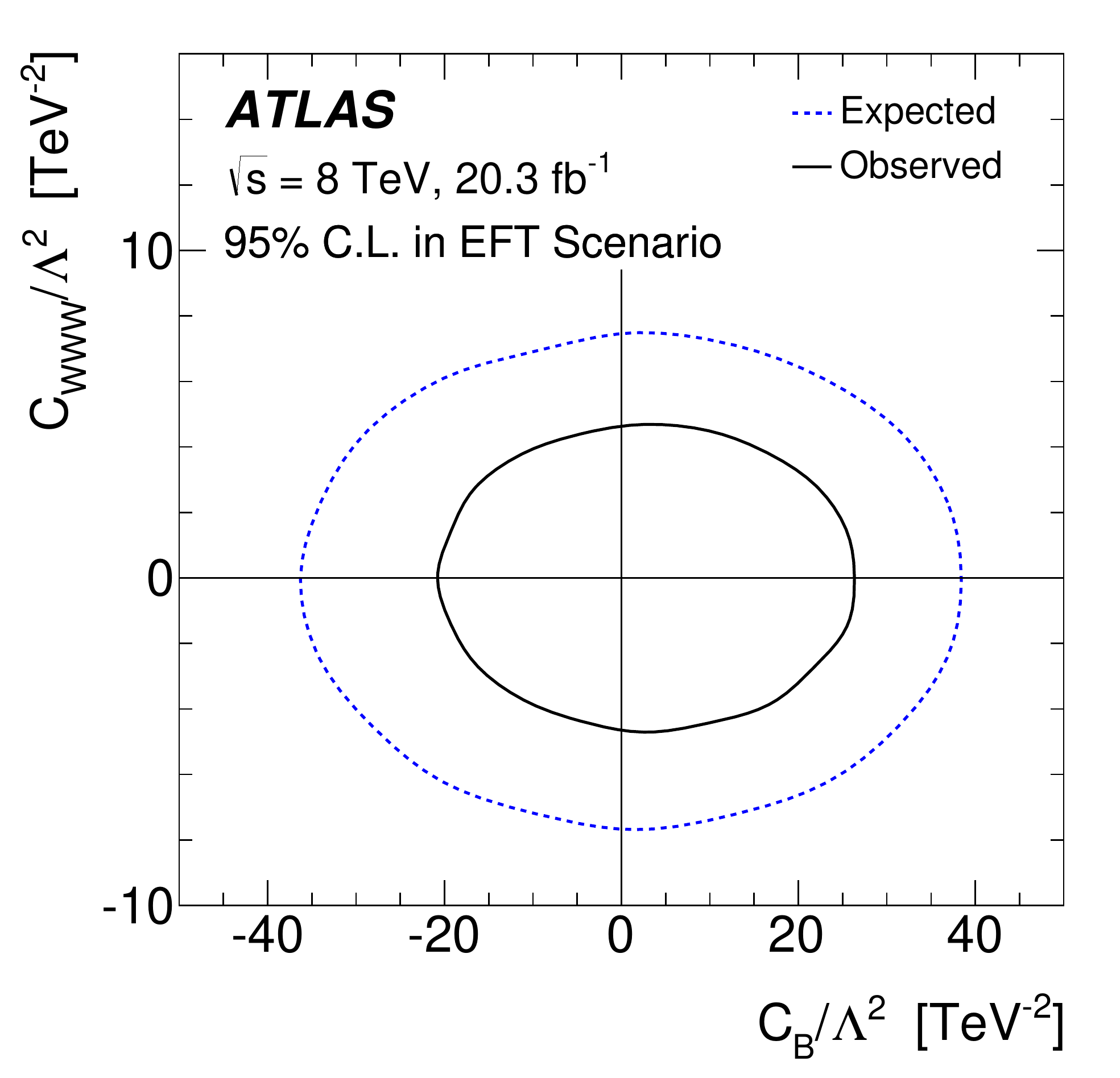}}
	{\includegraphics[width=0.49\textwidth]{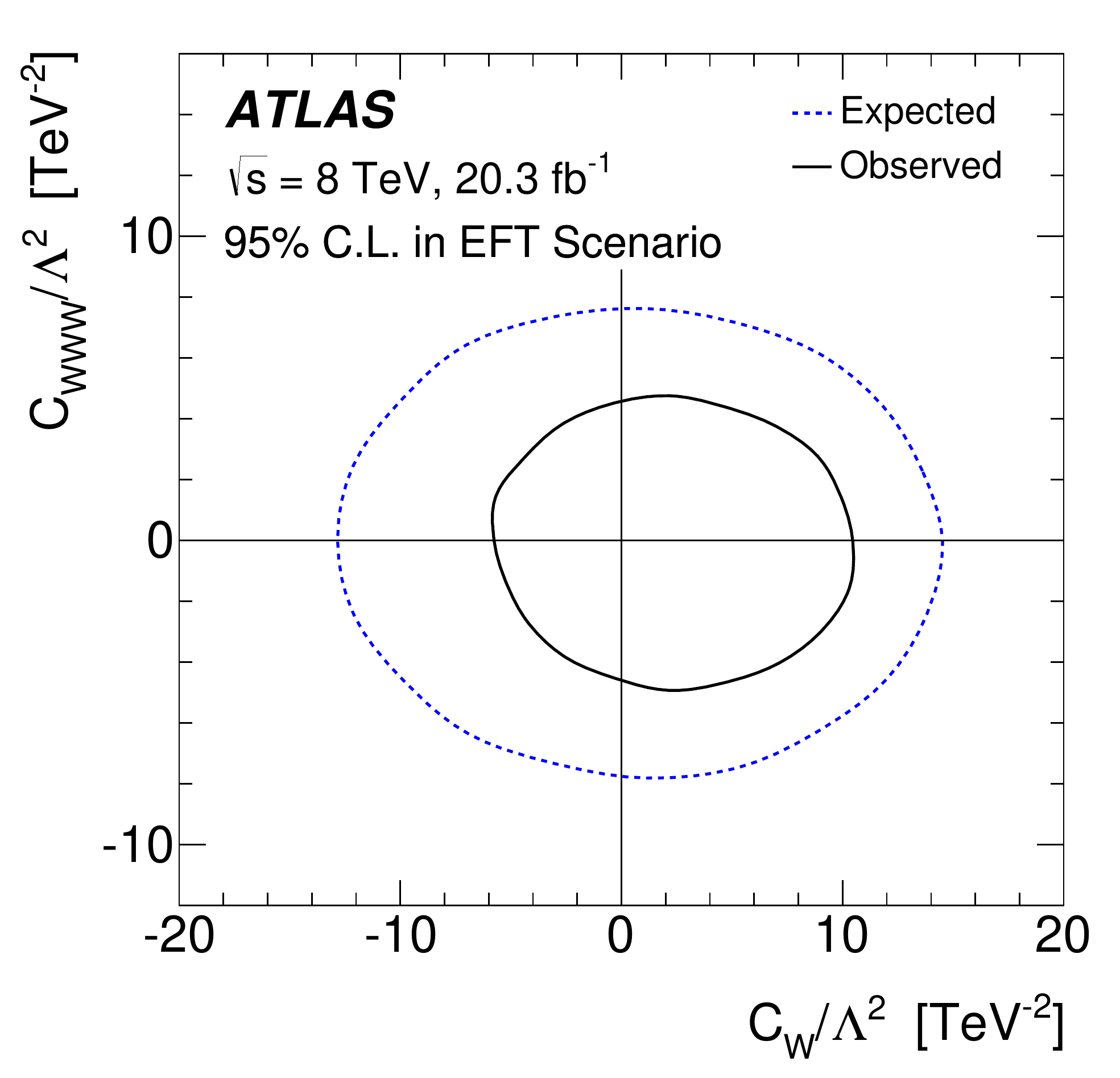}}
	\caption{The expected and observed 95\% confidence-level contours for limits in the plane of two simultaneously non-zero parameters in the effective field theory framework. In each case, only the two effective field theory couplings under study are allowed to differ from zero.}
	\label{fig:atgc_2D_limits_EFT}
\end{figure}

\FloatBarrier
\section{Conclusions}
\label{sec:summary}

The \WW production cross section in $pp$ collisions at $\sqrt{s}=8~\TeV$ is measured using data corresponding to an integrated luminosity of 20.3~\ifb\ collected by the ATLAS detector at the LHC during 2012. The measurement is conducted in three dilepton final states ($e\mu$, $ee$ and $\mu\mu$) that are all accompanied by missing transverse momentum due to the neutrinos produced in the leptonic $W$ decays. Fiducial and total cross sections are measured and limits on anomalous triple gauge couplings are derived. The uncertainty of the fiducial cross-section measurement is dominated by systematic uncertainties due to reconstruction and background estimation, while the total cross-section uncertainty is subject to significant contributions from the modelling of the extrapolation from the fiducial to the full phase space. 

The measured total \WW production cross section is measured to be \WZtotXsec $\pm \WZtotXsecStatErrUp$(stat) $^{+\WZtotXsecSysErrUp}_{\WZtotXsecSysErrDw}$(syst) $\pm\WZtotXsecLumiErrUp$(lumi)~pb, from the combination of the three analysed final states extrapolated to the full phase space. This is about 1.4 standard deviations higher than the NNLO prediction of 63.2 $^{+1.6}_{-1.4}$(scale) $\pm1.2$(PDF) pb. 

The fiducial cross sections for the three final states are about two standard deviations higher than the partial NLO predictions. However, the difference is reduced by taking into account higher-order effects that increase the cross-section prediction by $5$--$10\%$. The measured fiducial cross sections are found to be consistent with predictions that include both the NNLO and resummed QCD corrections up to NNLL accuracy. Differential cross sections are measured in the fiducial region using events in the $e\mu$ final state. The shapes of the measured unfolded differential cross section distributions agree with the predictions at the level of 15\%, 
and the discrepancy is mainly caused by the overall normalization offset. Larger deviations at the 20\% level are observed in the unfolded distribution of the leading lepton \pt\ for large transverse momenta. All measured cross sections are consistent within 1.5-2 standard deviations with the predictions.

The distribution of the transverse momentum of the leading lepton, $\pt^{\mathrm{lead}}$, is used to investigate anomalous triple-gauge-boson coupling parameters. The data show no indications of anomalous couplings and are fully compatible with the SM, hence limits on these parameters are set. The derived limits are better than expected due to a deficit in data for large momenta of the leading lepton. Due to the increased integrated luminosity and the higher centre-of-mass energy, the limits reported here are more stringent than those previously published by the ATLAS Collaboration using data taken at $\sqrt{s}=7$ TeV. They are also competitive with the results obtained at the LEP collider~\cite{schael:2013ita}. The limits can also be compared with the limits observed by the CMS experiment~\cite{Khachatryan:2015sga}, which computed limits based on the dilepton invariant mass distribution $m_{\ell\ell}$, and which also reported better observed limits than expected limits. The confidence interval on $C_{WWW}/\Lambda^{2}$ and $C_{B}/\Lambda^{2}$ derived in this analysis are similar, or up to 20-30\% more restrictive than those observed by the CMS experiment~\cite{Khachatryan:2015sga}. The limits derived on $C_{W}/\Lambda^{2}$ cover a complementary range around zero compared to the bounds by the CMS experiment.
\clearpage

\section*{Acknowledgements}


We thank CERN for the very successful operation of the LHC, as well as the
support staff from our institutions without whom ATLAS could not be
operated efficiently.

We acknowledge the support of ANPCyT, Argentina; YerPhI, Armenia; ARC, Australia; BMWFW and FWF, Austria; ANAS, Azerbaijan; SSTC, Belarus; CNPq and FAPESP, Brazil; NSERC, NRC and CFI, Canada; CERN; CONICYT, Chile; CAS, MOST and NSFC, China; COLCIENCIAS, Colombia; MSMT CR, MPO CR and VSC CR, Czech Republic; DNRF and DNSRC, Denmark; IN2P3-CNRS, CEA-DSM/IRFU, France; GNSF, Georgia; BMBF, HGF, and MPG, Germany; GSRT, Greece; RGC, Hong Kong SAR, China; ISF, I-CORE and Benoziyo Center, Israel; INFN, Italy; MEXT and JSPS, Japan; CNRST, Morocco; FOM and NWO, Netherlands; RCN, Norway; MNiSW and NCN, Poland; FCT, Portugal; MNE/IFA, Romania; MES of Russia and NRC KI, Russian Federation; JINR; MESTD, Serbia; MSSR, Slovakia; ARRS and MIZ\v{S}, Slovenia; DST/NRF, South Africa; MINECO, Spain; SRC and Wallenberg Foundation, Sweden; SERI, SNSF and Cantons of Bern and Geneva, Switzerland; MOST, Taiwan; TAEK, Turkey; STFC, United Kingdom; DOE and NSF, United States of America. In addition, individual groups and members have received support from BCKDF, the Canada Council, CANARIE, CRC, Compute Canada, FQRNT, and the Ontario Innovation Trust, Canada; EPLANET, ERC, FP7, Horizon 2020 and Marie Sk{\l}odowska-Curie Actions, European Union; Investissements d'Avenir Labex and Idex, ANR, R{\'e}gion Auvergne and Fondation Partager le Savoir, France; DFG and AvH Foundation, Germany; Herakleitos, Thales and Aristeia programmes co-financed by EU-ESF and the Greek NSRF; BSF, GIF and Minerva, Israel; BRF, Norway; the Royal Society and Leverhulme Trust, United Kingdom.

The crucial computing support from all WLCG partners is acknowledged
gratefully, in particular from CERN and the ATLAS Tier-1 facilities at
TRIUMF (Canada), NDGF (Denmark, Norway, Sweden), CC-IN2P3 (France),
KIT/GridKA (Germany), INFN-CNAF (Italy), NL-T1 (Netherlands), PIC (Spain),
ASGC (Taiwan), RAL (UK) and BNL (USA) and in the Tier-2 facilities
worldwide.

\printbibliography
\clearpage

\appendix
\section{Tables of differential cross sections}
\subsection{Differential cross section measurements}

\begin{sidewaystable}[!ht]
\footnotesize
  \centering
  {
    \sisetup{round-mode = figures, round-precision = 2,
      table-format = 2.5,
      table-number-alignment = center,table-space-text-post=\si{\percent} 
    }
    \begin{tabular}{|l| 
      S[round-mode = figures, round-precision = 2,
      table-format = 2.1,
      table-number-alignment = center]
      @{\,}|
      S[round-mode = figures, round-precision = 2,
      table-format = 2.1, table-number-alignment = center]
      @{\,}|
      S[round-mode = figures, round-precision = 2,
      table-format = 2.1, table-number-alignment = center]
      @{\,}|
      S[round-mode = figures, round-precision = 2,
      table-format = 2.1, table-number-alignment = center]
      @{\,}|
      S[round-mode = figures, round-precision = 2,
      table-format = 2.1, table-number-alignment = center]
      @{\,}|
      S[round-mode = figures, round-precision = 2,
      table-format = 2.1, table-number-alignment = center]
      @{\,}|
      S[round-mode = figures, round-precision = 2,
      table-format = 2.1, table-number-alignment = center]
      @{\,}|
      S[round-mode = figures, round-precision = 2,
      table-format = 2.1, table-number-alignment = center]
      @{\,}|
      S[round-mode = figures, round-precision = 2,
      table-format = 2.1, table-number-alignment = center]
      @{\,}|
      S[round-mode = figures, round-precision = 2,
      table-format = 2.1, table-number-alignment = center]
      @{\,}|}
      \hline\hline
$\pt^{\mathrm{lead}}$ [\GeV] & \multicolumn{1}{c|}{25--30}           & \multicolumn{1}{c|}{30--35}          & \multicolumn{1}{c|}{35--40}          & \multicolumn{1}{c|}{40--50}          & \multicolumn{1}{c|}{50--60}          & \multicolumn{1}{c|}{60--70}          & \multicolumn{1}{c|}{70--80}          & \multicolumn{1}{c|}{80--100}         & \multicolumn{1}{c|}{100--150}         & \multicolumn{1}{c|}{150--500}        \\
\hline
\multicolumn{11}{|c|}{Differential cross sections}\\\hline
 Results [fb/\GeV]    & \multicolumn{1}{c|}{\num[round-precision=3]{3.48998}} & \multicolumn{1}{c|}{\num[round-precision=3]{7.04841}} & \multicolumn{1}{c|}{\num[round-precision=3]{10.67684}} & \multicolumn{1}{c|}{\num[round-precision=3]{9.30288}} & \multicolumn{1}{c|}{\num[round-precision=3]{6.64604}} & \multicolumn{1}{c|}{\num[round-precision=3]{3.96929}} & \multicolumn{1}{c|}{\num[round-precision=3]{2.34890}}  & \multicolumn{1}{c|}{\num[round-precision=3]{1.20408}} & \multicolumn{1}{c|}{\num[round-precision=3]{0.36314}} & \multicolumn{1}{c|}{\num[round-precision=3]{0.008674}} \\

 Total Unc. 	 &	22.7794\%	 &	16.311\%	 &	11.952\%	 &	9.43332\%	 &	7.63034\%	 &	12.5447\%	 &	16.0131\%	 &	14.8223\%	 &13.435\%	 &	30.1599\%	 \\
\hline
 Stat. Unc.	 &	11.7\%	 &	7.56\%	 &	5.68\%	 &	4.04\%	 &	4.79\%	 &	6.09\%	 &	7.97\%	 &	7.94\%	 &	8.61\%	 &	23.8\%	 \\
 Syst. Unc.	 &	6.67689\%	 &	6.26846\%	 &	5.30726\%	 &	5.17865\%	 &	5.42605\%	 &	5.57742\%	 &	5.49739\%	 &	6.34765\%	 &5.98911\%	 &	7.91409\%	 \\
 Bkg. Unc. 	 &	18.2945\%	 &	12.971\%	 &	9.15231\%	 &	6.76344\%	 &	2.40468\%	 &	9.49789\%	 &	12.6844\%	 &	10.7739\%	 &8.43673\%	 &	16.8654\%	 \\
\hline
\hline
  \end{tabular}
  }
  \caption{Unfolded cross sections measured differentially in bins of the transverse momentum of the leading lepton, $\pt^{\mathrm{lead}}$, and related uncertainties. The systematic uncertainty includes uncertainties from event reconstruction and theoretical uncertainties in the signal acceptance, including the luminosity. Uncertainties in the estimated background contributions are shown separately. Systematic, background and statistical uncertainties are added in quadrature to obtain the total uncertainty in each bin. The correlation matrix for the total uncertainty is made available separately. }
  \label{tab::Analysis.Results.Unfolding.8TeV.pt1}
  {
    \sisetup{round-mode = figures, round-precision = 2,
      table-format = 2.5,
      table-number-alignment = center,table-space-text-post=\si{\percent} 
    }
  \begin{tabular}{|l|
  S[round-mode = figures, round-precision = 2,
table-format = 2.1,
table-number-alignment = center]
@{\,}|
S[round-mode = figures, round-precision = 2,
table-format = 2.1, table-number-alignment = center]
@{\,}|
S[round-mode = figures, round-precision = 2,
table-format = 2.1, table-number-alignment = center]
@{\,}|
S[round-mode = figures, round-precision = 2,
table-format = 2.1, table-number-alignment = center]
@{\,}|
S[round-mode = figures, round-precision = 2,
table-format = 2.1, table-number-alignment = center]
@{\,}|
S[round-mode = figures, round-precision = 2,
table-format = 2.1, table-number-alignment = center]
@{\,}|
S[round-mode = figures, round-precision = 2,
table-format = 2.1, table-number-alignment = center]
@{\,}|
S[round-mode = figures, round-precision = 2,
table-format = 2.1, table-number-alignment = center]
@{\,}|
S[round-mode = figures, round-precision = 2,
table-format = 2.1, table-number-alignment = center]
@{\,}|
S[round-mode = figures, round-precision = 2,
table-format = 2.1, table-number-alignment = center]
@{\,}|}
    \hline\hline
    \ptll\  [\GeV]               & \multicolumn{1}{c|}{0--25  }        & \multicolumn{1}{c|}{25--30 }         & \multicolumn{1}{c|}{30--35  }        & \multicolumn{1}{c|}{35--40    }      & \multicolumn{1}{c|}{40--50 }         & \multicolumn{1}{c|}{50--60  }        & \multicolumn{1}{c|}{60--70 }         & \multicolumn{1}{c|}{70--80 }         & \multicolumn{1}{c|}{80--500  }      \\
    \hline
    \multicolumn{10}{|c|}{Differential cross sections}\\\hline
    Results [fb/\GeV]    & \multicolumn{1}{c|}{\num[round-precision=3]{1.69389}} & \multicolumn{1}{c|}{\num[round-precision=3]{5.60392}} & \multicolumn{1}{c|}{\num[round-precision=3]{6.10728}}  & \multicolumn{1}{c|}{\num[round-precision=3]{6.58806}} & \multicolumn{1}{c|}{\num[round-precision=3]{6.80873}} & \multicolumn{1}{c|}{\num[round-precision=3]{6.84233}}  & \multicolumn{1}{c|}{\num[round-precision=3]{4.84052}} & \multicolumn{1}{c|}{\num[round-precision=3]{3.00434}} & \multicolumn{1}{c|}{\num[round-precision=3]{0.06005}} \\
    
 Total Unc. 	 &	28.8799\%	 &	13.4522\%	 &	14.2339\%	 &	13.5344\%	 &	10.0121\%	 &	10.7696\%	 &	9.41525\%	 &	10.0588\%	 &13.0397\%	 \\
\hline
 Stat. Unc.	 &	8.94\%	 &	8.51\%	 &	7.69\%	 &	7.17\%	 &	4.56\%	 &	4.29\%	 &	5.2\%	 &	6.96\%	 &	7.43\%	 \\
 Syst. Unc.	 &	6.40085\%	 &	5.60583\%	 &	5.79711\%	 &	5.3124\%	 &	5.21913\%	 &	5.1266\%	 &	5.57284\%	 &	5.46573\%	 &7.79602\%	 \\
 Bkg. Unc. 	 &	26.7\%	 &	8.85\%	 &	10.5\%	 &	10.2\%	 &	7.22\%	 &	8.46\%	 &	5.52\%	 &	4.78\%	 &	7.41\%	 \\
\hline
\hline
  \end{tabular}
}
  \caption{Unfolded cross sections measured differentially in bins of the transverse momentum of the dilepton system, \ptll, and related uncertainties. The systematic uncertainty includes uncertainties from event reconstruction and theoretical uncertainties in the signal acceptance, including the luminosity. Uncertainties in the estimated background contributions are shown separately. Systematic, background and statistical uncertainties are added in quadrature to obtain the total uncertainty in each bin. The correlation matrix for the total uncertainty is made available separately. }
   \label{tab::Analysis.Results.Unfolding.8TeV.pt2l}
  {
    \sisetup{round-mode = figures, round-precision = 2,
      table-format = 2.5,
      table-number-alignment = center,table-space-text-post=\si{\percent} 
    }
    \begin{tabular}{|l|
      S[round-mode = figures, round-precision = 2,
      table-format = 2.1,
      table-number-alignment = center]
      @{\,}|
      S[round-mode = figures, round-precision = 2,
      table-format = 2.1, table-number-alignment = center]
      @{\,}|
      S[round-mode = figures, round-precision = 2,
      table-format = 2.1, table-number-alignment = center]
      @{\,}|
      S[round-mode = figures, round-precision = 2,
      table-format = 2.1, table-number-alignment = center]
      @{\,}|
      S[round-mode = figures, round-precision = 2,
      table-format = 2.1, table-number-alignment = center]
      @{\,}|
      S[round-mode = figures, round-precision = 2,
      table-format = 2.1, table-number-alignment = center]
      @{\,}|
      S[round-mode = figures, round-precision = 2,
      table-format = 2.1, table-number-alignment = center]
      @{\,}|
      S[round-mode = figures, round-precision = 2,
      table-format = 2.1, table-number-alignment = center]
      @{\,}|
      S[round-mode = figures, round-precision = 2,
      table-format = 2.1, table-number-alignment = center]
      @{\,}|
      S[round-mode = figures, round-precision = 2,
      table-format = 2.1, table-number-alignment = center]
      @{\,}|
      S[round-mode = figures, round-precision = 2,
      table-format = 2.1, table-number-alignment = center]
      @{\,}|
      S[round-mode = figures, round-precision = 2,
      table-format = 2.1, table-number-alignment = center]
      @{\,}|
      S[round-mode = figures, round-precision = 2,
      table-format = 2.1, table-number-alignment = center]
      @{\,}|
      S[round-mode = figures, round-precision = 2,
      table-format = 2.1, table-number-alignment = center]
      @{\,}|
      }
      \hline\hline
      $m_{\ell\ell}$ [\GeV] & \multicolumn{1}{l|}{10--20}                 & \multicolumn{1}{l|}{20--30}                & \multicolumn{1}{l|}{30--40}                 & \multicolumn{1}{l|}{40--50}               & \multicolumn{1}{l|}{50--60}             & \multicolumn{1}{l|}{60--70}                 & \multicolumn{1}{l|}{70--85}               & \multicolumn{1}{l|}{85--100}                & \multicolumn{1}{l|}{100--115}              & \multicolumn{1}{l|}{115--135}              & \multicolumn{1}{l|}{135--155}            & \multicolumn{1}{l|}{155--175}            & \multicolumn{1}{l|}{175--210}            & \multicolumn{1}{l|}{210--650}           \\
                      
      \hline

      \multicolumn{15}{|c|}{Differential cross sections}\\\hline
      Results [fb/\GeV]     & \multicolumn{1}{c|}{\num[round-precision=3]{1.19347}} & \multicolumn{1}{c|}{\num[round-precision=3]{1.40042}} & \multicolumn{1}{c|}{\num[round-precision=3]{1.84326}}  & \multicolumn{1}{c|}{\num[round-precision=3]{2.62435}} & \multicolumn{1}{c|}{\num[round-precision=3]{3.49617}} & \multicolumn{1}{c|}{\num[round-precision=3]{3.32439}} & \multicolumn{1}{c|}{\num[round-precision=3]{3.45215}}  & \multicolumn{1}{c|}{\num[round-precision=3]{2.92048}} & \multicolumn{1}{c|}{\num[round-precision=3]{1.85733}}  & \multicolumn{1}{c|}{\num[round-precision=3]{1.626185}} & \multicolumn{1}{c|}{\num[round-precision=3]{1.22331}} & \multicolumn{1}{c|}{\num[round-precision=3]{0.761027}} & \multicolumn{1}{c|}{\num[round-precision=3]{0.497945}} & \multicolumn{1}{c|}{\num[round-precision=3]{0.050070}} \\ 
 Total Unc. 	 &	12.2606\%	 &	11.7582\%	 &	11.8852\%	 &	21.7069\%	 &	12.1758\%	 &	12.2445\%	 &	9.4283\%	 &	8.78993\%	 &13.009\%	 &	13.6378\%	 &	11.316\%	 &	23.5211\%	 &	20.5905\%	 &	16.1243\%	 \\
\hline
 Stat. Unc.	 &	10.1\%	 &	9.98\%	 &	8.88\%	 &	7.84\%	 &	7\%	 &	7.48\%	 &	5.49\%	 &	5.63\%	 &	7.78\%	 &	6.75\%	 &	7.58\%	 &	10.4\%	 &8.84\%	 &	7.57\%	 \\
 Syst. Unc.	 &	5.77905\%	 &	5.84504\%	 &	5.65061\%	 &	5.50772\%	 &	5.57516\%	 &	5.53148\%	 &	5.57251\%	 &	5.5093\%	 &5.75604\%	 &	5.29571\%	 &	5.44449\%	 &	5.6932\%	 &	5.63994\%	 &	6.10633\%	 \\
 Bkg. Unc. 	 &	3.68\%	 &	2.23\%	 &	5.53\%	 &	19.4\%	 &	8.32\%	 &	8.09\%	 &	5.25\%	 &	3.89\%	 &	8.71\%	 &	10.6\%	 &	6.53\%	 &	20.3\%	 &17.8\%	 &	12.8\%	 \\
\hline
\hline
    \end{tabular}
  }
  \caption{Unfolded cross sections measured differentially in bins of the invariant mass of the dilepton system, $m_{\ell\ell}$, and related uncertainties. The systematic uncertainty includes uncertainties from event reconstruction and theoretical uncertainties in the signal acceptance, including the luminosity. Uncertainties in the estimated background contributions are shown separately. Systematic, background and statistical uncertainties are added in quadrature to obtain the total uncertainty in each bin. The correlation matrix for the total uncertainty is made available separately. }
   \label{tab::Analysis.Results.Unfolding.8TeV.M2l}
\end{sidewaystable}

\begin{sidewaystable}[!ht]
\footnotesize
  \centering
  {
    \sisetup{round-mode = figures, round-precision = 2,
      table-format = 2.5,
      table-number-alignment = center,table-space-text-post=\si{\percent} 
    }
    \begin{tabular}{|l|
      S[round-mode = figures, round-precision = 2,
      table-format = 2.1,
      table-number-alignment = center]
      @{\,}|
      S[round-mode = figures, round-precision = 2,
      table-format = 2.1, table-number-alignment = center]
      @{\,}|
      S[round-mode = figures, round-precision = 2,
      table-format = 2.1, table-number-alignment = center]
      @{\,}|
      S[round-mode = figures, round-precision = 2,
      table-format = 2.1, table-number-alignment = center]
      @{\,}|
      S[round-mode = figures, round-precision = 2,
      table-format = 2.1, table-number-alignment = center]
      @{\,}|
      S[round-mode = figures, round-precision = 2,
      table-format = 2.1, table-number-alignment = center]
      @{\,}|
      S[round-mode = figures, round-precision = 2,
      table-format = 2.1, table-number-alignment = center]
      @{\,}|
      S[round-mode = figures, round-precision = 2,
      table-format = 2.1, table-number-alignment = center]
      @{\,}|
      S[round-mode = figures, round-precision = 2,
      table-format = 2.1, table-number-alignment = center]
      @{\,}|
      }
      \hline\hline
      $\Delta\phi_{\ell\ell}$                         & \multicolumn{1}{c|}{0--$\pi/8$ }    & \multicolumn{1}{c|}{$\pi/8$--$\pi/4$  }   & \multicolumn{1}{c|}{$\pi/4$--$3\pi/8$ }&\multicolumn{1}{c|}{ $3\pi/8$--$\pi/2$} & \multicolumn{1}{c|}{$\pi/2$--$5\pi/8$} & \multicolumn{1}{c|}{$5\pi/8$--$3\pi/4$ }& \multicolumn{1}{c|}{$3\pi/4$--$7\pi/8$} & \multicolumn{1}{c|}{$7\pi/8$--$\pi$} \\
      \hline
      \multicolumn{9}{|c|}{Differential cross sections}\\\hline
      Results [fb]      &\multicolumn{1}{c|}{\num[round-precision=3]{47.313}} &\multicolumn{1}{c|}{\num[round-precision=3]{70.6934}} &\multicolumn{1}{c|}{\num[round-precision=3]{81.6484}} &\multicolumn{1}{c|}{\num[round-precision=3]{101.977}} &\multicolumn{1}{c|}{\num[round-precision=3]{140.38}} &\multicolumn{1}{c|}{\num[round-precision=3]{192.876}} &\multicolumn{1}{c|}{\num[round-precision=3]{217.425}} &\multicolumn{1}{c|}{\num[round-precision=3]{98.7955}} \\  

 Total Unc. 	 &	11.2675\%	 &	8.98473\%	 &	9.53634\%	 &	14.6303\%	 &	9.458\%	 &	7.51091\%	 &	7.97371\%	 &	15.7181\%	 \\
\hline
 Stat. Unc.	 &	8.04\%	 &	6.29\%	 &	6.07\%	 &	5.37\%	 &	4.62\%	 &	3.89\%	 &	3.96\%	 &	7.25\%	 \\
 Syst. Unc.	 &	6.03865\%	 &	5.48903\%	 &	5.3205\%	 &	5.32884\%	 &	5.06048\%	 &	5.41384\%	 &	5.22418\%	 &	5.86444\%	 \\
 Bkg. Unc. 	 &	5.23\%	 &	3.31\%	 &	5.07\%	 &	12.5\%	 &	6.52\%	 &	3.45\%	 &	4.53\%	 &	12.7\%	 \\
\hline
\hline
    \end{tabular}
  }
  \caption{Unfolded cross sections measured differentially in bins of the difference in azimuth angle between the decay leptons, $\Delta\phi_{\ell\ell}$, and related uncertainties. The systematic uncertainty includes uncertainties from event reconstruction and theoretical uncertainties in the signal acceptance, including the luminosity. Uncertainties in the estimated background contributions are shown separately. Systematic, background and statistical uncertainties are added in quadrature to obtain the total uncertainty in each bin. The correlation matrix for the total uncertainty is made available separately. }
   \label{tab::Analysis.Results.Unfolding.8TeV.Dphi2l}

  {
    \sisetup{round-mode = figures, round-precision = 2,
      table-format = 2.5,
      table-number-alignment = center,table-space-text-post=\si{\percent} 
    }
    \begin{tabular}{|l|  S[round-mode = figures, round-precision = 2,
      table-format = 2.1,
      table-number-alignment = center]
      @{\,}|
      S[round-mode = figures, round-precision = 2,
      table-format = 2.1, table-number-alignment = center]
      @{\,}|
      S[round-mode = figures, round-precision = 2,
      table-format = 2.1, table-number-alignment = center]
      @{\,}|
      S[round-mode = figures, round-precision = 2,
      table-format = 2.1, table-number-alignment = center]
      @{\,}|
      S[round-mode = figures, round-precision = 2,
      table-format = 2.1, table-number-alignment = center]
      @{\,}|
      S[round-mode = figures, round-precision = 2,
      table-format = 2.1, table-number-alignment = center]
      @{\,}|
      S[round-mode = figures, round-precision = 2,
      table-format = 2.1, table-number-alignment = center]
      @{\,}|
      S[round-mode = figures, round-precision = 2,
      table-format = 2.1, table-number-alignment = center]
      @{\,}|
      S[round-mode = figures, round-precision = 2,
      table-format = 2.1, table-number-alignment = center]
      @{\,}|
      S[round-mode = figures, round-precision = 2,
      table-format = 2.1, table-number-alignment = center]
      @{\,}|
      S[round-mode = figures, round-precision = 2,
      table-format = 2.1, table-number-alignment = center]
      @{\,}|
      }
      \hline\hline
      \absyll & \multicolumn{1}{c|}{0--0.2  }       & \multicolumn{1}{c|}{0.2--0.4 }        & \multicolumn{1}{c|}{0.4--0.6  }       &\multicolumn{1}{c|}{ 0.6--0.8  }       & \multicolumn{1}{c|}{0.8--1     }       & \multicolumn{1}{c|}{1--1.2   }        & \multicolumn{1}{c|}{1.2--1.4   }      & \multicolumn{1}{c|}{1.4--1.6  }       & \multicolumn{1}{c|}{1.6--1.8}         & \multicolumn{1}{c|}{1.8--2  }          & \multicolumn{1}{c|}{2--2.5   }      \\

      \hline
      \multicolumn{12}{|c|}{Differential cross sections}\\\hline
      Results [fb]         & \multicolumn{1}{c|}{\num[round-precision=3]{237.411}} & \multicolumn{1}{c|}{\num[round-precision=3]{222.909}} & \multicolumn{1}{c|}{\num[round-precision=3]{232.468}}  & \multicolumn{1}{c|}{\num[round-precision=3]{207.213}} & \multicolumn{1}{c|}{\num[round-precision=3]{216.154}} & \multicolumn{1}{c|}{\num[round-precision=3]{193.908}} & \multicolumn{1}{c|}{\num[round-precision=3]{145.746}}  & \multicolumn{1}{c|}{\num[round-precision=3]{126.300}} & \multicolumn{1}{c|}{\num[round-precision=3]{101.754}}  & \multicolumn{1}{c|}{\num[round-precision=3]{97.437}}  & \multicolumn{1}{c|}{\num[round-precision=3]{35.5271}} \\
 Total Unc. 	 &	8.88678\%	 &	9.21857\%	 &	8.88552\%	 &	9.53225\%	 &	8.29104\%	 &	8.79858\%	 &	16.3252\%	 &	14.2288\%	 &19.1137\%	 &	11.4586\%	 &	24.5744\%	 \\
\hline
 Stat. Unc.	 &	5.23\%	 &	5.75\%	 &	5.44\%	 &	5.86\%	 &	5.54\%	 &	5.69\%	 &	6.89\%	 &	7.27\%	 &	8.32\%	 &	8.41\%	 &	8.54\%	 \\
 Syst. Unc.	 &	5.35093\%	 &	5.54175\%	 &	5.46117\%	 &	5.54167\%	 &	5.1351\%	 &	5.2664\%	 &	5.32703\%	 &	5.35257\%	 &5.54546\%	 &	5.57383\%	 &	5.3806\%	 \\
 Bkg. Unc. 	 &	4.79\%	 &	4.61\%	 &	4.42\%	 &	5.07\%	 &	3.42\%	 &	4.14\%	 &	13.8\%	 &	11.1\%	 &	16.3\%	 &	5.48\%	 &	22.4\%	 \\
\hline
\hline
    \end{tabular}
  }
  \caption{Unfolded cross sections measured differentially in bins of the absolute value of the rapidity of the dilepton system,  $\left|y_{\ell\ell}\right|$, and related uncertainties. The systematic uncertainty includes uncertainties from event reconstruction and theoretical uncertainties in the signal acceptance, including the luminosity. Uncertainties in the estimated background contributions are shown separately. Systematic, background and statistical uncertainties are added in quadrature to obtain the total uncertainty in each bin. The correlation matrix for the total uncertainty is made available separately. }
  \label{tab::Analysis.Results.Unfolding.8TeV.AbsY2l}

  {
    \sisetup{round-mode = figures, round-precision = 2,
      table-format = 2.5,
      table-number-alignment = center,table-space-text-post=\si{\percent} 
    }
    \begin{tabular}{|l| S[round-mode = figures, round-precision = 2,
      table-format = 2.1,
      table-number-alignment = center]
      @{\,}|
      S[round-mode = figures, round-precision = 2,
      table-format = 2.1, table-number-alignment = center]
      @{\,}|
      S[round-mode = figures, round-precision = 2,
      table-format = 2.1, table-number-alignment = center]
      @{\,}|
      S[round-mode = figures, round-precision = 2,
      table-format = 2.1, table-number-alignment = center]
      @{\,}|
      S[round-mode = figures, round-precision = 2,
      table-format = 2.1, table-number-alignment = center]
      @{\,}|
      S[round-mode = figures, round-precision = 2,
      table-format = 2.1, table-number-alignment = center]
      @{\,}|
      S[round-mode = figures, round-precision = 2,
      table-format = 2.1, table-number-alignment = center]
      @{\,}|
      S[round-mode = figures, round-precision = 2,
      table-format = 2.1, table-number-alignment = center]
      @{\,}|
      S[round-mode = figures, round-precision = 2,
      table-format = 2.1, table-number-alignment = center]
      @{\,}|
      }
      \hline\hline
      \abscostheta & \multicolumn{1}{c|}{0--0.1 }         & \multicolumn{1}{c|}{0.1--0.2  }        & \multicolumn{1}{c|}{0.2--0.3 }         & \multicolumn{1}{c|}{0.3--0.4   }       & \multicolumn{1}{c|}{0.4--0.5    }      & \multicolumn{1}{c|}{0.5--0.6  }        & \multicolumn{1}{c|}{0.6--0.7  }        & \multicolumn{1}{c|}{0.7--0.8  }        & \multicolumn{1}{c|}{0.8--1   }       \\

      \hline
      \multicolumn{10}{|c|}{Differential cross sections}\\\hline
      Results [fb]          & \multicolumn{1}{c|}{\num[round-precision=3]{460.293}} & \multicolumn{1}{c|}{\num[round-precision=3]{419.064}} & \multicolumn{1}{c|}{\num[round-precision=3]{469.952}}  & \multicolumn{1}{c|}{\num[round-precision=3]{402.454}} & \multicolumn{1}{c|}{\num[round-precision=3]{371.482}} & \multicolumn{1}{c|}{\num[round-precision=3]{402.756}} & \multicolumn{1}{c|}{\num[round-precision=3]{404.576}}  & \multicolumn{1}{c|}{\num[round-precision=3]{387.064}} & \multicolumn{1}{c|}{\num[round-precision=3]{211.296}} \\
 Total Unc. 	 &	10.1626\%	 &	18.3031\%	 &	8.39574\%	 &	10.0069\%	 &	15.1251\%	 &	10.4763\%	 &	10.1849\%	 &	12.1523\%	 &14.3308\%	 \\
\hline
 Stat. Unc.	 &	5.42\%	 &	5.67\%	 &	5.1\%	 &	5.75\%	 &	6.13\%	 &	5.63\%	 &	5.54\%	 &	5.37\%	 &	5.26\%	 \\
 Syst. Unc.	 &	5.37383\%	 &	5.28853\%	 &	5.6136\%	 &	5.52617\%	 &	5.19638\%	 &	5.5035\%	 &	5.22063\%	 &	5.12251\%	 &5.3696\%	 \\
 Bkg. Unc. 	 &	6.71\%	 &	16.6\%	 &	3.59\%	 &	6.04\%	 &	12.9\%	 &	6.98\%	 &	6.79\%	 &	9.62\%	 &	12.2\%	 \\
\hline
\hline
    \end{tabular}
  }
  \caption{Unfolded cross sections measured differentially in bins of the observable \abscostheta\ and related uncertainties. The systematic uncertainty includes uncertainties from event reconstruction and theoretical uncertainties in the signal acceptance, including the luminosity. Uncertainties in the estimated background contributions are shown separately. Systematic, background and statistical uncertainties are added in quadrature to obtain the total uncertainty in each bin. The correlation matrix for the total uncertainty is made available separately. }
  \label{tab::Analysis.Results.Unfolding.8TeV.AbsCosTheta}
\end{sidewaystable}

\clearpage

\subsection{Normalised differential cross sections}

\begin{sidewaystable}[!ht]
\scriptsize
  \centering
  {
    \sisetup{round-mode = figures, round-precision = 2,
      table-format = 2.5,
      table-number-alignment = center,table-space-text-post=\si{\percent} 
    }
  \begin{tabular}{|l| 
    S[round-mode = figures, round-precision = 2,
    table-format = 2.1,    table-number-alignment = center    ]
    @{\,}|
    S[round-mode = figures, round-precision = 2,
    table-format = 2.1, table-number-alignment = center]
    @{\,}|
    S[round-mode = figures, round-precision = 2,
    table-format = 2.1, table-number-alignment = center]
    @{\,}|
    S[round-mode = figures, round-precision = 2,
    table-format = 2.2, table-number-alignment = center]
    @{\,}|
    S[round-mode = figures, round-precision = 2,
    table-format = 2.2, table-number-alignment = center]
    @{\,}|
    S[round-mode = figures, round-precision = 2,
    table-format = 2.1, table-number-alignment = center]
    @{\,}|
    S[round-mode = figures, round-precision = 2,
    table-format = 2.1, table-number-alignment = center]
    @{\,}|
    S[round-mode = figures, round-precision = 2,
    table-format = 2.1, table-number-alignment = center]
    @{\,}|
    S[round-mode = figures, round-precision = 2,
    table-format = 2.1, table-number-alignment = center]
    @{\,}|
    S[round-mode = figures, round-precision = 2,
    table-format = 2.1, table-number-alignment = center]
    @{\,}|}
    \hline\hline
$\pt^{\mathrm{lead}}$ [\GeV] & \multicolumn{1}{c|}{25--30}           & \multicolumn{1}{c|}{30--35}          & \multicolumn{1}{c|}{35--40}          & \multicolumn{1}{c|}{40--50}          & \multicolumn{1}{c|}{50--60}          & \multicolumn{1}{c|}{60--70}          & \multicolumn{1}{c|}{70--80}          & \multicolumn{1}{c|}{80--100}         & \multicolumn{1}{c|}{100--150}         & \multicolumn{1}{c|}{150--500}        \\
\hline
    \multicolumn{11}{|c|}{Normalised differential cross sections}\\\hline
Results [\GeV$^{-1}$]       & \multicolumn{1}{c|}{\num[round-precision=3]{0.0093309}} & \multicolumn{1}{c|}{\num[round-precision=3]{0.0188449}} & \multicolumn{1}{c|}{\num[round-precision=3]{0.0285461}}  & \multicolumn{1}{c|}{\num[round-precision=3]{0.0248725}} & \multicolumn{1}{c|}{\num[round-precision=3]{0.0177691}} & \multicolumn{1}{c|}{\num[round-precision=3]{0.0106125}} & \multicolumn{1}{c|}{\num[round-precision=3]{0.0062801}} & \multicolumn{1}{c|}{\num[round-precision=3]{0.0032193}} & \multicolumn{1}{c|}{\num[round-precision=3]{0.0009709}} & \multicolumn{1}{c|}{\num[round-precision=3]{0.0000232}} \\

    Total Unc.  &25.3002\% &18.5611\% &8.32472\% &5.01081\% &5.09204\% &8.44533\% &13.1005\% &10.6025\% &10.9737\% &27.6609\% \\
\hline
    Stat. Unc. &11.5\% &7.23\% &5.38\% &3.62\% &4.48\% &5.85\% &7.79\% &7.72\% &8.5\% &23.6\% \\
    Syst. Unc. &3.34117\% &3.15493\% &2.20057\% &0.666108\% &0.791581\% &1.32442\% &1.39435\% &1.75514\% &2.13806\% &5.0057\% \\
    Bkg. Unc.  &22.3\% &16.9\% &5.96\% &3.4\% &2.3\% &5.95\% &10.4\% &7.13\% &6.6\% &13.4\% \\
\hline
\hline
  \end{tabular}
  }
  \caption{Unfolded cross sections measured differentially, normalised to unity and scaled with the inverse bin width in bins of the transverse momentum of the leading lepton, $\pt^{\mathrm{lead}}$, and related uncertainties. The systematic uncertainty includes uncertainties from event reconstruction and theoretical uncertainties in the signal acceptance, including the luminosity. Uncertainties in the estimated background contributions are shown separately. Systematic, background and statistical uncertainties are added in quadrature to obtain the total uncertainty in each bin. The correlation matrix for the total uncertainty is made available separately. }
  \label{tab::Analysis.Results.Unfolding.8TeV.pt1.Normalised}

  {
    \sisetup{round-mode = figures, round-precision = 2,
      table-format = 2.5,
      table-number-alignment = center,table-space-text-post=\si{\percent} 
    }
  \begin{tabular}{|l|
    S[round-mode = figures, round-precision = 2,
    table-format = 2.1,table-number-alignment = center]
    @{\,}|
S[round-mode = figures, round-precision = 2,
table-format = 2.1, table-number-alignment = center]
@{\,}|
S[round-mode = figures, round-precision = 2,
table-format = 2.1, table-number-alignment = center]
@{\,}|
S[round-mode = figures, round-precision = 2,
table-format = 2.1, table-number-alignment = center]
@{\,}|
S[round-mode = figures, round-precision = 2,
table-format = 2.2, table-number-alignment = center]
@{\,}|
S[round-mode = figures, round-precision = 2,
table-format = 2.2, table-number-alignment = center]
@{\,}|
S[round-mode = figures, round-precision = 2,
table-format = 2.1, table-number-alignment = center]
@{\,}|
S[round-mode = figures, round-precision = 2,
table-format = 2.1, table-number-alignment = center]
@{\,}|
S[round-mode = figures, round-precision = 2,
table-format = 2.1, table-number-alignment = center]
@{\,}|
S[round-mode = figures, round-precision = 2,
table-format = 2.1, table-number-alignment = center]
@{\,}|}
    \hline\hline
    \ptll\  [\GeV]               & \multicolumn{1}{c|}{0--25  }        & \multicolumn{1}{c|}{25--30 }         & \multicolumn{1}{c|}{30--35  }        & \multicolumn{1}{c|}{35--40    }      & \multicolumn{1}{c|}{40--50 }         & \multicolumn{1}{c|}{50--60  }        & \multicolumn{1}{c|}{60--70 }         & \multicolumn{1}{c|}{70--80 }         & \multicolumn{1}{c|}{80--500  }      \\
    \hline
 \multicolumn{10}{|c|}{Normalised differential cross sections}\\\hline
 Results [\GeV$^{-1}$]                     & \multicolumn{1}{c|}{\num[round-precision=3]{0.0045289}} & \multicolumn{1}{c|}{\num[round-precision=3]{0.0149828}} & \multicolumn{1}{c|}{\num[round-precision=3]{0.0163287}} & \multicolumn{1}{c|}{\num[round-precision=3]{0.0176140}} & \multicolumn{1}{c|}{\num[round-precision=3]{0.0182041}} & \multicolumn{1}{c|}{\num[round-precision=3]{0.0182939}} & \multicolumn{1}{c|}{\num[round-precision=3]{0.0129418}} & \multicolumn{1}{c|}{\num[round-precision=3]{0.0080325}} & \multicolumn{1}{c|}{\num[round-precision=3]{0.0001605}} \\
    Total Unc.  &30.5059\% &11.6511\% &12.1\% &10.9041\% &6.59367\% &7.52868\% &6.48235\% &7.52\% &9.64457\% \\
\hline
    Stat. Unc. &8.16\% &8.27\% &7.46\% &6.95\% &4.24\% &4.12\% &5.02\% &6.8\% &7.24\% \\
    Syst. Unc. &2.91247\% &2.19046\% &1.95\% &1.5003\% &0.848999\% &0.902109\% &1.58142\% &1.58\% &3.62891\% \\
    Bkg. Unc.  &29.3\% &7.93\% &9.32\% &8.27\% &4.98\% &6.23\% &3.78\% &2.8\% &5.23\% \\
\hline
\hline
  \end{tabular}
  }
   \caption{Unfolded cross sections measured differentially, normalised to unity and scaled with the inverse bin width in bins of the transverse momentum of the dilepton system, \ptll, and related uncertainties. The systematic uncertainty includes uncertainties from event reconstruction and theoretical uncertainties in the signal acceptance, including the luminosity. Uncertainties in the estimated background contributions are shown separately. Systematic, background and statistical uncertainties are added in quadrature to obtain the total uncertainty in each bin. The correlation matrix for the total uncertainty is made available separately. }
  \label{tab::Analysis.Results.Unfolding.8TeV.pt2l.Normalised}
  {
    \sisetup{round-mode = figures, round-precision = 2,
      table-format = 2.5,
      table-number-alignment = center,table-space-text-post=\si{\percent} 
    }
  \begin{tabular}{|l|
    S[round-mode = figures, round-precision = 2,
    table-format = 2.1,table-number-alignment = center]
@{\,}|
S[round-mode = figures, round-precision = 2,
table-format = 2.1, table-number-alignment = center]
@{\,}|
S[round-mode = figures, round-precision = 2,
table-format = 2.1, table-number-alignment = center]
@{\,}|
S[round-mode = figures, round-precision = 2,
table-format = 2.1, table-number-alignment = center]
@{\,}|
S[round-mode = figures, round-precision = 2,
table-format = 2.1, table-number-alignment = center]
@{\,}|
S[round-mode = figures, round-precision = 2,
table-format = 2.1, table-number-alignment = center]
@{\,}|
S[round-mode = figures, round-precision = 2,
table-format = 2.2, table-number-alignment = center]
@{\,}|
S[round-mode = figures, round-precision = 2,
table-format = 2.1, table-number-alignment = center]
@{\,}|
S[round-mode = figures, round-precision = 2,
table-format = 2.1, table-number-alignment = center]
@{\,}|
S[round-mode = figures, round-precision = 2,
table-format = 2.1, table-number-alignment = center]
@{\,}|
S[round-mode = figures, round-precision = 2,
table-format = 2.1, table-number-alignment = center]
@{\,}|
S[round-mode = figures, round-precision = 2,
table-format = 2.1, table-number-alignment = center]
@{\,}|
S[round-mode = figures, round-precision = 2,
table-format = 2.1, table-number-alignment = center]
@{\,}|
S[round-mode = figures, round-precision = 2,
table-format = 2.1, table-number-alignment = center]
@{\,}|
}
        \hline\hline
      $m_{\ell\ell}$ [\GeV] & \multicolumn{1}{l|}{10--20}                 & \multicolumn{1}{l|}{20--30}                & \multicolumn{1}{l|}{30--40}                 & \multicolumn{1}{l|}{40--50}               & \multicolumn{1}{l|}{50--60}             & \multicolumn{1}{l|}{60--70}                 & \multicolumn{1}{l|}{70--85}               & \multicolumn{1}{l|}{85--100}                & \multicolumn{1}{l|}{100--115}              & \multicolumn{1}{l|}{115--135}              & \multicolumn{1}{l|}{135--155}            & \multicolumn{1}{l|}{155--175}            & \multicolumn{1}{l|}{175--210}            & \multicolumn{1}{l|}{210--650}           \\
\hline
 \multicolumn{15}{|c|}{Normalised differential cross sections}\\\hline
Results [\GeV$^{-1}$]                     &  \multicolumn{1}{c|}{\num[round-precision=3]{0.0031909}} & \multicolumn{1}{c|}{\num[round-precision=3]{0.0037442}} & \multicolumn{1}{c|}{\num[round-precision=3]{0.0049282}} & \multicolumn{1}{c|}{\num[round-precision=3]{0.0070165}} & \multicolumn{1}{c|}{\num[round-precision=3]{0.0093475}} & \multicolumn{1}{c|}{\num[round-precision=3]{0.0088882}} & \multicolumn{1}{c|}{\num[round-precision=3]{0.0092298}} & \multicolumn{1}{c|}{\num[round-precision=3]{0.0078083}} & \multicolumn{1}{c|}{\num[round-precision=3]{0.0049658}} & \multicolumn{1}{c|}{\num[round-precision=3]{0.0043478}} & \multicolumn{1}{c|}{\num[round-precision=3]{0.0032707}} & \multicolumn{1}{c|}{\num[round-precision=3]{0.0020347}} & \multicolumn{1}{c|}{\num[round-precision=3]{0.0013313}} & \multicolumn{1}{c|}{\num[round-precision=3]{0.0001344}} \\ 
    Total Unc.  &10.4389\% &11.3001\% &8.96509\% &17.8122\% &8.11193\% &8.46029\% &5.48\% &5.73838\% &10.7447\% &9.91\% &9.1374\% &22.7262\% &23.9094\% &18.7019\% \\
\hline
    Stat. Unc. &10\% &9.83\% &8.71\% &7.59\% &6.74\% &7.13\% &5.15\% &5.34\% &7.58\% &6.49\% &7.42\% &10.3\% &8.7\% &7.45\% \\
    Syst. Unc. &1.99459\% &1.66075\% &1.36308\% &1.14856\% &1.51093\% &2.0854\% &0.94\% &1.82661\% &2.58588\% &1\% &2.08607\% &1.84613\% &1.34826\% &1.63248\% \\
    Bkg. Unc.  &2.11\% &5.33\% &1.66\% &16\% &4.24\% &4.03\% &1.6\% &1\% &7.17\% &7.42\% &4.9\% &20.1\% &22.2\% &17\% \\
\hline
\hline
  \end{tabular}
  }
  \caption{Unfolded cross sections measured differentially, normalised to unity and scaled with the inverse bin width in bins of the invariant mass of the dilepton system, $m_{\ell\ell}$, and related uncertainties. The systematic uncertainty includes uncertainties from event reconstruction and theoretical uncertainties in the signal acceptance, including the luminosity. Uncertainties in the estimated background contributions are shown separately. Systematic, background and statistical uncertainties are added in quadrature to obtain the total uncertainty in each bin. The correlation matrix for the total uncertainty is made available separately. }
  \label{tab::Analysis.Results.Unfolding.8TeV.M2l.Normalised}
\end{sidewaystable}

\begin{sidewaystable}[!htbp]
\footnotesize
  \centering
  {
    \sisetup{round-mode = figures, round-precision = 2,
      table-format = 2.5,
      table-number-alignment = center,table-space-text-post=\si{\percent} 
    }
  \begin{tabular}{|l|
  S[round-mode = figures, round-precision = 2,
table-format = 2.1,table-number-alignment = center]
@{\,}|
S[round-mode = figures, round-precision = 2,
table-format = 2.1, table-number-alignment = center]
@{\,}|
S[round-mode = figures, round-precision = 2,
table-format = 2.2, table-number-alignment = center]
@{\,}|
S[round-mode = figures, round-precision = 2,
table-format = 2.2, table-number-alignment = center]
@{\,}|
S[round-mode = figures, round-precision = 2,
table-format = 2.2, table-number-alignment = center]
@{\,}|
S[round-mode = figures, round-precision = 2,
table-format = 2.2, table-number-alignment = center]
@{\,}|
S[round-mode = figures, round-precision = 2,
table-format = 2.2, table-number-alignment = center]
@{\,}|
S[round-mode = figures, round-precision = 2,
table-format = 2.1, table-number-alignment = center]
@{\,}|
S[round-mode = figures, round-precision = 2,
table-format = 2.1, table-number-alignment = center]
@{\,}|
}
    \hline\hline
      $\Delta\phi_{\ell\ell}$                         & \multicolumn{1}{c|}{0--$\pi/8$ }    & \multicolumn{1}{c|}{$\pi/8$--$\pi/4$  }   & \multicolumn{1}{c|}{$\pi/4$--$3\pi/8$ }&\multicolumn{1}{c|}{ $3\pi/8$--$\pi/2$} & \multicolumn{1}{c|}{$\pi/2$--$5\pi/8$} & \multicolumn{1}{c|}{$5\pi/8$--$3\pi/4$ }& \multicolumn{1}{c|}{$3\pi/4$--$7\pi/8$} & \multicolumn{1}{c|}{$7\pi/8$--$\pi$} \\
\hline
 \multicolumn{9}{|c|}{Normalised differential cross sections}\\\hline

    Results       &\multicolumn{1}{c|}{\num[round-precision=3]{0.126675}} &\multicolumn{1}{c|}{\num[round-precision=3]{0.189273}} &\multicolumn{1}{c|}{\num[round-precision=3]{0.218603}} &\multicolumn{1}{c|}{\num[round-precision=3]{0.27303}} &\multicolumn{1}{c|}{\num[round-precision=3]{0.375849}} &\multicolumn{1}{c|}{\num[round-precision=3]{0.516401}} &\multicolumn{1}{c|}{\num[round-precision=3]{0.582129}} &\multicolumn{1}{c|}{\num[round-precision=3]{0.264513}} \\ 

    Total Unc.  &7.97\% &6.80124\% &11.2\% &9.85051\% &4.94\% &4.14\% &3.92062\% &11.9\% \\
\hline
    Stat. Unc. &7.78\% &6.13\% &5.85\% &5.11\% &4.32\% &3.45\% &3.45\% &6.76\% \\
    Syst. Unc. &1.35\% &1.05802\% &0.97\% &0.766551\% &0.58\% &0.6\% &0.842912\% &1.59\% \\
    Bkg. Unc.  &1.11\% &2.74\% &9.53\% &8.39\% &2.32\% &2.2\% &1.66\% &9.73\% \\
\hline
\hline
  \end{tabular}
  }
  \caption{Unfolded cross sections measured differentially, normalised to unity and scaled with the inverse bin width in bins of the difference in azimuth angle between the decay leptons, $\Delta\phi_{\ell\ell}$, and related uncertainties. The systematic uncertainty includes uncertainties from event reconstruction and theoretical uncertainties in the signal acceptance, including the luminosity. Uncertainties in the estimated background contributions are shown separately. Systematic, background and statistical uncertainties are added in quadrature to obtain the total uncertainty in each bin. The correlation matrix for the total uncertainty is made available separately. }
  \label{tab::Analysis.Results.Unfolding.8TeV.Dphi2l.Normalised}

  {
    \sisetup{round-mode = figures, round-precision = 2,
      table-format = 2.5,
      table-number-alignment = center,table-space-text-post=\si{\percent} 
    }
  \begin{tabular}{|l|  S[round-mode = figures, round-precision = 2,
table-format = 2.2,
table-number-alignment = center]
@{\,}|
S[round-mode = figures, round-precision = 2,
table-format = 2.2, table-number-alignment = center]
@{\,}|
S[round-mode = figures, round-precision = 2,
table-format = 2.2, table-number-alignment = center]
@{\,}|
S[round-mode = figures, round-precision = 2,
table-format = 2.2, table-number-alignment = center]
@{\,}|
S[round-mode = figures, round-precision = 2,
table-format = 2.2, table-number-alignment = center]
@{\,}|
S[round-mode = figures, round-precision = 2,
table-format = 2.2, table-number-alignment = center]
@{\,}|
S[round-mode = figures, round-precision = 2,
table-format = 2.2, table-number-alignment = center]
@{\,}|
S[round-mode = figures, round-precision = 2,
table-format = 2.2, table-number-alignment = center]
@{\,}|
S[round-mode = figures, round-precision = 2,
table-format = 2.2, table-number-alignment = center]
@{\,}|
S[round-mode = figures, round-precision = 2,
table-format = 2.1, table-number-alignment = center]
@{\,}|
S[round-mode = figures, round-precision = 2,
table-format = 2.1, table-number-alignment = center]
@{\,}|
}
    \hline\hline
      \absyll & \multicolumn{1}{c|}{0--0.2  }       & \multicolumn{1}{c|}{0.2--0.4 }        & \multicolumn{1}{c|}{0.4--0.6  }       &\multicolumn{1}{c|}{ 0.6--0.8  }       & \multicolumn{1}{c|}{0.8--1     }       & \multicolumn{1}{c|}{1--1.2   }        & \multicolumn{1}{c|}{1.2--1.4   }      & \multicolumn{1}{c|}{1.4--1.6  }       & \multicolumn{1}{c|}{1.6--1.8}         & \multicolumn{1}{c|}{1.8--2  }          & \multicolumn{1}{c|}{2--2.5   }      \\
\hline
 \multicolumn{12}{|c|}{Normalised differential cross sections}\\\hline
 Results                    & \multicolumn{1}{c|}{\num[round-precision=3]{0.6347495}} & \multicolumn{1}{c|}{\num[round-precision=3]{0.595975}} & \multicolumn{1}{c|}{\num[round-precision=3]{0.6215345}} & \multicolumn{1}{c|}{\num[round-precision=3]{0.554009}} & \multicolumn{1}{c|}{\num[round-precision=3]{0.577915}} & \multicolumn{1}{c|}{\num[round-precision=3]{0.518437}} & \multicolumn{1}{c|}{\num[round-precision=3]{0.3896705}} & \multicolumn{1}{c|}{\num[round-precision=3]{0.337678}} & \multicolumn{1}{c|}{\num[round-precision=3]{0.272053}} & \multicolumn{1}{c|}{\num[round-precision=3]{0.260511}} & \multicolumn{1}{c|}{\num[round-precision=3]{0.0949862}}  \\

    Total Unc.  &9.30155\% &5.78678\% &9.18427\% &6.19494\% &5.50523\% &5.54326\% &13.005\% &10.3007\% &15.2089\% &9.3055\% &21.0003\% \\
\hline
    Stat. Unc. &4.86\% &5.36\% &5.14\% &5.52\% &5.23\% &5.39\% &6.72\% &7.09\% &8.05\% &8.24\% &8.38\% \\
    Syst. Unc. &0.730068\% &1.02883\% &0.763217\% &0.83006\% &0.692892\% &0.6772\% &0.840952\% &0.947629\% &1.04809\% &1.17444\% &1.36528\% \\
    Bkg. Unc.  &7.89\% &1.89\% &7.58\% &2.68\% &1.58\% &1.07\% &11.1\% &7.52\% &12.9\% &4.16\% &19.2\% \\
\hline
\hline
  \end{tabular}
  }
  \caption{Unfolded cross sections measured differentially, normalised to unity and scaled with the inverse bin width in bins of the rapidity of the dilepton system,  $\left|y_{\ell\ell}\right|$, and related uncertainties. The systematic uncertainty includes uncertainties from event reconstruction and theoretical uncertainties in the signal acceptance, including the luminosity. Uncertainties in the estimated background contributions are shown separately. Systematic, background and statistical uncertainties are added in quadrature to obtain the total uncertainty in each bin. The correlation matrix for the total uncertainty is made available separately. }
  \label{tab::Analysis.Results.Unfolding.8TeV.AbsY2l.Normalised}

  {
    \sisetup{round-mode = figures, round-precision = 2,
      table-format = 2.5,
      table-number-alignment = center,table-space-text-post=\si{\percent} 
    }
  \begin{tabular}{|l| S[round-mode = figures, round-precision = 2,
table-format = 2.2,
table-number-alignment = center]
@{\,}|
S[round-mode = figures, round-precision = 2,
table-format = 2.2, table-number-alignment = center]
@{\,}|
S[round-mode = figures, round-precision = 2,
table-format = 2.2, table-number-alignment = center]
@{\,}|
S[round-mode = figures, round-precision = 2,
table-format = 2.2, table-number-alignment = center]
@{\,}|
S[round-mode = figures, round-precision = 2,
table-format = 2.2, table-number-alignment = center]
@{\,}|
S[round-mode = figures, round-precision = 2,
table-format = 2.2, table-number-alignment = center]
@{\,}|
S[round-mode = figures, round-precision = 2,
table-format = 2.2, table-number-alignment = center]
@{\,}|
S[round-mode = figures, round-precision = 2,
table-format = 2.2, table-number-alignment = center]
@{\,}|
S[round-mode = figures, round-precision = 2,
table-format = 2.2, table-number-alignment = center]
@{\,}|
}
    \hline\hline
      \abscostheta & \multicolumn{1}{c|}{0--0.1 }         & \multicolumn{1}{c|}{0.1--0.2  }        & \multicolumn{1}{c|}{0.2--0.3 }         & \multicolumn{1}{c|}{0.3--0.4   }       & \multicolumn{1}{c|}{0.4--0.5    }      & \multicolumn{1}{c|}{0.5--0.6  }        & \multicolumn{1}{c|}{0.6--0.7  }        & \multicolumn{1}{c|}{0.7--0.8  }        & \multicolumn{1}{c|}{0.8--1   }       \\
\hline
 \multicolumn{10}{|c|}{Normalised differential cross sections}\\\hline
 Results                    & \multicolumn{1}{c|}{\num[round-precision=3]{1.23884}} & \multicolumn{1}{c|}{\num[round-precision=3]{1.12293}} & \multicolumn{1}{c|}{\num[round-precision=3]{1.25413}} & \multicolumn{1}{c|}{\num[round-precision=3]{1.07891}} & \multicolumn{1}{c|}{\num[round-precision=3]{0.993448}} & \multicolumn{1}{c|}{\num[round-precision=3]{1.07428}} & \multicolumn{1}{c|}{\num[round-precision=3]{1.07517}} & \multicolumn{1}{c|}{\num[round-precision=3]{1.03173}} & \multicolumn{1}{c|}{\num[round-precision=3]{0.565281}}  \\
    Total Unc.  &5.9503\% &14.1001\% &5.14\% &5.89\% &10.8003\% &6.11066\% &6.89\% &13.5001\% &17.3\% \\
\hline
    Stat. Unc. &5.13\% &5.25\% &4.76\% &5.43\% &5.73\% &5.35\% &5.27\% &5.15\% &4.96\% \\
    Syst. Unc. &0.722496\% &0.49366\% &0.7\% &0.75\% &0.664831\% &0.715681\% &0.87\% &0.712531\% &0.83\% \\
    Bkg. Unc.  &2.91\% &13\% &1.81\% &2.15\% &9.14\% &2.86\% &4.35\% &12.5\% &16.5\% \\
\hline
\hline
  \end{tabular}
  }
  \caption{Unfolded cross sections measured differentially, normalised to unity and scaled with the inverse bin width in bins of the observable, \abscostheta, and related uncertainties. The systematic uncertainty includes uncertainties from event reconstruction and theoretical uncertainties in the signal acceptance, including the luminosity. Uncertainties in the estimated background contributions are shown separately. Systematic, background and statistical uncertainties are added in quadrature to obtain the total uncertainty in each bin. The correlation matrix for the total uncertainty is made available separately. }
  \label{tab::Analysis.Results.Unfolding.8TeV.AbsCosTheta.Normalised}
\end{sidewaystable}
\clearpage

\subsection{Bin-to-bin correlation matrices for the differential measurements}

\begin{table}[h!]
  \centering
\footnotesize
 {
  \begin{tabular}{|r
    |S[round-mode = figures, round-precision = 2,
    table-format = 1.3, table-number-alignment = center]
    @{\,}|S[round-mode = figures, round-precision = 2,
    table-format = 1.3, table-number-alignment = center]
    @{\,}|S[round-mode = figures, round-precision = 2,
    table-format = 1.2, table-number-alignment = center]
    @{\,}|S[round-mode = figures, round-precision = 2,
    table-format = 1.2, table-number-alignment = center]
    @{\,}|S[round-mode = figures, round-precision = 2,
    table-format = 1.2, table-number-alignment = center]
    @{\,}|S[round-mode = figures, round-precision = 2,
    table-format = 1.2, table-number-alignment = center]
    @{\,}|S[round-mode = figures, round-precision = 2,
    table-format = 1.2, table-number-alignment = center]
    @{\,}|S[round-mode = figures, round-precision = 2,
    table-format = 1.2, table-number-alignment = center]
    @{\,}|S[round-mode = figures, round-precision = 2,
    table-format = 1.3, table-number-alignment = center]
    @{\,}|S[round-mode = figures, round-precision = 2,
    table-format = 1.2, table-number-alignment = center]
    @{\,}|}
    \hline
    $\pt^{\mathrm{lead}}$ [\GeV] & \multicolumn{1}{c|}{25--30}           & \multicolumn{1}{c|}{30--35}          & \multicolumn{1}{c|}{35--40}          & \multicolumn{1}{c|}{40--50}          & \multicolumn{1}{c|}{50--60}          & \multicolumn{1}{c|}{60--70}          & \multicolumn{1}{c|}{70--80}          & \multicolumn{1}{c|}{80--100}         & \multicolumn{1}{c|}{100--150}         & \multicolumn{1}{c|}{150--500}        \\
  \hline
  \hline
\text{  25--\hphantom{0}30}&1	&0.134542	&0.0913856	&0.144466	&0.201013	&0.108193	&0.093555	&0.125829	&0.124912	&0.0924576 \\
\text{30--\hphantom{0}35}  &0.134542	&1	&0.0597336	&0.170643	&0.241116	&0.115384	&0.0906569	&0.138861	&0.147256	&0.0777681 \\
\text{35--\hphantom{0}40}  &0.0913856	&0.0597336	&1	&0.216938	&0.316312	&0.198246	&0.150651	&0.192957	&0.220977	&0.0882091 \\
\text{40--\hphantom{0}50}  &0.144466	&0.170643	&0.216938	&1	&0.39377	&0.269048	&0.215193	&0.282275	&0.298717	&0.150317 \\
\text{50--\hphantom{0}60}  &0.201013	&0.241116	&0.316312	&0.39377	&1	&0.329182	&0.280535	&0.392204	&0.425087	&0.210905 \\
\text{60--\hphantom{0}70}  &0.108193	&0.115384	&0.198246	&0.269048	&0.329182	&1	&0.158541	&0.24888	&0.292884	&0.172176 \\
\text{70--\hphantom{0}80}  &0.093555	&0.0906569	&0.150651	&0.215193	&0.280535	&0.158541	&1	&0.187896	&0.24995	&0.143889 \\
\text{80--100}             &0.125829	&0.138861	&0.192957	&0.282275	&0.392204	&0.24888	&0.187896	&1	&0.327677	&0.214662 \\
\text{100--150}  &          0.124912	&0.147256	&0.220977	&0.298717	&0.425087	&0.292884	&0.24995	&0.327677	&1	&0.213062 \\
\text{150--500}  &          0.0924576	&0.0777681	&0.0882091	&0.150317	&0.210905	&0.172176	&0.143889	&0.214662	&0.213062	&1 \\
    \hline
  \end{tabular}
  }
  \caption{Correlation matrix for the total uncertainties for the unnormalised unfolded distribution of the leading lepton \pt, including all sources of systematic and
statistical uncertainties.}
    \label{tab::Systematic.Correlation.Matricies.Unnormalised.pT1}
  \end{table}

\begin{table}[h!]
  \centering
\footnotesize
  {
  \begin{tabular}{|r|
  S[round-mode = figures, round-precision = 2,
table-format = 1.3,
table-number-alignment = center]
@{\,}|
S[round-mode = figures, round-precision = 2,
table-format = 1.2, table-number-alignment = center]
@{\,}|
S[round-mode = figures, round-precision = 2,
table-format = 1.2, table-number-alignment = center]
@{\,}|
S[round-mode = figures, round-precision = 2,
table-format = 1.2, table-number-alignment = center]
@{\,}|
S[round-mode = figures, round-precision = 2,
table-format = 1.2, table-number-alignment = center]
@{\,}|
S[round-mode = figures, round-precision = 2,
table-format = 1.2, table-number-alignment = center]
@{\,}|
S[round-mode = figures, round-precision = 2,
table-format = 1.2, table-number-alignment = center]
@{\,}|
S[round-mode = figures, round-precision = 2,
table-format = 1.2, table-number-alignment = center]
@{\,}|
S[round-mode = figures, round-precision = 2,
table-format = 1.3, table-number-alignment = center]
@{\,}|
S[round-mode = figures, round-precision = 2,
table-format = 1.3, table-number-alignment = center]
@{\,}|}
    \hline
    \ptll\  [\GeV]               & \multicolumn{1}{c|}{0--25  }        & \multicolumn{1}{c|}{25--30 }         & \multicolumn{1}{c|}{30--35  }        & \multicolumn{1}{c|}{35--40    }      & \multicolumn{1}{c|}{40--50 }         & \multicolumn{1}{c|}{50--60  }        & \multicolumn{1}{c|}{60--70 }         & \multicolumn{1}{c|}{70--80 }         & \multicolumn{1}{c|}{80--500  }      \\
  \hline
  \hline
\text{  0--\hphantom{0}25}	& 1	&0.0835804	&0.0818019	&0.0851709	&0.124122	&0.102359	&0.132214	&0.119732	&0.13508 \\
\text{ 25--\hphantom{0}30}	& 0.0840369	&1	&0.140225	&0.15922	&0.236879	&0.197697	&0.230641	&0.233894	&0.240162 \\
 \text{30--\hphantom{0}35}	& 0.0821927	&0.140388	&1	&0.134536	&0.237679	&0.20776	&0.241777	&0.22083	&0.24472 \\
 \text{35--\hphantom{0}40}	& 0.085598	&0.159555	&0.134647	&1	&0.194033	&0.202099	&0.226399	&0.22107	&0.2231 \\
\text{ 40--\hphantom{0}50}	& 0.124114	&0.235292	&0.236224	&0.192777	&1	&0.261258	&0.344046	&0.326157	&0.372406 \\
\text{ 50--\hphantom{0}60}	& 0.102581	&0.197955	&0.207837	&0.20206	&0.262183	&1	&0.285991	&0.288988	&0.349784 \\
\text{ 60--\hphantom{0}70}	& 0.131126	&0.224722	&0.236236	&0.22117	&0.340515	&0.281503	&1	&0.327783	&0.446096 \\
\text{ 70--\hphantom{0}80}	& 0.120879	&0.234367	&0.220892	&0.220847	&0.329785	&0.288859	&0.339853	&1	&0.425535 \\
\text{ 80--500}           	& 0.136264	&0.24085	&0.245009	&0.223103	&0.376269	&0.349896	&0.461136	&0.426084	&1 \\
\hline
\end{tabular}
}
\caption{Correlation matrix for the total uncertainties for the unnormalised unfolded \ptll~distribution, including all sources of systematic and
statistical uncertainties.}
    \label{tab::Systematic.Correlation.Matricies.Unnormalised.pT2l}
\end{table}

  \begin{table}[h!]
  \centering
 \footnotesize
  {
   \begin{tabular}{|r|
  S[round-mode = figures, round-precision = 2,
table-format = 1.3, table-number-alignment = center]
@{\,}|
S[round-mode = figures, round-precision = 2,
table-format = 1.3, table-number-alignment = center]
@{\,}|
S[round-mode = figures, round-precision = 2,
table-format = 1.3, table-number-alignment = center]
@{\,}|
S[round-mode = figures, round-precision = 2,
table-format = 1.3, table-number-alignment = center]
@{\,}|
S[round-mode = figures, round-precision = 2,
table-format = 1.3, table-number-alignment = center]
@{\,}|
S[round-mode = figures, round-precision = 2,
table-format = 1.3, table-number-alignment = center]
@{\,}|
S[round-mode = figures, round-precision = 2,
table-format = 1.3, table-number-alignment = center]
@{\,}|
S[round-mode = figures, round-precision = 2,
table-format = 1.4, table-number-alignment = center]
@{\,}|
S[round-mode = figures, round-precision = 2,
table-format = 1.3, table-number-alignment = center]
@{\,}|
S[round-mode = figures, round-precision = 2,
table-format = 1.3, table-number-alignment = center]
@{\,}|
S[round-mode = figures, round-precision = 2,
table-format = 1.3, table-number-alignment = center]
@{\,}|
S[round-mode = figures, round-precision = 2,
table-format = 1.3, table-number-alignment = center]
@{\,}|
S[round-mode = figures, round-precision = 2,
table-format = 1.3, table-number-alignment = center]
@{\,}|
S[round-mode = figures, round-precision = 2,
table-format = 1.4, table-number-alignment = center]
@{\,}|
}
    \hline
      $m_{\ell\ell}$ [\GeV] & \multicolumn{1}{l|}{10}                 & \multicolumn{1}{l|}{20}                & \multicolumn{1}{l|}{30}                 & \multicolumn{1}{l|}{40}               & \multicolumn{1}{l|}{50}             & \multicolumn{1}{l|}{60}                 & \multicolumn{1}{l|}{70}               & \multicolumn{1}{l|}{85}                & \multicolumn{1}{l|}{100}              & \multicolumn{1}{l|}{115}              & \multicolumn{1}{l|}{135}            & \multicolumn{1}{l|}{155}            & \multicolumn{1}{l|}{175}            & \multicolumn{1}{l|}{210}           \\
                           &        \multicolumn{1}{r|}{ --20   }       & \multicolumn{1}{r|}{ --30      }     & \multicolumn{1}{r|}{ --40  }         & \multicolumn{1}{r|}{ --50     }      &\multicolumn{1}{r|}{ --60 }         & \multicolumn{1}{r|}{ --70 }          & \multicolumn{1}{r|}{ --85   }        & \multicolumn{1}{r|}{ --100 }         & \multicolumn{1}{r|}{ --115     }     &\multicolumn{1}{r|}{ --135    }      &\multicolumn{1}{r|}{ --155    }      & \multicolumn{1}{r|}{ --175  }        &\multicolumn{1}{r|}{ --210    }      & \multicolumn{1}{r|}{ --650  }       \\
\hline
 \hline

     \text{     10--\hphantom{0}20}	& 1	&0.191001	&0.230356	&0.123688	&0.228499	&0.193943	&0.259555	&0.321657	&0.151642	&0.183523	&0.236907	&0.1041	&0.137815	&0.172649 \\
     \text{     20--\hphantom{0}30}	& 0.191213	&1	&0.203835	&0.129185	&0.230161	&0.21217	&0.33193	&0.331848	&0.227781	&0.198462	&0.235807	&0.127901	&0.145434	&0.201939 \\
     \text{    30--\hphantom{0}40} 	& 0.228705	&0.202431	&1	&0.105054	&0.250308	&0.231377	&0.306599	&0.312753	&0.233789	&0.194289	&0.235591	&0.127134	&0.12348	&0.173442 \\
     \text{ 40--\hphantom{0}50}    	& 0.121707	&0.127453	&0.104499	&1	&0.101401	&0.133333	&0.166359	&0.179331	&0.13045	&0.109828	&0.127998	&0.0711743	&0.0812416	&0.0927904 \\
     \text{     50--\hphantom{0}60}	& 0.224209	&0.226515	&0.248595	&0.101366	&1	&0.18711	&0.308999	&0.328538	&0.218579	&0.200335	&0.236087	&0.119804	&0.148185	&0.165764 \\
     \text{60--\hphantom{0}70}     	& 0.171444	&0.191427	&0.217061	&0.130634	&0.182441	&1	&0.265014	&0.260667	&0.258064	&0.197772	&0.211517	&0.129257	&0.132918	&0.157876 \\
     \text{ 70--\hphantom{0}85}    	& 0.258257	&0.330411	&0.306659	&0.166638	&0.309808	&0.271798	&1	&0.382737	&0.306965	&0.260403	&0.313007	&0.152582	&0.184794	&0.22486 \\
     \text{  85--100}              	& 0.296108	&0.309723	&0.300008	&0.177009	&0.323432	&0.26376	&0.376275	&1	&0.226172	&0.250643	&0.340565	&0.163653	&0.1989	&0.247886 \\
     \text{  100--115}             	& 0.133087	&0.204115	&0.218041	&0.127235	&0.212152	&0.257847	&0.297819	&0.222975	&1	&0.178446	&0.183214	&0.11597	&0.139596	&0.161345 \\
     \text{ 115--135}              	& 0.183079	&0.197963	&0.19459	&0.110081	&0.201015	&0.20337	&0.260528	&0.255437	&0.184438	&1	&0.170329	&0.127407	&0.129186	&0.165021 \\
     \text{   135--155}            	& 0.232122	&0.231664	&0.233698	&0.127962	&0.236184	&0.220455	&0.311683	&0.349307	&0.191919	&0.169414	&1	&0.0845525	&0.158484	&0.189133 \\
     \text{  155--175}             	& 0.10288	&0.126597	&0.12673	&0.07126	&0.120043	&0.133496	&0.152364	&0.167084	&0.120369	&0.127127	&0.0847772	&1	&0.069636	&0.101842 \\
     \text{     175--210}          	& 0.137852	&0.14538	&0.123876	&0.0815351	&0.148918	&0.137674	&0.185023	&0.203708	&0.145369	&0.129272	&0.159638	&0.0698774	&1	&0.118822 \\
     \text{  210--650}             	& 0.171971	&0.20117	&0.173614	&0.0930517	&0.166421	&0.163415	&0.224933	&0.253655	&0.167881	&0.164967	&0.190214	&0.102075	&0.118679	&1 \\

\hline
\end{tabular}
}
\caption{Correlation matrix for the total uncertainties for the unnormalised unfolded $m_{\ell\ell}$ distribution, including all sources of systematic and
statistical uncertainties.}
    \label{tab::Systematic.Correlation.Matricies.Unnormalised.M2l}
\end{table}

\begin{table}[h!]
  \centering
\footnotesize
  {
  \begin{tabular}{|r|
  S[round-mode = figures, round-precision = 2,
table-format = 2.3,
table-number-alignment = center]
@{\,}|
S[round-mode = figures, round-precision = 2,
table-format = 2.3, table-number-alignment = center]
@{\,}|
S[round-mode = figures, round-precision = 2,
table-format = 2.3, table-number-alignment = center]
@{\,}|
S[round-mode = figures, round-precision = 2,
table-format = 2.3, table-number-alignment = center]
@{\,}|
S[round-mode = figures, round-precision = 2,
table-format = 2.3, table-number-alignment = center]
@{\,}|
S[round-mode = figures, round-precision = 2,
table-format = 2.3, table-number-alignment = center]
@{\,}|
S[round-mode = figures, round-precision = 2,
table-format = 2.3, table-number-alignment = center]
@{\,}|
S[round-mode = figures, round-precision = 2,
table-format = 2.3, table-number-alignment = center]
@{\,}|
S[round-mode = figures, round-precision = 2,
table-format = 2.3, table-number-alignment = center]
@{\,}|
}
    \hline
      $\Delta\phi_{\ell\ell}$                         & \multicolumn{1}{c|}{0--$\pi/8$ }    & \multicolumn{1}{c|}{$\pi/8$--$\pi/4$  }   & \multicolumn{1}{c|}{$\pi/4$--$3\pi/8$ }&\multicolumn{1}{c|}{ $3\pi/8$--$\pi/2$} & \multicolumn{1}{c|}{$\pi/2$--$5\pi/8$} & \multicolumn{1}{c|}{$5\pi/8$--$3\pi/4$ }& \multicolumn{1}{c|}{$3\pi/4$--$7\pi/8$} & \multicolumn{1}{c|}{$7\pi/8$--$\pi$} \\
    \hline
\hline
\hline
\text{0--\hphantom{0}$\pi/8$}                  	&1	&0.349359	&0.340203	&0.228053	&0.331656	&0.433498	&0.382975	&0.238476 \\
\text{$\pi/8$--\hphantom{0}$\pi/4$}            	&0.348626	&1	&0.375819	&0.245492	&0.362414	&0.476214	&0.416574	&0.250825 \\
\text{$\pi/4$--$3\pi/8$}                       	&0.339912	&0.376106	&1	&0.232188	&0.322002	&0.447681	&0.400087	&0.22151 \\
\text{ $3\pi/8$--\hphantom{0}$\pi/2$}          	&0.227333	&0.245165	&0.231877	&1	&0.219115	&0.303511	&0.275727	&0.159575 \\
\text{$\pi/2$--$5\pi/8$}                       	&0.330598	&0.361929	&0.321556	&0.219118	&1	&0.451535	&0.425549	&0.240501 \\
\text{$5\pi/8$--$3\pi/4$ }                     	&0.430723	&0.47442	&0.446291	&0.30328	&0.451139	&1	&0.574742	&0.332369 \\
\text{$3\pi/4$--$7\pi/8$}                      	&0.381059	&0.415441	&0.399173	&0.275616	&0.425346	&0.574886	&1	&0.319634 \\
\text{$7\pi/8$--\hphantom{0}$\pi$\hphantom{/0}}	&0.238212	&0.250873	&0.221456	&0.159647	&0.24062	&0.332767	&0.319881	&1 \\
\hline
\end{tabular}
}
\caption{Correlation matrix for the total uncertainties for the unnormalised unfolded $\Delta\phi_{\ell\ell}$ distribution, including all sources of systematic and
statistical uncertainties.}
    \label{tab::Systematic.Correlation.Matricies.Unnormalised.dphi2l}
\end{table}

 \begin{table}[h!]
  \centering
\footnotesize
  {
    \begin{tabular}{|r|  
      S[round-mode = figures, round-precision = 2,
      table-format = 1.3,
      table-number-alignment = center]
      @{\,}|
      S[round-mode = figures, round-precision = 2,
      table-format = 1.3, table-number-alignment = center]
      @{\,}|
      S[round-mode = figures, round-precision = 2,
      table-format = 1.3, table-number-alignment = center]
      @{\,}|
      S[round-mode = figures, round-precision = 2,
      table-format = 1.3, table-number-alignment = center]
      @{\,}|
      S[round-mode = figures, round-precision = 2,
      table-format = 1.3, table-number-alignment = center]
      @{\,}|
      S[round-mode = figures, round-precision = 2,
      table-format = 1.3, table-number-alignment = center]
      @{\,}|
      S[round-mode = figures, round-precision = 2,
      table-format = 1.3, table-number-alignment = center]
      @{\,}|
      S[round-mode = figures, round-precision = 2,
      table-format = 1.3, table-number-alignment = center]
      @{\,}|
      S[round-mode = figures, round-precision = 2,
      table-format = 1.3, table-number-alignment = center]
      @{\,}|
      S[round-mode = figures, round-precision = 2,
      table-format = 1.3, table-number-alignment = center]
      @{\,}|
      S[round-mode = figures, round-precision = 2,
      table-format = 1.3, table-number-alignment = center]
      @{\,}|
      }
      \hline
      \absyll & \multicolumn{1}{c|}{0--0.2  }       & \multicolumn{1}{c|}{0.2--0.4 }        & \multicolumn{1}{c|}{0.4--0.6  }       &\multicolumn{1}{c|}{ 0.6--0.8  }       & \multicolumn{1}{c|}{0.8--1     }       & \multicolumn{1}{c|}{1--1.2   }        & \multicolumn{1}{c|}{1.2--1.4   }      & \multicolumn{1}{c|}{1.4--1.6  }       & \multicolumn{1}{c|}{1.6--1.8}         & \multicolumn{1}{c|}{1.8--2  }          & \multicolumn{1}{c|}{2--2.5   }      \\
      \hline
      \hline
      \text{0\hphantom{.0}--0.2}	&1	&0.45712	&0.491001	&0.460715	&0.46101	&0.443834	&0.224607	&0.261485	&0.211709	&0.347042	&0.14925 \\
      \text{0.2--0.4}           	&0.45531	&1	&0.44145	&0.482168	&0.443872	&0.430578	&0.235854	&0.245284	&0.217174	&0.344407	&0.135408 \\
      \text{0.4--0.6}           	&0.491143	&0.443374	&1	&0.434966	&0.447215	&0.435795	&0.230272	&0.253816	&0.210375	&0.341946	&0.140518 \\
      \text{0.6--0.8}           	&0.458115	&0.481419	&0.432328	&1	&0.41854	&0.416125	&0.221857	&0.255037	&0.200376	&0.333348	&0.143657 \\
      \text{0.8--1.0}           	&0.46113	&0.446085	&0.447179	&0.421457	&1	&0.410177	&0.233061	&0.25834	&0.212027	&0.342302	&0.143131 \\
      \text{1.0--1.2}           	&0.443914	&0.43281	&0.435715	&0.41915	&0.410132	&1	&0.203145	&0.25057	&0.218781	&0.334054	&0.143942 \\
      \text{1.2--1.4}           	&0.224223	&0.236276	&0.229808	&0.222578	&0.232522	&0.20266	&1	&0.128093	&0.114827	&0.184659	&0.0752544 \\
      \text{1.4--1.6}           	&0.261529	&0.246473	&0.253773	&0.256774	&0.258313	&0.25057	&0.128377	&1	&0.121375	&0.194583	&0.0977603 \\
      \text{1.6--1.8}           	&0.211389	&0.217771	&0.209984	&0.20128	&0.21159	&0.218327	&0.114886	&0.121139	&1	&0.158012	&0.0806512 \\
      \text{1.8--2.0}           	&0.347267	&0.347699	&0.341946	&0.337705	&0.342359	&0.334175	&0.185542	&0.194651	&0.158639	&1	&0.125638 \\
      \text{2.0--2.5}           	&0.149281	&0.135958	&0.140508	&0.144479	&0.143131	&0.143955	&0.0753924	&0.0977684	&0.0807879	&0.125618	&1 \\
      \hline
    \end{tabular}
  }
  \caption{Correlation matrix for the total uncertainties for the unnormalised unfolded \absyll~distribution, including all sources of systematic and
statistical uncertainties.}
  \label{tab::Systematic.Correlation.Matricies.Unnormalised.absy2l}
\end{table}

\begin{table}[h!]
  \centering
 \footnotesize
  {
    \begin{tabular}{|r|
      S[round-mode = figures, round-precision = 2,
     table-format = 1.2,
      table-number-alignment = center]
      @{\,}|
      S[round-mode = figures, round-precision = 2,
      table-format = 1.2, table-number-alignment = center]
      @{\,}|
      S[round-mode = figures, round-precision = 2,
      table-format = 1.3, table-number-alignment = center]
      @{\,}|
      S[round-mode = figures, round-precision = 2,
      table-format = 1.3, table-number-alignment = center]
      @{\,}|
      S[round-mode = figures, round-precision = 2,
      table-format = 1.2, table-number-alignment = center]
      @{\,}|
      S[round-mode = figures, round-precision = 2,
      table-format = 1.3, table-number-alignment = center]
      @{\,}|
      S[round-mode = figures, round-precision = 2,
      table-format = 1.3, table-number-alignment = center]
      @{\,}|
      S[round-mode = figures, round-precision = 2,
      table-format = 1.2, table-number-alignment = center]
      @{\,}|
      S[round-mode = figures, round-precision = 2,
      table-format = 1.3, table-number-alignment = center]
      @{\,}|
      }
      \hline
      \abscostheta & \multicolumn{1}{c|}{0--0.1 }         & \multicolumn{1}{c|}{0.1--0.2  }        & \multicolumn{1}{c|}{0.2--0.3 }         & \multicolumn{1}{c|}{0.3--0.4   }       & \multicolumn{1}{c|}{0.4--0.5    }      & \multicolumn{1}{c|}{0.5--0.6  }        & \multicolumn{1}{c|}{0.6--0.7  }        & \multicolumn{1}{c|}{0.7--0.8  }        & \multicolumn{1}{c|}{0.8--1   }       \\
      \hline
      \hline
      \text{ 0\hphantom{.0}--0.1}	&1	&0.195388	&0.439621	&0.355276	&0.230662	&0.328325	&0.325133	&0.263835	&0.221578 \\
      \text{ 0.1--0.2}           	&0.195313	&1	&0.239974	&0.212612	&0.135821	&0.1886	&0.179286	&0.140131	&0.123416 \\
      \text{0.2--0.3}            	&0.439276	&0.239955	&1	&0.435878	&0.294734	&0.420543	&0.395765	&0.317773	&0.285654 \\
      \text{0.3--0.4}            	&0.355252	&0.212705	&0.436237	&1	&0.246535	&0.341816	&0.328202	&0.277404	&0.232667 \\
      \text{0.4--0.5}            	&0.230593	&0.135836	&0.294826	&0.24646	&1	&0.226811	&0.212115	&0.178213	&0.149293 \\
      \text{0.5--0.6}            	&0.32842	&0.188724	&0.421048	&0.341956	&0.226945	&1	&0.315608	&0.259366	&0.222786 \\
      \text{0.6--0.7}            	&0.325035	&0.179331	&0.395973	&0.328114	&0.212135	&0.315402	&1	&0.24618	&0.210029 \\
      \text{0.7--0.8}            	&0.263885	&0.140204	&0.318095	&0.277484	&0.178295	&0.259346	&0.246307	&1	&0.171884 \\
      \text{0.8--1.0}            	&0.221518	&0.123429	&0.285742	&0.232602	&0.149293	&0.222664	&0.210012	&0.171811	&1 \\
      \hline
    \end{tabular}
  }
  \caption{Correlation matrix for the total uncertainties for the unnormalised unfolded \abscostheta~distribution, including all sources of systematic and
statistical uncertainties.}
  \label{tab::Systematic.Correlation.Matricies.Unnormalised.abscostheta}
\end{table}
\clearpage
\newpage
\subsection{Bin-to-bin correlation matrices for the normalised differential measurements}

\begin{table}[h!]
  \centering
 \footnotesize
  {
  \begin{tabular}{|r
    |S[round-mode = figures, round-precision = 2,
    table-format = 2.3, table-number-alignment = center]
    @{\,}|S[round-mode = figures, round-precision = 2,
    table-format = 2.3, table-number-alignment = center]
    @{\,}|S[round-mode = figures, round-precision = 2,
    table-format = 2.3, table-number-alignment = center]
    @{\,}|S[round-mode = figures, round-precision = 2,
    table-format = 2.3, table-number-alignment = center]
    @{\,}|S[round-mode = figures, round-precision = 2,
    table-format = 2.3, table-number-alignment = center]
    @{\,}|S[round-mode = figures, round-precision = 2,
    table-format = 2.3, table-number-alignment = center]
    @{\,}|S[round-mode = figures, round-precision = 2,
    table-format = 2.3, table-number-alignment = center]
    @{\,}|S[round-mode = figures, round-precision = 2,
    table-format = 2.3, table-number-alignment = center]
    @{\,}|S[round-mode = figures, round-precision = 2,
    table-format = 2.3, table-number-alignment = center]
    @{\,}|S[round-mode = figures, round-precision = 2,
    table-format = 2.3, table-number-alignment = center]
    @{\,}|}
    \hline
    $\pt^{\mathrm{lead}}$ [\GeV] & \multicolumn{1}{c|}{25--30}           & \multicolumn{1}{c|}{30--35}          & \multicolumn{1}{c|}{35--40}          & \multicolumn{1}{c|}{40--50}          & \multicolumn{1}{c|}{50--60}          & \multicolumn{1}{c|}{60--70}          & \multicolumn{1}{c|}{70--80}          & \multicolumn{1}{c|}{80--100}         & \multicolumn{1}{c|}{100--150}         & \multicolumn{1}{c|}{150--500}        \\
    \hline
    \hline

    {25--\hphantom{0}30}      &  1	&0.743244	&-0.659934	&-0.585009	&0.350158	&-0.630403	&-0.466807	&-0.543547	&0.122237	&-0.253344 \\
    {30--\hphantom{0}35}      &  0.742356	&1	&-0.708039	&-0.596108	&0.348638	&-0.648044	&-0.493483	&-0.565866	&0.111844	&-0.288154 \\
    {35--\hphantom{0}40}      &  -0.659026	&-0.707968	&1	&0.346687	&-0.390487	&0.453527	&0.308714	&0.328888	&-0.225382	&0.158147 \\
    {40--\hphantom{0}50}      &  -0.585155	&-0.596309	&0.34676	&1	&-0.504055	&0.38835	&0.136744	&0.223954	&-0.178444	&0.151621 \\
    {50--\hphantom{0}60}      &  0.350219	&0.348762	&-0.390584	&-0.504051	&1	&-0.45309	&-0.205956	&-0.297333	&-0.0485598	&-0.219938 \\
    {60--\hphantom{0}70}      &  -0.630576	&-0.648382	&0.453728	&0.388349	&-0.453098	&1	&0.188077	&0.340752	&-0.0800655	&0.272454 \\
    {70--\hphantom{0}80}      &  -0.46699	&-0.49396	&0.309019	&0.136745	&-0.205967	&0.18808	&1	&0.352619	&-0.258152	&0.0204098 \\
    { 80--100}                &  -0.543752	&-0.566392	&0.329199	&0.223955	&-0.297347	&0.340757	&0.352618	&1	&-0.117393	&0.213042 \\
    { 100--150}               &  0.122276	&0.111991	&-0.225709	&-0.178431	&-0.0485599	&-0.0800598	&-0.258119	&-0.117379	&1	&0.0866612 \\
    { 150--500}               &  -0.255328	&-0.299351	&0.165696	&0.15157	&-0.220425	&0.272468	&0.0203957	&0.212922	&0.0868535	&1 \\
\hline
  \end{tabular}
  }
  \caption{Correlation matrix for the total uncertainties for the normalised unfolded leading lepton \pt~distribution, including all sources of systematic and
statistical uncertainties.}
    \label{tab::Systematic.Correlation.Matricies.Normalised.pT1}
  \end{table}

\begin{table}[h!]
  \centering
 \footnotesize
  {
  \begin{tabular}{|r|
    S[round-mode = figures, round-precision = 2,
    table-format = 2.3,table-number-alignment = center]
@{\,}|
    S[round-mode = figures, round-precision = 2,
    table-format = 2.3, table-number-alignment = center]
    @{\,}|
    S[round-mode = figures, round-precision = 2,
    table-format = 2.3, table-number-alignment = center]
    @{\,}|
    S[round-mode = figures, round-precision = 2,
    table-format = 2.4, table-number-alignment = center]
    @{\,}|
    S[round-mode = figures, round-precision = 2,
    table-format = 2.3, table-number-alignment = center]
    @{\,}|
    S[round-mode = figures, round-precision = 2,
    table-format = 2.3, table-number-alignment = center]
    @{\,}|
    S[round-mode = figures, round-precision = 2,
    table-format = 2.3, table-number-alignment = center]
    @{\,}|
    S[round-mode = figures, round-precision = 2,
    table-format = 2.4, table-number-alignment = center]
    @{\,}|
    S[round-mode = figures, round-precision = 2,
    table-format = 2.3, table-number-alignment = center]
    @{\,}|
    S[round-mode = figures, round-precision = 2,
    table-format = 2.3, table-number-alignment = center]
    @{\,}|}
    \hline
    \ptll\  [\GeV]               & \multicolumn{1}{c|}{0--25  }        & \multicolumn{1}{c|}{25--30 }         & \multicolumn{1}{c|}{30--35  }        & \multicolumn{1}{c|}{35--40    }      & \multicolumn{1}{c|}{40--50 }         & \multicolumn{1}{c|}{50--60  }        & \multicolumn{1}{c|}{60--70 }         & \multicolumn{1}{c|}{70--80 }         & \multicolumn{1}{c|}{80--500  }      \\
     \hline
  \hline

 {0--\hphantom{0}25}    &1	&-0.203809	&-0.417981	&-0.722061	&-0.629649	&-0.817134	&-0.233299	&-0.0780548	&0.0164206 \\
 {25--\hphantom{0}30}   &-0.203803	&1	&0.40795	&0.116331	&-0.19156	&0.119915	&-0.372659	&-0.261404	&-0.210925 \\
 {30--\hphantom{0}35}   &-0.418003	&0.407994	&1	&0.252236	&-0.0475611	&0.345603	&-0.313607	&-0.249736	&-0.164235 \\
 {35--\hphantom{0}40}   &-0.72209	&0.116342	&0.252234	&1	&0.371348	&0.603991	&0.0963432	&-0.00715498	&-0.146638 \\
 {40--\hphantom{0}50}   &-0.629659	&-0.19157	&-0.0475604	&0.371346	&1	&0.425221	&0.302711	&0.12123	&-0.0500793 \\
 {50--\hphantom{0}60}   &-0.817113	&0.119911	&0.345574	&0.603949	&0.425211	&1	&0.0721124	&-0.0296049	&-0.109044 \\
 {60--\hphantom{0}70}   &-0.233278	&-0.37261	&-0.313548	&0.0963277	&0.302689	&0.0721097	&1	&0.0801906	&0.0484276 \\
 {70--\hphantom{0}80}   &-0.078059	&-0.261433	&-0.249736	&-0.00715505	&0.121232	&-0.0296075	&0.0802059	&1	&-0.021394 \\
 {80--500}              &0.0164214	&-0.210947	&-0.164231	&-0.146636	&-0.0500795	&-0.109055	&0.0484387	&-0.0213935	&1 \\
\hline
\end{tabular}
}
\caption{Correlation matrix for the total uncertainties for the normalised unfolded \ptll~distribution, including all sources of systematic and
statistical uncertainties.}
    \label{tab::Systematic.Correlation.Matricies.Normalised.pT2l}
\end{table}

  \begin{table}[h!]
\hspace{-1.5cm}

\footnotesize
  {
 \centering
 \begin{tabular}{|r|
   S[round-mode = figures, round-precision = 2,
   table-format = 1.4, table-number-alignment = center]
   @{\,}|
S[round-mode = figures, round-precision = 2,
table-format = 1.3, table-number-alignment = center]
@{\,}|
S[round-mode = figures, round-precision = 2,
table-format = 1.5, table-number-alignment = center]
@{\,}|
S[round-mode = figures, round-precision = 2,
table-format = 1.3, table-number-alignment = center]
@{\,}|
S[round-mode = figures, round-precision = 2,
table-format = 1.3, table-number-alignment = center]
@{\,}|
S[round-mode = figures, round-precision = 2,
table-format = 1.4, table-number-alignment = center]
@{\,}|
S[round-mode = figures, round-precision = 2,
table-format = 1.3, table-number-alignment = center]
@{\,}|
S[round-mode = figures, round-precision = 2,
table-format = 1.4, table-number-alignment = center]
@{\,}|
S[round-mode = figures, round-precision = 2,
table-format = 1.3, table-number-alignment = center]
@{\,}|
S[round-mode = figures, round-precision = 2,
table-format = 1.4, table-number-alignment = center]
@{\,}|
S[round-mode = figures, round-precision = 2,
table-format = 1.5, table-number-alignment = center]
@{\,}|
S[round-mode = figures, round-precision = 2,
table-format = 1.3, table-number-alignment = center]
@{\,}|
S[round-mode = figures, round-precision = 2,
table-format = 1.3, table-number-alignment = center]
@{\,}|
S[round-mode = figures, round-precision = 2,
table-format = 1.3, table-number-alignment = center]
@{\,}|
}
    \hline
      $m_{\ell\ell}$ [\GeV] & \multicolumn{1}{l|}{10}                 & \multicolumn{1}{l|}{20}                & \multicolumn{1}{l|}{30}                 & \multicolumn{1}{l|}{40}               & \multicolumn{1}{l|}{50}             & \multicolumn{1}{l|}{60}                 & \multicolumn{1}{l|}{70}               & \multicolumn{1}{l|}{85}                & \multicolumn{1}{l|}{100}              & \multicolumn{1}{l|}{115}              & \multicolumn{1}{l|}{135}            & \multicolumn{1}{l|}{155}            & \multicolumn{1}{l|}{175}            & \multicolumn{1}{l|}{210}           \\
                           &        \multicolumn{1}{r|}{ --20   }       & \multicolumn{1}{r|}{--30      }     & \multicolumn{1}{r|}{--40  }         & \multicolumn{1}{r|}{--50     }      &\multicolumn{1}{r|}{ --60 }         & \multicolumn{1}{r|}{--70 }          & \multicolumn{1}{r|}{--85   }        & \multicolumn{1}{r|}{--100 }         & \multicolumn{1}{r|}{--115     }     &\multicolumn{1}{r|}{ --135    }      &\multicolumn{1}{r|}{ --155    }      & \multicolumn{1}{r|}{--175  }        &\multicolumn{1}{r|}{ --210    }      & \multicolumn{1}{r|}{--650  }       \\
\hline
 \hline
    \text{ 10--\hphantom{0}20}   &    1	&-0.0254037	&-0.0111842	&-0.074481	&-0.0694196	&-0.0931438	&-0.101977	&0.00422657	&-0.127159	&-0.0518698	&-0.0409186	&0.0160891	&0.0633366	&0.0488489 \\
    \text{ 20--\hphantom{0}30}   &    -0.0254055	&1	&-0.150505	&-0.430935	&-0.278349	&-0.273215	&-0.0600889	&0.0418988	&0.0183117	&-0.318231	&0.0210882	&-0.0256703	&0.391507	&0.404701 \\
    \text{ 30--\hphantom{0}40}   &    -0.0111843	&-0.1505	&1	&0.063066	&0.0652375	&0.00671487	&-0.0484994	&-0.0589278	&-0.0133265	&0.00886546	&-0.000935922	&-0.0747935	&-0.174902	&-0.144134 \\
\text{     40--\hphantom{0}50}   &    -0.0744752	&-0.430892	&0.0630631	&1	&0.369241	&0.367973	&0.0385444	&-0.138035	&-0.0789238	&0.486181	&-0.0776059	&-0.052562	&-0.797289	&-0.791135 \\
     \text{50--\hphantom{0}60}   &    -0.0693788	&-0.278216	&0.0652188	&0.369185	&1	&0.107166	&-0.0124068	&-0.155311	&0.0144638	&0.177923	&-0.0455458	&-0.151788	&-0.486745	&-0.463119 \\
     \text{60--\hphantom{0}70}   &    -0.0930118	&-0.272912	&0.00671022	&0.367826	&0.107146	&1	&-0.0942941	&-0.165216	&0.0482481	&0.140856	&0.0195322	&-0.175846	&-0.444433	&-0.411248 \\
     \text{70--\hphantom{0}85}   &    -0.101981	&-0.060089	&-0.0485001	&0.0385454	&-0.012408	&-0.0943164	&1	&-0.20364	&0.0666845	&-0.111873	&0.0464722	&-0.221593	&-0.147826	&-0.108502 \\
\text{     85--100}   &               0.00421825	&0.0418359	&-0.0588742	&-0.137967	&-0.155272	&-0.165197	&-0.203591	&1	&-0.102725	&-0.19776	&-0.00953563	&-0.0597196	&0.0983379	&0.118248 \\
     \text{100--115}  &               -0.126953	&0.0182882	&-0.0133154	&-0.0788834	&0.0144601	&0.0482463	&0.0666622	&-0.102738	&1	&-0.412482	&0.30168	&-0.591795	&-0.0538352	&0.115643 \\
     \text{115--135}  &               -0.0518722	&-0.318231	&0.00886562	&0.486198	&0.177948	&0.140905	&-0.111873	&-0.197838	&-0.412667	&1	&-0.387939	&0.380626	&-0.450682	&-0.578693 \\
     \text{135--155}  &               -0.040898	&0.0210791	&-0.000935666	&-0.0775941	&-0.0455477	&0.0195383	&0.0464661	&-0.00953989	&0.301803	&-0.387874	&1	&-0.523618	&-0.0335799	&0.124153 \\
     \text{155--175}  &               0.0160876	&-0.025667	&-0.0747885	&-0.0525623	&-0.151841	&-0.176003	&-0.221579	&-0.0597928	&-0.592418	&0.380598	&-0.523775	&1	&0.154849	&-0.0875969 \\
     \text{175--210}  &               0.0633359	&0.391485	&-0.1749	&-0.79732	&-0.486893	&-0.444766	&-0.147823	&0.0984356	&-0.0538833	&-0.450668	&-0.0335894	&0.154857	&1	&0.751475 \\
    \text{ 210--650}  &               0.0488507	&0.404696	&-0.144136	&-0.791171	&-0.463223	&-0.411471	&-0.108501	&0.118329	&0.11572	&-0.57869	&0.124182	&-0.0876036	&0.751496	&1 \\
\hline
\end{tabular}
}
\caption{Correlation matrix for the total uncertainties for the normalised unfolded $m_{\ell\ell}$ distribution, including all sources of systematic and
statistical uncertainties.}
    \label{tab::Systematic.Correlation.Matricies.Normalised.M2l}
\end{table}

\clearpage

\begin{table}[h!]
  \centering
 \footnotesize
  {
  \begin{tabular}{|r|
    S[round-mode = figures, round-precision = 2,
    table-format = 2.3, table-number-alignment = center]
    @{\,}|
S[round-mode = figures, round-precision = 2,
table-format = 2.3, table-number-alignment = center]
@{\,}|
S[round-mode = figures, round-precision = 2,
table-format = 2.3, table-number-alignment = center]
@{\,}|
S[round-mode = figures, round-precision = 2,
table-format = 2.3, table-number-alignment = center]
@{\,}|
S[round-mode = figures, round-precision = 2,
table-format = 2.3, table-number-alignment = center]
@{\,}|
S[round-mode = figures, round-precision = 2,
table-format = 2.3, table-number-alignment = center]
@{\,}|
S[round-mode = figures, round-precision = 2,
table-format = 2.3, table-number-alignment = center]
@{\,}|
S[round-mode = figures, round-precision = 2,
table-format = 2.3, table-number-alignment = center]
@{\,}|
S[round-mode = figures, round-precision = 2,
table-format = 2.3, table-number-alignment = center]
@{\,}|
}
    \hline
      $\Delta\phi_{\ell\ell}$                         & \multicolumn{1}{c|}{0--$\pi/8$ }    & \multicolumn{1}{c|}{$\pi/8$--$\pi/4$  }   & \multicolumn{1}{c|}{$\pi/4$--$3\pi/8$ }&\multicolumn{1}{c|}{ $3\pi/8$--$\pi/2$} & \multicolumn{1}{c|}{$\pi/2$--$5\pi/8$} & \multicolumn{1}{c|}{$5\pi/8$--$3\pi/4$ }& \multicolumn{1}{c|}{$3\pi/4$--$7\pi/8$} & \multicolumn{1}{c|}{$7\pi/8$--$\pi$} \\
    \hline
\hline

\text{0--\hphantom{0}$\pi/8$}                  &1	&-0.0709753	&-0.0521466	&0.0238657	&-0.0337191	&-0.112386	&-0.164851	&-0.0551079 \\
\text{$\pi/8$--\hphantom{0}$\pi/4$}            &-0.0709751	&1	&0.24369	&-0.272114	&-0.134743	&0.0654979	&-0.0762035	&-0.249719 \\
\text{$\pi/4$--$3\pi/8$}                       &-0.0521466	&0.243691	&1	&-0.702675	&-0.320928	&0.362786	&0.139981	&-0.407202 \\
\text{ $3\pi/8$--\hphantom{0}$\pi/2$}          &0.0238656	&-0.272114	&-0.702674	&1	&0.229946	&-0.454462	&-0.271638	&0.164946 \\
\text{$\pi/2$--$5\pi/8$}                       &-0.0337191	&-0.134744	&-0.320928	&0.229947	&1	&-0.243424	&-0.131285	&-0.210851 \\
\text{$5\pi/8$--$3\pi/4$}                      &-0.112386	&0.0654979	&0.362786	&-0.454462	&-0.243424	&1	&-0.00773089	&-0.409949 \\
\text{$3\pi/4$--$7\pi/8$}                      &-0.16485	&-0.0762034	&0.139981	&-0.271638	&-0.131285	&-0.00773089	&1	&-0.421748 \\
\text{$7\pi/8$--\hphantom{0}$\pi$\hphantom{/0}}&-0.0551079	&-0.24972	&-0.407202	&0.164946	&-0.210851	&-0.409949	&-0.421748	&1 \\
\hline
\end{tabular}
}
\caption{Correlation matrix for the total uncertainties for the normalised unfolded $\Delta\phi_{\ell\ell}$ distribution, including all sources of systematic and
statistical uncertainties.}
    \label{tab::Systematic.Correlation.Matricies.Normalised.dphi2l}
\end{table}

 \begin{table}[h!]
  \centering
 \footnotesize
  {
   \begin{tabular}{|r|  
     S[round-mode = figures, round-precision = 2,
     table-format = 1.4,table-number-alignment = center]
     @{\,}|
S[round-mode = figures, round-precision = 2,
table-format = 1.4, table-number-alignment = center]
@{\,}|
S[round-mode = figures, round-precision = 2,
table-format = 1.4, table-number-alignment = center]
@{\,}|
S[round-mode = figures, round-precision = 2,
table-format = 1.4, table-number-alignment = center]
@{\,}|
S[round-mode = figures, round-precision = 2,
table-format = 1.4, table-number-alignment = center]
@{\,}|
S[round-mode = figures, round-precision = 2,
table-format = 1.4, table-number-alignment = center]
@{\,}|
S[round-mode = figures, round-precision = 2,
table-format = 1.4, table-number-alignment = center]
@{\,}|
S[round-mode = figures, round-precision = 2,
table-format = 1.4, table-number-alignment = center]
@{\,}|
S[round-mode = figures, round-precision = 2,
table-format = 1.4, table-number-alignment = center]
@{\,}|
S[round-mode = figures, round-precision = 2,
table-format = 1.4, table-number-alignment = center]
@{\,}|
S[round-mode = figures, round-precision = 2,
table-format = 2.3, table-number-alignment = center]
@{\,}|
}
    \hline
      \absyll & \multicolumn{1}{c|}{0--0.2  }       & \multicolumn{1}{c|}{0.2--0.4 }        & \multicolumn{1}{c|}{0.4--0.6  }       &\multicolumn{1}{c|}{ 0.6--0.8  }       & \multicolumn{1}{c|}{0.8--1     }       & \multicolumn{1}{c|}{1--1.2   }        & \multicolumn{1}{c|}{1.2--1.4   }      & \multicolumn{1}{c|}{1.4--1.6  }       & \multicolumn{1}{c|}{1.6--1.8}         & \multicolumn{1}{c|}{1.8--2  }          & \multicolumn{1}{c|}{2--2.5   }      \\
\hline
\hline
\text{0\hphantom{.0}--0.2}  &1	&0.0371596	&0.616569	&0.0267759	&0.144529	&0.00660197	&-0.387196	&-0.598602	&-0.583199	&0.0879905	&-0.747626 \\
\text{0.2--0.4}             &0.0371594	&1	&0.0865706	&0.0690781	&-0.0703944	&-0.0782533	&-0.280334	&-0.134451	&-0.260712	&-0.118056	&-0.126667 \\
\text{0.4--0.6}             &0.616567	&0.0865705	&1	&0.113467	&0.140736	&0.037521	&-0.57014	&-0.47269	&-0.689513	&0.00075758	&-0.61372 \\
\text{0.6--0.8}             &0.0267755	&0.0690775	&0.113466	&1	&-0.0935497	&-0.0633695	&-0.383221	&-0.0261756	&-0.331286	&-0.160329	&-0.0328752 \\
\text{0.8--1.0}             &0.144528	&-0.0703946	&0.140737	&-0.0935508	&1	&-0.106031	&-0.192847	&-0.190791	&-0.247304	&-0.0546151	&-0.228507 \\
\text{1.0--1.2}             &0.00660197	&-0.0782537	&0.0375212	&-0.0633706	&-0.106032	&1	&-0.196226	&-0.066438	&-0.127869	&-0.0855447	&-0.0643779 \\
\text{1.2--1.4}             &-0.387194	&-0.280335	&-0.570142	&-0.383228	&-0.192847	&-0.196225	&1	&0.0541498	&0.637521	&0.0891532	&0.168966 \\
\text{1.4--1.6}             &-0.598606	&-0.134453	&-0.472698	&-0.0261764	&-0.190793	&-0.0664384	&0.0541505	&1	&0.275809	&-0.231083	&0.634671 \\
\text{1.6--1.8}             &-0.583195	&-0.260713	&-0.689519	&-0.331296	&-0.247304	&-0.127868	&0.637521	&0.275803	&1	&-0.0278053	&0.44421 \\
\text{1.8--2.0}             &0.0879909	&-0.118058	&0.000757595	&-0.160336	&-0.0546157	&-0.085545	&0.0891544	&-0.231081	&-0.0278059	&1	&-0.216228 \\
\text{2.0--2.5}             &-0.747635	&-0.126671	&-0.613738	&-0.0328771	&-0.228512	&-0.0643785	&0.16897	&0.634672	&0.444223	&-0.21623	&1 \\
\hline
\end{tabular}
}
\caption{Correlation matrix for the total uncertainties for the normalised unfolded \absyll~distribution, including all sources of systematic and
statistical uncertainties.}
    \label{tab::Systematic.Correlation.Matricies.Normalised.absy2l}
\end{table}

 \begin{table}[h!]
  \centering
 \footnotesize
  {
  \begin{tabular}{|r| 
S[round-mode = figures, round-precision = 2,
table-format = 2.3,
table-number-alignment = center]
@{\,}|
S[round-mode = figures, round-precision = 2,
table-format = 1.4, table-number-alignment = center]
@{\,}|
S[round-mode = figures, round-precision = 2,
table-format = 2.3, table-number-alignment = center]
@{\,}|
S[round-mode = figures, round-precision = 2,
table-format = 2.3, table-number-alignment = center]
@{\,}|
S[round-mode = figures, round-precision = 2,
table-format = 2.3, table-number-alignment = center]
@{\,}|
S[round-mode = figures, round-precision = 2,
table-format = 2.3, table-number-alignment = center]
@{\,}|
S[round-mode = figures, round-precision = 2,
table-format = 1.4, table-number-alignment = center]
@{\,}|
S[round-mode = figures, round-precision = 2,
table-format = 2.3, table-number-alignment = center]
@{\,}|
S[round-mode = figures, round-precision = 2,
table-format = 2.3, table-number-alignment = center]
@{\,}|
}
    \hline
      \abscostheta & \multicolumn{1}{c|}{0--0.1 }         & \multicolumn{1}{c|}{0.1--0.2  }        & \multicolumn{1}{c|}{0.2--0.3 }         & \multicolumn{1}{c|}{0.3--0.4   }       & \multicolumn{1}{c|}{0.4--0.5    }      & \multicolumn{1}{c|}{0.5--0.6  }        & \multicolumn{1}{c|}{0.6--0.7  }        & \multicolumn{1}{c|}{0.7--0.8  }        & \multicolumn{1}{c|}{0.8--1   }       \\
  \hline
  \hline
 \text{0\hphantom{.0}--0.1}    & 1	&0.303543	&-0.24937	&0.0474565	&0.26557	&0.0646275	&-0.24647	&-0.465765	&-0.288937 \\
\text{ 0.1--0.2} &               0.303543	&1	&-0.348642	&0.233438	&0.744261	&0.382582	&0.00262248	&-0.789597	&-0.830774 \\
 \text{0.2--0.3} &               -0.249371	&-0.348642	&1	&-0.266545	&-0.33649	&-0.256381	&-0.0231025	&0.270195	&0.217172 \\
 \text{0.3--0.4} &               0.0474565	&0.233438	&-0.266545	&1	&0.196988	&0.0135321	&-0.227913	&-0.351371	&-0.213896 \\
 \text{0.4--0.5} &               0.26557	&0.744261	&-0.336489	&0.196988	&1	&0.333173	&-0.0712106	&-0.746273	&-0.740839 \\
 \text{0.5--0.6} &               0.0646274	&0.382582	&-0.256381	&0.0135321	&0.333173	&1	&-0.0762605	&-0.435572	&-0.427168 \\
 \text{0.6--0.7} &               -0.246471	&0.00262248	&-0.0231025	&-0.227913	&-0.0712107	&-0.0762605	&1	&0.16767	&-0.262256 \\
 \text{0.7--0.8} &               -0.465765	&-0.789597	&0.270195	&-0.351371	&-0.746274	&-0.435573	&0.16767	&1	&0.614904 \\
 \text{0.8--1.0} &               -0.288937	&-0.830775	&0.217172	&-0.213896	&-0.740839	&-0.427169	&-0.262256	&0.614904	&1 \\
\hline
\end{tabular}
}
\caption{Correlation matrix for the total uncertainties for the normalised unfolded \abscostheta~distribution, including all sources of systematic and
statistical uncertainties.}
    \label{tab::Systematic.Correlation.Matricies.Normalised.abscostheta}
\end{table}

\clearpage

\newpage 

\begin{flushleft}
{\Large The ATLAS Collaboration}

\bigskip

G.~Aad$^\textrm{\scriptsize 87}$,
B.~Abbott$^\textrm{\scriptsize 115}$,
J.~Abdallah$^\textrm{\scriptsize 153}$,
O.~Abdinov$^\textrm{\scriptsize 11}$,
R.~Aben$^\textrm{\scriptsize 109}$,
M.~Abolins$^\textrm{\scriptsize 92}$,
O.S.~AbouZeid$^\textrm{\scriptsize 160}$,
H.~Abramowicz$^\textrm{\scriptsize 155}$,
H.~Abreu$^\textrm{\scriptsize 154}$,
R.~Abreu$^\textrm{\scriptsize 118}$
% ...
\bigskip
\\
% ...
$^{11}$ Institute of Physics, Azerbaijan Academy of Sciences, Baku, Azerbaijan\\
% ...
$^{87}$ CPPM, Aix-Marseille Universit{\'e} and CNRS/IN2P3, Marseille, France\\
% ...
$^{92}$ Department of Physics and Astronomy, Michigan State University, East Lansing MI, United States of America\\
% ...
$^{109}$ Nikhef National Institute for Subatomic Physics and University of Amsterdam, Amsterdam, Netherlands\\
% ...
$^{115}$ Homer L. Dodge Department of Physics and Astronomy, University of Oklahoma, Norman OK, United States of America\\
% ...
$^{118}$ Center for High Energy Physics, University of Oregon, Eugene OR, United States of America\\
% ...
$^{153}$ Institute of Physics, Academia Sinica, Taipei, Taiwan\\
$^{154}$ Department of Physics, Technion: Israel Institute of Technology, Haifa, Israel\\
$^{155}$ Raymond and Beverly Sackler School of Physics and Astronomy, Tel Aviv University, Tel Aviv, Israel\\
% ...
$^{160}$ Department of Physics, University of Toronto, Toronto ON, Canada\\
% ...
\end{flushleft}

 
\end{document}